\documentclass[journal,a4paper]{IEEEtran}
\usepackage[cmex10]{amsmath}
\usepackage[mathscr]{euscript}
\usepackage[T1]{fontenc}
\usepackage[utf8]{inputenc} 
\usepackage{algorithmic}
\usepackage{amsfonts,amssymb,amsthm,amstext}
\usepackage{balance} 	% Force a column break in bibliography
\usepackage{booktabs} 	% fixes poor spacing in tables and provides \toprule, \midrule, \bottomrule
\usepackage{cancel,cite,color,dsfont,enumerate}
\usepackage{empheq}
\usepackage{graphicx}   % provides \includegraphics{...} to include graphics (pdf format)
\usepackage[geometry]{ifsym}
\usepackage{ifthen}
\usepackage{makeidx,multicol}
\usepackage{mleftright}       % fix to wrong spacing of \left-,
\mleftright                   % \middle- \right-commands 
\usepackage{orcidlink}
\usepackage{pgfplots}
\usepackage{pict2e}
\usepackage{relsize}
\usepackage{shadethm,soul}
\usepackage{textcomp}
\usepackage[normalem]{ulem}
\usepackage{url,verbatim,xcolor,xfrac,xr}
\usepackage{hyperref}
\hypersetup{hyperindex}
\hypersetup{breaklinks}
\hypersetup{hidelinks=true}
\hypersetup{
pdftitle	={A Spectral Representation Of The Simple Hypothesis Testing Problem},
pdfauthor	={Barış Nakiboğlu},
pdfsubject	={-},
pdfkeywords	={-}, 
}
\theoremstyle{plain}

\newtheorem{lemma}{Lemma} 
\newtheorem{theorem}{Theorem} 
\newtheorem{corollary}{Corollary}

\newtheorem*{conjecture*}{Conjecture}
\theoremstyle{definition}
\newtheorem{definition}{Definition} 
\newtheorem{assumption}{Assumption}

\newtheorem{remark}{Remark}
\newtheorem*{remark*}{Remark}
\theoremstyle{plain}
\newcommand{\ones}[0]			{\mathbf{1}}

\newcommand{\DEF}[0]			{{:=}}

\newcommand{\AC}[0]            {{\prec}}

\newcommand{\sss}[1]	       {\mathit{2}^{#1}}
\NewDocumentCommand\numbers{m !o}{\IfNoValueTF{#2}{\mathbb{#1}}{\mathbb{#1}_{^{#2}}}}
\newcommand{\set} [1]		{{\mathscr{#1}}}
\newcommand{\alg}[1]		{{\mathcal{#1}}}

\NewDocumentCommand\rvaux{m m m}{{\mathsf#1}\IfNoValueTF{#2}{}{_{#2}}\IfNoValueTF{#3}{}{^{#3}}}
\NewDocumentCommand\rv{>{\SplitArgument{2}{;}}m !o}{\rvaux #1\IfNoValueTF{#2}{}{\left(#2\right)}}

\newcommand{\msr}[1]       {{\it    {#1}}}
\newcommand{\cnst}[1]      {{\mathit{#1}}}

\newcommand{\bigo} [1]     {{\cnst{O}\left({#1}\right)}}
\newcommand{\smallo}[1]    {{\cnst{o}\left({#1}\right)}}

\NewDocumentCommand\lmea{o m}{\IfNoValueTF{#1}{\alg{L}(#2)}{\alg{L}^{\!^{#1}}\!\!(#2)}}
\NewDocumentCommand\smea{o m}{\IfNoValueTF{#1}{\alg{M}(#2)}{\alg{M}^{\!^{#1}}\!\!(#2)}}
\newcommand{\pmea}[1]        {\alg{P}({#1})}
\newcommand{\tmea}[1]        {\alg{T}({#1})}
\newcommand{\tmead}[1]       {\alg{T}^{d}({#1})}

\NewDocumentCommand\ldis{o m}{\IfNoValueTF{#1}{\set{L}(#2)}{\set{L}^{\!^{#1}}\!\!(#2)}}
\NewDocumentCommand\sdis{o m}{\IfNoValueTF{#1}{\set{M}(#2)}{\set{M}^{\!^{#1}}\!\!(#2)}}

\DeclareMathOperator*{\essup}{ess\,sup}
\DeclareMathOperator*{\esinf}{ess\,inf}

\DeclareMathOperator{\Tr}{Tr}
\NewDocumentCommand\trace{o}{\IfNoValueTF{#1}{\Tr}{\Tr\left[{#1}\right]}} %Trace

\newcommand{\supp}[1]      {\mathfrak{S}_{#1}}

\newcommand{\inte}[1]		{\mathtt{int}\left({#1}\right)}

\newcommand{\domain}[1]   	{\mathtt{dom}\left({#1}\right)}

\NewDocumentCommand\dual{m !o}{\IfNoValueTF{#2}{{#1}^{\star}}{{#1}_{#2}^{\star}}}
\NewDocumentCommand\polar{m !o}{\IfNoValueTF{#2}{{#1}^{\circ}}{{#1}_{#2}^{\circ}}}
\NewDocumentCommand\AEcoefficient{o m !o}{\IfNoValueTF{#3}{\kappa_{#2}}{\kappa_{#2,#3}}\IfNoValueTF{#1}{}{\left(#1\right)}}
\NewDocumentCommand\TEcoefficient{o m !o}{\IfNoValueTF{#3}{\gamma_{#2}}{\gamma_{#2,#3}}\IfNoValueTF{#1}{}{\left(#1\right)}}

\NewDocumentCommand\normal{o m}{\alg{N}_{#2} \IfNoValueTF{#1}{} {\left({#1}\right)}}

\NewDocumentCommand\tangent{o m}{\alg{T}_{#2} \IfNoValueTF{#1}{} {\left({#1}\right)}}

\NewDocumentCommand\pushover{o m m}{\alg{T}_{#2,#3} \IfNoValueTF{#1}{} {\left({#1}\right)}}
\NewDocumentCommand\projected{m m !o}{\IfNoValueTF{#3}{\mathtt{P}_{#2}^{-1}\left({#1}\right)}{\mathtt{P}_{#2,#3}^{-1}\left({#1}\right)}}

\newcommand{\dif}[1]        {\mathrm{d}{#1}}  
\newcommand{\lder}[1]       {\mathrm{D}_{\!-}{#1}}  
\newcommand{\rder}[1]       {\mathrm{D}_{\!+}{#1}}  
\newcommand{\der}[2]        {\tfrac{\dif{#1}}{\dif{#2}}}  
\newcommand{\pder}[2]       {\tfrac{\partial{#1}}{\partial{#2}}}  

\newcommand{\abs}[1]           {{\left\lvert{#1}\right\lvert}}
\newcommand{\abp}[1]           {{\left\lvert{#1}\right\lvert^{+}}}

\newcommand{\norm}[2][]        {\left\lVert{#2}\right\lVert_{#1}}

\NewDocumentCommand\potential{!o}{\mathrm{\Phi}\IfNoValueTF{#1}{}{\!\left({#1}\right)}}
\NewDocumentCommand\pdpotential{m !o}{\mathrm{\Phi}_{#1}'\IfNoValueTF{#2}{}{\!\left({#2}\right)}}
\newcommand{\IND}[1]           {{\mathds{1}_{\{#1\}}}}    

\newcommand{\jnd}[0]           {\jmath}

\newcommand{\tin}[0]           {\cnst{t}}
\newcommand{\mlx}[0]           {\cnst{m}}
\newcommand{\blx}[0]           {\cnst{n}}
\newcommand{\tlx}[0]           {\cnst{T}}

\newcommand{\pxs}[0]		{\bf P}
\newcommand{\exs}[0]		{\bf E}
\newcommand{\vxs}[0]		{\bf V}
\newcommand{\covxs}[0]		{\bf cov}
\newcommand{\txs}[0]		{\bf T}

\NewDocumentCommand\PX{o m !o}{
	\IfNoValueTF{#1}{{\pxs}_{\!}\left(\IfNoValueTF{#3}{#2}{#2\middle\vert#3}\right)}
	{{\pxs}_{\!#1}\left(\IfNoValueTF{#3}{#2}{#2\middle\vert#3}\right)}}			%Probability
\NewDocumentCommand\EX{o m !o}{
	\IfNoValueTF{#1}{{\exs}_{\!}\left[\IfNoValueTF{#3}{#2}{#2\middle\vert#3}\right]}
	{{\exs}_{#1\!}\left[\IfNoValueTF{#3}{#2}{#2\middle\vert#3}\right]}}			%Expectation
\NewDocumentCommand\VX{o m !o}{
	\IfNoValueTF{#1}{{\vxs}\!\left[\IfNoValueTF{#3}{#2}{#2\middle\vert#3}\right]}
	{{\vxs}_{#1\!}\left[\IfNoValueTF{#3}{#2}{#2\middle\vert#3}\right]}}			%Variance
\NewDocumentCommand\COVX{o m m !o}{
	\IfNoValueTF{#1}{{\covxs}\!\left[\IfNoValueTF{#4}{#2,#3}{#2,#3\middle\vert#4}\right]}
	{{\covxs}_{#1\!}\left[\IfNoValueTF{#4}{#2,#3}{#2,#3\middle\vert#4}\right]}}	 %CoVariance
\NewDocumentCommand\TX{o m !o}{
	\IfNoValueTF{#1}{{\txs}\!\left[\IfNoValueTF{#3}{#2}{#2\middle\vert#3}\right]}
	{{\txs}_{#1\!}\left[\IfNoValueTF{#3}{#2}{#2\middle\vert#3}\right]}}			 %CenteredThirdMoment

\newcommand{\agf}[0]    {\cnst{A}}
\NewDocumentCommand\AGF{o m m !o}{\agf_{#2}\!\left(
	\IfNoValueTF{#1}{\IfNoValueTF{#4}{\!#3\!}{\!#3;\!#4\!}}
	{\IfNoValueTF{#4}{\!#3,\!#1\!}{\!#3,\!#1;\!#4\!}}\right)}		%
\NewDocumentCommand\pmf{o m !o}{p_{#2}\IfNoValueTF{#1}{}{^{#1}}\!\IfNoValueTF{#3}{~}{\left(#3\right)}}
\NewDocumentCommand\pdf{o m !o}{f_{#2}\IfNoValueTF{#1}{}{^{#1}}\!\IfNoValueTF{#3}{~}{\left(#3\right)}}
\NewDocumentCommand\ccdf{o m !o !o}{\overline{F}_{#2}\IfNoValueTF{#1}{}{^{#1}}\!\IfNoValueTF{#3}{}{\IfNoValueTF{#4}{\left(#3\right)}{\left(#3\middle\vert#4\right)}}}
\NewDocumentCommand\cdf{o m !o !o}{F_{#2}\IfNoValueTF{#1}{}{^{#1}}\!\IfNoValueTF{#3}{}{\IfNoValueTF{#4}{\left(#3\right)}{\left(#3\middle\vert#4\right)}}}
\NewDocumentCommand\chf{o m !o !o}{\varphi_{#2}\IfNoValueTF{#1}{}{^{#1}}\!\IfNoValueTF{#3}{}{\IfNoValueTF{#4}{\left(#3\right)}{\left(#3\middle\vert#4\right)}}}
\NewDocumentCommand\mgf{o m !o !o}{M_{#2}\IfNoValueTF{#1}{}{^{#1}}\!\IfNoValueTF{#3}{}{\IfNoValueTF{#4}{\left(#3\right)}{\left(#3\middle\vert#4\right)}}}
\NewDocumentCommand\cgf{o m !o !o}{K_{#2}\IfNoValueTF{#1}{}{^{#1}}\!\IfNoValueTF{#3}{}{\IfNoValueTF{#4}{\left(#3\right)}{\left(#3\middle\vert#4\right)}}}
\NewDocumentCommand\crf{o m !o !o}{K_{#2}^{*}\IfNoValueTF{#1}{}{^{#1}}\!\IfNoValueTF{#3}{}{\IfNoValueTF{#4}{\left(#3\right)}{\left(#3\middle\vert#4\right)}}}
\NewDocumentCommand\prp{o m !o !o}{\tau_{#2}\IfNoValueTF{#1}{}{^{#1}}\!\IfNoValueTF{#3}{}{\IfNoValueTF{#4}{\left(#3\right)}{\left(#3\middle\vert#4\right)}}}
\NewDocumentCommand\prf{o m !o !o}{L_{#2}\IfNoValueTF{#1}{}{^{#1}}\!\IfNoValueTF{#3}{}{\IfNoValueTF{#4}{\left(#3\right)}{\left(#3\middle\vert#4\right)}}}

\NewDocumentCommand\cumulant{m !o !o}{\kappa_{#1}\IfNoValueTF{#2}{}{\IfNoValueTF{#3}{^{#2}}{^{#2}\left(#3\right)}}}
\NewDocumentCommand\cramer{m !o !o}{\mA_{#1}\IfNoValueTF{#2}{}{\IfNoValueTF{#3}{^{#2}}{^{#2}\left(#3\right)}}}
\NewDocumentCommand\petrov{m !o !o}{\mB_{#1}\IfNoValueTF{#2}{}{\IfNoValueTF{#3}{^{#2}}{^{#2}\left(#3\right)}}}

\NewDocumentCommand\scf{o m !o}{H_{#2}\IfNoValueTF{#1}{}{^{#1\!}}\!\IfNoValueTF{#3}{}{\left(#3\right)}}
\NewDocumentCommand\acumulant{o m !o}{\varGamma_{#2}\IfNoValueTF{#1}{}{^{#1\!}}\!\IfNoValueTF{#3}{}{\left(#3\right)}}

\NewDocumentCommand\dmgf{m m !o}{M\IfNoValueTF{#3}{_{#1}^{#2'\!}}{_{#1\!}^{#2'\!}\left(#3\right)}}
\NewDocumentCommand\dcgf{m m !o}{K\IfNoValueTF{#3}{_{#1}^{#2'\!}}{_{#1\!}^{#2'\!}\left(#3\right)}}
\NewDocumentCommand\expf{m o m}{\mathbf{#1}\!\IfNoValueTF{#2}{\left({#3}\right)}{\left({#2}\middle\vert{#3}\right)}}

\newcommand{\llr}[0]    {{\rv{L}}}

\newcommand{\ylr}[0]	{\ell}

\NewDocumentCommand\jacobian{o m}{\cnst{J}\IfNoValueTF{#1}{_{#2}}{_{#2}^{#1}}}
\NewDocumentCommand\shm{o m}{\cnst{H}\IfNoValueTF{#1}{_{#2}}{_{#2}^{#1}}}

\NewDocumentCommand\chiD{o m m}{
	\IfNoValueTF{#1}{\cnst{\chi}^{2}\!\left({#2}\middle\Vert{#3}\right)}
	{{\chi}^{#1}       \!\left({#2}\middle\Vert{#3}\right)}}
\NewDocumentCommand\CchiD{o m m m}{
	\IfNoValueTF{#1}{\cnst{\chi}^{2}\!\left({#2}\middle\Vert{#3}\middle\vert{#4}\right)}
	{{\chi}^{#1}       \!\left({#2}\middle\Vert{#3}\middle\vert{#4}\right)}}
\newcommand{\RD}[3]		{\cnst{D}_{#1}          \!\left({#2}\middle\Vert{#3}\right)}

\newcommand{\KLD}[2]	{\cnst{D}				\!\left({#1}\middle\Vert{#2}\right)} 
\newcommand{\CKLD}[3]	{\cnst{D}				\!\left({#1}\middle\Vert{#2}\middle\vert{#3}\right)} 

\NewDocumentCommand\entropy{o m !o}{
	\IfNoValueTF{#1}{\cnst{h}_{\!}\left(\IfNoValueTF{#3}{#2}{#2\middle\vert#3}\right)}
	{\cnst{h}_{#1\!}\left(\IfNoValueTF{#3}{#2}{#2\middle\vert#3}\right)}}

\NewDocumentCommand\MI{o m m !o}{
	\IfNoValueTF{#1}{\cnst{I}\!\left(\IfNoValueTF{#4}{#2;#3}{#2;#3\middle\vert{#4}}\right)}
	{\cnst{I}_{#1\!\!}\left(\IfNoValueTF{#4}{#2;#3}{#2;#3\middle\vert{#4}}\right)}}

\NewDocumentCommand\LMI{m o m m !o}{
	\IfNoValueTF{#2}{\cnst{I}^{#1}\!\left(\IfNoValueTF{#5}{#3;#4}{#3;#4\middle\vert{#5}}\right)}
	{\cnst{I}_{#2}^{#1}\!\left(\IfNoValueTF{#5}{#3;#4}{#3;#4\middle\vert{#5}}\right)}}
\NewDocumentCommand\Loptimalset{m !o}{
	\IfNoValueTF{#2}{\Pi^{#1}}
	{\Pi^{#1,#2}}}

\newcommand{\rfm}[0]			{\msr{\nu}}

\newcommand{\cset}[0]			{\set{A}}

\newcommand{\GMRdummy}[0]		{\cnst{\varPsi}}
\NewDocumentCommand\GMR{o m !o}{
\IfNoValueTF{#1}{\IfNoValueTF{#3}{\GMRdummy\!\left({#2}\right)}{\GMRdummy{#3}\!\left({#2}\right)}}{\IfNoValueTF{#3}{\GMRdummy_{#1}\!\left({#2}\right)}{\GMRdummy_{#1}\!\!{#3}\!\left({#2}\right)}}}

\newcommand{\GQF}[1]			{\cnst{Q}\left({#1}\right)}
\newcommand{\IGQF}[1]			{\cnst{Q}^{-1}\left({#1}\right)}

\newcommand{\GCD}[1]			{\cnst{\Phi}\left({#1}\right)}
\newcommand{\IGCD}[1]			{\cnst{\Phi}^{-1}\left({#1}\right)}
\newcommand{\GD}[1]				{\cnst{\varphi}\left({#1}\right)}

\newcommand{\rno}[0]          {{\cnst{\alpha}}}

\newcommand{\oev}[0]           {{\set{E}}}

\newcommand{\dinp}[0]          {{\cnst{x}}}
\newcommand{\inp}[0]           {{\rv{X}}}
\newcommand{\inpS}[0]          {{\set{X}}}
\newcommand{\inpA}[0]          {{\alg{X}}}

\newcommand{\dout}[0]          {{\cnst{y}}}
\newcommand{\out}[0]           {{\rv{Y}}}
\newcommand{\outS}[0]          {{\set{Y}}}
\newcommand{\outA}[0]          {{\alg{Y}}}

\newcommand{\dsta}[0]          {{\cnst{z}}}
\newcommand{\sta}[0]           {{\rv{Z}}}

\newcommand{\mA}[0]				{\msr{a}}    
\newcommand{\amn}[1]			{{\mA_{#1}}}
\newcommand{\ama}[2]			{{\mA_{#1}^{#2}}}

\newcommand{\mB}[0]				{\msr{b}}    
\newcommand{\bmn}[1]			{{\mB_{#1}}}
\newcommand{\bma}[2]			{{\mB_{#1}^{#2}}}

\newcommand{\mQ}[0]				{\msr{q}}

\newcommand{\mS}[0]				{\msr{s}}

\newcommand{\mW}[0]				{\msr{w}}

\makeatletter
\DeclareRobustCommand{\bigplus}{%
	\mathop{\vphantom{\sum}\mathpalette\@bigplus\relax}\slimits@
}
\newcommand{\@bigplus}[2]{\vcenter{\hbox{\make@bigplus{#1}}}}
\newcommand{\make@bigplus}[1]{%
	\sbox\z@{$\m@th#1\sum$}%
	\setlength{\unitlength}{\wd\z@}%
	\begin{picture}(1.4,1.4)
	\linethickness{.17ex}
	\Line(.7,.14)(.7,1.26)
	\Line(.14,.7)(1.26,.7)
	\end{picture}%
}
\DeclareRobustCommand{\bigtimes}{%
	\mathop{\vphantom{\sum}\mathpalette\@bigtimes\relax}\slimits@
}
\newcommand{\@bigtimes}[2]{\vcenter{\hbox{\make@bigtimes{#1}}}}
\newcommand{\make@bigtimes}[1]{%
	\sbox\z@{$\m@th#1\sum$}%
	\setlength{\unitlength}{\wd\z@}%
	\begin{picture}(1,1)
	\linethickness{.17ex}
	\Line(.1,.1)(.9,.9)
	\Line(.1,.9)(.9,.1)
	\end{picture}%
}
\makeatother

\NewDocumentCommand\maux{m m m}{{\mathit#1}\IfNoValueTF{#2}{}{_{#2}}\IfNoValueTF{#3}{}{^{#3}}}
\NewDocumentCommand\tmargument{m m}{\left(#1{;}#2\right)}
\NewDocumentCommand\faux{m m m}{{\mathit#1}\IfNoValueTF{#2}{}{_{#2}}\IfNoValueTF{#3}{}{^{#3}}}

\NewDocumentCommand\m{>{\SplitArgument{2}{;}}m !o}{\maux #1\IfNoValueTF{#2}{}{\left(#2\right)}}
\NewDocumentCommand\tm{>{\SplitArgument{2}{;}}m >{\SplitArgument{1}{;}} o}{\maux #1\IfNoValueTF{#2}{}{\tmargument#2}}
\NewDocumentCommand\f{>{\SplitArgument{2}{;}}m !o}{\faux #1\IfNoValueTF{#2}{}{\left(#2\right)}}

\NewDocumentCommand\cdualaux{m m m}{{\color{blue}\widetilde{\mathit#1}}\IfNoValueTF{#2}{}{_{#2}}\IfNoValueTF{#3}{}{^{#3}}}
\NewDocumentCommand\cdual{>{\SplitArgument{2}{;}}m !o}{\cdualaux#1\IfNoValueTF{#2}{}{\left(#2\right)}}
\NewDocumentCommand\impf{o !o}{\IfNoValueTF{#1}{\mathrm{F}\IfNoValueTF{#2}{}{\left(#2\right)}}{\mathrm{#1}\IfNoValueTF{#2}{}{\left(#2\right)}}}
\NewDocumentCommand\pdimpf{o m  !o}{\IfNoValueTF{#1}{\mathrm{F}_{#2}'\IfNoValueTF{#3}{}{\!\left({#3}\right)}}{\mathrm{#1}_{#2}'\IfNoValueTF{#3}{}{\!\left({#3}\right)}}}

\NewDocumentCommand\metricaux{m m}{\left(#1,#2\right)}
\NewDocumentCommand\metric{m >{\SplitArgument{1}{;}} !o}{{\it#1}\IfNoValueTF{#2}{}{\metricaux#2}}

\newcommand{\opnorm}[2][]{\left\vert\kern-0.21em\left\vert\kern-0.21em\left\vert#2\right\vert\kern-0.21em\right\vert\kern-0.21em\right\vert_{#1}}

\NewDocumentCommand\fav{m m}{\varkappa\left({#1},{#2}\right)}
\NewDocumentCommand\mdv{m m}{\overline{\varkappa}\left({#1},{#2}\right)}

\NewDocumentCommand\LCextension{m o}{\IfNoValueTF{#2}{\underline{\!~#1}}{\underline{\!~#1}\!\left({#2}\right)}}		
\NewDocumentCommand\UCextension{m o}{\IfNoValueTF{#2}{\overline{\!~#1}}{\overline{\!~#1}\!\left({#2}\right)}}		

\NewDocumentCommand\lsccl{m o}{\IfNoValueTF{#2}{\underline{\mathtt{cl}#1}}{\underline{\mathtt{cl}#1}\!\left({#2}\right)}}		
\NewDocumentCommand\usccl{m o}{\IfNoValueTF{#2}{\overline{\mathtt{cl}#1}}{\overline{\mathtt{cl}#1}\!\left({#2}\right)}}		

\NewDocumentCommand\sci{m o}{\IfNoValueTF{#2}{#1^{\star}}{{#1}^{\star}\!\left({#2}\right)}}		
\NewDocumentCommand\lsci{m o}{\IfNoValueTF{#2}{#1^{\wedge}}{{#1}^{\wedge}\!\left({#2}\right)}}		
\NewDocumentCommand\usci{m o}{\IfNoValueTF{#2}{#1^{\vee}}{{#1}^{\vee}\!\left({#2}\right)}}		

\NewDocumentCommand\gcdf{m o m m m}{F\IfNoValueTF{#2}{_{#1}}{_{#1}^{#2}}\!\left({#3}\middle\vert_{#5}^{#4}\!\right)}	
\NewDocumentCommand\scq{o m m m}{   F\IfNoValueTF{#1}{_{\ylr}}{_{\ylr\!}^{#1\!}}\left({#2}\middle\vert_{#4}^{#3}\!\right)}		
\newcommand{\lscq}[3]{\scq[\wedge]{#1}{#2}{#3}}		
\newcommand{\uscq}[3]{\scq[\vee]{#1}{#2}{#3}}		

\NewDocumentCommand\scqs{o m m m}{\cset\IfNoValueTF{#1}{_{\ylr}}{_{\ylr\!}^{#1}}\!\left({#2}\middle\vert_{#4}^{#3}\right)}		
\newcommand{\lscqs}[3]{\scqs[\wedge]{#1}{#2}{#3}}		
\newcommand{\uscqs}[3]{\scqs[\vee]{#1}{#2}{#3}}

\NewDocumentCommand\kbm{o m !o}{\mu_{#2}\IfNoValueTF{#1}{}{^{#1}}\!\IfNoValueTF{#3}{}{\left(#3\right)}}
\newcommand{\SGBdomain}[2] {\set{I}_{#2}^{#1}} 
 
\NewDocumentCommand\SGB{o m m m}{\mu\IfNoValueTF{#1}{}{^{#1\!}}\left({#2}\middle\vert_{#4}^{#3}\!\right)}
\NewDocumentCommand\NPDd{o m m m}{\beta\IfNoValueTF{#1}{}{^{#1\!}}\left({#2}\middle\Vert_{#4}^{#3}\!\right)}
\NewDocumentCommand\NPD{o m m m}{\beta\IfNoValueTF{#1}{}{^{#1\!}}\left({#2}\middle\vert_{#4}^{#3}\!\right)}

\NewDocumentCommand\ent{o m m m}{\cnst{\hslash}\IfNoValueTF{#1}{}{^{#1\!}}\left({#2}\middle\vert_{#4}^{#3}\!\right)}

\renewcommand{\div}[3]	{\cnst{d}				\left({#1}\middle\vert_{#3}^{#2}\!\right)}

\NewDocumentCommand\XTE{o m m m}{E_{e}\IfNoValueTF{#1}{}{^{#1\!}}\left({#2}\middle\vert_{#4}^{#3}\!\right)}
\NewDocumentCommand\XTC{o m m m}{E_{c}\IfNoValueTF{#1}{}{^{#1\!}}\left({#2}\middle\vert_{#4}^{#3}\!\right)}
\NewDocumentCommand\XT{o m m m}{E\IfNoValueTF{#1}{}{^{#1\!}}\left({#2}\middle\vert_{#4}^{#3}\!\right)}

\NewDocumentCommand\SD{o m m m}{
	\IfNoValueTF{#1}{{\zeta_{#2}\!\left(\!{#3}\middle\Vert{#4}\right)}}
	{\zeta_{#2}^{#1}			\!\left(\!{#3}\middle\Vert{#4}\right)}}		
\NewDocumentCommand\SDS{o m m m}{
	\IfNoValueTF{#1}{{\cset_{#2}	\!\left(\!{#3}\middle\Vert{#4}\right)}}
	{\cset_{#2}^{#1}			\!\left(\!{#3}\middle\Vert{#4}\right)}}

\newif\ifnullhyperlink
\nullhyperlinkfalse
\newif\iffreshbib
\freshbibfalse
\newif\ifextra
\extratrue 
\newif\ifshowproof
\showprooftrue 
\begin{document}
\ifnullhyperlink \begin{NoHyper}\fi
\title{A Spectral Representation Of\\The Simple Hypothesis Testing Problem
} 	
\author{Bar\i\c{s} Nakibo\u{g}lu,~\IEEEmembership{Member,~IEEE,} 
\thanks{B. Nakibo\u{g}lu is with 
		the Department of Electrical and Electronics Engineering
Middle East Technical University 06800 Ankara, T\"{u}rkiye
(\orcidlink{0000-0001-7737-5423}
\href{https://orcid.org/0000-0001-7737-5423}{0000-0001-7737-5423}).}% <-this % stops a space
\thanks{This paper was presented in part at 
the \href{https://www.aparat.com/v/sefhpq9}{14th Iran Workshop on Communication and Information Theory}, 
April 2026,	and at 
the IEEE International Symposium on Information Theory, 
(ISIT) Guangzhou, China, July 2026 
[DOI:\href{https://dx.doi.org/10.1109/ISIT62367.2026.11653959}{10.1109/ISIT62367.2026.11653959}].}% <-this % stops a space
}
\IEEEaftertitletext{ 
\flushright
\vspace{-2\baselineskip}
{\it 
Can{\i}m anneannem Fatma Canbaz'{\i}n an{\i}s{\i}na adanm{\i}{\c{s}}t{\i}r. \hspace{1.7cm}~
\\
Dedicated to the memory of my dear grandmother Fatma Canbaz.}\quad~\\
~\\
}
\maketitle
\begin{abstract}	
The minimum Type II error probability (or volume)
of simple hypothesis tests with randomized detectors 
is expressed as a Riemann integral of the (cumulative) distribution 
function of the likelihood ratio for all non-negative 
Type I error probability (or volume) values:
$\beta(\varepsilon)\!=\!\int_{0}^{\infty}
|F(\sfrac{1}{\tau})\!-\!\varepsilon|^{+}d\tau$
for all $\varepsilon\!\geq\!0$. 
The derivation relies on convex conjugation (the Legendre transform)
and level set integration, and is sufficiently general to extend to 
tests between $\sigma$-finite measures. 
Measure change identities relating the $\beta(\cdot)$ functions of
different hypothesis testing problems are established.
Approximations for Type II and Type I volumes are derived under 
two different hypotheses concerning the Kolmogorov distance to 
normality for the log-likelihood ratio distributions.
In both the Central Limit Theorem and the large deviations regimes,
the resulting non-asymptotic expressions recover and extend 
state-of-the-art characterizations of optimal performance in 
the memoryless case, and improve and generalize them in 
the Markovian case.
Finally, The distinction between the implications of 
the Central Limit Theorem and the Berry--Esseen Theorem 
for the asymptotic simple hypothesis testing problem is 
clarified.
\end{abstract}		
%\begin{IEEEkeywords}
%Hypothesis testing, 
%convex conjugation, %Legendre transform, 
%level set integration, %tail expectation formula, survival integral, 
%Gaussian Mills ratio, 
%Kolmogorov Distance, %Berry--Esseen theorem, 
%%discrete (countable state space) Markov chains,
%tilting.
%\end{IEEEkeywords}
{\hypersetup{hidelinks=true}\tableofcontents} 
\section{Introduction}\label{sec:Introduction}
The following elementary decision problem illustrates our 
principal novelty regarding the Neyman--Pearson lemma.
%\subsection[An Elementary Decision Problem]{An Instructive Elementary Decision Problem}\label{sec:InitialExample}
Let \(\out\) be a zero mean unit variance Gaussian random variable.
What is the infimum of the lengths of Borel sets \(\oev\)
satisfying \(\PX{\out\!\notin\!\oev}\!\leq\!\varepsilon\), 
i.e, what is the value of \(\f{\beta}[\varepsilon]\) defined
in the following? 
\begin{align}
	\notag
	\f{\beta}[\varepsilon]
	&\DEF\inf\left\{\int_{\oev}\!\!\dif{\dout}:\oev\!\in\!\mathcal{B}\text{~and~}\!
	\int_{\numbers{R}\setminus\oev}\!\!\GD{\dout}\dif{\dout}\!\leq\!\varepsilon\right\}
	&
	&\forall \varepsilon\!\in\![0,1],	
\end{align}
where \(\mathcal{B}\) is the set of all Borel subsets of \(\numbers{R}\) 
and \(\GD{\cdot}\) is the Gaussian probability density function:
\begin{align}
	\label{eq:def:GD}
	\GD{\dout}
	&\DEF\tfrac{1}{\sqrt{2\pi}}e^{-\sfrac{\dout^{2}}{2}}
	&
	&\forall \dout\in\numbers{R}.
\end{align}
The answer is intuitively obvious: we first determine
the threshold \(\f{\phi}[\varepsilon]\) for \(\GD{\cdot}\) 
such that \(\PX{\GD{\out}\!\leq\!\f{\phi}[\varepsilon]}\!=\!\varepsilon\), 
then \(\f{\beta}[\varepsilon]\) is the length of the set 
of all \(\dout\)'s satisfying  \(\GD{\dout}\!>\!\f{\phi}[\varepsilon]\).
The threshold \(\f{\phi}[\varepsilon]\) is 
the \(\varepsilon\)-quantile for the random variable \(\GD{\out}\), 
i.e., \(\f{\phi}[\varepsilon]\!=\!\cdf[-1]{\GD{\out}}[\varepsilon]\).
\begin{align}
	\notag
	\cdf[-1]{\GD{\out}}[\varepsilon]
	&\DEF\inf\left\{\tau:\cdf{\GD{\out}}[\tau]\geq \varepsilon\right\}
	&
	&\forall \varepsilon\!\in\!(0,1),
	\\
	\label{eq:GaussLebesgue:LRQ}
	&=\GD{\IGQF{\tfrac{\varepsilon}{2}}}
	&
	&\forall \varepsilon\!\in\!(0,1),	
\end{align}
where\footnote{Note that \(\cdf[-1]{\GD{\out}}\!:\!(0,1)\!\mapsto\!\left(0,\sfrac{1}{\sqrt{2\pi}}\right)\) 
	is the inverse of the function \(\cdf{\GD{\out}}\).}
\(\cdf{\GD{\out}}[\cdot]\) is the cumulative distribution function of 
the random variable \(\GD{\out}\) 
and \(\IGQF{\cdot}\) is the inverse of the Gaussian \(Q\)-function
defined as 
\begin{align}
	\label{eq:def:GQF}
	\GQF{\tau}
	&\DEF \int_{\tau}^{\infty}\GD{\dout}\dif{\dout}
	&
	&\forall \tau\in\numbers{R}.	
	\intertext{Thus \(\left\{\dout:\abs{\dout}\!<\!\IGQF{\sfrac{\varepsilon}{2}}\right\}\) is 
		the set of all \(\dout\)'s satisfying the condition 
		\(\GD{\dout}\!>\!\f{\phi}[\varepsilon]\), and consequently}
	\label{eq:GaussLebesgue:Beta}
	\f{\beta}[\varepsilon]
	&\!=\!2\IGQF{\sfrac{\varepsilon}{2}}
	&
	&\forall \varepsilon\!\in\!(0,1).	
\end{align}
What we have observed until now, is implied by Neyman--Pearson lemma
and there is nothing new conceptually. 
The expressions in 
\eqref{eq:GaussLebesgue:LRQ}
and 
\eqref{eq:GaussLebesgue:Beta}
for \(\cdf[-1]{\GD{\out}}[\cdot]\) and \(\f{\beta}[\cdot]\)
allows us to 
establish alternative characterizations for \(\f{\beta}[\cdot]\)
and to relate \(\cdf[-1]{\GD{\out}}[\cdot]\)  and \(\f{\beta}[\cdot]\).

First\begin{subequations}\label{eq:GaussLebesgue} note that 
\(\pder{}{\tau}\GQF{\tau}\!=\!-\GD{\tau}\), 
\eqref{eq:GaussLebesgue:LRQ},
and \eqref{eq:GaussLebesgue:Beta}
imply
\begin{align}
	\notag
	\pdimpf[\beta]{}[\varepsilon]
	&\!=\!\tfrac{-1}{\GD{\IGCD{\frac{\varepsilon}{2}}}} 
	\\
	\label{eq:GaussLebesgue:Beta:derivative}
	&\!=\!\tfrac{-1}{\cdf[-1]{\GD{\out}}[\varepsilon]}
	&
	&\forall\varepsilon\!\in\!(0,1).
	\intertext{Then using 
		\(\f{\beta}[1]\!-\!\f{\beta}[\varepsilon]\!=\!\int_{\varepsilon}^{1}\!\pdimpf[\beta]{}[\tau]\dif{\tau}\) 
		and	\(\f{\beta}[1]=0\), we get}
	\label{eq:GaussLebesgue:Beta:IOSD}
	\f{\beta}[\varepsilon]
	&\!=\!\int_{\varepsilon}^{1}\!\tfrac{1}{\cdf[-1]{\GD{\out}}[\tau]}\dif{\tau}
	&
	&\forall\varepsilon\!\in\!(0,1).
\end{align}
On the other hand, \(\f{\beta}[\cdot]\) is convex 
because \(\pdimpf[\beta]{}[\cdot]\) is increasing. 
Furthermore, \(\f{\beta}[\cdot]\) can be described 
as the convex conjugate of its convex conjugate
on \((0,1)\). Consequently, it has the following 
dual characterization	
\begin{align}
	\label{eq:GaussLebesgue:Beta:DualCharacterization}
	\f{\beta}[\varepsilon]
	&=\sup\nolimits_{\gamma\geq 0}
	\f{\hbar}[\gamma]-\gamma\varepsilon
	&
	&\forall\varepsilon\!\in\!(0,1),
	\\
	\label{eq:GaussLebesgue:Beta:DualCharacterization:max}
	&=\f{\hbar}[\gamma_{\varepsilon}]-\gamma_{\varepsilon}\varepsilon
	&
	&\forall\varepsilon\!\in\!(0,1),
\end{align}
where \(\f{\hbar}[\gamma]\!=\!\EX{\sfrac{1}{\GD{\out}}\!\wedge\!\gamma}\)
and \(\gamma_{\varepsilon}\!=\!\tfrac{1}{\cdf[-1]{\GD{\out}}[\varepsilon]}\). 
Recall that the expected value of any non-negative random variable \(\inp\)
equals to the integral of its complementary distribution function,
i.e., \(\EX{\inp}\!=\!\int_{0}^{\infty}\!\PX{\inp\!>\!\tau}\dif{\tau}\),
by \cite[(1.30)]{gallagerSP}, \cite[p.247]{shiryaev2016probability1}. %\cite[p.208]{shiryaev}
Thus, {\color{white} \cite{shiryaev2019probability2}.}
\begin{align}
	\notag
	\f{\hbar}[\gamma]
	&\!=\!\int_{0}^{\infty} \PX{\left[\tfrac{1}{\GD{\out}}\wedge\gamma \right]\!>\!\tau}\dif{\tau}	
	&
	&\forall \gamma\geq0,
	\\
	\notag 
	&\!=\!\int_{0}^{\gamma} 
	\PX{\tfrac{1}{\GD{\out}}\!>\!\tau}\dif{\tau}
	&
	&\forall \gamma\geq0.		
\end{align}
Hence, \eqref{eq:GaussLebesgue:Beta:DualCharacterization:max},
the fact that \(\PX{(\sfrac{1}{\GD{\out}})\!>\!\tau}\) is 
non-increasing in \(\tau\),
and \(\PX{\sfrac{1}{\GD{\out}}\!>\!\gamma_{\varepsilon}}\!=\!\varepsilon\), 
imply
%Hence, \eqref{eq:GaussLebesgue:Beta:DualCharacterization}, 
%\eqref{eq:GaussLebesgue:Beta:DualCharacterization:max},
%and the monotonicity of \(\PX{\tfrac{1}{\GD{\out}}\!>\!\tau}\) 
%as a function of \(\tau\) imply
\begin{align}
	\label{eq:GaussLebesgue:Beta:SI}
	\f{\beta}[\varepsilon]
	&\!=\!\int_{0}^{\gamma_{\varepsilon}}\!
	\left[\PX{\tfrac{1}{\GD{\out}}\!>\!\tau}\!-\!\varepsilon\right]
	\dif{\tau}
	&
	&&\forall\varepsilon\!\in\!(0,1),	
	\\
	\label{eq:GaussLebesgue:Beta:CI}
	&\!=\!\int_{0}^{\infty}\!
	\abp{\PX{\tfrac{1}{\GD{\out}}\!>\!\tau}\!-\!\varepsilon}
	\!\dif{\tau}
	&
	&&\forall\varepsilon\!\in\!(0,1),
\end{align}
where \(\abp{\dinp}\!=\!\max\{0,\dinp\}\).	
\end{subequations}

The above problem is a simple hypothesis testing problem 
with deterministic detectors between 
the standard Gaussian measure on \(\numbers{R}\)
and the Lebesgue measure on \(\numbers{R}\);
and \(\GD{\dout}\) is the likelihood-ratio of the outcome \(\dout\).
Then \(\f{\beta}[\varepsilon]\) is the infimum 
type II error volume (length)
when type I error probability 
is less than or equal to \(\varepsilon\).

We call the infimum of type II error volume
(i.e., missed detection volume, power) 
as a function of type I error volume 
(i.e., false alarm volume, significance level) 
the deterministic-detector Neyman--Pearson function 
(dNP-function) and denote it by \(\NPDd{\cdot}{\m{q}}{\m{w}}\)
when the measures associated with the alternative  and null hypotheses
are \(\m{q}\) and \(\m{w}\), respectively.
\begin{subequations}\label{eq:def:NPD}If 
	the randomized detectors are permitted then
	we call it the randomized-detector Neyman--Pearson function 
	(rNP-function) and denote it by \(\NPD{\cdot}{\m{q}}{\m{w}}\).
	That is for any pair of \(\sigma\)-finite measures \((\m{w},\m{q})\) 
	on the measurable space \((\outS,\outA)\), 
	dNP-function \(\NPDd{\cdot}{\m{q}}{\m{w}}\!:\!\numbers{R}\!\mapsto\![0,\infty]\) 
	and 
	rNP-function \(\NPD{\cdot}{\m{q}}{\m{w}}\!:\!\numbers{R}\!\mapsto\![0,\infty]\) 
	are defined as 
	\begin{align}
		\label{eq:def:NPDd}
		\NPDd{\varepsilon}{\m{q}}{\m{w}}
		&\DEF \inf\limits_{\rv{T}\in\tmead{\outA}:\int \rv{T} \dif{\m{w}}\leq \varepsilon} \int (1-\rv{T}) \dif{\m{q}}
		&
		&\forall\varepsilon\!\in\!\numbers{R},	
		\\
		\label{eq:def:NPDr}
		\NPD{\varepsilon}{\m{q}}{\m{w}}
		&\DEF \inf\limits_{\rv{T}\in\tmea{\outA}:\int \rv{T} \dif{\m{w}}\leq \varepsilon} \int (1-\rv{T}) \dif{\m{q}}
		&
		&\forall\varepsilon\!\in\!\numbers{R},	
	\end{align}
	where \(\tmead{\outA}\) is  the set of all \(\outA\)-measurable 
	deterministic tests on \(\outS\)
	(i.e., \(\outA\)-measurable functions of the form 
	\(\rv{T}\!:\!\outS\!\mapsto\!\{0,1\}\)), 
	and \(\tmea{\outA}\) is the set of all \(\outA\)-measurable 
	randomized tests on \(\outS\)
	(i.e., \(\outA\)-measurable functions of the form 
	\(\rv{T}\!:\!\outS\!\mapsto\![0,1]\)).	
\end{subequations}			
Then \(\f{\beta}[\cdot]\!=\!\NPDd{\cdot}{\m{q}}{\m{w}}\) 
on \([0,\infty)\) for the case when 
\(\m{q}\) is the Lebesgues measure and
\(\m{w}\) is the standard Gaussian measure on \(\numbers{R}\)
by definition. 
Furthermore,  \(\f{\beta}[\cdot]\!=\!\NPD{\cdot}{\m{q}}{\m{w}}\)
on \([0,\infty)\) by \eqref{eq:GaussLebesgue:Beta} because
\(\NPD{\cdot}{\m{q}}{\m{w}}\) is the 
convex envelope of \(\NPDd{\cdot}{\m{q}}{\m{w}}\)
and \(\f{\beta}[\cdot]\) is convex.

The extension of definitions of \(\NPD{\cdot}{\m{q}}{\m{w}}\) and 
\(\NPDd{\cdot}{\m{q}}{\m{w}}\) from the customary setting
in which both \(\m{w}\) and \(\m{q}\) are probability measures,
to the case when both \(\m{w}\) and \(\m{q}\) are \(\sigma\)-finite
measures results in  a subtlety: for some \(\sigma\)-finite 
measures \(\m{w}\) and \(\m{q}\), both \(\NPD{\varepsilon}{\m{q}}{\m{w}}\) 
and \(\NPDd{\varepsilon}{\m{q}}{\m{w}}\) are infinite for all 
finite \(\varepsilon\) values. We deem those cases to be 
void cases. 
We will see that non-void cases are those cases satisfying
the following condition: 
\begin{align}
	\label{eq:assumption:nonvoid}
	\m{w}[\der{\m{q;\sim}}{\m{w}}\!\geq\!1]
	+
	\m{q}[\der{\m{w;\sim}}{\m{q}}\!>\!1]
	&<\infty,			
\end{align}
where \(\m{q;\sim}\) is 
the \(\m{w}\)-absolutely continuous component of \(\m{q}\) 
and \(\m{w;\sim}\) is 
the \(\m{q}\)-absolutely continuous component of \(\m{w}\),
\cite[\S31]{halmos}.  

If \(\m{w}\) and \(\m{q}\) are \(\sigma\)-finite 
\begin{subequations}\label{eq:Beta}measures, then
%\begin{empheq}[box=\fbox]{align}
\begin{align}
\label{eq:Beta:CI}
\NPD{\varepsilon}{\m{q}}{\m{w}}
&\!=\!\int_{0}^{\infty}\!
\abp{\m{w}[\der{\m{q;\sim}}{\m{w}}\!>\!\tau]\!-\!\varepsilon}
\!\dif{\tau}
&
&\forall\varepsilon\!\geq\!0,
\\
\label{eq:Beta:CI:DF} 
&\!=\!\int_{0}^{\infty}\!
\abp{\m{w}[\der{\m{w;\sim}}{\m{q;\sim}}\!\leq\!\tfrac{1}{\tau}]\!-\!\varepsilon}
\!\dif{\tau}
&
&\forall\varepsilon\!\geq\!0,
\end{align}
%\end{empheq}
To the best of our knowledge, 
the characterization of \(\NPD{\cdot}{\m{q}}{\m{w}}\) given 
in \eqref{eq:Beta} has not been reported before, 
even for probability measures \(\m{w}\) and \(\m{q}\) satisfying \(\m{w}\!=\!\m{w;\sim}\) and \(\m{q}\!=\!\m{q;\sim}\). 
Furthermore, all of the identities given in \eqref{eq:GaussLebesgue} 
for \(\NPD{\cdot}{\m{q}}{\m{w}}\), hold with appropriate modifications
in all non-void cases, 
i.e., when \(\m{w}\) and \(\m{q}\) satisfy \eqref{eq:assumption:nonvoid}, 
see \eqref{eq:thm:SNPL} of Theorem \ref{thm:SNPL}. 
We provide an account of related prior results in 
\S\ref{sec:PriorWorkOnOneShot}.
\end{subequations}		

Evidently, neither \eqref{eq:GaussLebesgue:LRQ} 
nor \eqref{eq:GaussLebesgue:Beta} is true for arbitrary 
\(\sigma\)-finite measures ---or for arbitrary probability measures---
\(\m{w}\) and \(\m{q}\).
In fact, \(\NPD{\cdot}{\m{q}}{\m{w}}\) is not differentiable
for some \(\varepsilon\) values for certain \(\m{w}\) and \(\m{q}\).  
However, \(\NPD{\cdot}{\m{q}}{\m{w}}\) is convex and finite on
\((0,\infty)\) in all non-void cases. 
Thus \(\NPD{\cdot}{\m{q}}{\m{w}}\) has directional derivatives 
everywhere on \((0,\infty)\). Furthermore, 
the left and right derivatives of \(\NPD{\cdot}{\m{q}}{\m{w}}\) 
are \(\sfrac{-1}{\lscq{\cdot}{\m{w}}{\m{q}}}\) and
\(\sfrac{-1}{\uscq{\cdot}{\m{w}}{\m{q}}}\), where
\(\lscq{\cdot}{\m{w}}{\m{q}}\) and \(\uscq{\cdot}{\m{w}}{\m{q}}\) 
are the lower semicontinuous and the upper semicontinuous 
quantiles for the likelihood-ratio, 
see \eqref{eq:def:lr} and \eqref{eq:def:LRQ} in the following 
for formal definitions.
When the distribution function \(\scq{\cdot}{\m{w}}{\m{q}}\)
is constant on a subinterval of \((0,\infty)\), the 
inverse function in the usual sense 
does not exist; that is why we have to work with   
the semicontinuous inverses of 
\(\scq{\cdot}{\m{w}}{\m{q}}\), i.e.,  likelihood-ratio quantiles
\(\lscq{\cdot}{\m{w}}{\m{q}}\) and \(\uscq{\cdot}{\m{w}}{\m{q}}\).
Whenever \(\scq{\cdot}{\m{w}}{\m{q}}\) is invertible
\(\scq[-1]{\cdot}{\m{w}}{\m{q}}
=\lscq{\cdot}{\m{w}}{\m{q}}=\uscq{\cdot}{\m{w}}{\m{q}}\)
holds, as one would expect.
%We discuss semicontinuous inverses of monotonic functions 
%and likelihood-ratio quantiles in more detail in \S\ref{sec:LikelihoodRatioQuantiles}.

The function \(\f{\hbar}[\cdot]\) in \eqref{eq:GaussLebesgue:Beta:DualCharacterization} 
is the primitive entropy spectrum of the Lebesgue measure on \(\numbers{R}\) with respect to the
standard Gaussian measure on \(\numbers{R}\).
For arbitrary \(\sigma\)-finite measures \(\m{q}\) and \(\m{w}\) 
on the same measurable space, the primitive entropy spectrum 
of \(\m{q}\) with respect to \(\m{w}\) is defined as 
\begin{align}
\label{eq:def:EntropySpectrum}
\ent{\gamma}{\m{q}}{\m{w}}
&\DEF
\displaystyle{\int}\left(\der{\m{q;\sim}}{\m{w}}\wedge\gamma\right)\dif{\m{w}}
&
&\forall\gamma\!\in\![0,\infty).
\intertext{Evidently, the primitive entropy spectrum has 
the following variational characterization}
\label{eq:def:EntropySpectrum:Variational}
\ent{\gamma}{\m{q}}{\m{w}}
&=\inf\nolimits_{\oev\in\outA}
\m{q}[\oev]\!+\!\gamma\m{w}[\outS\!\setminus\!\oev]
&
&\forall\gamma\!\in\![0,\infty).	
\end{align}

We obtain the spectral representation of \(\ent{\cdot}{\m{q}}{\m{w}}\),
using the following level set integration, which is 
a consequence of integration by parts,
\cite[Theorem 2.9.3]{bogachev}: for any 
measure\footnote{All measures are assumed to be non-signed 
	and countably-additive.} \(\m{\mu}\) and \(\m{\mu}\)-measurable function 
\(\f{f}\) satisfying \(\esinf_{\m{\mu}}\f{f}\geq0\),
\begin{align}
\label{eq:LevelSetIntegration}%
\int \f{f}\dif{\m{\mu}}
&=\int_{0}^{\infty}\m{\mu}[\f{f}>\tau]\dif{\tau}.
\end{align}
In the parlance of probability theory
(i.e., when \(\m{\mu}\) is a probability measure) 
\(\EX{\f{f}}\!=\!\int_{0}^{\infty}\!\PX{\f{f}\!>\!\tau}\dif{\tau}\)
for any non-negative random variable \(\f{f}\). 
In \cite[\S 1.5]{gallagerSP}, Gallager defines 
the expectation of a random variable using this 
identity, i.e., the tail expectation formula, 
yielding a deceptively simple, yet fully rigorous 
construction aligned with measure-theoretic foundations.

What one learns from \eqref{eq:GaussLebesgue} that is not explicit in
\eqref{eq:GaussLebesgue:Beta} is 
how \(\f{\beta}[\cdot]\) as a function is related to
\(\f{\hbar}[\cdot]\) and \(\cdf[-1]{\GD{\out}}[\cdot]\) 
as functions.
Analogously Theorem \ref{thm:SNPL}, in the following,
describes how \(\NPD{\cdot}{\m{q}}{\m{w}}\) as a function 
is related to \(\ent{\cdot}{\m{q}}{\m{w}} \), 
\(\lscq{\cdot}{\m{w}}{\m{q}}\), 
and  \(\uscq{\cdot}{\m{w}}{\m{q}}\) as functions
for  a given pair of \(\sigma\)-finite measures \(\m{q}\)
and \(\m{w}\).
In other words, Theorem \ref{thm:SNPL} makes 
the implications of the Neyman--Pearson lemma explicit 
and extends the lemma  and its consequences 
to randomized decisions between \(\sigma\)-finite
measures.

As pointed out in \cite{strassen62} and later employed in \cite{csiszarL71}, 
lossless source coding provides an operational justification 
for hypothesis testing problems where
the measure associated with the alternative hypothesis is a 
\(\sigma\)-finite measure rather than a probability measure. 
Specifically, finding the minimum cardinality of a subset 
of a countably infinite set that contains all but 
\(\varepsilon\) of the probability mass of a discrete 
probability measure \(\m{w}\) is a hypothesis 
testing problem against the counting measure \(\m{\sharp}\) 
with deterministic tests. In this context, 
\(\NPDd{\varepsilon}{\m{\sharp}}{\m{w}}\) is the 
cardinality of any such minimum cardinality sets. 
While the counting measure \(\m{\sharp}\) 
is not finite for countably infinite sets, 
it remains \(\sigma\)-finite for any countable set.

Furthermore,  for the lossless source coding problem,
the randomized-detector Neyman--Pearson function 
determines the value of the deterministic-detector Neyman--Pearson function.
Note that for \begin{subequations}\label{eq:LosslessSourceCoding}those 
\(\varepsilon\) values for which a minimum-cardinality-set-\(\oev\) 
satisfies \(\m{w}[\outS\!\setminus\!\oev]\!=\!\varepsilon\), we have
\(\NPDd{\varepsilon}{\m{\sharp}}{\m{w}}\!=\!\NPD{\varepsilon}{\m{\sharp}}{\m{w}}\).
On the other hand, if \(\m{w}[\outS\!\setminus\!\oev]\!<\!\varepsilon\)
for a minimum-cardinality-set-\(\oev\),  
then \(\NPD{\varepsilon}{\m{\sharp}}{\m{w}}\!<\!
\NPDd{\varepsilon}{\m{\sharp}}{\m{w}}\!<
\!\NPD{\varepsilon}{\m{\sharp}}{\m{w}}\!+\!1\).
Thus, using the fact that \(\NPDd{\cdot}{\m{\sharp}}{\m{w}}\) 
is an integer valued function, we conclude
\begin{align}
	\label{eq:LosslessSourceCoding:Equivalence}
	\NPDd{\varepsilon}{\m{\sharp}}{\m{w}}
	&\!=\!\lceil\NPD{\varepsilon}{\m{\sharp}}{\m{w}}\rceil	
	&
	&\forall\varepsilon\geq0,
	\\
	\label{eq:LosslessSourceCoding:Exact-original}
	&\!=\!\left\lceil \int_{0}^{\infty}\!
	\abp{\m{w}[\der{\m{\sharp}_{\sim}}{\m{w}}\!>\!\tau]\!-\!\varepsilon}
	\!\dif{\tau}\right\rceil	
	&
	&\forall\varepsilon\geq0,
	\\
	\label{eq:LosslessSourceCoding:Exact}
	&\!=\!\left\lceil
	\int_{0}^{\infty}\!
	\abp{\m{w}[\der{\m{w}}{\m{\sharp}}\!\leq\!\tfrac{1}{\tau}]\!-\!\varepsilon}
	\!\dif{\tau}
	\right\rceil	
	&
	&\forall\varepsilon\geq0,		
\end{align}
\eqref{eq:LosslessSourceCoding:Exact-original} 
follows from\end{subequations}~		
\eqref{eq:Beta:CI}, and
\eqref{eq:LosslessSourceCoding:Exact} follows 
from \eqref{eq:Beta:CI:DF}.
Note that, in general \(\NPDd{\cdot}{\m{q}}{\m{w}}\)
is bounded but not necessarily determined by \(\NPD{\cdot}{\m{q}}{\m{w}}\), 
see Remark \ref{rem:NPDvsNPDdWorstcase}.

In the remainder of this section, 
we first briefly review prior work related to
\(\NPD{\cdot}{\m{q}}{\m{w}}\), \(\uscq{\cdot}{\m{w}}{\m{q}}\), 
and \(\ent{\cdot}{\m{q}}{\m{w}}\) in \S\ref{sec:PriorWorkOnOneShot}.
Next, we discuss prior work on approximations of rNP-functions in 
\S\ref{sec:PriorWorkOnApproximations}. 
We then highlight our main contributions in \S\ref{sec:MainContributions}
and conclude the introduction with outlining the remainder 
of the paper in \S\ref{sec:outline}.

\subsection{Prior Work On \(\NPD{\cdot}{\m{q}}{\m{w}}\), 
	\(\uscq{\cdot}{\m{w}}{\m{q}}\), and 
	\(\ent{\cdot}{\m{q}}{\m{w}}\)}\label{sec:PriorWorkOnOneShot}
The simple hypothesis testing problem is a fundamental problem 
that can be motivated by various applications; our interest
in the problem is from the perspective of information transmission
problems \cite{csiszarkorner,gallager,coverthomas,han,polyanskiyWu}.
%An exhaustive survey of prior results is neither feasible nor helpful for
%our purposes; nevertheless, the author would welcome references to any prior work 
%establishing the identities in \eqref{eq:thm:SNPL} and \eqref{eq:thm:SNPL:DF}.

The importance of hypothesis testing problem, and of 
the NP-functions in particular, 
was already clear for information transmission problems 
by the time of Strassen's seminal work \cite{strassen62}.
Furthermore, as we have recalled above the lossless source coding problem 
provides a concrete operational justification for considering
\(\NPD{\cdot}{\m{q}}{\m{w}}\) for the case when \(\m{q}\) is 
a \(\sigma\)-finite measure, which is
not finite. The hypothesis testing problem recognized to play 
a crucial role for understanding the behavior
of the error probability in the channel coding problem,
not only at rates close to capacity \cite{strassen62},
but also at other rates 
\cite{shannonGB67A,haroutunian68,augustin68,haroutunian77,augustin78}.

The hypothesis testing problem is now a key tool 
for analyzing and understanding information transmission problems,
\cite[Ch. 1]{csiszarkorner},
\cite[Ch. 3 and Appendix 5.A]{gallager},
\cite[Ch. 5 and 11]{coverthomas},
\cite[Ch. 1 and 4]{han},
\cite[Ch. 10, 11, and Part III]{polyanskiyWu}. 
%{\color{magenta}\cite[\S 2.3]{polyanskiythesis},\cite[\S 2.1]{tan14}, \cite[\S 14.1]{polyanskiyWu}}

The interest in finite-length (i.e., non-asymptotic) analysis
and exact asymptotic analysis  
in information transmission problems revived the interest
in \(\NPD{\cdot}{\m{q}}{\m{w}}\),  
\cite{hayashi09B,polyanskiythesis,polyanskiyPV10,
	polyanskiy13,moulin17,yavasKE24,tomamichelT13,tan14}. 
In particular, the maximum number of messages that a
code on the channel \(\tm{W}\!:\!\inpS\!\mapsto\!\pmea{\outA}\) 
with maximum error probability \(\varepsilon\) can have,
i.e., \(\f{M}[\tm{W},\varepsilon]\), 
is bounded above as,  
\begin{align}
	\label{eq:MetaConverse}
	\f{M}[\tm{W},\varepsilon]
	&\leq \inf\nolimits_{\m{q}\in\pmea{\outA}}\sup\nolimits_{\dinp\in\inpS}
	\tfrac{1}{\NPD{\varepsilon}{~\m{q}}{\tm{W}(\dinp)}},
\end{align}
see  \cite[Theorem 31]{polyanskiyPV10}. 
This upper bound is studied more closely in \cite{polyanskiy13}.
In addition new divergences between probability measures,
such as the hypothesis testing divergence 
in \cite[p. 7044]{tomamichelT13}, \cite[(2.6)]{tan14}, 
are introduced using \(\NPD{\cdot}{\m{q}}{\m{w}}\)
so as to analyze information transmission 
problems in non-asymptotic and refined asymptotic regimes,
see \cite{tan14,wangCR09,wangR12,tomamichelH13,tomamichelT13}. 

The maximum number of messages is also bounded from below 
for any probability mass function \(\m{p}\) on the input set
\(\inpS\) and positive threshold \(\gamma\) in terms of 
the likelihood ratio distribution as a result of 
the maximum-code-theorem, see \cite[Theorem 2.1]{augustin78}, as 
\begin{align}
	\label{eq:MaximalCode}	
	\f{M}[\tm{W},\varepsilon]
	&\geq \gamma\left(\varepsilon-
	\scq{\gamma}{\m{p}\otimes\tm{W}}{\m{p}\otimes (\m{p}\tm{W})}\right),
\end{align}
where \(\m{p}\tm{W}\) is output marginal of the joint distribution 
\(\m{p}\otimes\tm{W}\) generated by the input distribution \(\m{p}\) 
and the channel \(\tm{W}\).

The likelihood-ratio quantiles \(\lscq{\cdot}{\m{w}}{\m{q}}\)
and \(\uscq{\cdot}{\m{w}}{\m{q}}\), defined as the semi-continuous
inverses of likelihood ratio distribution \(\scq{\cdot}{\m{w}}{\m{q}}\),
determine the sub-differential of  \(\NPD{\cdot}{\m{q}}{\m{w}}\), as we will see 
in \eqref{eq:thm:SNPL:leftderivative} and \eqref{eq:thm:SNPL:rightderivative}, 
and have seen in \eqref{eq:GaussLebesgue:Beta:derivative}
for a specific case.
The upper-semicontinuous likelihood-ratio quantile \(\uscq{\cdot}{\m{w}}{\m{q}}\) 
and its natural logarithm (i.e., \(\ln\uscq{\cdot}{\m{w}}{\m{q}}\),  
the log-likelihood quantile) 
have been introduced and analyzed by many authors under various names. 
\(\uscq{\cdot}{\m{w}}{\m{q}}\) is denoted by \(\mu\) in \cite[(2.5)]{strassen62}.
Inspired by \cite[Ch. IV]{han}, 
\(\ln\uscq{\cdot}{\m{w}}{\m{q}}\) is called 
information spectrum divergence in \cite[(2.9)]{tan14},
the classical entropic information spectrum in \cite[(2)]{tomamichelH13},
the relative entropy information spectrum (and the divergence spectrum) 
in \cite[p. 7044]{tomamichelT13}. 
Furthermore, \(\NPD{\cdot}{\m{q}}{\m{w}}\) has been bounded in terms of 
\(\uscq{\cdot}{\m{w}}{\m{q}}\) 
for any probability measures \(\m{w}\) and \(\m{q}\)
provided that \(\m{w}\) is absolutely continuous  in \(\m{q}\), 
i.e., \(\m{w}\AC\m{q}\), see
\cite[(2.67), (2.69)]{polyanskiythesis},
\cite[(103)]{polyanskiyPV10},
\cite[Lemma 2]{tomamichelT13}, 
\cite[Lemma 2.4]{tan14}, 
\cite[Lemma 12]{tomamichelH13}:
\begin{align}
	\label{eq:Tan-2-4}
	\!\!\tfrac{1}{\uscq{\varepsilon}{\m{w}}{\m{q}}}
	\!\geq\!
	\NPD{\varepsilon}{\m{q}}{\m{w}}
	&\!\geq\! 
	\tfrac{\delta}{\uscq{\varepsilon+\delta}{\m{w}}{\m{q}}}
	&&\forall\varepsilon\!\in\!(0,1),\delta\!\in\!(0,1\!-\!\varepsilon).	
\end{align}
The upper and lower bounds in 
\eqref{eq:Tan-2-4} is the state-of-the-art relation between 
likelihood-ratio quantiles and \(\NPD{\cdot}{\m{q}}{\m{w}}\), prior
to Theorem \ref{thm:SNPL}. 
It is worth noting that although the upper semicontinuous quantile 
\(\uscq{\cdot}{\m{w}}{\m{q}}\) appear more prominently in information
theory; when used without any qualifiers, what is meant by quantile 
in probability theory, usually, is the lower semicontinuous quantile,
see \cite[p. 427]{shiryaev2016probability1}.%\cite[p. 355]{shiryaev}

For probability measures \(\m{w}\) and \(\m{q}\),
the primitive entropy spectrum of \(\ent{\cdot}{\m{w}}{\m{q}}\)
is related to the primitive divergence spectrum 
\(\div{\cdot}{\m{w}}{\m{q}}\), employed in 
the spectral representation of \(\f{f}\)-divergences
\cite{csiszar63,csiszar67A,morimoto63,aliS66} 
reported in \cite{osterreicherV93,lieseV06,reidW11, liese12,guntuboyinaSS14}:
%\cite[Theorem 12]{reidW11} Implicitly assumes there are noatoms 
\begin{align}
	\div{\gamma}{\m{w}}{\m{q}}
	&=1\wedge\gamma-\ent{\gamma}{\m{w}}{\m{q}}	
	&
	&\forall \gamma\in[0,\infty).	
\end{align}
Note that \(\div{\gamma}{\m{w}}{\m{q}}\) is the \(\f{f}\)-divergence
\(\RD{\f{f}}{\m{w}}{\m{q}}\) 
corresponding to the function 
\(\f{f;\gamma}[\dinp]\!=\!(1\wedge\gamma)-(\dinp\wedge\gamma)\)
for each \(\gamma\in[0,\infty)\). 
The quantity \(\tfrac{1}{1+\gamma}\ent{\gamma}{\m{w}}{\m{q}}\) is called 
Bayes loss in \cite[(66)]{lieseV06},
posterior Bayes risk in \cite[p.734]{reidW11},
and a posteriori loss in \cite[(2.25)]{liese12}.
The primitive entropy spectrum \(\ent{\cdot}{\m{w}}{\m{q}}\)
is denoted by \(\f{\psi;{\m{w},\m{q}}}[\cdot]\) 
in  \cite[\S III.A]{guntuboyinaSS14}. 
The left and right derivatives of \(\ent{\cdot}{\m{w}}{\m{q}}\)
were determined in \cite[Lemma 3.2]{guntuboyinaSS14} for 
the case when \(\m{w}\) and \(\m{q}\) are probability measures.
%\cite[\S III]{osterreicherV93}
%%\cite[(66)]{lieseV06}% 

For any channel \(\tm{W}\!:\!\inpS\!\mapsto\!\pmea{\outA}\),
probability mass function \(\m{p}\) on \(\inpS\),
and integer \(M\geq 2\), there exists a code 
with \(M\) messagages whose average error probability 
\(\varepsilon\) satisfies
\begin{align}
	\label{eq:DependenceTestingBound}
	\varepsilon
	&\leq \ent{\tfrac{M-1}{2}}{\m{p}\otimes\tm{W}}{\m{p}\otimes (\m{p}\tm{W})},
\end{align}
see \cite[(76)]{polyanskiyPV10} and \eqref{eq:lem:PES:integral:r}.
A similar dependence test bound was suggested for the lossless source coding problem
in \cite[Theorem 3]{chenEK20},
establishing an upper bound to the error probability when source code can have 
at most \(M\) distinct outputs for any probability mass function \(\m{w}\)
\begin{align}
	\label{eq:DependenceTestingBound2}
	\varepsilon
	&\leq \ent{\tfrac{1}{M}}{\m{w}}{\m{\sharp}}.
\end{align}	

Motivated by an expression equivalent to \cite[(76)]{polyanskiyPV10}
and inspired by \(\f{f}\)-divergences \cite{csiszar67A}, the divergence
\(E_{\gamma}(\m{w}\Vert\m{q})\) was defined in \cite[(2.141)]{polyanskiythesis}
for all probability measures \(\m{w}\) and \(\m{q}\) such that
\(\m{w}\) is absolutely continuous in \(\m{q}\), i.e., \(\m{w}\AC\m{q}\), 
see also \cite{sasonV16,liuthesis,liuCV17}. 
%in \cite[(18)]{liuCV17}
One can show for all probability measures \(\m{w}\) and \(\m{q}\)
satisfying the hypothsis \(\m{w}\AC\m{q}\), that
\begin{align}
\label{eq:def:DivergenceSpectrumAlternative}
E_{\gamma}(\m{w}\Vert\m{q})
&\!=\!1\!-\!\ent{\gamma}{\m{w}}{\m{q}}
&
&\forall\gamma\!\in\![0,\infty).\!		
\end{align}
For all probability measures \(\m{w}\) and \(\m{q}\)
satisfying \(\m{w}\!\sim\!\m{q}\), i.e., 
satisfying both \(\m{w}\AC\m{q}\) and \(\m{q}\AC\m{w}\), 
the directional derivatives
of \(E_{\gamma}(\m{w}\Vert\m{q})\) as a function of \(\gamma\)
were determined and an integral representation of the Kulllback-Leibler divergence 
\(\KLD{\m{w}}{\m{q}}\) in terms of \(E_{\gamma}(\m{w}\Vert\m{q})\), 
similar to the more general ones for \(\f{f}\)-divergences 
in terms of \(\div{\gamma}{\m{w}}{\m{q}}\) in
\cite{osterreicherV93,lieseV06,reidW11, 
	liese12,guntuboyinaSS14}, was
established in \cite[Theorem 21]{polyanskiythesis}.
Later this result is extended all \(\f{f}\)-divergences 
with twice differentiable \(\f{f}\)'s relying on \cite{lieseV06},
see \cite[Proposition 3]{sasonV16}.

The following measure change bound, recently established 
in \cite[Proposition 1]{LiuFG26} for probability measure 
\(\m{w}\) and  \(\m{q}\) satisfying \(\m{w}\AC\m{q}\),
follows from \eqref{eq:def:EntropySpectrum:Variational}
and \eqref{eq:def:DivergenceSpectrumAlternative}:
\begin{align}
\label{LiuFanGunduz}
E_{\gamma}(\m{w}\Vert\m{q})+\gamma\m{q}[\oev]
&\geq\m{w}[\oev]
&
&\forall \oev\!\in\!\outA,\gamma\!>\!0.
\end{align}

None of the identities of given in  \eqref{eq:thm:SNPL} of 
Theorem \ref{thm:SNPL} was reported before in this form, 
even for the case of arbitrary probability measures 
\(\m{w}\) and \(\m{q}\). However, there are a number of 
closely related prior results.

The prior results closest to Theorem \ref{thm:SNPL} 
are \cite{AudenaertMF12} and \cite{elkayamF16},
which we became aware of after the initial arXiv and ISIT 
submissions of the present work 
\cite{nakiboglu26A,nakiboglu26C,nakiboglu26B}. 
For the quantum formulation of the problem under the assumption 
that underlying Hilbert space is finite dimensional,
\cite[Proposition 3.2]{AudenaertMF12} characterizes rNP-function
\(\NPD{\cdot}{\m{q}}{\m{w}}\) as its own bidual for convex 
conjugation.\footnote{The dual characterization of the rNP-function 
in \cite[Proposition 3.2]{AudenaertMF12} has been modified, 
since \cite{nakiboglu26A}, to adopt primitive entropy spectrum 
form, see \cite[Proposition 6.1]{chengL26} 
and \cite[Lemma 1]{LizarribarVK26B}.} 
Thus proposition \cite[Proposition 3.2]{AudenaertMF12} implies
\eqref{eq:thm:SNPL:DualCharacterization} and
\eqref{eq:thm:SNPL:DualCharacterization:max}
for the case when \(\m{w}\) and \(\m{q}\)
are probability mass functions on a finite set \(\outS\). 
Later \cite{elkayamF16} set out to express \(\NPD{1\!-\!\alpha}{\m{q}}{\m{w}}\)
as the Legendre transform (i.e., convex conjugate) of a function for the case when	
\(\m{w}\) and \(\m{q}\) are probability measures
on a countable set \(\outS\), i.e., \(\m{w}\) and \(\m{q}\) 
are probability mass functions. 
\cite[Lemmas 1 and 2]{elkayamF16}
establish expressions that are equivalent to  
\eqref{eq:thm:SNPL:DualCharacterization}, 
\eqref{eq:thm:SNPL:DualCharacterization:max}, and
\eqref{eq:thm:SNPL:SI}.
More recently, a characterization of \(\NPD{\cdot}{\m{w}}{\m{q}}\) that is 
equivalent to \eqref{eq:thm:SNPL:DualCharacterization} and
\eqref{eq:thm:SNPL:DualCharacterization:max} of Theorem \ref{thm:SNPL} 
was established for the relative information 
spectrum,\footnote{The relative information spectrum 
	in \cite{verdu24,verdu25} is equal to \(\scq{\exp(\cdot)}{\m{w}}{\m{q}}\).} 
see \cite[(4.1) and (8.35)]{verdu24}, \cite{verdu25},
for the case when \(\m{w}\) and \(\m{q}\) are probability measures. 
A relation equivalent to \eqref{eq:thm:SNPL:DualCharacterization:max} of 
Theorem \ref{thm:SNPL} was reported in
\cite[(2.147) of Theorem 21]{polyanskiythesis}
for probability measures \(\m{w}\) and \(\m{q}\) 
satisfying \(\m{w}\!\sim\!\m{q}\). 

Working under implicit assumption that 
\(\NPD{\cdot}{\m{q}}{\m{w}}\!=\!\NPDd{\cdot}{\m{q}}{\m{w}}\),
which is guaranteed to hold when \(\m{w;\sim}\) does not have any 
atoms, an 
expressions equivalent to \eqref{eq:thm:SNPL:DualCharacterization} 
of Theorem \ref{thm:SNPL} and \eqref{eq:Beta:ConvexConjugate} 
were reported in \cite[Theorem 21]{reidW11}, see
\cite[(43)]{reidW11}  and \cite[(42)]{reidW11}.
The conjugacy relation established in \eqref{eq:Beta:ConvexConjugate} 
is not present in their framework because of their parameterization. 
Nevertheless \cite[(42) and (43)]{reidW11} are deemed
to be ``dual'' relations in \cite[Theorem 21]{reidW11}.

A closely related result was also reported in the framework
of lossless source coding.
\cite{kontoyiannisV14} characterizes the optimal performance 
of one-to-one but not necessarily prefix free variable length 
binary codes in terms of a quantity denoted by 
\(\f{R;;*}[\blx,\epsilon]\), see \cite[(5)]{kontoyiannisV14}.
The value of \(\f{R;;*}[\blx,\varepsilon]\) is
determined by \(\NPDd{\varepsilon}{\m{\sharp}}{\m{w}}\):
%\begin{subequations}
\begin{align}
\label{eq:KontoyiannisVerdu}
\blx \f{R;;*}[\blx,\varepsilon]
&=\lceil\log_{2}\left(1+\NPDd{\varepsilon}{\m{\sharp}}{\m{w}}\right)\rceil-1.	
%\\
%&=\lceil\log\left(1+\lceil\NPD{\varepsilon}{\m{\sharp}}{\m{w}}\rceil\right)\rceil-1,
%\\
%&=\lceil\log\left(1+\lceil\NPD{\varepsilon}{\m{\sharp}}{\m{w}}\rceil\right)\rceil-1.
\end{align}		
%\end{subequations}
Thus the value of \(\f{R;;*}[\blx,\varepsilon]\) is determined by 
\(\NPD{\varepsilon}{\m{\sharp}}{\m{w}}\), as a result of 
\eqref{eq:LosslessSourceCoding:Equivalence}.
\cite[Theorem 7]{kontoyiannisV14} characterizes 
\(\f{R;;*}[\varepsilon]=\f{R;;*}[1,\varepsilon]\)
for \(\varepsilon\) values satisfying
\(\varepsilon\!=\!\m{w}[\der{\m{w}}{\m{\sharp}}\!\leq\!\gamma]\)
for some \(\gamma\!\geq\!0\). 
Evidently, not all \(\varepsilon\) values in \((0,1)\) interval satisfy that
condition. This, however, does not prevent \cite[Theorem 7]{kontoyiannisV14}
from characterizing \(\f{R;;*}[\cdot]\) and 
\(\NPDd{\cdot}{\m{\sharp}}{\m{w}}\) as functions on \((0,1)\) 
as long as all \(\m{w}[\{\dout\}]\) values are distinct for all 
\(\dout\) in \(\outS\). 
When multiple \(\dout\)'s have the same \(\m{w}[\{\dout\}]\) value,
however, \cite[Theorem 7]{kontoyiannisV14} is mute about the behavior of 
\(\f{R;;*}[\cdot]\) and 
\(\NPDd{\cdot}{\m{\sharp}}{\m{w}}\) for the values of \(\varepsilon\) 
corresponding to  these \(\dout\)'s.
Consider for example the case when \(\m{w}\) is the uniform distribution 
on a finite set or when \(\m{w}\) is weighted sum of uniform distributions 
on disjoint sets.
Evidently,  \eqref{eq:LosslessSourceCoding} and \eqref{eq:KontoyiannisVerdu}
characterize \(\f{R;;*}[\blx,\varepsilon]\) completely even in such cases.
{\color{white}  \cite{Yushkevich53}}

\subsection{Prior Work On Approximations}\label{sec:PriorWorkOnApproximations}
In many applications one is interested in the behavior 
of \(\NPD{\cdot}{\m{q}}{\m{w}}\) on a sub-interval of 
the positive real line under various structural 
assumptions on \(\m{q}\) and \(\m{w}\). 
In this subsection, we consider the prior work on the problem only
for the memoryless and discrete Markov cases. 
In the memoryless case \(\m{w}\) and \(\m{q}\) are both 
product measures of the form 
\(\m{w}\!=\!\otimes_{\tin=1}^{\blx}\m{w;\tin}\)
and
\(\m{q}\!=\!\otimes_{\tin=1}^{\blx}\m{q;\tin}\),
where \(\m{w;\tin}\) and \(\m{q;\tin}\) are measures on 
the same measurable space \((\outS_{\tin},\outA_{\tin})\). 
The memoryless case is called stationary 
if \((\m{w;\tin},\m{q;\tin})\!=\!(\m{w;1},\m{q;1})\) 
for all \(\tin\). 
In the time-homogeneous discrete Markov case 
\(\m{w}\) and \(\m{q}\) are both given in terms of 
an initial \(\sigma\)-finite measure on a countable set \(\set{S}\)
and a discrete \(\sigma\)-finite transition measure from \(\set{S}\)
to \(\set{S}\); that is \(\m{w}\) and \(\m{q}\) are of the form 
\(\m{w}\!=\!\m{w;0}\otimes \tm{W;;\otimes\blx}\)
and
\(\m{q}\!=\!\m{q;0}\otimes \tm{Q;;\otimes\blx}\).
All of the  results mentioned in this subsection 
are derived for the case when \(\m{w}\) is a 
probability measure.
In the following, we deem an approximation non-asymptotic 
only when both the approximation error term and the condition 
on \(\blx\) (and on other quantities if any) are explicit 
in the statement of the result.

For the stationary memoryless 
case\footnote{Although \cite[Theorem 1.1]{strassen62} is 
	stated for the case when \(\outS_{1}\) is a finite set, proof 
	does not rely on that assumption and the extension for arbitrary 
	measurable space is explicitly mentioned in \cite[\S5]{strassen62}.} 
\cite[Theorem 1.1]{strassen62} determined
\(\ln\NPDd{\varepsilon}{\m{q}}{\m{w}}\) 
---and \(\ln\NPD{\varepsilon}{\m{q}}{\m{w}}\) within the proof--- 
up to an \(\smallo{1}\) approximation error term for every 
\(\varepsilon\!\in\!(0,1)\). 
The result for \(\ln\NPD{\varepsilon}{\m{q}}{\m{w}}\) is extended 
to the possibly non-stationary memoryless case under appropriate 
hypotheses with the additional assumption of \(\m{q}\) being
a probability measure, see \cite[Theorem 18]{moulin17}.
For the time-homogeneous discrete Markov case with a finite state
space \(\set{S}\), assuming \(\tm{W}\) corresponds to an irreducible 
aperiodic Markov chain and loglikelihood ratios are non-lattice 
the same result is proved in \cite{nagaevA02} for the case
when \(\m{q}\) is Markov probability measure.

For the possibly non-stationary memoryless case,
\cite[Theorem 3.1]{strassen62} provides an approximation 
for  \(\ln\NPD{\cdot}{\m{q}}{\m{w}}\) on \((\delta,1\!-\!\delta)\) 
whose approximation error is bounded from above uniformly by \(\sfrac{140}{\delta^{8}}\) 
provided that \(\sqrt{\blx}\geq\sfrac{140}{\delta^{8}}\),
where \(\delta\) is constrained by certain average moment conditions
due to the approximation error term in Berry--Esseen Theorem. 
In most cases of interest for any positive \(\delta\),
non-asymptotic  approximation for \(\ln\NPD{\cdot}{\m{q}}{\m{w}}\) on \((\delta,1\!-\!\delta)\)
by \cite[Theorem 3.1]{strassen62} has a uniform \(\bigo{1}\) approximation error. 
An equally tight approximation was established later in 
\cite[Lemma 14]{polyanskiythesis}, \cite[\S III.B]{elkayamF16}.
For the lossless source coding problem, i.e.,
when \(\m{q}\!=\!\m{\sharp}\), a qualitatively equivalent approximation 
is obtained for the memoryless case 
in\footnote{Note that for the lossless source coding problem
	\(\blx\f{R;;*}[\blx,\varepsilon]\) is essentially, 
	\(\ln\NPD{\cdot}{\m{\sharp}}{\m{w}}\)
	as a result of \eqref{eq:LosslessSourceCoding} and
	\eqref{eq:KontoyiannisVerdu}}
both stationary \cite[(44)]{kontoyiannisV14}
and non-stationary \cite[Theorems 5 and 6]{GavalakisK21}
cases. 
For the time-homogeneous discrete Markov case with 
a finite state-space assuming that \(\tm{W}\) corresponds 
to an irreducible aperiodic Markov chain, 
the lower bound of the approximation 
is established in \cite[Theorem 20]{kontoyiannisV14}
and the upper bound of the approximation is conjectured 
in \cite{kontoyiannisV14}.
More recently, for the memoryless case an asymptotic approximation 
for \(\ln\NPD{\varepsilon_{\blx}}{\m{q}}{\m{w}}\) 
was established under the hypothesis 
\(\abs{\IGQF{\varepsilon_{\blx}}}=\smallo{\sqrt{\blx}}\)
with an approximation  error term of the form 
\(\bigo{\tfrac{(\IGQF{\varepsilon_{\blx}})^{4}}{\blx}}+\bigo{1}\), 
see \cite[Theorem 4]{yavasKE24}.

In approximations for \(\ln\NPD{\varepsilon}{\m{q}}{\m{w}}\), 
the leading term is \(-\KLD{\m{w}}{\m{q}}\) both for the memoryless 
and Markovian cases where \(\KLD{\m{w}}{\m{q}}\) is the
Kullaback-Leibler divergence. Recall that for any probability 
measure \(\m{w}\) and \(\sigma\)-finite measure \(\m{q}\)
on the same measurable space satisfying \(\m{w}\AC\m{q}\), 
the Kullaback-Leibler divergence 
\(\KLD{\m{w}}{\m{q}}\) is defined as
\begin{align}
	\label{eq:def:KLD}
	\KLD{\m{w}}{\m{q}}
	&\DEF\EX[\m{w}]{\ln \der{\m{w}}{\m{q}}},			
\end{align}	
where \(\EX[\m{w}]{\cdot}\) is the expected value
for random variables on the probability space 		
\((\outS,\outA,\m{w})\), whenever the expectation 
on the right hand side exists.{\color{white}\cite{vazquezFKL18}}

For the stationary memoryless case, Csiszar and 
Longo\footnote{There is a typo in \cite{csiszarL71}, 
	see \cite[Remark 3.7]{nakiboglu20F}.} 
determined the behavior of \(\NPDd{e^{-\blx E }}{\m{w}}{\m{q}}\) 
for any fixed \(E\) value in \cite{csiszarL71}, 
with the help of \cite[Theorem 1.1]{strassen62}.
In particular,
\(\ln\NPDd{e^{-\blx E }}{\m{w}}{\m{q}}\) for  \(E\!<\!\KLD{\m{w;1}}{\m{q;1}}\)  case
and
\(\ln(1-\NPDd{e^{-\blx E }}{\m{w}}{\m{q}})\) for  \(E\!>\!\KLD{\m{w;1}}{\m{q;1}}\)  case,
were determined in \cite{csiszarL71} up to an \(\smallo{1}\) approximation error 
term
for a finite measure \(\m{q}\). The Kullback--Leibler divergence
\(\KLD{\m{w;1}}{\m{q;1}}\) acts as a threshold for 
the change in behavior of \(\NPDd{e^{-\blx E }}{\m{w}}{\m{q}}\)
with increasing \(\blx\): 
if  \(E\!<\!\KLD{\m{w;1}}{\m{q;1}}\), then  
\(\NPDd{e^{-\blx E }}{\m{w}}{\m{q}}\) converges to \(0\);
if  \(E\!>\!\KLD{\m{w;1}}{\m{q;1}}\), on the other hand,
\(\NPDd{e^{-\blx E }}{\m{w}}{\m{q}}\)
converges to  \(1\).
An equally tight bound is obtained via saddle point approximation 
in \cite[Theorem 2]{vazquezFKL18},
for the case when \(\m{w}\!\sim\!\m{q}\)  and \(\m{q}\) 
is probability measure.

The total variation norm of a signed measure \(\m{v}\)
on the measurable space \((\outS,\outA)\) is defined as
\begin{align}
	\label{eq:def:TotalVariationNorm}
	\norm{\m{v}}
	&\DEF \sup\nolimits_{\oev\in\outA} 
	\m{v}[\oev]\!+\!\abs{\m{v}[\outS\!\setminus\!\oev]}.
\end{align}
In \cite{csiszarL71}, a measure change argument is employed 
via tilting and the value of \(E\) is set to 
\(\KLD{\m{v}}{\m{q}}\) for \(\m{v}=\sfrac{\m{v;\rno}}{\norm{\m{v;\rno}}}\) 
where \(\m{v;\rno}\) is the tilted measure 
defined in \eqref{eq:def:TiltedMeasure}.
For all \(\rno\in(0,1)\), a non-asymptotic approximation for 
\(\ln\NPDd{e^{-\KLD{\m{v}}{\m{q}}+\cramer{\blx}}}{\m{w}}{\m{q}}\) 
with \(\bigo{\sqrt{\blx}}\) approximation error term
is established in \cite[Theorem 5]{shannonGB67A} 
assuming \(\abs{\cramer{\blx}}\!=\!\bigo{\sqrt{\blx}}\) 
for the possibly non-stationary memoryless case,
prior to \cite{csiszarL71}, 
by invoking Chebyshev's inequality instead 
of \cite[Theorem 1.1]{strassen62}. 
More recently non-asymptotic approximations for 
\(\ln\NPDd{e^{-\KLD{\m{v}}{\m{q}}+\cramer{\blx}}}{\m{w}}{\m{q}}\) 
with \(\bigo{1}\) approximation error terms have been be obtained
under \(\abs{\cramer{\blx}}\!=\!\bigo{\sqrt{\blx}}\) hypothesis
by employing Berry--Esseen Theorem instead of Chebyshev's inequality
for the memoryless case 
see  \cite{nakiboglu19-ISIT,nakiboglu20F,lunguK24,theocharousK25,theocharousGK26}.
For \(\rno\!\in\!(0,1)\) case, an asymptotic approximation of similar nature 
were derived earlier in \cite{altugW14A} relying on  strong large deviation 
results of \cite{chagantyS93,bahadurR60}, rather than Berry--Esseen Theorem.
For \(\rno\!>\!1\) case, a \(\bigo{1}\) approximation for
\(\ln(1\!-\!\NPDd{e^{-\KLD{\m{v}}{\m{q}}+\cramer{\blx}}}{\m{w}}{\m{q}})\) 
was established in \cite{chengN20A} relying on Berry--Esseen Theorem.

\subsection{Main Contributions}\label{sec:MainContributions}
One of the main contributions of this work is the explicit
characterization of implications of Neyman--Pearson lemma
in Theorem \ref{thm:SNPL} for the relations between 
\(\ent{\cdot}{\m{q}}{\m{w}} \), \(\lscq{\cdot}{\m{w}}{\m{q}}\), 
\(\uscq{\cdot}{\m{w}}{\m{q}}\) and \(\NPD{\cdot}{\m{q}}{\m{w}}\)
as functions for \(\sigma\)-finite measures \(\m{w}\) and \(\m{q}\) 
satisfying \eqref{eq:assumption:nonvoid}.
Resulting identities in \eqref{eq:thm:SNPL} and \eqref{eq:thm:SNPL:DF},
have not been reported either in this form or with this generality
before to the best our knowledge. 
In particular, the identities in \eqref{eq:Beta}
(i.e., \eqref{eq:thm:SNPL:CI:DF} and \eqref{eq:thm:SNPL:CI}), 
\eqref{eq:thm:SNPL:IOSD:1}, and \eqref{eq:thm:SNPL:IOSD:2}  
have not appeared before.
However, as we have already pointed
out statements equivalent to \eqref{eq:thm:SNPL:DualCharacterization}, 
\eqref{eq:thm:SNPL:DualCharacterization:max}, and
\eqref{eq:thm:SNPL:SI} have appeared 
in \cite{reidW11,polyanskiythesis,elkayamF16,verdu24,verdu25}, as discussed 
in \S \ref{sec:PriorWorkOnOneShot}. 
Furthermore, for a different parametrization convex conjugate of 
\(\NPD{\cdot}{\m{q}}{\m{w}}\) was characterized in \cite{elkayamF16}
for the case when \(\m{w}\) and \(\m{q}\) are probability mass functions
on a countable set.

Theorem \ref{thm:MeasureChange} establishes sufficient conditions 
for expressing \(\NPD{\cdot}{\m{q}}{\m{w}}\) either
as \(\NPD{\f{g}[\cdot]}{\m{q}}{\m{v}}\)
or as \(\norm{\m{q;\sim}}-\NPD{\f{g}[\cdot]}{\m{q}}{\m{v}}\) 
for some  an invertible function \(\f{g}[\cdot]\).  
Such a characterization has not been reported before
and works quite well in simplifying the derivations
when employed with the monotonicity properties 
of \(\NPD{\cdot}{\m{q}}{\m{w}}\) described in 
Lemma \ref{lem:NPD:Elementary} and tilting, 
as we demonstrate while establishing 
Theorem \ref{thm:TKDtoN} and  
Corollary \ref{cor:DiscreteMarkov:TKDtoN}.

Note that with\begin{subequations}\label{eq:Beta:L}~the changes 
of variable \(\tau\!=\!e^{\tin}\) and \(\tau\!=\!e^{-\tin}\),
\eqref{eq:Beta:CI} and \eqref{eq:Beta:CI:DF} become
\begin{align}
	\label{eq:Beta:CI:L}
	\NPD{\varepsilon}{\m{q}}{\m{w}}
	&\!=\!\int_{-\infty}^{\infty}e^{\tin}
	\abp{\m{w}[\ln\der{\m{q;\sim}}{\m{w}}\!>\!\tin]\!-\!\varepsilon}
	\!\dif{\tin}
	&
	&\forall\varepsilon\!\geq\!0,
	\\
	\label{eq:Beta:CI:DF:L} 
	&\!=\!\int_{-\infty}^{\infty}e^{-\tin}
	\abp{\m{w}[\ln\der{\m{w;\sim}}{\m{q;\sim}}\!\leq\!\tin]\!-\!\varepsilon}
	\!\dif{\tin}
	&
	&\forall\varepsilon\!\geq\!0.
\end{align}
If\end{subequations}~the Kolmogorov distance of 	
the log-likelihood ratio distribution, i.e., 
\(\m{w}[\ln\der{\m{w;\sim}}{\m{q;\sim}}\!\leq\!\cdot]\),
to a Gaussian distribution is small, 
then one can approximate \(\NPD{\cdot}{\m{q}}{\m{w}}\) 
using that Gaussian distribution with the help of 
\eqref{eq:Beta:CI:DF:L}. 
Theorem \ref{thm:KDtoN} pursues this idea 
to characterize \(\NPD{\cdot}{\m{q}}{\m{w}}\) 
up to a bounded explicit multiplicative error term uniformly 
on all closed subsets of \((0,1)\). 
For the case when \(\m{w}\) and \(\m{q}\) are probability 
mass functions on a countable set a similar and
qualitatively equivalent one-shot bound
was derived in \cite[\S III.B]{elkayamF16}.

The customary approach in bounding \(\NPD{\varepsilon}{\m{q}}{\m{w}}\),  
adopted explicitly or implicitly in much of the prior work (some of which 
is cited in \S\ref{sec:PriorWorkOnApproximations}), 
first invokes \eqref{eq:Tan-2-4} and then bounds
likelihood-ratio quantile \(\uscq{\cdot}{\m{w}}{\m{q}}\). 
With the help of \eqref{eq:Beta:CI:DF:L}, we not only 
avoid using likelihood-ratio quantiles completely,
but also obtain tighter results with briefer analysis.
Furthermore, it should be noted that Theorem \ref{thm:KDtoN}
can also be proved 
by first expressing \(\NPD{\cdot}{\m{q}}{\m{w}}\)
in terms of likelihood ratio quantiles \(\uscq{\cdot}{\m{w}}{\m{q}}\)
using\footnote{Whenever \(\norm{\m{q;\sim}}\) is finite, 
	\eqref{eq:thm:SNPL:IOSD:2} can be used instead of \eqref{eq:thm:SNPL:IOSD:1}.} 
\eqref{eq:thm:SNPL:IOSD:1} and then bounding
\(\uscq{\cdot}{\m{w}}{\m{q}}\) by invoking
\cite[(2.34)]{tan14}. %\cite[Proposition 2.1]{tan14} %\cite[(33)]{tomamichelH13},%\cite[Lemma 5]{tomamichelT13},

Kolmogorov distance to normality hypothesis 
for the log-likelihood ratio distribution
is pursued with tilting in Theorem \ref{thm:TKDtoN}, 
for \(\m{v}=\gamma\m{v;\rno}\) 
and appropriately bounded \(\tau\) to characterize 
\(\NPD{\tau e^{-\KLD{\m{v}}{\m{q}}}}{\m{w}}{\m{q}}\) 
for \(\rno\in(0,1)\) 
and \(\norm{\m{w;\sim}}\!-\!\NPD{\tau e^{-\KLD{\m{v}}{\m{q}}}}{\m{w}}{\m{q}}\) 
for \(\rno>1\), upto a bounded multiplicative error term
by invoking Theorem \ref{thm:MeasureChange} and 
Theorem \ref{thm:KDtoN}. Theorem \ref{thm:TKDtoN} 
or a qualitatively equivalent bound has not appeared 
before to the best of our knowledge.

In the memoryless case, 
hypothesis of Theorems \ref{thm:KDtoN} and \ref{thm:TKDtoN} 
are implied by the Berry--Esseen Theorem
and Corollaries \ref{cor:Memoryless:KDtoN}
and \ref{cor:Memoryless:TKDtoN} follow.
Corollary \ref{cor:Memoryless:KDtoN} is qualitatively 
equivalent to the previously known best results of 
this kind both for hypothesis testing
\cite[Theorem 3.1]{strassen62},
\cite[Lemma 14]{polyanskiythesis}, 
\cite[\S III.B]{elkayamF16}
and for lossless source coding
\cite[(44), Theorems 17 and 18]{kontoyiannisV14},
\cite[Theorems 5 and 6]{GavalakisK21},
discussed in \S\ref{sec:PriorWorkOnApproximations}.
Likewise, Corollary \ref{cor:Memoryless:TKDtoN}
recovers previously known best results 
of its kind for both for hypothesis testing
\cite{nakiboglu19-ISIT,nakiboglu20F,chengN20A,lunguK24}
and for lossless source coding
\cite{theocharousK25,theocharousGK26},
discussed in \S\ref{sec:PriorWorkOnApproximations}.
Corollary \ref{cor:Memoryless:TKDtoN}
demonstrates for the \(\rno\!\in\!(0,1)\) case, 
in contrast to prior work,  
neither \(\m{w}\) nor \(\m{q}\) need to
be finite measure for these approximations 
to hold.{\color{white}
\cite{nagaev57,nagaev61,Nummelin,vere-Jones67,seneta}}

For the time-homogeneous discrete Markov case if
\(\tm{W}\) corresponds to an irreducible aperiodic 
Markov chain satisfying Dobrushin--Doeblin condition, 
see \cite[(1.1)]{nagaev57}, \cite[(0.1)]{nagaev61},
then the hypothesis of Theorem \ref{thm:KDtoN}
follows from the functional Berry--Esseen Theorem 
of Nagaev \cite[Theorem 1]{nagaev61} and 
Corollary \ref{cor:DiscreteMarkov:KDtoN} follows.
For the lossless source coding problem, 
Corollary \ref{cor:DiscreteMarkov:KDtoN}
generalizes the lower bound established 
in \cite[Theorem 20]{kontoyiannisV14}
and establishes the improvement 
of \cite[Theorem 19]{kontoyiannisV14}
conjectured\footnote{\!\cite{kontoyiannisV14} points out
	that \cite[Theorems 19]{kontoyiannisV14}
	can be improved but only by ``extending the involved proof 
	of'' \cite[Theorems 17]{kontoyiannisV14}.}
in \cite{kontoyiannisV14}, 
without making the finite state-space
assumption of \cite{kontoyiannisV14}.
The approximation error of
Corollary \ref{cor:DiscreteMarkov:KDtoN} 
for \(\ln\NPD{\cdot}{\m{q}}{\m{w}}\) is \(\bigo{1}\),
which is worse than \(\smallo{1}\) approximation error
derived in \cite{nagaevA02} under the additional
finite state-space and non-lattice loglikelihood ratio
assumptions. 
Without non-lattice loglikelihood ratio assumption
\cite[Theorem 9.6]{watanabeH17} with 
\(\smallo{\sqrt{\blx}}\) approximation error,
which is qualitatively equivalent to \cite{Yushkevich53},
is the best approximation we are aware of, even  
for the finite state-space case.
%\cite[Lemma 9.5]{watanabeH17} and \cite[Theorem 9.6]{watanabeH17}.

Corollary \ref{cor:DiscreteMarkov:TKDtoN} establishes 
for the time-homogeneous discrete Markov case with a countable state-space 
the approximation corresponding to 
the one reported in Corollary \ref{cor:Memoryless:TKDtoN}
for the memoryless case, relying on the concept of 
\(R\)-positivity \cite{Nummelin,vere-Jones67,seneta} 
and Doob's \(\f{h}\)-transform. 
Corollary \ref{cor:DiscreteMarkov:TKDtoN} improves previously 
known best approximations in the same regime
both for hypothesis testing
\cite{watanabeH17,Koopmans60,NakagawaK93B,Natarajan85}
and lossless source coding 
\cite{Vasek80,DavissonLS81,Natarajan85},
which are, except for \cite{Koopmans60}, 
derived under the additional finite state-space assumption.

Corollaries \ref{cor:Memoryless:KDtoN} and \ref{cor:DiscreteMarkov:KDtoN}
are non-asymptotic results following from the Berry--Esseen Theorems;
they can be interpreted as third order non-asymptotic Stein's lemmas for the memoryless
and discrete Markovian cases. 
Theorem \ref{thm:NPD:CentralLimitTheorem} establishes 
the second order Stein's lemma for
the case when  log-likelihood-ratio of a standard 
simple hypothesis testing problem satisfies 
the Central Limit Theorem,
using only Theorem \ref{thm:KDtoN} 
and the fact that convergence in distribution 
to the normal distribution is metrized by 
the Kolmogorov distance.
For the Markov case the fact that Central Limit Theorem 
implies the second order Stein's lemma was previously
pointed out in \cite{watanabeH17}, as well,
see \cite[Corollary 7.2 and Theorem 9.6]{watanabeH17}.
Theorem \ref{thm:NPD:CentralLimitTheorem} establishes 
the same result far more generally.

\subsection{Outline}\label{sec:outline}
In \S\ref{sec:Preliminaries}, we present preliminary 
results that are later employed in our analysis.
We start by recalling certain well-known properties of
convex functions in \S\ref{sec:Convexity}. 
We introduce the semicontinuous inverses of 
monotonic functions and explain their construction intuitively
in \S\ref{sec:SemicontinuousInverses}.
Then we introduce the likelihood-ratio distribution 
\(\scq{\cdot}{\m{w}}{\m{q}}\) and its semicontinous inverses, 
i.e., the likelihood-ratio quantiles \(\lscq{\cdot}{\m{w}}{\m{q}}\) 
and \(\uscq{\cdot}{\m{w}}{\m{q}}\), and review certain elementary 
properties of the likelihood-ratio quantiles in 
\S\ref{sec:LikelihoodRatioQuantiles}. 
We conclude the preliminaries section by deriving 
elementary properties of the primitive entropy 
spectrum \(\ent{\cdot}{\m{w}}{\m{q}}\) in \S\ref{sec:Entropy}.

We start \S\ref{sec:SRandMC}, by 
discussing certain elementary observations about
\(\NPDd{\cdot}{\m{q}}{\m{w}}\) and \(\NPD{\cdot}{\m{q}}{\m{w}}\).
Then in \S \ref{sec:LikelihoodRatioTest}, we obtain an 
upper bound on \(\NPD{\cdot}{\m{q}}{\m{w}}\) by analyzing 
performance of a randomized likelihood-ratio test.
In \S\ref{sec:ASpectralNeymanPearsonLemma}, we 
prove Theorem \ref{thm:SNPL}, which is essentially 
the Neyman--Pearson lemma for \(\sigma\)-finite measures, 
with the additional identities we discussed above
including \eqref{eq:Beta}. 
In \S\ref{sec:MeasureChangeIdentities}, we prove 
Theorem \ref{thm:MeasureChange}, which establishes 
sufficient conditions for expressing 
\(\NPD{\cdot}{\m{q}}{\m{w}}\) in terms of 
\(\NPD{\cdot}{\m{q}}{\m{v}}\) for an 
auxiliary measure \(\m{v}\).

In \S\ref{sec:Approximations}, we first 
review certain well-known properties of Gaussian Mills ratio
and then derive Theorems \ref{thm:KDtoN} and \ref{thm:TKDtoN}
under small Kolmogorov distance to normality hypothesis both
without and with tilting, respectively.

\S\ref{sec:Applications} is devoted to applications.
In \S\ref{sec:Memoryless}, we consider the memoryless case
and establish Corollaries \ref{cor:Memoryless:KDtoN}
and \ref{cor:Memoryless:TKDtoN}. 
In \S\ref{sec:DiscreteMarkov}, we consider time-homogeneous discrete
Markov case with possibly countably infinite number of states
and establish Corollaries \ref{cor:DiscreteMarkov:KDtoN}
and \ref{cor:DiscreteMarkov:TKDtoN}. 
In \S\ref{sec:Asymptotic}, we introduce the 
standard simple asymptotic hypothesis testing problem, 
in its complete measure theoretic generality 
and show that the Central Limit Theorem for 
the log-likelihood-ratio implies 
the second order Stein's lemma
in Theorem \ref{thm:NPD:CentralLimitTheorem}.

In \S\ref{sec:Discussion}, we discuss 
the broader implications of the spectral approach 
proposed in this paper for information transmission problems,
and highlight several interesting technical questions
that have became more manageable as a result of this work.

\section{Preliminaries}\label{sec:Preliminaries}
\subsection{Convex Functions on the Real Line}\label{sec:Convexity}
\begin{assumption}\label{assumption:HasAConjugate}
The function 
\(\f{f}:\numbers{R}\mapsto\numbers{R}\cup\{\infty\}\)
takes at least one finite value and has an affine minorant, 
i.e., 
\(\exists\dinp_{0}\!\in\!\numbers{R}\) such that 
\(\f{f}[\dinp_{0}]\!<\!\infty\)
and 
\(\exists\mA,\mB\!\in\!\numbers{R}\) such that 
\(\f{f}[\dinp]\geq\mA\dinp+\mB\) holds for all 
\(\dinp\in\numbers{R}\).
(\(\f{f}\) in not necessarily convex.)  	
\end{assumption}
For any function \(\f{f}\) satisfying Assumption \ref{assumption:HasAConjugate}, 
subdiffential of \(\f{f}\) at \(\dinp\) is,
\cite[(E.1.4.1)]{hiriart-urrutyLemarechal}, 
\begin{align}
	\label{eq:def:Subdifferential}
	\f{\partial f}[\dinp]
	&\DEF\left\{\mS\!\in\!\numbers{R}:
	\f{f}[\dout]\!\geq\!
	\f{f}[\dinp]\!+\!\mS(\dout\!-\!\dinp)
	\text{~for all~}\dout\!\in\!\numbers{R}\right\}.
\end{align}
If  \(\f{f}\) is convex and \(\dinp\) is in the interior 
of the interval on which 
\(\f{f}\) is finite, then \(\f{\partial f}[\dinp]\) is the 
closed interval between 
the left and right derivatives of \(\f{f}\) at \(\dinp\)
by \cite[Theorem I.4.1.1]{hiriart-urrutyLemarechal-I}:
\begin{align}
	\label{eq:Subdifferential}
	\f{\partial f}[\dinp]
	&\DEF [\f{\lder{f}}[\dinp],\f{\rder{f}}[\dinp]]
	&
	&\forall \dinp\!\in\!\inte{\domain{\f{f}}}.		
\end{align}
If \(\f{f}\) is closed (i.e., lower semicontinuous) 
and \(\mA\) is the left end point of 
the domain then \(\f{\partial f}[\mA]=(-\infty,\rder{\f{f}}(\mA)]\),
see \cite[p.22]{hiriart-urrutyLemarechal-I}.
If \(\f{f}\) is closed and \(\mB\) is the right end point of 
the domain then \(\f{\partial f}[\mB]=[\lder{\f{f}}(\mB),\infty)\).

%For all \(\dinp_{0}\!\in\!\inte{\domain{\f{f}}}\), 
%subdifferential \(\f{\partial f}[\dinp]\) converges 
%increasingly to 
%\(\f{\lder{f}}[\dinp]\) when \(\dinp\!\uparrow\!\dinp_{0}\)
%and
%decreasingly to 
%\(\f{\rder{f}}[\dinp]\) when \(\dinp\!\downarrow\!\dinp_{0}\),
%see \cite[Theorem I.4.2.1]{hiriart-urrutyLemarechal-I}.
%Thus \(\f{f}\), in fact its subdifferential \(\f{\partial f}\),
%uniquely determines a \(\sigma\)-finite Borel measure 
%on \(\inte{\domain{\f{f}}}\) via the following relations,
%\cite{liese12},
%\begin{align}
%	\notag	
%	\f{\rho;f}[\{\mA\}]	
%	&\!=\!\f{\rder{f}}[\mA]\!-\!\f{\lder{f}}[\mA]
%	&
%	&\forall \mA\in \inte{\domain{\f{f}}},
%	\\
%	\notag	
%	\f{\rho;f}[{(\mA,\mB)}]	
%	&\!=\!\f{\lder{f}}[\mB]\!-\!\f{\rder{f}}[\mA]
%	&
%	&\forall \mA,\mB\in \inte{\domain{\f{f}}}:\mA<\mB.
%\end{align}
%Evidently, the subdifferential \(\f{\partial f}\) of \(\f{f}\)
%is  determined by \(\f{\rho;f}\), up to an 
%additive constant.

If \(\f{f}:\numbers{R}\mapsto\numbers{R}\cup\{\infty\}\) 
is a closed (i.e., lower semicontinuous) convex 
function and \([\mA,\mB]\subset\domain{\f{f}}\), then  
\begin{align}
	\label{eq:ConvexFunctionIR}	
	\f{f}[\dinp]
	&=\f{f}[\mA]+\int_{\mA}^{\dinp} \f{f'}[\tau]\dif{\tau}
	&
	&\forall \dinp\in[\mA,\mB],			
\end{align}	
where \( \f{f'}\) can be either 
\(\lder{\f{f}}\) or \(\rder{\f{f}}\), 
see \cite[Remark I.4.2.5]{hiriart-urrutyLemarechal-I}.
%\cite[Theorem D.2.3.4]{hiriart-urrutyLemarechal} considers only real valued functions.
%\cite[Appendix B.2]{dudley} C1 required.
Furthermore,
\cite[Theorem I.4.2.4]{hiriart-urrutyLemarechal-I}
%\cite[Theorem D.2.3.3]{hiriart-urrutyLemarechal} considers only real valued functions.
asserts that for any 
\(\mA,\mB\in\domain{\f{f}}\)
such that \(\mA<\mB\),
there exists a
\(\dinp\in(\mA,\mB)\) such that
\begin{align}
	\label{eq:ConvexMeanValueTheorem}	
	\f{f}[\mB]
	&=\f{f}[\mA]+(\mB-\mA)\f{f'}
	&
	&\f{f'}\in\f{\partial f}[\dinp].	
\end{align}

For any function \(\f{f}\) satisfying Assumption
\ref{assumption:HasAConjugate}, 
the convex conjugate \(\f{f;;*}\) of \(\f{f}\) is
defined as, 
\begin{align}
	\label{eq:def:ConvexConjugate}	
	\f{f;;*}[\lambda]
	&\DEF\sup\{\lambda\dinp-\f{f}[\dinp]:\dinp\!\in\!\domain{\f{f}}\}
	&
	&\forall \lambda\!\in\!\numbers{R},
\end{align}	
see \cite[\S I.6]{hiriart-urrutyLemarechal-I}.%\cite[Definition E.1.1.1]{hiriart-urrutyLemarechal},

\subsection[The Semicontinuous Inverses]{The Semicontinuous Inverses of Monotonic Functions}\label{sec:SemicontinuousInverses}
Inverse of a function, in the usual sense, is defined
only for bijective (i.e., one-to-one and onto) functions. 
For non-bijective monotone functions 
the generalized (i.e., semi-continuous) 
inverses constitute a partial remedy, 
see \cite{klementMP99,vicenik99,fengWTK12,falknerT12,
	embrechtsH13,fortelle20,wacker23}. 
\begin{definition}\label{def:SCInverse}
For any non-decreasing function 
\(\f{f}\!:\!\numbers{R}\!\mapsto\!\overline{\numbers{R}}\)
the u.s.c. (i.e., upper-semi-continuous) inverse 
\(\lsci{\f{f}}\!:\!\numbers{R}\!\mapsto\!\overline{\numbers{R}}\) 
and the l.s.c. (i.e., lower-semi-continuous) inverse 
\(\usci{\f{f}}\!:\!\numbers{R}\!\mapsto\!\overline{\numbers{R}}\) 
are defined as
\begin{subequations}
\label{eq:def:SCInverse:nondecreasing}	
\begin{align}
\label{eq:def:SCInverse:USC:nondecreasing}	
\usci{\f{f}}[\dout]
&\DEF\inf\left\{\dinp\!\in\!\numbers{R}:\f{f}[\dinp]>\dout\right\}
&
&\forall \dout\in\numbers{R},
\\
\label{eq:def:SCInverse:LSC:nondecreasing}	
\lsci{\f{f}}[\dout]
&\DEF\inf\left\{\dinp\!\in\!\numbers{R}:\f{f}[\dinp]\geq\dout\right\}
&
&\forall \dout\in\numbers{R}.	
\end{align}		
\end{subequations}		
For any non-increasing function \(\f{g}\!:\!\numbers{R}\!\mapsto\!\overline{\numbers{R}}\)
the u.s.c. inverse 
\(\usci{\f{g}}\!:\!\numbers{R}\!\mapsto\!\overline{\numbers{R}}\) 
and the l.s.c. inverse 
\(\lsci{\f{g}}\!:\!\numbers{R}\!\mapsto\!\overline{\numbers{R}}\) 
are defined as
\begin{subequations}
\label{eq:def:SCInverse:nonincreasing}	
\begin{align}
\label{eq:def:SCInverse:USC:nonincreasing}	
\usci{\f{g}}[\dout]
&\DEF\inf\left\{\dinp\!\in\!\numbers{R}:\f{g}[\dinp]<\dout\right\}
&
&\forall \dout\in\numbers{R},
\\
\label{eq:def:SCInverse:LSC:nonincreasing}	
\lsci{\f{g}}[\dout]
&\DEF\inf\left\{\dinp\!\in\!\numbers{R}:\f{g}[\dinp]\leq\dout\right\}
&
&\forall \dout\in\numbers{R}.			
\end{align}			
\end{subequations}				
\end{definition}
\begin{remark}
	Recall that \(\inf\emptyset=\infty\), \(\sup\emptyset=-\infty\), and 
	the inequality \(\inf\cset\leq \sup\cset\) holds only for 
	\(\cset\neq\emptyset\) case; in that case 
	\(\inf\cset\leq \dinp\leq\sup\cset\) for all \(\dinp\in\cset\),
	as well.
\end{remark}
Customarily, only the non-decreasing functions are studied 
because corresponding claims for the non-increasing ones 
are evident. 
In addition, the u.s.c. inverse 
\(\usci{\f{f}}\) is usually called the right continuous inverse
and the l.s.c. inverse 
\(\lsci{\f{f}}\) is usually called the left continuous inverse. 
Excellent overviews and derivations of some of the fundamental 
results and some applications can be found in 
\cite{embrechtsH13,fortelle20,wacker23}.

The definitions above are describing an operation
that is quite easy to grasp intuitively.
Let \(\f{h}:\numbers{R}\mapsto\overline{\numbers{R}}\) be
a monotonic function. 
We extend the graph of \(\f{h}[\cdot]\) by drawing vertical 
lines between 
\((\dinp,\lim\nolimits_{\gamma\uparrow\dinp}\f{h}[\gamma])\) 
and \((\dinp,\lim\nolimits_{\gamma\downarrow\dinp}\f{h}[\gamma])\) 
at the points 
where\footnote{For any extended real valued monotonic function 
	both the right limit and the left limit exist at every point in 
	\(\numbers{R}\), by \cite[Theorem 4.29]{rudin}.}
the left limit \(\lim\nolimits_{\gamma\uparrow\dinp}\f{h}[\gamma]\) 
is not equal to the right limit \(\lim\nolimits_{\gamma\downarrow\dinp}\f{h}[\gamma]\); 
those lines are not present in the graph of \(\f{h}[\cdot]\) 
to ensure that \(\f{h}[\dinp]\) has a single value for each 
\(\dinp\in\numbers{R}\). 
Note that extended graph is connected.
Then the inversion operation simply takes the mirror image 
of the extended graph with respect to \(\dout\!=\!\dinp\) line,
i.e., with respect to the graph of the identity function,
and replace any vertical lines in the mirror image 
(i.e., horizontal lines in the original graph) 
with upper most point for \(\usci{\f{h}}(\cdot)\) and
the lower most point for \(\lsci{\f{h}}(\cdot)\).
Then following three observations are evident from 
the description of the semicontinuous inversion given above:
\begin{itemize}
\item The semicontinuous inverses
\(\usci{\f{h}}(\cdot)\) and \(\lsci{\f{h}}(\cdot)\)
would be the same if replace \(\f{h}[\cdot]\)
with its lower semicontinuous 
closure \(\lsccl{\f{h}}(\cdot)\),
or 	
with its upper semicontinuous closure \(\usccl{\f{h}}(\cdot)\),	
because \(\f{h}[\cdot]\), \(\lsccl{\f{h}}(\cdot)\), and 
\(\usccl{\f{h}}(\cdot)\) share the same the extended graph,
where 
the lower semicontinuous closure \(\lsccl{\f{h}}\) 
of \(\f{h}\) and 
the upper semicontinuous closure \(\usccl{\f{h}}\) 
of \(\f{h}\) are defined as
\begin{align}
	\notag	
	\lsccl{\f{h}}(\dinp)
	&\DEF
	(\lim\nolimits_{\gamma\uparrow\dinp}\f{h}[\gamma])
	\wedge
	(\lim\nolimits_{\gamma\downarrow\dinp}\f{h}[\gamma])
	&
	&\forall \dinp\in\numbers{R},
	\\
	\notag	
	\usccl{\f{h}}(\dinp)
	&\DEF
	(\lim\nolimits_{\gamma\uparrow\dinp}\f{h}[\gamma])
	\vee
	(\lim\nolimits_{\gamma\downarrow\dinp}\f{h}[\gamma])
	&
	&\forall \dinp\in\numbers{R},	
\end{align}
see \cite[Definition B.1.2.4]{hiriart-urrutyLemarechal} for 
a more general definition.
\item 
\(\lsci{\f{h}}(\cdot)\) is lower semicontinuous on \(\numbers{R}\)
and 
\(\usci{\f{h}}(\cdot)\) is upper semicontinuous on \(\numbers{R}\).

\item The semicontinuous inversion is involutive 
in the sense that
the  u.s.c. inverses of  both 
\(\lsci{\f{h}}(\cdot)\)
and
\(\usci{\f{h}}(\cdot)\)
equal to \(\usccl{\f{h}}(\cdot)\)
and
the  l.s.c. inverses of  both 
\(\lsci{\f{h}}(\cdot)\)
and
\(\usci{\f{h}}(\cdot)\)
equal to \(\lsccl{\f{h}}(\cdot)\).	
\end{itemize}
%We will not invoke any of the above observations or any other 
%result about semicontinuous inverses explicitly in the following. 
%Instead, we will establish those properties of the semicontinuous 
%inverses that will be necessary for our purposes 
%for the case of  likelihood-ratio quantiles
%in Lemma \ref{lem:LRQ} in the following.

\subsection[The Likelihood-Ratio Quantiles]{The Likelihood-Ratio Quantiles}\label{sec:LikelihoodRatioQuantiles}
Recall that the Lebesgue decomposition theorem holds 
not only for finite measures but also for \(\sigma\)-finite measures, 
see \cite[Theorem 5.5.3]{dudley} and \cite[\S32, Theorem C]{halmos}.
As a consequence, Radon-Nikodym Theorem for finite measures
%,see e.g., \cite[Theorem 5.5.4]{dudley},
generalizes to the \(\sigma\)-finite measures,
see \cite[\S3.2]{bogachev}, \cite[\S31, Theorem B]{halmos}.
Thus for any two \(\sigma\)-finite measures \(\m{w}\) and \(\rfm\)  
satisfying   \(\m{w}\AC\rfm\),  
there is a non-negative \(\rfm\)-measurable function  \(\der{\m{w}}{\rfm}\),
called the Radon-Nikodym derivative of \(\m{w}\) with respect to \(\rfm\). 
\begin{definition}\label{def:lr}
For any pair of \(\sigma\)-finite measures \((\m{w},\m{q})\) 
on the same measurable space, say \((\outS,\outA)\),
the likelihood-ratio \(\f{\ylr;\m{w},\m{q}}[\cdot]\)
and the likelihood-ratio\begin{subequations}\label{eq:def:lr} 
distribution \(\scq{\cdot}{\m{w}}{\m{q}}\) are defined as
\begin{align}
\label{eq:def:lrf}
\f{\ylr;\m{w},\m{q}}
&\DEF
\begin{cases}
\infty 
&\text{if~\(\der{\m{q}}{(\!\m{w}+\m{q}\!)}\!=\!0\),}
\\
\der{\m{w;\sim}}{\m{q}}
&\text{if~\(\der{\m{q}}{(\!\m{w}+\m{q}\!)}\!>\!0\),}
\end{cases}		
\\
\label{eq:def:lrcdf}
\scq{\tau}{\m{w}}{\m{q}} 
&\DEF \int \IND{\f{\ylr;\m{w},\m{q}}\leq\tau} \dif{\m{w}}
&
&\forall \tau\!\in\!\numbers{R},	
\end{align}
where \(\m{w;\sim}\) is the \(\m{q}\)-absolutely~\end{subequations} 
continuous component of \(\m{w}\). 	
\end{definition}
The likelihood-ratio distribution \(\scq{\cdot}{\m{w}}{\m{q}}\) 
is ``insidiously self-referential'' as 
\cite[p. 57]{coverthomas} deems entropy to be 
while discussing asymptotic equipartition property. 
Nevertheless, \(\scq{\cdot}{\m{w}}{\m{q}}\) is a 
well--defined monotonic function for any pair of 
\(\sigma\)-finite measures \((\m{w},\m{q})\)
on a common measurable space.
\begin{subequations}\label{eq:lrf:Elementary:ScalarMultiplication}
	Furthermore,
	\begin{align}
		\label{eq:lrf:Elementary:ScalarMultiplication:mW}	
		\scq{\tau}{\lambda\m{w}}{\m{q}}
		&=\lambda\scq{\tfrac{\tau}{\lambda}}{\m{w}}{\m{q}}	
		&
		&\forall\tau\!\in\!\numbers{R},
		\\	
		\label{eq:lrf:Elementary:ScalarMultiplication:mQ}	
		\scq{\tau}{\m{w}}{\lambda\m{q}}
		&=\scq{\lambda\tau}{\m{w}}{\m{q}}
		&
		&\forall\tau\!\in\!\numbers{R},
	\end{align} 
	where \(\lambda\) is positive constant, i.e., \(\lambda\!\in\!(0,\infty)\).
\end{subequations}

If replace \(\m{w}\!+\!\m{q}\) with any other  
\(\sigma\)-finite reference measure \(\rfm\) satisfying 
\((\!\m{w}\!+\!\m{q}\!)\AC\rfm\), then 
resulting likelihood-ratio is equal to the one given in
\eqref{eq:def:lrf} \((\!\m{w}\!+\!\m{q}\!)\)-a.e.;
hence the resulting \(\scq{\cdot}{\m{w}}{\m{q}}\) 
will be equal to the one given in \eqref{eq:def:lrcdf}.
Furthermore, with the understanding 
that \(\tfrac{1}{0}\) and \(\tfrac{1}{\infty}\)
stand for \(\infty\) and \(0\), respectively, we have
\begin{align}
	\label{eq:lrf}
	\f{\ylr;\m{q},\m{w}}
	&=\tfrac{1}{\f{\ylr;\m{w},\m{q}}}
	&
	&(\m{w}\!+\!\m{q})\text{-a.e.}.	
\end{align}
If \(\m{w}\) is a probability measure, then \(\scq{\cdot}{\m{w}}{\m{q}}\) 
is the cumulative distribution of 
the extended real valued random variable \(\f{\ylr;\m{w},\m{q}}\) 
in the probability space \((\outS,\outA,\m{w})\), 
as its name suggests.
However, we do not assume either \(\m{w}\) or \(\m{q}\) to be a
probability measure; we only assume both of them to be \(\sigma\)-finite
measures. Note that the likelihood-ratio distribution  \(\scq{\cdot}{\m{w}}{\m{q}}\)
depends only on the absolutely continuous components of 
\(\m{w}\) and \(\m{q}\) in each other, 
as an immediate consequence of Definition \ref{def:lr}:
\begin{align}
	\label{eq:lrcdf:ACequivalence}
	\scq{\cdot}{\m{w}}{\m{q}} 
	&=\scq{\cdot}{\m{w;\sim}}{\m{q;\sim}}, 	
\end{align}
where \(\m{q;\sim}\) is the \(\m{w}\)-absolutely 
continuous component of \(\m{q}\) and 
\(\m{w;\sim}\) is the \(\m{q}\)-absolutely component of 
\(\m{w}\). 
Furthermore, for any \(\sigma\)-finite measure \(\m{\mu}\AC\m{q}\) 
satisfying \(\der{\m{\mu}}{\m{q}}\!\leq\!1\) \(\m{q;\sim}\)-a.e., 
the inequality \(\f{\ylr;\m{w},\m{\mu}}\!\geq\!\f{\ylr;\m{w},\m{q}}\) 
holds \(\m{q}\)-almost-everywhere. Thus 
\begin{align}
\label{eq:lrcdf:Elementary:Monotonicity}
\scq{\cdot}{\m{w}}{\m{\mu}}
&\!\leq\!\scq{\cdot}{\m{w}}{\m{q}}
&
&\forall\m{\mu}:
\der{\m{\mu}}{\m{q}}\!\leq\!1\quad\m{q;\sim}\text{-a.e}.
\end{align}
We call the cases when the
likelihood-ratio distribution is a function of the form
\(\scq{\cdot}{\m{w}}{\m{q}}:\numbers{R}\!\mapsto\!\{0,\lambda\}\) 
to be the trivial case.
The case when \(\scq{\tau}{\m{w}}{\m{q}}\!=\!0\)
for all \(\tau\!\in\!\numbers{R}\) is called the singular 
case because it happens iff 
\(\m{w}\) and \(\m{q}\) are mutually singular, i.e., \(\m{w}\!\perp\!\m{q}\).
Recall that 
\(\m{w}\perp\m{q}\Leftrightarrow\norm{\m{w;\sim}}\!=\!0
\Leftrightarrow\norm{\m{q;\sim}}\!=\!0\).
We call the case when the
likelihood-ratio distribution is a surjective (onto) function 
of the form
\(\scq{\cdot}{\m{w}}{\m{q}}:\numbers{R}\!\mapsto\!\{0,\lambda\}\) 
to be the trivial-non-singular case.
Note that the trivial-non-singular case happens iff 
\(\m{w}\not\perp\m{q}\) and
\(\der{\m{w;\sim}}{\m{q}}\!=\!\zeta\) holds \(\m{w;\sim}\)-a.e. 
some \(\zeta\!\in\!(0,\infty)\). Note that in the trivial-non-singular 
case \(\norm{\m{w;\sim}}\!=\!\lambda\) and 
\(\norm{\m{q;\sim}}\!=\!\sfrac{\lambda}{\zeta}\)

Although the likelihood-ratio distribution \(\scq{\cdot}{\m{w}}{\m{q}}\) 
is well-defined, it might not have an inverse in the
usual sense. Nevertheless, the monotonicity of \(\scq{\cdot}{\m{w}}{\m{q}}\) 
ensures that it has semicontinuous inverses, called likelihood-ratio quantiles.

\begin{definition}\label{def:LRQ}
For any pair of \(\sigma\)-finite measures \((\m{w},\m{q})\) on 
the same measurable space, say \((\outS,\outA)\), 
the likelihood-ratio quantiles (LRQs)
\(\lscq{\cdot}{\m{w}}{\m{q}}:\numbers{R}\!\mapsto\!\overline{\numbers{R}}\) 
and 
\(\uscq{\cdot}{\m{w}}{\m{q}}:\numbers{R}\!\mapsto\!\overline{\numbers{R}}\) 
are defined as
\begin{subequations}
	\label{eq:def:LRQ}	
	\begin{align}
		\label{eq:def:LRQ:LSC}
		\lscq{\varepsilon}{\m{w}}{\m{q}}
		&\DEF \inf\left\{\gamma\in\numbers{R}:
		\scq{\gamma}{\m{w}}{\m{q}}
		\geq\varepsilon\right\}
		&
		&\forall\varepsilon\!\in\!\numbers{R},
		\\
		\label{eq:def:LRQ:USC}
		\uscq{\varepsilon}{\m{w}}{\m{q}}
		&\DEF \inf\left\{\gamma\in\numbers{R}:
		\scq{\gamma}{\m{w}}{\m{q}}
		>\varepsilon\right\}
		&
		&\forall\varepsilon\!\in\!\numbers{R}.
	\end{align}			
\end{subequations}
\end{definition}	
Note that the likelihood-ratio quantiles are nothing but the semicontinuous 
inverses of likelihoods ratio distribution \(\scq{\cdot}{\m{w}}{\m{q}}\); 
this is why we imitate the notation in \cite{klementMP99,fortelle20} 
and denote them with \(\lscq{\cdot}{\m{w}}{\m{q}}\) and 
\(\uscq{\cdot}{\m{w}}{\m{q}}\).

As a result of \eqref{eq:lrcdf:ACequivalence}, we have
\begin{align}
	\label{eq:LRQ:ACequivalence}
	\uscq{\cdot}{\m{w}}{\m{q}} 
	&=\uscq{\cdot}{\m{w;\sim}}{\m{q;\sim}} 	
	&
	&\text{and}
	&
	\lscq{\cdot}{\m{w}}{\m{q}} 
	&=\lscq{\cdot}{\m{w;\sim}}{\m{q;\sim}}. 			
\end{align}
Furthermore, evident inclusion relations for the sets 
we are talking the infimum over in \eqref{eq:def:LRQ} imply,
\begin{align}
	\label{eq:LRQ:Monotonicity}
	\lscq{\varepsilon}{\m{w}}{\m{q}}
	&\!\leq\!\uscq{\varepsilon}{\m{w}}{\m{q}}
	\!\leq\!\lscq{\varepsilon\!+\!\delta}{\m{w}}{\m{q}}
	&
	&\forall\varepsilon\!\in\!\numbers{R},\delta\!>\!0.\!
\end{align} 

Let the LRQ-halflines\footnote{We call 
	\(\lscqs{\varepsilon}{\m{w}}{\m{q}}\) and \(\uscqs{\varepsilon}{\m{w}}{\m{q}}\) 
	halflines with the understanding that 
	\(\emptyset\) and \(\numbers{R}\) are  halflines, as well.} 
\(\lscqs{\varepsilon}{\m{w}}{\m{q}}\)
and 
\(\uscqs{\varepsilon}{\m{w}}{\m{q}}\)
be 
\begin{subequations}\label{eq:def:LRQ:halfline}	
\begin{align}
\label{eq:def:LRQM:halfline}
\lscqs{\varepsilon}{\m{w}}{\m{q}}
&\DEF \left\{\gamma\in\numbers{R}:
\scq{\gamma}{\m{w}}{\m{q}}
\geq\varepsilon\right\}
&
&\forall\varepsilon\!\in\!\numbers{R},
\\
\label{eq:def:LRQP:halfline}
\uscqs{\varepsilon}{\m{w}}{\m{q}}
&\DEF \left\{\gamma\in\numbers{R}:
\scq{\gamma}{\m{w}}{\m{q}}
>\varepsilon\right\}
&
&\forall\varepsilon\!\in\!\numbers{R}.
\end{align}		
\end{subequations}
Then
\begin{align}
\label{eq:LRQ:halfline:Monotonicity}
\lscqs{\varepsilon\!+\!\delta}{\m{w}}{\m{q}}\!\subset\!
\uscqs{\varepsilon}{\m{w}}{\m{q}}
&\!\subset\!\lscqs{\varepsilon}{\m{w}}{\m{q}}	
&
&\forall\varepsilon\!\in\!\numbers{R},\delta\!>\!0.\!
\end{align}

Since \(\scq{\cdot}{\m{w}}{\m{q}}\) is non-decreasing, both 
\(\lscqs{\varepsilon}{\m{w}}{\m{q}}\) and \(\uscqs{\varepsilon}{\m{w}}{\m{q}}\) 
are halflines directed towards \(\infty\).
Thus
\begin{subequations}
\label{eq:LRQ:halfline:CB}	
\begin{align}
\label{eq:LRQM:halfline:CB}
\inf\lscqs{\varepsilon}{\m{w}}{\m{q}}
&=\sup\left(\numbers{R}
\setminus\lscqs{\varepsilon}{\m{w}}{\m{q}}\right),
\\		
\label{eq:LRQP:halfline:CB}
\inf\uscqs{\varepsilon}{\m{w}}{\m{q}}
&=\sup\left(\numbers{R}
\setminus\uscqs{\varepsilon}{\m{w}}{\m{q}}
\right).
\end{align}		
\end{subequations}		
As a result of \eqref{eq:def:LRQ}, \eqref{eq:def:LRQ:halfline},
and \eqref{eq:LRQ:halfline:CB}, we have the following alternative
characterization of LRQs 
\begin{subequations}\label{eq:LRQ}	
\begin{align}
\label{eq:LRQ:LSC}
\lscq{\varepsilon}{\m{w}}{\m{q}}
&\!=\!\sup\left\{\gamma\!\in\!\numbers{R}:
\scq{\gamma}{\m{w}}{\m{q}}
\!<\!\varepsilon\right\}
&
&\forall\varepsilon\!\in\!\numbers{R},
\\	
\label{eq:LRQ:USC}
\uscq{\varepsilon}{\m{w}}{\m{q}}
&\!=\!\sup\left\{\gamma\!\in\!\numbers{R}:
\scq{\gamma}{\m{w}}{\m{q}}
\!\leq\!\varepsilon\right\}
&
&\forall\varepsilon\!\in\!\numbers{R}.		
\end{align}			
\end{subequations}
\begin{lemma}\label{lem:LRQ}
For any pair of \(\sigma\)-finite measures \((\m{w},\m{q})\) on 
the same measurable space \((\outS,\outA)\), 
LRQs \(\lscq{\cdot}{\m{w}}{\m{q}}\) and \(\uscq{\cdot}{\m{w}}{\m{q}}\), 
defined in \eqref{eq:def:LRQ},
are\begin{subequations}\label{eq:lem:LRQ}~ non-decreasing functions on \(\numbers{R}\) satisfying 
\begin{align}
\label{eq:lem:LRQ:leftlimit}
\lim\nolimits_{\tau\uparrow\varepsilon}
\scq[-1]{\tau}{\m{w}}{\m{q}}
&\!=\!
\lscq{\varepsilon}{\m{w}}{\m{q}}
&
&\forall\varepsilon\!\in\!\numbers{R},
\\
\label{eq:lem:LRQ:rightlimit}
\lim\nolimits_{\tau\downarrow\varepsilon}
\scq[-1]{\tau}{\m{w}}{\m{q}}
&\!=\!\uscq{\varepsilon}{\m{w}}{\m{q}}
&
&\forall\varepsilon\!\in\!\numbers{R},
\\
\label{eq:lem:LRQUpperBound}  
\m{w;\sim}[\f{\ylr;\m{w},\m{q}}
\!\leq\!\lscq{\varepsilon}{\m{w}}{\m{q}}]
&\!\geq\!\varepsilon
&
&\forall\varepsilon\leq \norm{\m{w;\sim}},
\\
\label{eq:lem:LRQLowerBound}
\m{w}[\f{\ylr;\m{w},\m{q}}
\!<\!\uscq{\varepsilon}{\m{w}}{\m{q}}]
&\!\leq\!\varepsilon
&
&\forall\varepsilon\geq0,
\\
\label{eq:lem:LRQ:AEequality}
\uscq{\varepsilon}{\m{w}}{\m{q}}
&=\lscq{\varepsilon}{\m{w}}{\m{q}}
&
&\forall\varepsilon\!\in\!\numbers{R}\!\setminus\!\set{D},
\\
\label{eq:lem:LRQ:StrictlyMonone}
\left[\lscq{\varepsilon}{\m{w}}{\m{q}},\uscq{\varepsilon}{\m{w}}{\m{q}}\right]
&\!\subset\!(0,\infty)
&
&\forall\varepsilon\!\in\!(0,\norm{\m{w;\sim}}),
\end{align}				
for\end{subequations}~some countable set \(\set{D}\subset[0,\norm{\m{w;\sim}}]\), 
where \(\scq[-1]{\cdot}{\m{w}}{\m{q}}\) in 
\eqref{eq:lem:LRQ:leftlimit} and \eqref{eq:lem:LRQ:rightlimit} 
can be either \(\lscq{\cdot}{\m{w}}{\m{q}}\) or \(\uscq{\cdot}{\m{w}}{\m{q}}\).
\end{lemma}	
Proof of Lemma \ref{lem:LRQ} is presented in Appendix \ref{sec:Proofs:LikelihoodRatioQuantiles}.
Let us proceed with elaborating on some of the implications of Lemma \ref{lem:LRQ}, 
briefly. If \(\lscq{\varepsilon}{\m{w}}{\m{q}}\neq\uscq{\varepsilon}{\m{w}}{\m{q}}\)
for an \(\varepsilon\in\numbers{R}\), then 
\([\lscq{\varepsilon}{\m{w}}{\m{q}},\uscq{\varepsilon}{\m{w}}{\m{q}})\neq\emptyset\)
by \eqref{eq:LRQ:Monotonicity}.
Furthermore, additivity of  \(\m{w}\) implies 
\begin{align}
\notag
\m{w}[\f{\ylr;\m{w},\m{q}}\!\leq\!\lscq{\varepsilon}{\m{w}}{\m{q}}]
\leq 
\scq{\gamma}{\m{w}}{\m{q}}
&\leq
\m{w}[\f{\ylr;\m{w},\m{q}}\!<\!\uscq{\varepsilon}{\m{w}}{\m{q}}]	
\end{align}
for all \(\gamma\) in \([\lscq{\varepsilon}{\m{w}}{\m{q}},\uscq{\varepsilon}{\m{w}}{\m{q}})\).
Then \eqref{eq:lem:LRQUpperBound} and \eqref{eq:lem:LRQLowerBound} imply
\begin{align}
	\label{eq:LRQDistinctLeftAndRightLimits}
	\hspace{-.19cm}
	\scq{\gamma}{\m{w}}{\m{q}}
	&\!=\!\varepsilon
	&
	&\forall
	\varepsilon\!\in\![0,\norm{\m{w;\sim}}],\!~
	\gamma\!\in\![\lscq{\varepsilon}{\m{w}}{\m{q}},\uscq{\varepsilon}{\m{w}}{\m{q}}).	
\end{align}
The behavior of \(\lscq{\cdot}{\m{w}}{\m{q}}\) and
\(\uscq{\cdot}{\m{w}}{\m{q}}\) are 
interesting only on \([0,\norm{\m{w;\sim}}]\).
What happens outside that range is determined by 
\eqref{eq:LRQ:Monotonicity},
\eqref{eq:lem:LRQ:leftlimit},
\eqref{eq:lem:LRQ:rightlimit},
and the values of \(\lscq{\cdot}{\m{w}}{\m{q}}\) and
\(\uscq{\cdot}{\m{w}}{\m{q}}\) on the boundary
\begin{subequations}\label{eq:LRQ:Obvious}
\begin{align}
\label{eq:LRQ:Obvious:Zero:LSC}	
\lscq{0}{\m{w}}{\m{q}}
&\!=\!-\infty,
\\
\label{eq:LRQ:Obvious:Zero:USC}	
\uscq{0}{\m{w}}{\m{q}}
&\!=\!\esinf\nolimits_{\m{w;\sim}}\f{\ylr;\m{w},\m{q}},
\\
\label{eq:LRQ:Obvious:Singular:LSC}	
\lscq{\norm{\m{w;\sim}}}{\m{w}}{\m{q}}
&\!=\!\essup\nolimits_{\m{w;\sim}}\f{\ylr;\m{w},\m{q}},
\\
\label{eq:LRQ:Obvious:Singular:USC}	
\uscq{\norm{\m{w;\sim}}}{\m{w}}{\m{q}}
&\!=\!\infty.							
\end{align}\end{subequations}

The effect of certain elementary modifications of \(\m{w}\) 
and \(\m{q}\) on the likelihood ratio distribution was 
described \eqref{eq:lrf:Elementary:ScalarMultiplication} and 
\eqref{eq:lrcdf:Elementary:Monotonicity}.
Lemma \ref{lem:LRQ:Elementary}, in the following, 
describes the effect of the very same modifications on the LRQs.
For the case when \(\m{w}\) and \(\m{q}\) are probability measures,
\eqref{eq:lem:LRQ:Elementary:Monotonicity:mQ} follows from 
similar identities reported before 
see \cite[Lemma 3]{tomamichelT13} and \cite[Lemma 2.2]{tan14}. 
\begin{lemma}\label{lem:LRQ:Elementary}
For any pair of \(\sigma\)-finite measures \((\m{w},\m{q})\) on 
the same measurable space \((\outS,\outA)\)
\begin{subequations}\label{eq:lem:LRQ:Elementary}the 
LRQs defined in \eqref{eq:def:LRQ} satisfy
\begin{align}
\label{eq:lem:LRQ:Elementary:ScalarMultiplication:mW}
\scq[-1]{\cdot}{\lambda\m{w}}{\m{q}}
&\!=\!\lambda\scq[-1]{\tfrac{\cdot}{\lambda}}{\m{w}}{\m{q}}
&
&\forall \lambda\!\in\!(0,\infty),
\\
\label{eq:lem:LRQ:Elementary:ScalarMultiplication:mQ}
\scq[-1]{\cdot}{\m{w}}{\lambda\m{q}}
&\!=\!\tfrac{1}{\lambda}\scq[-1]{\cdot}{\m{w}}{\m{q}}
&
&\forall \lambda\!\in\!(0,\infty),
\\
\label{eq:lem:LRQ:Elementary:Monotonicity:mQ}
\scq[-1]{\cdot}{\m{w}}{\m{\mu}}
&\!\geq\!\scq[-1]{\cdot}{\m{w}}{\m{q}}
&
&\forall\m{\mu}:\der{\m{\mu}}{\m{q}}\!\leq\!1\quad\m{q;\sim}\text{-a.e},
\end{align}		
where \(\m{q;\sim}\) is\end{subequations}~the 
\(\m{w}\)-absolutely component of \(\m{q}\).		
\end{lemma}
Each relation in Lemma \ref{lem:LRQ:Elementary} follows from 
the corresponding relation in
\eqref{eq:lrf:Elementary:ScalarMultiplication} 
or \eqref{eq:lrcdf:Elementary:Monotonicity}, 
as an immediate consequence of the definition of 
LRQs in \eqref{eq:def:LRQ}. 

\subsection{The Primitive Entropy Spectrum}\label{sec:Entropy}
Recall that for any pair of \(\sigma\)-finite measures 
\((\m{w},\m{q})\) on the same measurable space, say \((\outS,\outA)\), 
the primitive entropy spectrum of \(\m{w}\) with respect to 
\(\m{q}\) is defined
\begin{subequations}\label{eq:EntropySpectrum} in \eqref{eq:def:EntropySpectrum} as 
\begin{align}
\label{eq:fed:EntropySpectrum}
\ent{\gamma}{\m{w}}{\m{q}}
&=\displaystyle{\int}\left(\der{\m{w;\sim}}{\m{q}}\wedge\gamma\right)\dif{\m{q}}
&
&\forall\gamma\!\in\![0,\infty).
\end{align}
Thus for all \(\sigma\)-finite measures \(\m{w}\) and \(\m{q}\), we have
\begin{align}
\label{eq:EntropySpectrum:Variational}
\ent{\gamma}{\m{w}}{\m{q}}
&=\inf\nolimits_{\oev\in\outA}
\m{w}[\oev]\!+\!\gamma\m{q}[\outS\!\setminus\!\oev]
&
&\forall\gamma\!\in\![0,\infty),
\\
\label{eq:EntropySpectrum:zero}	
\!
\ent{0}{\m{w}}{\m{q}}
&\!=\!0,
\\	
\label{eq:EntropySpectrum:ACequivalence}
\ent{\gamma}{\m{w}}{\m{q}}
&=\ent{\gamma}{\m{w;\sim}}{\m{q;\sim}}
&
&\forall\gamma\!\in\![0,\infty),\!
\\	
\label{eq:EntropySpectrum:Sum}	
\!
\ent{\gamma}{\m{w}}{\m{q}}
&\!=\!\m{w}[\f{\ylr;\m{w},\m{q}}\!\leq\!\gamma\!]
\!+\!\gamma\m{q}[\f{\ylr;\m{w},\m{q}}\!>\!\gamma\!]
&
&\forall\gamma\!\in\!(0,\infty),\!
\\
\label{eq:EntropySpectrum:Adjoint}
\!
\ent{\gamma}{\m{w}}{\m{q}}
&\!=\!\gamma\ent{\sfrac{1}{\gamma}}{\m{q}}{\m{w}}
&
&\forall\gamma\!\in\!(0,\infty),\!	
\end{align}	
\end{subequations}		
Note that \(\ent{\gamma}{\m{w}}{\m{q}}\) is non-decreasing
in \(\gamma\). Thus \eqref{eq:EntropySpectrum:Adjoint}
implies
\begin{align}
	\label{eq:EntropySpectrum:Bound}
	%\tfrac{\gamma\ent{1}{\m{w}}{\m{q}}}{1\wedge\gamma}
	%\!\geq\!
	%\ent{\gamma}{\m{w}}{\m{q}}
	%=\gamma\ent{\sfrac{1}{\gamma}}{\m{q}}{\m{w}}
	%&\!\geq\!\tfrac{\gamma\ent{1}{\m{w}}{\m{q}}}{1\vee\gamma}
	%&
	%&\forall\gamma\!\in\!(0,\infty).
	%\\
	(1\!\vee\!\gamma)
	\ent{1}{\m{w}}{\m{q}}
	\!\geq\!
	\ent{\gamma}{\m{w}}{\m{q}}
	&\!\geq\!
	(1\!\wedge\!\gamma)
	\ent{1}{\m{w}}{\m{q}}
	&
	&\forall\gamma\!\in\!(0,\infty).
\end{align}
Thus \begin{subequations}\label{eq:EntropySpectrum:DistinctionOfCases}	
for any \(\gamma\!\in\!(0,\infty)\), we have
\begin{align}
\label{eq:EntropySpectrum:Obvious}	
\ent{1}{\m{w}}{\m{q}}&=0
&
&\Longleftrightarrow
&
\ent{\gamma}{\m{w}}{\m{q}}
&=0,
\\
\label{eq:EntropySpectrum:FinitePositive}	
\ent{1}{\m{w}}{\m{q}}
&\in\!(0,\infty)
&
&\Longleftrightarrow
&
\ent{\gamma}{\m{w}}{\m{q}}
&\in\!(0,\infty),
\\
\label{eq:EntropySpectrum:Void}
\ent{1}{\m{w}}{\m{q}}
&=\infty
&
&\Longleftrightarrow
&
\ent{\gamma}{\m{w}}{\m{q}}
&=\infty.	
\end{align}
\end{subequations}		
We deem the case when  \(\ent{\gamma}{\m{w}}{\m{q}}\!=\!\infty\)
for all \(\gamma\!\in\!(0,\infty)\) to be the void case 
for primitive entropy spectrum \(\ent{\cdot}{\m{w}}{\m{q}}\). 
Thus a necessary and sufficient condition for the non-void
cases for \(\ent{\cdot}{\m{w}}{\m{q}}\) 
is \eqref{eq:assumption:nonvoid},
i.e., \(\ent{1}{\m{w}}{\m{q}}\!<\!\infty\). 
On the other hand the singular case \(\m{w}\perp\m{q}\) 
corresponds to \(\ent{1}{\m{w}}{\m{q}}\!=\!0\) case.
More generally in the trivial case
the entropy spectrum has the following 
simple expression
\(\ent{\gamma}{\m{w}}{\m{q}}\!=\!\norm{\m{w;\sim}}
\!\wedge\!(\gamma\norm{\m{q;\sim}})%=\lambda(1\wedge(\sfrac{\gamma}{\zeta}))
\)
for all \(\gamma\!\in\!(0,\infty)\).

Using the level set integration formula given in \eqref{eq:LevelSetIntegration},
we can express \(\ent{\gamma}{\m{w}}{\m{q}}\) as Riemann integrals in terms
of likelihood ratio distributions. To see why let us first express 
\(\m{w}[\f{\ylr;\m{w},\m{q}}\leq\gamma]\) as such an integral.
\begin{align}
\notag
\m{w}[\f{\ylr;\m{w},\m{q}}\leq\gamma]
&\!\mathop{=}^{(a)}\!\int\!
\IND{\f{\ylr;\m{w},\m{q}}\leq\gamma}\f{\ylr;\m{w},\m{q}} 
\dif{\m{q}},
\\
\notag
&\!\mathop{=}^{(b)}\!\int_{0}^{\infty}\!
\m{q}[\IND{\f{\ylr;\m{w},\m{q}}\leq\gamma}\f{\ylr;\m{w},\m{q}}\!>\!\tau]
\dif{\tau},
\\
\notag
&\!\mathop{=}^{(c)}\! \int_{0}^{\gamma}\!
\m{q}[\IND{\f{\ylr;\m{w},\m{q}}\leq\gamma}\f{\ylr;\m{w},\m{q}}\!>\!\tau]
\dif{\tau},
\\
\notag
&\!\mathop{=}^{}\! \int_{0}^{\gamma}\!
\m{q}[\gamma\geq\f{\ylr;\m{w},\m{q}}>\tau]
\dif{\tau},
\end{align}				
where 
\((a)\) follows from \eqref{eq:def:lr};
\((b)\) follows from \eqref{eq:LevelSetIntegration};
\((c)\) holds because 
\(\m{q}[\IND{\f{\ylr;\m{w},\m{q}}\leq\gamma}\f{\ylr;\m{w},\m{q}}\!>\!\tau]=0\)
for \(\tau\!\geq\!\gamma\). Then 
\eqref{eq:EntropySpectrum:Sum}~\begin{subequations}\label{eq:lem:PES:integral} 
	implies
\begin{align}
\label{eq:lem:PES:integral:r}
\ent{\gamma}{\m{w}}{\m{q}}
&=\int_{0}^{\gamma} \!\m{q}[\f{\ylr;\m{w},\m{q}}\!>\!\tau]\dif{\tau}
&
&\forall\gamma\!\in\!(0,\infty).
\intertext{Then \eqref{eq:EntropySpectrum:Adjoint} implies}
\notag	
\ent{\gamma}{\m{w}}{\m{q}}
&=\gamma\int_{0}^{\sfrac{1}{\gamma}} \!\m{w}[\f{\ylr;\m{q},\m{w}}\!>\!\mS]\dif{\mS}
&
&\forall\gamma\!\in\!(0,\infty).	
\intertext{Then applying the change of variable \(\mS=\sfrac{\tau}{\gamma}\), we get}
\label{eq:lem:PES:integral:p}
\ent{\gamma}{\m{w}}{\m{q}}
&=\int_{0}^{1} \!\m{w}[\f{\ylr;\m{q},\m{w}}\!>\!\tfrac{\tau}{\gamma}]\dif{\tau}
&
&\forall\gamma\!\in\!(0,\infty).
\end{align}
Note that integrands  
\(\m{q}[\f{\ylr;\m{w},\m{q}}\!>\!\tau]\) 
and
\(\m{w}[\f{\ylr;\m{q},\m{w}}\!>\!\sfrac{\tau}{\gamma}]\) 
in \eqref{eq:lem:PES:integral:r} 
and \eqref{eq:lem:PES:integral:p}
can be replaced by 
\(\m{q}[\f{\ylr;\m{w},\m{q}}\!\geq\!\tau]\) 
and
\(\m{w}[\f{\ylr;\m{q},\m{w}}\!\geq\!\sfrac{\tau}{\gamma}]\) 
because
monotonic functions have at most countably many discontinuities 
by \cite[Theorem 4.30]{rudin} 
and any countable set has zero Lebesgue measure. 
Thus using \eqref{eq:def:lr} and \eqref{eq:lrf}, we get 
\begin{align}
\label{eq:lem:PES:integral:r:DF}
\ent{\gamma}{\m{w}}{\m{q}}
&=\int_{0}^{\gamma}
\scq{\tfrac{1}{\tau}}{\m{q}}{\m{w}}\dif{\tau}
&
&\forall\gamma\!\in\!(0,\infty),
\\	
\label{eq:lem:PES:integral:p:DF}
\ent{\gamma}{\m{w}}{\m{q}}
&=\int_{0}^{1}\!\scq{\tfrac{\gamma}{\tau}}{\m{w}}{\m{q}}\dif{\tau}
&
&\forall\gamma\!\in\!(0,\infty).	
\end{align}
\end{subequations}
\begin{lemma}\label{lem:PES}
If a pair of \(\sigma\)-finite measures \((\m{w},\m{q})\)
satisfies \eqref{eq:assumption:nonvoid}, i.e., 
\(\ent{1}{\m{w}}{\m{q}}\!<\!\infty\),
then the primitive entropy spectrum \(\ent{\cdot}{\m{w}}{\m{q}}\) 
satisfies
\begin{subequations}\label{eq:lem:PES}
	\begin{align}
		\label{eq:lem:PES:zero}
		\lim\nolimits_{\gamma\downarrow 0}\ent{\gamma}{\m{w}}{\m{q}}
		&=\ent{0}{\m{w}}{\m{q}},
		\\
		\label{eq:lem:PES:infinity}
		\lim\nolimits_{\gamma\uparrow \infty}\ent{\gamma}{\m{w}}{\m{q}}
		&=\norm{\m{w;\sim}},
		%\ent{\infty}{\m{w}}{\m{q}}
		\\
		\label{eq:lem:PES:leftderivative}
		\lder{\ent{\gamma}{\m{w}}{\m{q}}}
		&=\m{q}[\f{\ylr;\m{w},\m{q}}\!\geq\!\gamma]
		&
		&\forall\gamma\!\in\!(0,\infty),
		\\
		\label{eq:lem:PES:rightderivative}
		\rder{\ent{\gamma}{\m{w}}{\m{q}}}
		&=\m{q}[\f{\ylr;\m{w},\m{q}}\!>\!\gamma]
		&
		&\forall\gamma\!\in\![0,\infty),	
	\end{align}		
	where \(\lder{\ent{\gamma}{\m{w}}{\m{q}}}\) and \(\rder{\ent{\gamma}{\m{w}}{\m{q}}}\)
	are left and right derivatives of \(\ent{\cdot}{\m{w}}{\m{q}}\) at \(\gamma\).
\end{subequations} 
Hence, either \(\ent{\cdot}{\m{w}}{\m{q}}\!=\!0\),
or \(\ent{\cdot}{\m{w}}{\m{q}}\)
is a non-decreasing, finite valued, continuous, and concave function,
which is monotonically increasing on \(\gamma\in[0,\essup_{\m{w;\sim}}\f{\ylr;\m{w},\m{q}}]\).
\end{lemma}
%Proof of Lemma \ref{lem:PES} is presented in Appendix \ref{sec:Proofs:PrimitiveEntropySpectrum}.

\begin{lemma}\label{lem:PES:Elementary}
Let \(\m{w;0}\), \(\m{w;1}\), \(\m{q;0}\), \(\m{q;1}\) be
\(\sigma\)-finite measures with
a common \(\sigma\)-finite reference measure \(\rfm\)
and \(\gamma_{0}\),  \(\gamma_{1}\) be non-negative real numbers.
If\begin{subequations}\label{eq:lem:PES:Elementary}~both 
\(\der{\m{w;0}}{\rfm}\leq\der{\m{w;1}}{\rfm}\) and 
\(\gamma_{0}\der{\m{q;0}}{\rfm}\leq\gamma_{1}\der{\m{q;1}}{\rfm}\) 
holds \(\rfm\)-a.e. then
\begin{align}
	\label{eq:lem:PES:Elementary:Monotonicty}
	\ent{\gamma_{0}}{\m{w;0}}{\m{q;0}}
	&\leq
	\ent{\gamma_{1}}{\m{w;1}}{\m{q;1}}.	
\end{align}
If \(\gamma_{\alpha}=\alpha\gamma_{1}+(1-\alpha)\gamma_{0}\),
\(\m{w;\alpha}=\alpha\m{w;1}+(1-\alpha)\m{w;0}\), and
\(\m{q;\alpha}=
\tfrac{\alpha\gamma_{1}}{\gamma_{\alpha}}\m{q;1}
+
\tfrac{(1-\alpha)\gamma_{0}}{\gamma_{\alpha}}\m{q;0}
\),
for some  \(\alpha\in[0,1]\), then
\begin{align}
	\label{eq:lem:PES:Elementary:Concavity}
	\alpha \ent{\gamma_{1}}{\m{w;1}}{\m{q;1}}
	+(1-\alpha)\ent{\gamma_{0}}{\m{w;0}}{\m{q;0}}
	&\leq \ent{\gamma_{\alpha}}{\m{w;\alpha}}{\m{q;\alpha}}.
\end{align}		
Hence \(\ent{\gamma}{\m{w}}{\m{q}}\) 
is non-decreasing and concave in the pair 
\((\m{w},\gamma\m{q})\). Furthermore, 
for any positive real \(\lambda\), i.e., \(\lambda\!\in\!(0,\infty)\),
\begin{align}
	\label{eq:lem:PES:Elementary:ScalarMultiplication:mW}
	\ent{\gamma}{\lambda\m{w}}{\m{q}}
	&\!=\!\lambda\ent{\tfrac{\gamma}{\lambda}}{\m{w}}{\m{q}}
	&
	&\forall \gamma\!\in\![0,\infty),
	\\
	\label{eq:lem:PES:Elementary:ScalarMultiplication:mQ}
	\ent{\gamma}{\m{w}}{\lambda\m{q}}
	&\!=\!\ent{\lambda\gamma}{\m{w}}{\m{q}}
	&
	&\forall \gamma\!\in\![0,\infty).
\end{align}
\end{subequations}				
\end{lemma}			
Proof of Lemmas \ref{lem:PES} and \ref{lem:PES:Elementary} is presented 
in Appendix \ref{sec:Proofs:PrimitiveEntropySpectrum}.	
	
\section[Spectral Representations and Measure Change]{The 
	Spectral Representations and A Measure Change Lemma 
	For Simple Hypothesis Testing}\label{sec:SRandMC}	
Our primary goal in this section is to establishing  
a spectral Neyman--Pearson lemma relating 
\(\NPD{\cdot}{\m{q}}{\m{w}}\) to \(\ent{\cdot}{\m{q}}{\m{w}}\),
\(\lscq{\cdot}{\m{w}}{\m{q}}\),  and \(\uscq{\cdot}{\m{w}}{\m{q}}\),
see Theorem \ref{thm:SNPL},
and a measure change lemma for the simple hypothesis testing 
problem expressing \(\NPD{\cdot}{\m{q}}{\m{w}}\) in terms of 
a \(\NPD{\cdot}{\m{q}}{\m{v}}\) for a \(\sigma\)-finite measure 
\(\m{v}\) satisfying certain conditions, 
see Theorem \ref{thm:MeasureChange}. 
To that end let us first point out certain elementary 
facts about rNP-function \(\NPD{\cdot}{\m{q}}{\m{w}}\) 
and dNP-function \(\NPDd{\cdot}{\m{q}}{\m{w}}\) 
defined in \eqref{eq:def:NPD}.
Note that
\begin{align}
\label{eq:NPDleqNPDd}
\NPD{\cdot}{\m{q}}{\m{w}}	
&\leq
\NPDd{\cdot}{\m{q}}{\m{w}}
&
&\forall \m{w},\m{q},			
\end{align}
because any deterministic detector is also randomized 
detector, i.e., \(\tmead{\outA}\!\subset\!\tmea{\outA}\).
Definitions in \eqref{eq:def:NPD} follow the framework of \cite{strassen62} 
and \cite{csiszarL71}, allowing \(\m{q}\)'s that are 
not probability measures and parameterizing \(\NPD{\cdot}{\m{q}}{\m{w}}\) 
with \(\varepsilon\) rather than \(1\!-\!\varepsilon\), unlike the majority
of the recent work on the subject 
\cite{polyanskiythesis,polyanskiyPV10,polyanskiy13,tomamichelT13,moulin17,yavasKE24,tan14},
i.e., \(\NPD{\varepsilon}{\m{q}}{\m{w}}=\beta_{1-\varepsilon}(\m{w}\Vert\m{q})\).
Extension to the case when \(\m{w}\) is a finite measure rather than a 
probability measures were already mentioned explicitly in \cite{strassen62} 
and \(\m{q}\) being a \(\sigma\)-finite measure is implicitly 
allowed in both \cite{strassen62} and \cite{csiszarL71}. 

Allowing both \(\m{w}\) and \(\m{q}\) to be \(\sigma\)-finite measures, however, 
introduces a nuance one needs to handle; that is  for certain \(\m{w}\) and \(\m{q}\)
choices \(\NPD{\varepsilon}{\m{q}}{\m{w}}\) is infinite for all finite \(\varepsilon\)
values, as we have mention in \S\ref{sec:Introduction}. 
We deem this case to be the void case for 
\(\NPD{\cdot}{\m{q}}{\m{w}}\).
Recall that we have called the case when \(\ent{1}{\m{w}}{\m{q}}\!=\!\infty\) 
to be the void case for \(\ent{\cdot}{\m{w}}{\m{q}}\) because in that case 
\(\ent{\gamma}{\m{w}}{\m{q}}\!=\!\infty\) for all positive \(\gamma\).
On the other hand, definitions of \(\NPD{\cdot}{\m{q}}{\m{w}}\) and 
\(\ent{\cdot}{\m{w}}{\m{q}}\) in \eqref{eq:def:NPD} and
\eqref{eq:def:EntropySpectrum} imply
\begin{align}
\NPD{\varepsilon}{\m{q}}{\m{w}}
+ \varepsilon	
&\geq 	\ent{1}{\m{w}}{\m{q}}
&
&\forall \varepsilon\geq 0.
\end{align}
Thus whenever the condition in \eqref{eq:assumption:nonvoid} is violated, i.e.,
whenever \(\ent{1}{\m{w}}{\m{q}}\!=\!\infty\), we are in the void case for 
\(\NPD{\cdot}{\m{q}}{\m{w}}\), as well. When \eqref{eq:assumption:nonvoid} holds,
i.e., \(\ent{1}{\m{w}}{\m{q}}\!<\!\infty\), the function 
\(\NPD{\cdot}{\m{q}}{\m{w}}\) is finite on \((0,\infty)\), as we will see in
Theorem \ref{thm:SNPL}.

Note that \(\NPD{\cdot}{\m{q;\sim}}{\m{w;\sim}}\!\leq\!\NPD{\cdot}{\m{q}}{\m{w}}\)
because
\(\int \rv{T} \dif{\m{w;\sim}}\!\leq\! \int \rv{T} \dif{\m{w}}\)
and
\(\int (1\!-\!\rv{T}) \dif{\m{q;\sim}}\!\leq\!\int (1\!-\!\rv{T}) \dif{\m{q}}\)
for all \(\rv{T}\!\in\!\tmea{\outA}\).
\begin{subequations}\label{eq:NPD:Obvious}		
On the other hand, 
\(\NPD{\cdot}{\m{q}}{\m{w}}\leq\NPD{\cdot}{\m{q;\sim}}{\m{w;\sim}}\)
because we can set \(\rv{T}\!=\!0\) whenever \(\f{\ylr;\m{w},\m{q}}\!=\!\infty\)
to ensure
\(\int \rv{T} \dif{\m{w;\sim}}\!=\! \int \rv{T} \dif{\m{w}}\)
and \(\rv{T}\!=\!1\) whenever \(\f{\ylr;\m{w},\m{q}}\!=\!0\),
to ensure
\(\int (1\!-\!\rv{T}) \dif{\m{q;\sim}}\!=\!\int (1\!-\!\rv{T}) \dif{\m{q}}\),
without changing the values of
\(\int \rv{T} \dif{\m{w;\sim}}\)
and
\(\int (1\!-\!\rv{T}) \dif{\m{q;\sim}}\).
Same argument applies to \(\NPDd{\cdot}{\m{q}}{\m{w}}\), as well.
Thus 
\begin{align}
\label{eq:NPD:ACequivalence}
\NPD{\cdot}{\m{q}}{\m{w}}
&\!=\!\NPD{\cdot}{\m{q;\sim}}{\m{w;\sim}},
\\
\label{eq:NPD:ACequivalence:d}
\NPDd{\cdot}{\m{q}}{\m{w}}
&\!=\!\NPDd{\cdot}{\m{q;\sim}}{\m{w;\sim}}.
\end{align}
Consequently, behavior of \(\NPDd{\cdot}{\m{q}}{\m{w}}\) and 
\(\NPD{\cdot}{\m{q}}{\m{w}}\) are interesting only 
on  \([0,\norm{\m{w;\sim}}]\).
What happens outside that range is determined 
by the monotonicity of \(\NPDd{\cdot}{\m{q}}{\m{w}}\) 
and \(\NPD{\cdot}{\m{q}}{\m{w}}\) and the values on the boundary:
\begin{align}
\label{eq:NPD:Obvious:Negative}	
\NPDd{\varepsilon}{\m{q}}{\m{w}}
=
~\NPD{\varepsilon}{\m{q}}{\m{w}}
&\!=\!\infty
&
&\forall\varepsilon<0,
\\
\label{eq:NPD:Obvious:Zero}	
\NPDd{0}{\m{q}}{\m{w}}
=
\NPD{0}{\m{q}}{\m{w}}
&\!=\!\norm{\m{q;\sim}},
&
&
\\
\label{eq:NPD:Obvious:Singular}	
\NPDd{\varepsilon}{\m{q}}{\m{w}}
=
~\NPD{\varepsilon}{\m{q}}{\m{w}}
&\!=0\!
&
&\forall\varepsilon\!\geq\!\norm{\m{w;\sim}},		
\end{align}						
\end{subequations}			
where \(\norm{\m{q;\sim}}\) can be infinite for the case 
when \(\m{q}\) is a \(\sigma\)-finite measure. 
When \(\m{q}\) is a finite measure both
\(\NPDd{\cdot}{\m{q}}{\m{w}}\) and \(\NPD{\cdot}{\m{q}}{\m{w}}\)
have a discontinuity at zero,
because \(\norm{\m{q;\sim}}\leq\norm{\m{q}}\).
%%%Note that \eqref{eq:NPD:Obvious:Negative}	
%%%holds because there are no tests either in \(\tmead{\outA}\)
%%%or in \(\tmea{\outA}\) satisfying \(\int \rv{T} \dif{\m{w}}\leq \varepsilon\)
%%%for a negative \(\varepsilon\),
%%%i.e., there are no-feasible tests for a negative \(\varepsilon\).
%%%
%%%If \(\int \rv{T} \dif{\m{w}}\!=\!0\), then \(\rv{T}\!=\!0\) 
%%%\(\m{w}\)-a.e., however, 
%%%we are free to choose the value of \(\rv{T}\)
%%%elsewhere, within \(\tmead{\outA}\) or \(\tmea{\outA}\).
%%%Evidently the equality \(\rv{T}\!=\!1\)
%%%\(\m{q;\perp}\)-a.e. should hold to minimize 
%%%\(\int (1\!-\!\rv{T}) \dif{\m{q}}\), 
%%%Thus \eqref{eq:NPD:Obvious:Zero} 
%%%follows from \eqref{eq:def:lr} and \eqref{eq:def:NPD},
%%%and \(\IND{\f{\ylr;\m{w},\m{q}}=0}\) is an optimal test.
%%%
%%%To prove \eqref{eq:NPD:Obvious:Singular}, 
%%%consider the non-randomized test \(\IND{\f{\ylr;\m{w},\m{q}}\!<\infty}\).
%%%Note that \(\IND{\f{\ylr;\m{w},\m{q}}\!<\infty}\!=\!1\) holds \(\m{q}\)-a.e.
%%%and \(\IND{\f{\ylr;\m{w},\m{q}}\!<\infty}\!=\!0\) holds \(\m{w;\perp}\)-a.e.;
%%%then \(\int (1\!-\!\IND{\f{\ylr;\m{w},\m{q}}\!<\!\infty})\dif{\m{q}}\!=\!0\)
%%%and \(\int \IND{\f{\ylr;\m{w},\m{q}}\!<\!\infty}\dif{\m{w}}=\norm{\m{w;\sim}}\).
%%%Then \eqref{eq:NPD:Obvious:Singular}	
%%%follows from the fact that both 
%%%\(\NPDd{\varepsilon}{\m{q}}{\m{w}}\) and \(\NPD{\varepsilon}{\m{q}}{\m{w}}\)
%%%are non-negative for all \(\varepsilon\).

Furthermore, the effect of certain elementary modifications 
of \(\m{w}\) and \(\m{q}\) on \(\NPD{\cdot}{\m{q}}{\m{w}}\) 
and  \(\NPDd{\cdot}{\m{q}}{\m{w}}\) given below follow from
definitions in \eqref{eq:def:NPD} and evident pointwise relation 
that are valid for all \(\rv{T}\!\in\!\tmea{\outA}\). 
\begin{lemma}\label{lem:NPD:Elementary}
For any pair of \(\sigma\)-finite measures \((\m{w},\m{q})\) on 
the same measurable space \((\outS,\outA)\),
\(\NPD{\cdot}{\m{q}}{\m{w}}\) defined
\begin{subequations}\label{eq:lem:NPD:Elementary}in 
\eqref{eq:def:NPDr} satisfies
\begin{align}
\label{eq:lem:NPD:Elementary:ScalarMultiplication:mQ}
\NPD{\cdot}{\lambda\m{q}}{\m{w}}
&\!=\!\lambda\NPD{\cdot}{\m{q}}{\m{w}}
&
&\forall \lambda\!\in\!(0,\infty),
\\
\label{eq:lem:NPD:Elementary:ScalarMultiplication:mW}
\NPD{\cdot}{\m{q}}{\lambda\m{w}}
&\!=\!\NPD{\tfrac{\cdot}{\lambda}}{\m{q}}{\m{w}}
&
&\forall \lambda\!\in\!(0,\infty),
\\
\label{eq:lem:NPD:Elementary:Monotonicity:mQ}
\NPD{\cdot}{\m{\mu}}{\m{w}}
&\!\leq\!\NPD{\cdot}{\m{q}}{\m{w}}
&
&\forall\m{\mu}:\der{\m{\mu}}{\m{q}}\!\leq\!1\quad\m{q;\sim}\text{-a.e},
\\	
\label{eq:lem:NDP:Elementary:Monotonicity:mW}
\NPD{\cdot}{\m{q}}{\m{\mu}}
&\!\geq\!\NPD{\cdot}{\m{q}}{\m{w}}
&
&\forall\m{\mu}:\der{\m{w;\sim}}{\m{\mu}}\leq 1\quad\m{\mu}\text{-a.e},	
\end{align}			
where \(\m{w;\sim}\) 
is\end{subequations}~the \(\m{q}\)-absolutely component of \(\m{w}\)	
and \(\m{q;\sim}\) is the \(\m{w}\)-absolutely component of \(\m{q}\).
Furthermore, \eqref{eq:lem:NPD:Elementary} hold as is
if \(\NPD{\cdot}{\cdot}{\cdot}\)'s are replaced by
\(\NPDd{\cdot}{\cdot}{\cdot}\)'s.			
\end{lemma}		

In the singular case, i.e., when \(\m{w}\perp\m{q}\), NP-functions satisfy
\(\NPDd{\varepsilon}{\m{q}}{\m{w}}\!=\!\NPD{\varepsilon}{\m{q}}{\m{w}}\!=0\!\)
for all \(\varepsilon\!\geq\!0\).
Note that  a trivial-non-singular case can simultaneously be 
a void case.
In the trivial-non-singular-non-void case, rNP-function has the
following closed form expression 
\begin{align}
	\label{eq:trivial-non-singular-non-void}	
	\NPD{\varepsilon}{\m{q}}{\m{w}}
	&=\norm{\m{q;\sim}}\abp{1\!-\!\tfrac{\varepsilon}{\norm{\m{w;\sim}}}}
	&
	&\forall \varepsilon\!\geq\!0.	
\end{align}

In the following, we will primarily focus on rNP-function 
\(\NPD{\cdot}{\m{q}}{\m{w}}\), nevertheless it is worth 
mentioning certain intuitive but subtle observations
about dNP-function \(\NPDd{\cdot}{\m{q}}{\m{w}}\)
and rNP-function  \(\NPD{\cdot}{\m{q}}{\m{w}}\).
If \(\m{w;\sim}\) (and hence \(\m{q;\sim}\)) does not have any 
atoms\footnote{An event \(\oev\!\in\!\outA\) is an atom of \(\m{w;\sim}\) iff
	\(\oev\) satisfies both \(\m{w;\sim}[\oev]\!>\!0\) and the condition 
	\(\m{w;\sim}[\oev\cap \set{S}]\!\in\!\{0,\m{w;\sim}[\oev]\}\) 
	for all \(\set{S}\in\outA\).}
then \(\NPDd{\cdot}{\m{q}}{\m{w}}\!=\!\NPD{\cdot}{\m{q}}{\m{w}}\). 
If the likelihood-ratio distribution \(\scq{\cdot}{\m{w}}{\m{q}}\) 
has a discontinuity at a value of \(\f{\ylr;\m{w},\m{q}}\) and 
if there are only finitely many outputs \(\dout\) taking 
that value, i.e., there exists a \(\gamma\) satisfying both
\begin{align}
	\notag	
	\lim\nolimits_{\tau\uparrow\gamma}\scq{\tau}{\m{w}}{\m{q}}
	&\neq \scq{\gamma}{\m{w}}{\m{q}}	
	&
	&\text{and}
	&
	\abs{\impf[\set{B}][\gamma|_{\m{q}}^{\m{w}}]}
	&<\infty,
\end{align}
where \(\impf[\set{B}][\gamma|_{\m{q}}^{\m{w}}]\!=\!\{\dout\!\in\!\outS:\f{\ylr;\m{w},\m{q}}\!=\!\gamma\}\).
Then the inequality in \eqref{eq:NPDleqNPDd} is strict 
for some \(\varepsilon\). 
If \(\impf[\set{B}][\gamma|_{\m{q}}^{\m{w}}]\) is not 
a finite set, then the issue is more nuanced, e.g.,  
\eqref{eq:NPDleqNPDd} can be satisfied as an equality
even for the discrete case.
Let both \(\m{w}\) and \(\m{q}\) be the geometric probability mass function 
with parameter \(\sfrac{1}{2}\) supported on positive integers. Then 
\(\NPD{\varepsilon}{\m{q}}{\m{w}}\!=\!\NPDd{\varepsilon}{\m{q}}{\m{w}}\!=\!\abp{1\!-\!\varepsilon}\)
for all \(\varepsilon\!\geq\!0\). 
See Remark \ref{rem:NPDvsNPDdWorstcase},
for a worst-case bound on 
\(\NPDd{\varepsilon}{\m{q}}{\m{w}}\!-\!\NPD{\varepsilon}{\m{q}}{\m{w}}\).

\subsection{Likelihood-Ratio Tests}\label{sec:LikelihoodRatioTest}
\begin{lemma}\label{lem:OLRTest}
For any pair of \(\sigma\)-finite measures \((\m{w},\m{q})\) 
satisfying \eqref{eq:assumption:nonvoid}, 
\(\varepsilon\!\in\!(0,\norm{\m{w;\sim}})\), and
\(\gamma\!\in\!\left[\sfrac{1}{\uscq{\varepsilon}{\m{w}}{\m{q}}},\sfrac{1}{\lscq{\varepsilon}{\m{w}}{\m{q}}}\right]\!\),
let \(\rv{T;\varepsilon,\gamma;\m{q},\m{w}}\) be
\begin{align}\label{eq:lem:OLRTest:def}
\hspace{-.2cm}
\rv{T;\varepsilon,\gamma;\m{q},\m{w}}
&\DEF
\begin{cases}
	1   
	&\text{if~}\f{\ylr;\m{q},\m{w}}\!>\!\gamma,
	\\ 	
	\tfrac{\varepsilon-\m{w}[\f{\ylr;\m{q},\m{w}}>\gamma]}{\m{w}[\f{\ylr;\m{q},\m{w}}=\gamma]}	
	&\text{if~\(\f{\ylr;\m{q},\m{w}}\!=\!\gamma\)~and~\(\m{w}[\f{\ylr;\m{q},\m{w}}\!=\!\gamma]\!>\!0\)},\!
	\\ 	
	0   
	&\text{otherwise.}		
\end{cases}			
\end{align}	
Then \(\rv{T;\varepsilon,\gamma;\m{q},\m{w}}\) is a randomized likelihood-ratio test,
\(\rv{T;\varepsilon,\gamma;\m{q},\m{w}}\!\in\!\tmea{\outA}\). 
Furthermore,\begin{subequations}\label{eq:lem:OLRTest}
\(\rv{T;\varepsilon,\gamma;\m{q},\m{w}}\) satisfies
\begin{align}
	\label{eq:lem:OLRTest:W}	
	\int \rv{T;\varepsilon,\gamma;\m{q},\m{w}} \dif{\m{w}}
	&\!=\!\varepsilon,
	\\
	\label{eq:lem:OLRTest:Q}	
	\int(1\!-\!\rv{T;\varepsilon,\gamma;\m{q},\m{w}})\dif{\m{q}}
	&\!=\!\ent{\gamma}{\m{q}}{\m{w}}-\gamma\varepsilon.
	%\\
	%\label{eq:lem:OLRTest:SI}
	%&\!=\!\int_{0}^{\gamma}\!
	%\left[\m{w}[\der{\m{q;\sim}}{\m{w}}\!>\!\tau]\!-\!\varepsilon\right]
	%\dif{\tau},
	%\\
	%\label{eq:lem:OLRTest:CI}
	%&\!=\!\int_{0}^{\infty}\!
	%\abp{\m{w}[\der{\m{q;\sim}}{\m{w}}\!>\!\tau]\!-\!\varepsilon}
	%\!\dif{\tau},
\end{align}			
\end{subequations}	
\end{lemma}
\begin{proof}
If \(\m{w}\perp\m{q}\), then \(\norm{\m{w;\sim}}\!=\!0\)
and the lemma is void. 
If \(\m{w}\not\perp\m{q}\) then 
and \(\varepsilon\in(0,\norm{\m{w;\sim}})\), then 
\(\left[\lscq{\varepsilon}{\m{w}}{\m{q}},\uscq{\varepsilon}{\m{w}}{\m{q}}\right]\)
is a non-empty subset of \((0,\infty)\) by Lemma \ref{lem:LRQ}.
Thus \eqref{eq:lem:LRQUpperBound},  \eqref{eq:lem:LRQLowerBound}, and \eqref{eq:lem:LRQ:StrictlyMonone}
imply 
%\(\m{w}[\f{\ylr;\m{w},\m{q}}\!<\!\sfrac{1}{\gamma}]\!\leq\!\varepsilon
%\!\leq\!\m{w}[\f{\ylr;\m{w},\m{q}}\!\leq\!\sfrac{1}{\gamma}]\). 
%Thus using \eqref{eq:lrf} and 
%\(\left[\lscq{\varepsilon}{\m{w}}{\m{q}},\uscq{\varepsilon}{\m{w}}{\m{q}}\right]\subset(0,\infty)\) 
%we get
\begin{align}
\notag
\m{w}[\f{\ylr;\m{w},\m{q}}\!<\!\sfrac{1}{\gamma}]
\!\leq\!\varepsilon&\!\leq\!
\m{w}[\f{\ylr;\m{w},\m{q}}\!\leq\!\sfrac{1}{\gamma}].
&
&
%\forall\gamma\!\in\!\left[\tfrac{1}{\uscq{\varepsilon}{\m{w}}{\m{q}}},\tfrac{1}{\lscq{\varepsilon}{\m{w}}{\m{q}}}\right].
%\end{align}
\intertext{Thus using \eqref{eq:lrf} and 
\(\left[\lscq{\varepsilon}{\m{w}}{\m{q}},\uscq{\varepsilon}{\m{w}}{\m{q}}\right]\subset(0,\infty)\) 
we get }
%	%\begin{align}
\notag
0\!\leq\!\varepsilon-\m{w}[\f{\ylr;\m{q},\m{w}}\!>\!\gamma]
&\!\leq\!
\m{w}[\f{\ylr;\m{q},\m{w}}\!=\!\gamma].
&
&
%\forall\gamma\!\in\!\left[\tfrac{1}{\uscq{\varepsilon}{\m{w}}{\m{q}}},\tfrac{1}{\lscq{\varepsilon}{\m{w}}{\m{q}}}\right].		
\end{align}
Then \(\rv{T;\varepsilon,\gamma;\m{q},\m{w}}\!\in\![0,1]\) and 
\(\rv{T;\varepsilon,\gamma;\m{q},\m{w}}\!\in\!\tmea{\outA}\). 
Furthermore,
\begin{align}
	\notag
	\int\!\rv{T;\varepsilon,\gamma;\m{q},\m{w}}\dif{\m{w}}
	&\!\mathop{=}^{(a)}\! 	
	\left(\varepsilon-\m{w}[\f{\ylr;\m{q},\m{w}}>\gamma]\right)
	\!+\!\int\!\IND{\f{\ylr;\m{q},\m{w}}>\gamma}\dif{\m{w}},
	\\
	\notag
	&\!\mathop{=}^{}\!
	\varepsilon.
	\\
	\notag
	\int\!(1\!-\!\rv{T;\varepsilon,\gamma;\m{q},\m{w}})\dif{\m{q}}
	&\!\mathop{=}^{(b)}\! 
	\left(\m{w}[\f{\ylr;\m{q},\m{w}}\geq\gamma]-\varepsilon\right)
	\gamma 
	\!+\!\int\!\IND{\f{\ylr;\m{q},\m{w}}<\gamma}\dif{\m{q}},
	\\
	\notag
	&\!\mathop{=}^{(c)}\!
	\ent{\gamma}{\m{q}}{\m{w}}-\gamma\varepsilon,
\end{align}		
where \((a)\) and \((b)\) follow the definitions
from \eqref{eq:def:lr} and \eqref{eq:lem:OLRTest:def};
\((c)\) follows \eqref{eq:EntropySpectrum:Sum},
with the roles of \(\m{w}\) and \(\m{q}\) interchanged.	
\end{proof}

\subsection{A Spectral Neyman--Pearson Lemma}\label{sec:ASpectralNeymanPearsonLemma}
\begin{theorem}\label{thm:SNPL}
If a pair of \(\sigma\)-finite measures \((\m{w},\m{q})\)
satisfies \eqref{eq:assumption:nonvoid}, i.e., 
\(\ent{1}{\m{w}}{\m{q}}\!<\!\infty\),
\begin{subequations}\label{eq:thm:SNPL}
then the restriction of \(\NPD{\cdot}{\m{q}}{\m{w}}\) 
to \([0,\infty)\) is a non-increasing, convex, and 
continuous function satisfying
\begin{align}
\label{eq:thm:SNPL:DualCharacterization}
\NPD{\varepsilon}{\m{q}}{\m{w}}
&=\sup\nolimits_{\gamma\geq 0}
\ent{\gamma}{\m{q}}{\m{w}}-\gamma\varepsilon
&
&\forall\varepsilon\!\in\!\numbers{R},
\\
\label{eq:thm:SNPL:DualCharacterization:max}
&=\ent{\gamma_{\varepsilon}}{\m{q}}{\m{w}}-\gamma_{\varepsilon}\varepsilon
&
&\forall\varepsilon\!>\!0,
\\
\label{eq:thm:SNPL:SI}
&\!=\!\int_{0}^{\gamma_{\varepsilon}}\!
\left[\m{w}[\der{\m{q;\sim}}{\m{w}}\!>\!\tau]\!-\!\varepsilon\right]
\dif{\tau}
&
&\forall\varepsilon\!\geq\!0,
\\
\label{eq:thm:SNPL:CI}
&\!=\!\int_{0}^{\infty}\!
\abp{\m{w}[\der{\m{q;\sim}}{\m{w}}\!>\!\tau]\!-\!\varepsilon}
\!\dif{\tau}
&
&\forall\varepsilon\!\in\!\numbers{R},
\\
\label{eq:thm:SNPL:leftderivative}	
\lder{\NPD{\varepsilon}{\m{q}}{\m{w}}}	
&\!=\!\tfrac{-1}{\lscq{\varepsilon}{\m{w}}{\m{q}}}
&
&\forall\varepsilon\!>\!0,
\\
\label{eq:thm:SNPL:rightderivative}
\rder{\NPD{\varepsilon}{\m{q}}{\m{w}}}	
&\!=\!\tfrac{-1}{\uscq{\varepsilon}{\m{w}}{\m{q}}}
&
&\forall\varepsilon\!\geq\!0,
\end{align}
for  any \(\gamma_{\varepsilon}\!\in\!\left[\tfrac{1}{\uscq{\varepsilon}{\m{w}}{\m{q}}},\tfrac{1}{(\lscq{\varepsilon}{\m{w}}{\m{q}}\vee0)}\right]\!\).
Furthermore, the restriction of \(\NPD{\cdot}{\m{q}}{\m{w}}\) to \((0,\norm{\m{w;\sim}})\)
is an invertible function with the image \((0,\norm{\m{q;\sim}})\)
and there exists a \(\rv{T;\varepsilon}\!\in\!\tmea{\outA}\) 
satisfying both 
\(\int\rv{T;\varepsilon} \dif{\m{w}}\!=\!\varepsilon\) and
\(\int(1\!-\!\rv{T;\varepsilon})\dif{\m{q}}\!=\!\NPD{\varepsilon}{\m{q}}{\m{w}}\)
for all \(\varepsilon\!\in\![0,\norm{\m{w;\sim}}]\).
\begin{align}
\label{eq:thm:SNPL:IOSD:1}
\NPD{\varepsilon}{\m{q}}{\m{w}}
&=\int_{\varepsilon}^{\norm{\m{w;\sim}}}
\f{g}[\tau]\dif{\tau}
&
&\forall\varepsilon\!\in\![0,\norm{\m{w;\sim}}],
\\
\label{eq:thm:SRBeta:FiniteAndNegative}
\partial\NPD{\varepsilon}{\m{q}}{\m{w}}
&\!\subset\!(-\infty,0)
&	
&\forall\varepsilon\!\in\!(0,\norm{\m{w;\sim}}),
\intertext{for any  \(\f{g}\!:\!(0,\norm{\m{w;\sim}})\!\mapsto\!(0,\infty)\) s.t.  \(\tfrac{1}{\uscq{\cdot}{\m{w}}{\m{q}}}\!\leq\!\f{g}[\cdot]\!\leq\!\tfrac{1}{\lscq{\cdot}{\m{w}}{\m{q}}}\). 
If \(\m{q;\sim}\) is a finite measure,  i.e.,\(\norm{\m{q;\sim}}<\infty\), then}
\addtocounter{equation}{1}
\label{eq:thm:SNPL:IOSD:2}
\NPD{\varepsilon}{\m{q}}{\m{w}}
&=\norm{\m{q;\sim}}-\int_{0}^{\varepsilon}\f{g}[\tau]\dif{\tau}
&
&\forall\varepsilon\!\geq\!0.		
\end{align}
\end{subequations}
Furthermore, \eqref{eq:thm:SNPL:DualCharacterization} and 
\eqref{eq:thm:SNPL:CI} hold even when
\eqref{eq:assumption:nonvoid} is violated, i.e., 
when \(\ent{1}{\m{w}}{\m{q}}\!=\!\infty\).
\end{theorem}			
In \eqref{eq:thm:SNPL:SI} and \eqref{eq:thm:SNPL:CI},
the `\(>\)' signs can be replaced by `\(\geq\)' signs
because monotonic functions have at most countably 
many discontinuities by \cite[Theorem 4.30]{rudin} 
and any countable set has zero Lebesgue measure.
\begin{subequations}\label{eq:thm:SNPL:DF}Hence
\eqref{eq:def:lr} and \eqref{eq:lrf} imply 
\begin{align}
\label{eq:thm:SNPL:SI:DF}
\NPD{\varepsilon}{\m{q}}{\m{w}}
&\!=\!\int_{0}^{\gamma_{\varepsilon}}\!
\left[\scq{\tfrac{1}{\tau}}{\m{w}}{\m{q}}\!-\!\varepsilon\right]
\dif{\tau}
&
&\forall\varepsilon\!\geq\!0,
\\
\label{eq:thm:SNPL:CI:DF} 
&\!=\!\int_{0}^{\infty}\!
\abp{\scq{\tfrac{1}{\tau}}{\m{w}}{\m{q}}\!-\!\varepsilon}
\!\dif{\tau}
&
&\forall\varepsilon\!\in\!\numbers{R}.
\end{align}	
\end{subequations}
As an immediate consequence of Theorem \ref{thm:SNPL},
we can express the convex conjugate 
\(\NPD[*]{\cdot}{\m{q}}{\m{w}}\) of \(\NPD{\cdot}{\m{q}}{\m{w}}\),
as defined in \eqref{eq:def:ConvexConjugate},
in terms of the primitive entropy spectrum,
as follows
\begin{align}
	\label{eq:Beta:ConvexConjugate}	
	\NPD[*]{\lambda}{\m{q}}{\m{w}}
	&=\begin{cases}
		-\ent{-\lambda}{\m{q}}{\m{w}}
		&\text{if~}\lambda\leq0,
		\\
		\infty 
		&\text{if~}\lambda>0.
	\end{cases}
\end{align}
On the other hand, 
the lower semicontinuous inverse 
\(\NPD[\wedge]{\cdot}{\m{q}}{\m{w}}\) is defined via 
\eqref{eq:def:SCInverse:LSC:nonincreasing},
because \(\NPD{\cdot}{\m{q}}{\m{w}}\) is non-increasing.
\(\NPD[\wedge]{\cdot}{\m{q}}{\m{w}}\) 
can be determined using injectivity of
\(\NPD{\cdot}{\m{q}}{\m{w}}\) on \([0,\norm{\m{w;\sim}}]\)
and \eqref{eq:NPD:Obvious}, to 
be\footnote{One can define upper semicontinuous inverse
	of \(\NPD{\cdot}{\m{q}}{\m{w}}\) using  
	\eqref{eq:def:SCInverse:USC:nonincreasing}, as well. 
	The value of semicontinuous inverses may 
	differ only at zero:
	the upper-semicontinuous inverse will have the value \(\infty\),
	and the lower-semicontinuous will have the value
	\(\norm{\m{w;\sim}}\) at zero.} 
\begin{align}
	\label{eq:Beta:SCInverse:LSC}		
	\NPD[\wedge]{\cdot}{\m{q}}{\m{w}}
	&=\NPD{\cdot}{\m{w}}{\m{q}},
\end{align}
where \(\NPD{\cdot}{\m{w}}{\m{q}}\) is defined via 
\eqref{eq:def:NPDr}, as 
\begin{subequations}\label{eq:NDPInversion}well. 
Thus\footnote{If \(\norm{\m{w;\sim}}\!<\!\infty\), then 
\(\norm{\m{w;\sim}}\) in \eqref{eq:NDPInversion:F}
and
\(0\) in \eqref{eq:NDPInversion:B} can be included in the intervals.
If \(\norm{\m{q;\sim}}\!<\!\infty\), then 
\(0\) in \eqref{eq:NDPInversion:F}
and
\(\norm{\m{q;\sim}}\) in \eqref{eq:NDPInversion:B} 
can be included in the intervals.}
\begin{align} 
\label{eq:NDPInversion:F}	
\NPD{\NPD{\varepsilon}{\m{q}}{\m{w}}}{\m{w}}{\m{q}}
&=\varepsilon
&
&\forall \varepsilon\in(0,\norm{\m{w;\sim}}),
\\
\label{eq:NDPInversion:B}	
\NPD{\NPD{\tau}{\m{w}}{\m{q}}}{\m{q}}{\m{w}}
&=\tau
&
&\forall \tau\in(0,\norm{\m{q;\sim}}). 		
\end{align}
That is the rNP-function \(\NPD{\cdot}{\m{q}}{\m{w}}\!:\!(0,\norm{\m{w;\sim}})\mapsto(0,\norm{\m{q;\sim}})\)
and the rNP-function\(\NPD{\cdot}{\m{w}}{\m{q}}\!:\!(0,\norm{\m{q;\sim}})\mapsto(0,\norm{\m{w;\sim}})\)
are \end{subequations}inverses of one another.	

If the likelihood-ratio distribution \(\scq{\cdot}{\m{w}}{\m{q}}\)
is constant on a subinterval of 
\((\uscq{0}{\m{w}}{\m{q}},\lscq{\norm{\m{w;\sim}}}{\m{w}}{\m{q}})\),
%see \eqref{eq:LRQ:Obvious},
%\((\esinf\nolimits_{\m{w;\sim}}\f{\ylr;\m{w},\m{q}},\essup\nolimits_{\m{w;\sim}}\f{\ylr;\m{w},\m{q}})\)
%\eqref{eq:LRQ:Obvious:Singular:LSC}	
%\eqref{eq:LRQ:Obvious:Zero:USC}	
%
then the restriction of \(\scq{\cdot}{\m{w}}{\m{q}}\) 
to \((\uscq{0}{\m{w}}{\m{q}},\lscq{\norm{\m{w;\sim}}}{\m{w}}{\m{q}})\) 
is not invertible and the left and right derivatives of 
\(\NPD{\cdot}{\m{q}}{\m{w}}\) differ at some point in \((0,\norm{\m{w;\sim}})\),
as a result of \eqref{eq:thm:SNPL:leftderivative} and \eqref{eq:thm:SNPL:rightderivative}.
On the other hand, any discontinuities of \(\scq{\cdot}{\m{w}}{\m{q}}\) 
in \((\uscq{0}{\m{w}}{\m{q}},\lscq{\norm{\m{w;\sim}}}{\m{w}}{\m{q}})\),
corresponds to a subinterval of \((0,\norm{\m{w;\sim}})\) 
on which \(\lscq{\cdot}{\m{w}}{\m{q}}\) and \(\uscq{\cdot}{\m{w}}{\m{q}}\) 
are equal to the same constant and 
\(\NPD{\cdot}{\m{q}}{\m{w}}\) is an affine function on the said interval
by  \eqref{eq:thm:SNPL:leftderivative} and \eqref{eq:thm:SNPL:rightderivative}.

\begin{proof}[Proof Theorem \ref{thm:SNPL}]
Let us first prove a lower bound on \(\NPD{\cdot}{\m{q}}{\m{w}}\)
\begin{align}
\notag	
\NPD{\varepsilon}{\m{q}}{\m{w}}
&\!\mathop{=}^{(a)}\! 
\inf\nolimits_{\rv{T}\!\in\!\tmea{\outA}}
\sup\nolimits_{\gamma\geq 0}
\int\!(1\!-\!\rv{T}) \dif{\m{q}}+
\gamma\left(\int\!\rv{T} \dif{\m{w}}\!-\!\varepsilon\right),
\\
\notag
&\!\mathop{\geq}^{(b)}\! 
\sup\nolimits_{\gamma \geq0}
\inf\nolimits_{\rv{T}\!\in\!\tmea{\outA}}
\int\!(1\!-\!\rv{T}) \dif{\m{q}}+
\gamma\left(\int\!\rv{T} \dif{\m{w}}\!-\!\varepsilon\right),
\\
\notag
&\!\mathop{\geq}^{(c)}\! 
\sup\nolimits_{\gamma\geq0}
-\gamma\varepsilon\!+\!
\inf\nolimits_{\rv{T}\!\in\!\tmea{\outA}}
\int\!\left[
(1\!-\!\rv{T})\der{\m{q;\sim}}{\m{w}}
\!+\!\gamma\rv{T}\right]\dif{\m{w}},
\\
\notag
&\!\mathop{=}^{(d)}\! 
\sup\nolimits_{\gamma\geq0}
-\gamma\varepsilon
\!+\!\int
\left(\der{\m{q;\sim}}{\m{w}}
\wedge \gamma\right)
\dif{\m{w}},
\\
\label{eq:thm:SNPL:1}
&\!\mathop{=}^{}\! 
\sup\nolimits_{\gamma\geq0}
\ent{\gamma}{\m{q}}{\m{w}}-\gamma\varepsilon.
\end{align}		
where 
\((a)\) follows from \eqref{eq:def:NPDr};
\((b)\) follows from max-min inequality;
\((c)\) holds because both \(\rv{T}\!\leq\!1\) 
and \(\der{\m{q;\sim}}{\m{q}}\!\leq\!1\) hold \(\m{q}\)-a.e.; 
\((d)\) follows from the fact that \(\rv{T}:\outS\mapsto[0,1]\)
for tests in \(\tmea{\outA}\).

If \(\varepsilon\in(0,\norm{\m{w;\sim}})\), then
\eqref{eq:thm:SNPL:DualCharacterization}  
and
\eqref{eq:thm:SNPL:DualCharacterization:max}
follow from  \eqref{eq:lem:OLRTest} of Lemma \ref{lem:OLRTest}
and \eqref{eq:thm:SNPL:1}.
Then \eqref{eq:thm:SNPL:SI} follows from
\eqref{eq:lem:PES:integral:r}
with the roles of \(\m{w}\) and \(\m{q}\) interchanged
and \eqref{eq:thm:SNPL:DualCharacterization:max}.
Furthermore, \eqref{eq:thm:SNPL:CI} follow from 
\eqref{eq:thm:SNPL:DualCharacterization} 
and \eqref{eq:thm:SNPL:SI} because
the integrand in \eqref{eq:thm:SNPL:SI}
is non-negative for all 
\(\tau\!<\!\sfrac{1}{\lscq{\varepsilon}{\m{w}}{\m{q}}}\)
by \eqref{eq:lrf} and \eqref{eq:lem:LRQUpperBound};
and non-positive for all 
\(\tau\!>\!\sfrac{1}{\uscq{\varepsilon}{\m{w}}{\m{q}}}\)
by \eqref{eq:lrf} and \eqref{eq:lem:LRQLowerBound}.

If \(\varepsilon\geq \norm{\m{w;\sim}}\), then 
\eqref{eq:thm:SNPL:DualCharacterization} and
\eqref{eq:thm:SNPL:DualCharacterization:max}, follow from 
\eqref{eq:NPD:Obvious:Singular} and \eqref{eq:thm:SNPL:1}
because \(\ent{0}{\m{q}}{\m{w}}=0\) by \eqref{eq:lem:PES:zero}.
Furthermore, \eqref{eq:thm:SNPL:CI}
follows from 
\(\m{w}[\der{\m{q;\sim}}{\m{w}}\!\geq\!0]\leq\norm{\m{w;\sim}}\).
For \(\varepsilon\!=\!\norm{\m{w;\sim}}\) case
\eqref{eq:thm:SNPL:SI}
follows from 
\eqref{eq:def:lr}, \eqref{eq:lrf},
\eqref{eq:LRQ:Obvious:Singular:LSC}	
and
\eqref{eq:LRQ:Obvious:Singular:USC}.
For \(\varepsilon>\norm{\m{w;\sim}}\) case,
\eqref{eq:thm:SNPL:SI} holds
because 
\(\lscq{\varepsilon}{\m{w}}{\m{q}}
\!=\!
\uscq{\varepsilon}{\m{w}}{\m{q}}
\!=\!\infty\).

For \(\varepsilon\!=\!0\) case 
\eqref{eq:thm:SNPL:DualCharacterization}
and
\eqref{eq:thm:SNPL:CI}
follow from applying
\eqref{eq:lem:PES:integral:r}
and
\eqref{eq:lem:PES:infinity}
of Lemma \ref{lem:PES},
to \(\ent{\gamma}{\m{q}}{\m{w}}\) rather than
\(\ent{\gamma}{\m{w}}{\m{q}}\).
For \(\varepsilon\!=\!0\) case \eqref{eq:thm:SNPL:SI}
follows from \eqref{eq:def:lr}, \eqref{eq:lrf},
\eqref{eq:LRQ:Obvious:Zero:LSC},
and \eqref{eq:LRQ:Obvious:Zero:USC}.	

\(\NPD{\cdot}{\m{q}}{\m{w}}\) is convex, non-increasing, and 
lower-semicontinous  because it is the supremum of such functions
by \eqref{eq:thm:SNPL:DualCharacterization}. 
Then \(\lim_{\varepsilon\downarrow0}
\NPD{\varepsilon}{\m{q}}{\m{w}}
=\NPD{0}{\m{q}}{\m{w}}\)
by   the lower-semicontinuity and monotonicity
of \(\NPD{\cdot}{\m{q}}{\m{w}}\).
Furthermore, \(\NPD{\cdot}{\m{q}}{\m{w}}\) is continuous
on \((0,\infty)\) because
any convex function is continuous on the interior of
the region that it is finite by \cite[Theorem 6.3.4]{dudley}. 
and 
\(\NPD{\varepsilon}{\m{q}}{\m{w}}<\infty\) for all \(\varepsilon>0\) 
by Lemma \ref{lem:OLRTest} and \eqref{eq:EntropySpectrum:Bound}.

One can prove the remaining assertions of the theorem,
via measure theoretic arguments 
using \eqref{eq:thm:SNPL:SI}.
We, however, we will rely on the conjugacy relation
implied by \eqref{eq:thm:SNPL:DualCharacterization} 
to prove those assertions. 
Let \(\f{f}[\cdot]:\numbers{R}\mapsto(-\infty,\infty]\) be
\begin{align}
\label{eq:thm:SNPL:2}
\f{f}[\dinp]
&\DEF\begin{cases}
	-\ent{-\dinp}{\m{q}}{\m{w}}	
	&\text{if~}\dinp\leq 0,
	\\
	\infty 
	&\text{if~}\dinp>0.
\end{cases}		
\end{align}
Then \(\f{f;;*}[\cdot]\!=\!\NPD{\cdot}{\m{q}}{\m{w}}\),
i.e., \(\NPD{\cdot}{\m{q}}{\m{w}}\) is the convex conjugate
of \(\f{f}[\cdot]\),
as a result of \eqref{eq:thm:SNPL:DualCharacterization}
and the definition convex conjugation given in  
\eqref{eq:def:ConvexConjugate}.
Furthermore, \(\f{f}[\cdot]\) is a closed convex function 
(i.e., lower semicontinuous and convex function not identically equal to \(\infty\)) by Lemma \ref{lem:PES}.
Then \(\NPD{\cdot}{\m{q}}{\m{w}}\) is a closed convex function 
by \cite[Theorem E.1.1.2]{hiriart-urrutyLemarechal}
and convex conjugate of \(\NPD{\cdot}{\m{q}}{\m{w}}\)
is \(\f{f}[\cdot]\), i.e., 
\(\NPD[*]{\cdot}{\m{q}}{\m{w}}=\f{f}[\cdot]\) by 
%\cite[Proposition B.2.5.1]{hiriart-urrutyLemarechal}
\cite[Theorem E.1.3.5]{hiriart-urrutyLemarechal}.
Furthermore, as a result of 
\cite[Corollary E.1.4.4]{hiriart-urrutyLemarechal},
\cite[Proposition I.6.1.2]{hiriart-urrutyLemarechal-I}.
\begin{align}
\notag
\dinp&\!\in\!\partial\NPD{\varepsilon}{\m{q}}{\m{w}}
&
&\Longleftrightarrow
&
\varepsilon&\!\in\!\partial\f{f}[\dinp].
\intertext{Thus as result of \eqref{eq:thm:SNPL:2} 
	for all \(\varepsilon>0\) we have}
\notag
-\lambda&\!\in\!\partial\NPD{\varepsilon}{\m{q}}{\m{w}}
&
&\Longleftrightarrow
&
\varepsilon&\!\in\!
\left[\rder{\ent{\lambda}{\m{q}}{\m{w}}},\lder{\ent{\lambda}{\m{q}}{\m{w}}}\right].
\intertext{Thus as result of Lemma \ref{lem:PES}
	for all \(\varepsilon>0\) we have}
\notag
-\lambda&\!\in\!\partial\NPD{\varepsilon}{\m{q}}{\m{w}}
&
&\Longleftrightarrow
&
\varepsilon&\!\in\!
\left[\m{w}[\der{\m{q;\sim}}{\m{w}}\!>\!\lambda],
\m{w}[\der{\m{q;\sim}}{\m{w}}\!\geq\!\lambda]\right].
\end{align}
Then for \(\varepsilon\!>\!0\)
\eqref{eq:thm:SNPL:leftderivative}	and
\eqref{eq:thm:SNPL:rightderivative}	
follow from \eqref{eq:lrf} and Lemma \ref{lem:LRQ}
and for \(\varepsilon\!=\!0\) \eqref{eq:thm:SNPL:rightderivative}   
follows from \eqref{eq:lem:LRQ:rightlimit} and 
because \(\rder{\NPD{\cdot}{\m{q}}{\m{w}}}\)
is continuous from the right by
\cite[Theorem I.4.2.1]{hiriart-urrutyLemarechal-I}.

\eqref{eq:thm:SRBeta:FiniteAndNegative}
follows from \eqref{eq:lem:LRQ:StrictlyMonone},
\eqref{eq:thm:SNPL:leftderivative},
and
\eqref{eq:thm:SNPL:rightderivative}.
Furthermore, 
\eqref{eq:thm:SNPL:IOSD:1} 
and \eqref{eq:thm:SNPL:IOSD:2} 
follow from \eqref{eq:ConvexFunctionIR}, \eqref{eq:NPD:Obvious},
\eqref{eq:thm:SNPL:leftderivative},
and
\eqref{eq:thm:SNPL:rightderivative}.
Note that 
\eqref{eq:thm:SNPL:leftderivative},
\eqref{eq:thm:SNPL:rightderivative},
\eqref{eq:thm:SNPL:IOSD:1},
and \eqref{eq:thm:SRBeta:FiniteAndNegative}
imply that \(\NPD{\cdot}{\m{q}}{\m{w}}\) is 
decreasing and injective on \([0,\norm{\m{w;\sim}}]\).

\eqref{eq:thm:SNPL:DualCharacterization}
continues to hold when \eqref{eq:assumption:nonvoid}
is violated because 
\(\ent{\varepsilon}{\m{w}}{\m{q}}\!=\!\infty\) 
for all \(\varepsilon\geq0\)
and 
\(\NPD{\gamma}{\m{q}}{\m{w}}\!=\!\infty\)
for all \(\gamma>0\) in that case. 
Then \eqref{eq:thm:SNPL:CI} holds 
as result of \eqref{eq:lem:PES:integral:r}
and \eqref{eq:EntropySpectrum:Void}.		
\end{proof}

The state-of-the-art bound
relating \(\uscq{\cdot}{\m{w}}{\m{q}}\) and  \(\NPD{\cdot}{\m{q}}{\m{w}}\)
for arbitrary probability measures \(\m{w}\) and \(\m{q}\), i.e.,
\eqref{eq:Tan-2-4}, can be improved slightly and extended 
to the case when \(\m{w}\) and \(\m{q}\) are \(\sigma\)-finite measures
using  Theorem \ref{thm:SNPL}.
Since \(\uscq{\cdot}{\m{w}}{\m{q}}\) is non-decreasing by Lemma \ref{lem:LRQ}, 
\eqref{eq:thm:SNPL:IOSD:1} implies
\begin{align}
	\label{eq:Tan-2-4-UB-Improved}
	\tfrac{\abp{\norm{\m{w;\sim}}-\varepsilon}}{\uscq{\varepsilon}{\m{w}}{\m{q}}}
	&\geq \NPD{\varepsilon}{\m{q}}{\m{w}}
	&
	&\forall \varepsilon>0.
\end{align}
On the other hand, since \(\NPD{\cdot}{\m{w}}{\m{q}}\) is convex
its supporting line any point and hence at \(\varepsilon\!+\!\delta\) 
lies below \(\NPD{\cdot}{\m{w}}{\m{q}}\) itself. Thus
for any \(\gamma\) in \(\left[\sfrac{1}{\uscq{\varepsilon+\delta}{\m{w}}{\m{q}}},\sfrac{1}{\lscq{\varepsilon+\delta}{\m{w}}{\m{q}}}\right]\),
\eqref{eq:thm:SNPL:leftderivative}, and \eqref{eq:thm:SNPL:rightderivative} 
imply 
\begin{align}
	\label{eq:Tan-2-4-LB-tangent}	
	\NPD{\varepsilon}{\m{q}}{\m{w}}
	&\!\geq\!\delta \gamma\!+\!\NPD{\varepsilon+\delta}{\m{w}}{\m{q}}
	&
	&\forall \delta\!\in\!(0,\norm{\m{w;\sim}}\!-\!\varepsilon).		
\end{align}
	
\begin{remark}\label{rem:NPDvsNPDdWorstcase}
Theorem \ref{thm:SNPL} characterizes \(\NPD{\cdot}{\m{q}}{\m{w}}\) 
rather than \(\NPDd{\cdot}{\m{q}}{\m{w}}\). 
Nevertheless, we can bound \(\NPDd{\cdot}{\m{q}}{\m{w}}\) 
in terms of \(\NPD{\cdot}{\m{q}}{\m{w}}\) 
using\begin{subequations}\label{eq:NPDvsNPDdWorstcase} 
Lemma \ref{lem:OLRTest}, deterministic likelihood ratio 
tests, \eqref{eq:lem:LRQLowerBound}, and \eqref{eq:NPDleqNPDd}:
\begin{align}
	\label{eq:NPDvsNPDdWorstcase:Bound} 
	\NPD{\varepsilon}{\m{q}}{\m{w}}	
	&\!\leq\!
	\NPDd{\varepsilon}{\m{q}}{\m{w}}
	\!\leq\! 
	\NPD{\varepsilon}{\m{q}}{\m{w}}	
	\!+\!\f{\zeta}[\varepsilon]
	&
	&\forall \varepsilon\!\in\!(0,\norm{\m{w;\sim}}),
	\\
	\label{eq:NPDvsNPDdWorstcase:Error} 
	\f{\zeta}[\varepsilon]
	&\DEF\inf\nolimits_{\mathcal{A}\in\mathfrak{B}}\max\nolimits_{\oev\in\mathcal{A}}\m{q}[\oev]
	&
	&\forall \varepsilon\!\in\!(0,\norm{\m{w;\sim}}),
\end{align}
where \(\mathfrak{B}\) is \end{subequations}the 		
set of all \(\outA\)-measurable finite partitions of the set 
\(\{\dout\!\in\!\outS\!:\!\der{\m{w;\sim}}{\m{q}}(\dout)\!=\!\uscq{\varepsilon}{\m{w}}{\m{q}}\}\).
If \(\m{w;\sim}\) does not have any atoms then \(\f{\zeta}[\varepsilon]\!=\!0\)
for all \(\varepsilon\!\in\!(0,\norm{\m{w;\sim}})\).
If  \(\outA\) is a countably separated \(\sigma\)-algebra, then all singletons are in \(\outA\)
by \cite[Theorem 6.5.7]{bogachev}. 
If in addition all atoms of \(\m{w;\sim}\) are singletons,
then \(\f{\zeta}[\varepsilon]\) has the following less daunting expression
\begin{align}
	\notag
	\f{\zeta}[\varepsilon]
	&\!=\!\max\left\{\m{q}[\{\dout\}]:\der{\m{w;\sim}}{\m{q}}(\dout)\!=\!\uscq{\varepsilon}{\m{w}}{\m{q}}\right\}
	&
	&\forall \varepsilon\!\in\!(0,\norm{\m{w;\sim}}).
\end{align}
The use of \(\max\) is justified because we are actually taking a supremum over countable set 
for which supremum is achieved even for the case when there are uncountably many 
\(\dout\)'s in \(\outS\) satisfying \(\der{\m{w;\sim}}{\m{q}}(\dout)\!=\!\uscq{\varepsilon}{\m{w}}{\m{q}}\)
by \cite[Theorem 4.30]{rudin}.
Note that \eqref{eq:NPDvsNPDdWorstcase} was already pointed by Strassen, 
see \cite[(2.7)]{strassen62} and the following unnumbered equation in \cite{strassen62}. 
It is wort mentioning that the upper bound in \eqref{eq:NPDvsNPDdWorstcase} is not tight 
even for deterministic likelihood ratio tests:
if \(\m{w}\) and \(\m{q}\) are geometric probability mass functions with 
parameter \(\sfrac{1}{2}\) supported on positive and non-negative integers, respectively,
then \(\der{\m{w}}{\m{q}}(\dout)\!=\!2\) hold \(\m{w}\)-a.e. and
\(\NPD{\varepsilon}{\m{q}}{\m{w}}\!=\!\NPDd{\varepsilon}{\m{q}}{\m{w}}\!=\!\abp{\tfrac{1-\varepsilon}{2}}\)
for all \(\varepsilon\!\geq\!0\); yet 
\(\f{\zeta}[\varepsilon]\!=\!\sfrac{1}{2}\) for all \(\varepsilon\!\in\!(0,1)\). 
Many of these observations, as well as some more advanced ideas going beyond
likelihood ratio tests are discussed in \cite[Appendix B]{polyanskiythesis}.		
\end{remark}
\subsection{Measure Change Identities}\label{sec:MeasureChangeIdentities}
\begin{assumption}\label{assumption:MeasureChange} 
	\(\m{w}\), \(\m{q}\), and \(\m{v}\) are \(\sigma\)-finite measures satisfying
	both \(\ent{1}{\m{w}}{\m{q}}\!\in\!(0,\infty)\) and \(\m{v}\sim\m{w;\sim}\), 
	and\begin{subequations}\label{eq:assumption:MeasureChange} 
		there exist a function \(\f{f}\) of the form
		\(\f{f}\!:\!(0,\infty)\!\mapsto\!(0,\infty)\)
		satisfying
		\begin{align}
			\label{eq:assumption:MeasureChange:RNDerivative}	
			\der{\m{v}}{\m{q}}
			&=\f{f}[\der{\m{w;\sim}}{\m{q}}]	
			&
			&\m{w;\sim}\text{-a.e.},
		\end{align}
		where
		\(\m{w;\sim}\) is the \(\m{q}\)-absolutely continuous component 
		of \(\m{w}\).
		Furthermore, \(\fav{\varepsilon}{\m{v}}\) 
		and \(\mdv{\varepsilon}{\m{v}}\) are defined in terms of
		\(\m{v}\) and \(\rv{T;\varepsilon,\gamma;\m{q},\m{w}}\) defined 
		in \eqref{eq:lem:OLRTest:def} 
		for any  \(\gamma\!\in\!\left[\sfrac{1}{\uscq{\varepsilon}{\m{w}}{\m{q}}},\sfrac{1}{\lscq{\varepsilon}{\m{w}}{\m{q}}}\right]\!\) 
		as
		\begin{align}
			\label{eq:assumption:MeasureChange:FalseAlarm}	
			\fav{\varepsilon}{\m{v}}
			&\DEF 
			\displaystyle{\int}
			\rv{T;\varepsilon,\gamma;\m{q},\m{w}}\dif{\m{v}}
			&
			&\forall\varepsilon\!\in\!\left(0,\norm{\m{w;\sim}}\right),
			\\
			\label{eq:assumption:MeasureChange:MissedDetection}	
			\mdv{\varepsilon}{\m{v}}
			&\DEF 
			\displaystyle{\int}
			\left(1\!-\!\rv{T;\varepsilon,\gamma;\m{q},\m{w}}\right)
			\dif{\m{v}}
			&
			&\forall\varepsilon\!\in\!\left(0,\norm{\m{w;\sim}}\right).
		\end{align}
	\end{subequations}
\end{assumption}
Note that \eqref{eq:assumption:MeasureChange:FalseAlarm}	
and \eqref{eq:assumption:MeasureChange:MissedDetection}	
imply
\begin{align}
	\label{eq:MeasureChange}
	\fav{\varepsilon}{\m{v}}
	+
	\mdv{\varepsilon}{\m{v}}
	&=\norm{\m{v}}
	&
	&\forall\varepsilon\!\in\!\left(0,\norm{\m{w;\sim}}\right).
\end{align}
The Csiszar-dual 
\(\cdual{f}\!:\!(0,\infty)\!\mapsto\!\numbers{R}\)
of a function \(\f{f}\) of the form 
\(\f{f}\!:\!(0,\infty)\!\mapsto\!\numbers{R}\)
is defined as, 
see \cite[(2)]{reidW11},
\cite[(29)]{topsoe04}.
\begin{align}
	\label{eq:def:CsiszarDual}	
	\cdual{f}[\dinp]
	&\DEF\dinp\f{f}[\tfrac{1}{\dinp}]
	&
	&\forall\dinp\!\in\!(0,\infty).
\end{align}
\begin{theorem}\label{thm:MeasureChange}
	Let measures \(\m{w}\), \(\m{q}\), \(\m{v}\) and
	function \(\f{f}[\cdot]\)
	satisfy\begin{subequations}\label{eq:thm:MeasureChange} 
		Assumption \ref{assumption:MeasureChange}. 
		Then the Csiszar-dual \(\cdual{f}\) 
		defined in \eqref{eq:def:CsiszarDual} satisfies
		\begin{align}
			\label{eq:thm:MeasureChange:RNDerivative} 	
			\der{\m{v}}{\m{w}}
			&=\cdual{f}[\der{\m{q;\sim}}{\m{w}}]	
			&
			&\m{q;\sim}\text{-a.e.}
			\intertext{If \(\f{f}[\cdot]\) is increasing and 
				bijective and \(\ent{1}{\m{v}}{\m{q}}\!<\!\infty\), then}
			\label{eq:thm:MeasureChange:Increasing}	
			\NPD{\varepsilon}{\m{q}}{\m{w}}
			&\!=\!\NPD{\fav{\varepsilon}{\m{v}}}{\m{q}}{\m{v}}
			&
			&\forall\varepsilon\!\in\!\left(0,\norm{\m{w;\sim}}\right).
			\intertext{If \(\f{f}[\cdot]\) is decreasing and 
				bijective and \(\norm{\m{q;\sim}}\!<\!\infty\), then}
			\label{eq:thm:MeasureChange:Decreasing}	
			\NPD{\varepsilon}{\m{q}}{\m{w}}
			&\!=\!\norm{\m{q;\sim}}\!-\!\NPD{\mdv{\varepsilon}{\m{v}}}{\m{q}}{\m{v}}
			&
			&
			\forall\varepsilon\!\in\!\left(0,\norm{\m{w;\sim}}\right).
			\intertext{If \(\cdual{f}[\cdot]\) is increasing and 
				bijective and \(\ent{1}{\m{v}}{\m{w}}\!<\!\infty\), then}
			\label{eq:thm:MeasureChange:CsiszarDual:Increasing}	
			\varepsilon
			&\!=\!\NPD{\mdv{\varepsilon}{\m{v}}}{\m{w}}{\m{v}}
			&
			&
			\forall\varepsilon\!\in\!\left(0,\norm{\m{w;\sim}}\right).
			\intertext{If \(\cdual{f}[\cdot]\) is decreasing and
				bijective, and \(\norm{\m{w;\sim}}\!<\!\infty\), then}
			\label{eq:thm:MeasureChange:CsiszarDual:Decreasing}	
			\varepsilon
			&\!=\!\norm{\m{w;\sim}}\!-\!\NPD{\fav{\varepsilon}{\m{v}}}{\m{w}}{\m{v}}
			&
			&\forall\varepsilon\!\in\!\left(0,\norm{\m{w;\sim}}\right).
		\end{align}
	\end{subequations}	
\end{theorem}
\begin{proof}[Proof of Theorem \ref{thm:MeasureChange}]
	First note that, 
	\begin{align}
		\notag
		\der{\m{v}}{\m{w}}
		&=\der{\m{q;\sim}}{\m{w}}\der{\m{v}}{\m{q;\sim}}
		&
		&\m{v}\text{-a.e.},
		\\
		\notag
		\tfrac{1}{\der{\m{q;\sim}}{\m{w}}}
		&=\der{\m{w;\sim}}{\m{q}}
		&
		&\m{v}\text{-a.e.}.
	\end{align}
	Then \eqref{eq:thm:MeasureChange:RNDerivative} 	
	follows from \eqref{eq:assumption:MeasureChange:RNDerivative} 
	and \eqref{eq:def:CsiszarDual}.
	On the other hand, 
	Lemma \ref{lem:OLRTest} and Theorem \ref{thm:SNPL} imply
	\begin{align}
		\label{eq:MeasureChange:MD}		
		\int(1\!-\!\rv{T;\varepsilon,\gamma;\m{q},\m{w}})\dif{\m{q}}
		&\!=\!\NPD{\varepsilon}{\m{q}}{\m{w}}
		&
		&\forall\varepsilon\!\in\!\left(0,\norm{\m{w;\sim}}\right),
		\\
		\label{eq:MeasureChange:FA}		
		\int \rv{T;\varepsilon,\gamma;\m{q},\m{w}}\dif{\m{w}}
		&\!=\!\varepsilon
		&
		&\forall\varepsilon\!\in\!\left(0,\norm{\m{w;\sim}}\right).
	\end{align}
	Furthermore, \(\gamma\!\in\!(0,\infty)\) 
	for all \(\varepsilon\!\in\!\left(0,\norm{\m{w;\sim}}\right)\) 
	by \eqref{eq:lem:LRQ:StrictlyMonone} and 
	Lemma \ref{lem:OLRTest}.
	
	When \(\f{f}[\cdot]\) is increasing and bijective
	\eqref{eq:lem:OLRTest:def}, 
	\eqref{eq:assumption:MeasureChange:RNDerivative},
	and
	\eqref{eq:assumption:MeasureChange:FalseAlarm}	
	imply \(\sfrac{1}{\f{f}[\sfrac{1}{\gamma}]}
	\in\!\left[\sfrac{1}{\uscq{\fav{\varepsilon}{\m{v}}}{\m{v}}{\m{q}}},\sfrac{1}{\lscq{\fav{\varepsilon}{\m{v}}}{\m{v}}{\m{q}}}\right]
	\) and
	\begin{align}
		\notag	
		\rv{T;\varepsilon,\gamma;\m{q},\m{w}}
		&=\rv{T;\fav{\varepsilon}{\m{v}},\sfrac{1}{\f{f}[\sfrac{1}{\gamma}]};\m{q},\m{v}}	
		&
		&\m{q}\text{-a.e.}.
	\end{align}
	Then \eqref{eq:thm:MeasureChange:Increasing} follows from
	\eqref{eq:MeasureChange:MD},
	Lemma \ref{lem:OLRTest},
	and
	Theorem \ref{thm:SNPL} whenever \(\ent{1}{\m{v}}{\m{q}}\) is finite.
	%The \(\varepsilon\!=\!0\) and \(\varepsilon\!\geq\!\norm{\m{w;\sim}}\) cases, 
	%are evident as a result of \(\m{v}\!\sim\!\m{w;\sim}\) and Theorem \ref{thm:SNPL}.
	
	When \(\f{f}[\cdot]\) is decreasing and bijective 
	\eqref{eq:lem:OLRTest:def}, 
	\eqref{eq:assumption:MeasureChange:RNDerivative},
	and
	\eqref{eq:assumption:MeasureChange:MissedDetection}	
	imply \(\sfrac{1}{\f{f}[\sfrac{1}{\gamma}]}
	\in\!\left[\sfrac{1}{\uscq{\mdv{\varepsilon}{\m{v}}}{\m{v}}{\m{q}}},\sfrac{1}{\lscq{\mdv{\varepsilon}{\m{v}}}{\m{v}}{\m{q}}}\right]
	\) and
	\begin{align}
		\notag	
		1-\rv{T;\varepsilon,\gamma;\m{q},\m{w}}
		&=\rv{T;\mdv{\varepsilon}{\m{v}},\sfrac{1}{\f{f}[\sfrac{1}{\gamma}]};\m{q},\m{v}}
		&
		&\m{q}\text{-a.e.}.	
	\end{align}
	Then \eqref{eq:thm:MeasureChange:Decreasing} follows from
	\eqref{eq:MeasureChange:MD},
	Lemma \ref{lem:OLRTest},
	and
	Theorem \ref{thm:SNPL} whenever \(\norm{\m{q;\sim}}\) is finite.
	%The \(\varepsilon\!=\!0\) and \(\varepsilon\!\geq\!\norm{\m{w;\sim}}\) cases, 
	%are evident as a result of \(\m{v}\!\sim\!\m{w;\sim}\) and Theorem \ref{thm:SNPL}.
	
	When \(\cdual{f}[\cdot]\) is increasing and bijective 
	\eqref{eq:lem:OLRTest:def}, 
	\eqref{eq:assumption:MeasureChange:MissedDetection},
	and
	\eqref{eq:thm:MeasureChange:RNDerivative} 
	imply \(\sfrac{1}{\cdual{f}[\gamma]}
	\in\!\left[\sfrac{1}{\uscq{\mdv{\varepsilon}{\m{v}}}{\m{v}}{\m{w}}},\sfrac{1}{\lscq{\mdv{\varepsilon}{\m{v}}}{\m{v}}{\m{w}}}\right]
	\) and
	\begin{align}
		\notag	
		1-\rv{T;\varepsilon,\gamma;\m{q},\m{w}}
		&=\rv{T;\mdv{\varepsilon}{\m{v}},\sfrac{1}{\cdual{f}[\gamma]};\m{w},\m{v}}	
		&
		&\m{w}\text{-a.e.}.	
	\end{align}
	Then \eqref{eq:thm:MeasureChange:CsiszarDual:Increasing} follows from
	\eqref{eq:MeasureChange:FA},
	Lemma \ref{lem:OLRTest},
	and
	Theorem \ref{thm:SNPL} whenever \(\ent{1}{\m{v}}{\m{w}}\) is finite.
	%The \(\varepsilon\!=\!0\) and \(\varepsilon\!=\!\norm{\m{w;\sim}}\) cases, 
	%cases, are evident as a result of \(\m{v}\!\sim\!\m{w;\sim}\) and Theorem \ref{thm:SNPL}.
	
	When \(\cdual{f}[\cdot]\) is decreasing and bijective 
	\eqref{eq:lem:OLRTest:def}, 
	\eqref{eq:assumption:MeasureChange:FalseAlarm},
	and
	\eqref{eq:thm:MeasureChange:RNDerivative} 
	imply \(\sfrac{1}{\cdual{f}[\gamma]}
	\in\!\left[\sfrac{1}{\uscq{\fav{\varepsilon}{\m{v}}}{\m{v}}{\m{w}}},\sfrac{1}{\lscq{\fav{\varepsilon}{\m{v}}}{\m{v}}{\m{w}}}\right]
	\) and
	\begin{align}
		\notag	
		\rv{T;\varepsilon,\gamma;\m{q},\m{w}}
		&=\rv{T;\fav{\varepsilon}{\m{v}},\sfrac{1}{\cdual{f}[\gamma]};\m{w},\m{v}}	
		&
		&\m{w}\text{-a.e.}.	
	\end{align}
	Then \eqref{eq:thm:MeasureChange:CsiszarDual:Decreasing} follows from
	\eqref{eq:MeasureChange:FA},
	Lemma \ref{lem:OLRTest},
	and
	Theorem \ref{thm:SNPL} whenever \(\norm{\m{w;\sim}}\) is finite.
	\begin{remark}
		There is a somewhat longer and indirect way to prove 
		\eqref{eq:thm:MeasureChange:Increasing}, 
		\eqref{eq:thm:MeasureChange:Decreasing}, 
		\eqref{eq:thm:MeasureChange:CsiszarDual:Increasing}, 
		and \eqref{eq:thm:MeasureChange:CsiszarDual:Decreasing},
		which is easier to confirm. Here is the proof of 
		\eqref{eq:thm:MeasureChange:Increasing} in this alternative way.
		\begin{align}
			\notag	
			\ent{\!\tfrac{1}{\f{f}[\sfrac{1}{\gamma}]}}{\m{q}}{\m{v}}
			&\!\mathop{=}^{(a)}
			\m{q}[\der{\m{v}}{\m{q}}\!>\!\f{f}[\sfrac{1}{\gamma}]]
			\!+\!\tfrac{1}{\f{f}[\sfrac{1}{\gamma}]}\m{v}[\der{\m{v}}{\m{q}}\!\leq\!\f{f}[\sfrac{1}{\gamma}]]
			\\
			\notag
			&\!\mathop{=}^{(b)}
			\m{q}[\der{\m{w;\sim}}{\m{q}}\!>\!\tfrac{1}{\gamma}]
			\!+\!\tfrac{1}{\f{f}[\sfrac{1}{\gamma}]}\m{v}[\der{\m{w;\sim}}{\m{q}}\!\leq\!\tfrac{1}{\gamma}]
			\\
			\notag
			&\!\mathop{=}^{(c)}
			\NPD{\varepsilon}{\m{q}}{\m{w}}
			+\tfrac{1}{\f{f}[\sfrac{1}{\gamma}]}\fav{\varepsilon}{\m{v}},
		\end{align}	
		where \((a)\) follows from the definition of primitive entropy spectrum
		in \eqref{eq:def:EntropySpectrum},
		\((b)\) follows from \eqref{eq:assumption:MeasureChange:RNDerivative}
		because \(\f{f}[\cdot]\) is increasing and invertible,
		\((c)\) follows from \eqref{eq:lem:OLRTest:def}, \eqref{eq:assumption:MeasureChange:FalseAlarm},
		and \eqref{eq:MeasureChange:MD}.
		Thus 
		\(\NPD{\fav{\varepsilon}{\m{v}}}{\m{q}}{\m{v}}\!\geq\!\NPD{\varepsilon}{\m{q}}{\m{w}}\) by 
		\eqref{eq:thm:SNPL:DualCharacterization}.
		On the other hand
		\(\NPD{\fav{\varepsilon}{\m{v}}}{\m{q}}{\m{v}}
		\!\leq\!\NPD{\varepsilon}{\m{q}}{\m{w}}\) by 
		\eqref{eq:def:NPDr},
		\eqref{eq:assumption:MeasureChange:FalseAlarm},
		and \eqref{eq:MeasureChange:MD}
		because \(\rv{T;\varepsilon,\gamma;\m{q},\m{w}}\!\in\!\tmea{\outA}\), i.e., 
		\(\rv{T;\varepsilon,\gamma;\m{q},\m{w}}\) is a randomized test, by Lemma \ref{lem:OLRTest}.
		Thus \eqref{eq:thm:MeasureChange:Increasing} holds for all \(\varepsilon\!\in\!(0,\norm{\m{w;\sim}})\).		
	\end{remark}\vspace{-.4cm}
\end{proof}

The tilted measures \(\m{v;\rno;\m{w},\m{q}}\), defined in the following,
satisfy the monotonicity assumptions of Theorem \ref{thm:MeasureChange}
on \(\f{f}[\cdot]\) and \(\cdual{f}[\cdot]\)
at the same time for all \(\rno\!\in\!\numbers{R}\!\setminus\!\{0,1\}\).
Usually, we will also assume \(\norm{\m{v;\rno}}\) to be finite to ensure 
finite primitive entropy hypotheses of Theorem \ref{thm:MeasureChange}.

\begin{definition}\label{def:Tilted} 
	For any pair of \(\sigma\)-finite measures \((\m{w},\m{q})\) 
	\begin{subequations}\label{eq:def:Tilted} 
		satisfying \(\m{w}\not\perp\m{q}\) and real number \(\rno\),
		the tilted measure \(\m{v;\rno;\m{w},\m{q}}\) is defined 
		as the following \(\m{w;\sim}\)-equivalent measure
		\begin{align}
			\label{eq:def:TiltedMeasure}
			\der{\m{v;\rno;\m{w},\m{q}}}{\m{q}}
			&\DEF (\der{\m{w;\sim}}{\m{q}})^{\rno}
			&
			&\m{q}\text{-a.e.}.
			\intertext{Domain of the tilted measure's finiteness \(\SGBdomain{\m{w}}{\m{q}}\) is defined as}
			\label{eq:def:Tilted:NormalizationFactor:domain} 
			\SGBdomain{\m{w}}{\m{q}}
			&\DEF\{\rno\in\numbers{R}:\norm{\m{v;\rno;\m{w},\m{q}}}<\infty\}.
		\end{align}
	\end{subequations} 
\end{definition}
Note that the tilted measure \(\m{v;\rno;\m{w},\m{q}}\) also satisfies 
\begin{align}
	\label{eq:TiltedMeasure}
	\der{\m{v;\rno;\m{w},\m{q}}}{\m{w}}
	&=(\der{\m{q;\sim}}{\m{w}})^{1-\rno}
	&
	&\m{w}\text{-a.e.}.	
\end{align}	
Furthermore, \(\ln \norm{\m{v;\rno;\m{w},\m{q}}}\) is a convex function 
as a result of H\"older's inequality \cite[Theorem 2.11.1]{bogachev}.
Thus \(\SGBdomain{\m{w}}{\m{q}}\) is an interval and \(\SGBdomain{\m{w}}{\m{q}}\)
is non-empty iff \(\ln \norm{\m{v;\rno;\m{w},\m{q}}}\) as a function 
of \(\rno\) satisfies Assumption \ref{assumption:HasAConjugate}. 
In the following, we will denote \(\m{v;\rno;\m{w},\m{q}}\) 
simply by \(\m{v;\rno}\), whenever \(\m{w}\) and \(\m{q}\)
are clear from the context.
\begin{remark}\label{rem:trivial-non-singular-tilting}
	In the trivial-non-singular case 
	\(\der{\m{w;\sim}}{\m{q}}\!=\!\zeta\) holds \(\m{w;\sim}\)-a.e.
	for some \(\zeta\!\in\!(0,\infty)\). 
	Thus \(\m{v;\rno}\!=\!\zeta^{\rno-1}\m{w;\sim}\!=\!\zeta^{\rno}\m{q;\sim}\) 
	holds \(\forall\rno\!\in\!\numbers{R}\).
	Furthermore, trivial-non-singular case is non-void iff
	\(\norm{\m{q;\sim}}\!\in\!(0,\infty)\), which is equivalent to
	\(\norm{\m{w;\sim}}\!\in\!(0,\infty)\) for the 
	trivial-non-singular case.
	In the trivial-non-singular-non-void case 
	\(\zeta\!=\!\sfrac{\norm{\m{w;\sim}}}{\norm{\m{q;\sim}}}\),
	and \(\fav{\varepsilon}{\m{v;\rno}}\!=\!\varepsilon\zeta^{\rno-1}\)
	for all \(\rno\!\in\!\numbers{R}\) and 
	\(\varepsilon\!\in\!(0,\norm{\m{w;\sim}})\).	
\end{remark}
Note that \(\m{v;\rno}\!\sim\!\m{q;\sim}\) holds 
by definition and \(\m{v;\rno}\) satisfies 
Assumption \ref{assumption:MeasureChange}
at least for all \(\rno\in\SGBdomain{\m{w}}{\m{q}}\). 
Furthermore,
\begin{enumerate}[(a)]
\item If \(\rno\!<\!0\), then \(\f{f}[\cdot]\)
is bijective and decreasing.
\item If \(\rno\!>\!0\), then \(\f{f}[\cdot]\) 
is bijective and increasing.
\item If \(\rno\!<\!1\), then \(\cdual{f}[\cdot]\)
is bijective and increasing.
\item If \(\rno\!>\!1\), then \(\cdual{f}[\cdot]\)
is bijective and decreasing.		
\end{enumerate}
Thus if \(\sigma\)-finite measure \(\m{w}\) and \(\m{q}\) 
satisfy \(\ent{1}{\m{w}}{\m{q}}\in(0,\infty)\) 
and \(\rno\in\SGBdomain{\m{w}}{\m{q}}\), 
then Theorem \ref{thm:MeasureChange} 
implies\begin{subequations}\label{eq:TiltedCharacterization}
for all \(\varepsilon\in(0,\norm{\m{w;\sim}})\), that  
\begin{align}
\label{eq:TiltedCharacterization:Q}
\NPD{\varepsilon}{\m{q}}{\m{w}}
&\!=\!\begin{cases}
	\norm{\m{q;\sim}}
	\!-\!\NPD{\mdv{\varepsilon}{\m{v;\rno}}}{\m{q}}{\m{v;\rno}}	
	&\text{if~\(\rno\!<\!0\),}	
	\\
	\mdv{\varepsilon}{\m{v;\rno}}
	&\text{if~\(\rno\!=\!0\),}
	\\
	\NPD{\fav{\varepsilon}{\m{v;\rno}}}{\m{q}}{\m{v;\rno}}	
	&\text{if~\(\rno\!>\!0\),}	
\end{cases}	
\\
\label{eq:TiltedCharacterization:W}	
\varepsilon
&\!=\!\begin{cases}
	\NPD{\mdv{\varepsilon}{\m{v;\rno}}}{\m{w}}{\m{v;\rno}}	
	&\text{if~\(\rno\!<\!1\),}
	\\
	\fav{\varepsilon}{\m{v;\rno}}
	&\text{if~\(\rno\!=\!1\),}
	\\
	\norm{\m{w;\sim}}\!-\!
	\NPD{\fav{\varepsilon}{\m{v;\rno}}}{\m{w}}{\m{v;\rno}}	
	&\text{if~\(\rno\!>\!1\).}			
\end{cases}
\intertext{Furthermore, invoking \eqref{eq:NDPInversion}, 
we can express the function 
\(\fav{\cdot}{\m{v;\rno}}:(0,\norm{\m{w;\sim}})\mapsto(0,\norm{\m{v;\rno}})\) 
in two distinct ways:}
\label{eq:TiltedCharacterization:VQ}	
\fav{\varepsilon}{\m{v;\rno}}
&\!=\!\begin{cases}
	\norm{\m{v;\rno}}
	-\NPD{\norm{\m{q;\sim}}\!-\!\NPD{\varepsilon}{\m{q}}{\m{w}}}{\m{v;\rno}}{\m{q}}	
	&\text{if~\(\rno\!<\!0\),}
	\\
	\norm{\m{v;\rno}}-\NPD{\varepsilon}{\m{q}}{\m{w}}
	&\text{if~\(\rno\!=\!0\),}
	\\
	\NPD{\NPD{\varepsilon}{\m{q}}{\m{w}}}{\m{v;\rno}}{\m{q}}
	&\text{if~\(\rno\!>\!0\),}			
\end{cases}
\\
\label{eq:TiltedCharacterization:VW}	
\fav{\varepsilon}{\m{v;\rno}}
&\!=\!\begin{cases}
	\norm{\m{v;\rno}}-\NPD{\varepsilon}{\m{v;\rno}}{\m{w}}	
	&\text{if~\(\rno\!<\!1\),}
	\\
	\varepsilon
	&\text{if~\(\rno\!=\!1\),}
	\\
	\NPD{\norm{\m{w;\sim}}\!-\!\varepsilon}{\m{v;\rno}}{\m{w}}	
	&\text{if~\(\rno\!>\!1\).}			
\end{cases}	
\end{align}			
\end{subequations}

Theorem \ref{thm:MeasureChange} and 
Definition \ref{def:Tilted}  are inspired by 
the information theorists approach to large deviations 
analysis that can be traced back to 
\cite{shannonGB67A,haroutunian68,augustin68,haroutunian77,augustin78}.
Customarily the tilted measure is defined as a probability measure
of the form \(\sfrac{\m{v;\rno}}{\norm{\m{v;\rno}}}\), 
e.g., \cite[Theorem 5 and (3.27)]{shannonGB67A} 
considering orders in \((0,1)\). 
The primary advantage of avoiding the scaling is that
not only memorylessness but also Markovian structure 
(and probably other structural assumption on \(\m{w}\) and \(\m{q}\)) 
are inherited by the tilted measure when there is no scaling, 
see \S\ref{sec:Memoryless} and \S\ref{sec:DiscreteMarkov}. 

\section{Approximations for randomized-NP-function}\label{sec:Approximations}
In this section, we approximate the rNP-function 
\(\NPD{\cdot}{\m{q}}{\m{w}}\) under hypotheses on 
the likelihood-ratio distribution that 
are implied by the Berry--Esseen type bounds non-asymptotically 
and Central Limit Theorem asymptotically. 
Since likelihood-ratios typically follow a multiplicative 
rather than additive structure 
(e.g., memoryless and Markovian models)
these assumptions are more naturally formulated and interpreted 
using the distribution of the log-likelihood ratio
rather than the likelihood ratio. 

In \S\ref{sec:LogLikelihoodRatio}, we first define 
the log-likelihood ratio and its distribution,
and then express \(\NPD{\cdot}{\m{q}}{\m{w}}\) 
in terms of the log-likelihood ratio distribution 
\(\gcdf{\llr}{\cdot}{\m{w}}{\m{q}}\) 
using Theorem \ref{thm:SNPL}. 
In \S\ref{sec:GaussianMillsRatio}, we define Gaussian Mills
ratio and review certain well-known bounds on it that will be 
helpful both in our analysis and in interpreting our results. 
Finally, in \S\ref{sec:KolmogorovDistanceToNormality} we
first approximate \(\NPD{\cdot}{\m{q}}{\m{w}}\) assuming 
\(\gcdf{\llr}{\cdot}{\m{w}}{\m{q}}\) is close to a 
Gaussian distribution in Kolmogorov distance and 
then bound \(\NPD{\cdot}{\m{w}}{\m{q}}\) assuming 
\(\gcdf{\llr}{\cdot}{\gamma\m{v;\rno}}{\m{w}}\) is close 
to a Gaussian distribution in Kolmogorov distance.

\subsection{The Log-Likelihood-Ratio}\label{sec:LogLikelihoodRatio}
\begin{definition}\label{def:llr}
For any pair of \(\sigma\)-finite measures \((\m{w},\m{q})\) 
on the same measurable space, say \((\outS,\outA)\),
the log-likelihood-ratio \(\f{\llr;\m{w},\m{q}}[\cdot]\)
and the log-likelihood-ratio\begin{subequations}\label{eq:def:llr} 
distribution \(\gcdf{\llr}{\cdot}{\m{w}}{\m{q}}\) are defined as
\begin{align}
\label{eq:def:llrf}
\f{\llr;\m{w},\m{q}}
&\DEF
\begin{cases}
\infty 
&\text{if~\(\der{\m{q}}{(\m{w}+\m{q})}\!=\!0\),}
\\
\ln\der{\m{w;\sim}}{\m{q}}
&\text{if~\(\der{\m{q}}{(\m{w}+\m{q})}\!>\!0\),}
\end{cases}		
\\
\label{eq:def:llrcdf}
\gcdf{\llr}{\tau}{\m{w}}{\m{q}}
&\DEF \int \IND{\f{\llr;\m{w},\m{q}}\leq\tau} \dif{\m{w}}
&
&\forall \tau\!\in\!\numbers{R},
\end{align}
where \(\m{w;\sim}\) is\end{subequations} 
the \(\m{q}\)-absolutely continuous component of \(\m{w}\). 	
\end{definition}
Note that definitions in 
\eqref{eq:def:lr} and \eqref{eq:def:llr} imply
\begin{align}
\label{eq:llr}
\f{\llr;\m{w},\m{q}}
&=\ln\f{\ylr;\m{w},\m{q}}
&
&(\m{w}\!+\!\m{q})\text{-a.e.,}
\\
\label{eq:llrcdf}
\gcdf{\llr}{\tau}{\m{w}}{\m{q}}
&=\scq{e^{\tau}}{\m{w}}{\m{q}}
&
&\forall \tau\!\in\!\numbers{R}.
\intertext{Thus identities 
	\eqref{eq:lrf} and \eqref{eq:lrcdf:ACequivalence} 
	imply}
\label{eq:llrf}
\f{\llr;\m{w},\m{q}}
&=-\f{\llr;\m{q},\m{w}}
&
&(\m{w}\!+\!\m{q})\text{-a.e.}
\\
\label{eq:llrcdf:ACequivalence}
\gcdf{\llr}{\tau}{\m{w}}{\m{q}}
&=\gcdf{\llr}{\tau}{\m{w;\sim}}{\m{q;\sim}}
&
&\forall \tau\!\in\!\numbers{R}.	
\end{align}
\begin{remark}\label{rem:LLRQvsLRQ}
\(\gcdf{\llr}[\wedge]{\cdot}{\m{w}}{\m{q}}
\!:\!\numbers{R}\!\mapsto\!\overline{\numbers{R}}\) 
and \(\gcdf{\llr}[\vee]{\cdot}{\m{w}}{\m{q}}
\!:\!\numbers{R}\!\mapsto\!\overline{\numbers{R}}\),
i.e., the log-likelihood ratio quantiles, 
can be defined using the definition of the likelihood ratio 
quantiles by replacing \(\scq{\cdot}{\m{w}}{\m{q}}\) 
with \(\gcdf{\llr}{\cdot}{\m{w}}{\m{q}}\)
in \eqref{eq:def:LRQ}. Then 
\(\gcdf{\llr}[\wedge]{\cdot}{\m{w}}{\m{q}}
\!=\!\ln\lscq{\cdot}{\m{w}}{\m{q}}\)
and \(\gcdf{\llr}[\vee]{\cdot}{\m{w}}{\m{q}}
\!=\!\ln\uscq{\cdot}{\m{w}}{\m{q}}\), 
with the understanding that
\(\ln(-\infty)\), \(\ln0\), and \(\ln\infty\) stand
for \(-\infty\), \(-\infty\), and \(\infty\),
respectively.
Then \eqref{eq:thm:SNPL:IOSD:1} implies
\begin{align}
	\label{eq:thm:SNPL:IOSD:1:L}
	\NPD{\varepsilon}{\m{q}}{\m{w}}
	&=\int_{\varepsilon}^{\norm{\m{w;\sim}}}
	e^{-\f{f}[\tau]}
	\dif{\tau}
	&
	&\forall\varepsilon\!\in\![0,\norm{\m{w;\sim}}],
\end{align}
for any  \(\f{f}\!:\!(0,\norm{\m{w;\sim}})\!\mapsto\!\numbers{R}\)  
s.t.  \(\gcdf{\llr}[\wedge]{\cdot}{\m{w}}{\m{q}}
\!\leq\!\f{f}[\cdot]\!\leq\!\gcdf{\llr}[\vee]{\cdot}{\m{w}}{\m{q}}\). 	
\end{remark}
For any pair of \(\sigma\)-finite measures \(\m{w}\) and \(\m{q}\)
---not necessarily  satisfying \eqref{eq:assumption:nonvoid}---
first invoking \eqref{eq:thm:SNPL:CI:DF}, i.e., \eqref{eq:Beta}, 
and then applying the change of variable 
\(\tau\!=\!e^{-\tin}\), we get 
\begin{align}
	\NPD{\varepsilon}{\m{q}}{\m{w}}
	%\notag
	%&\!=\!\int_{0}^{\infty}\!
	%\abp{\scq{\tfrac{1}{\tau}}{\m{w}}{\m{q}}-\varepsilon}
	%\!\dif{\tau}
	%&
	%&\forall\varepsilon\!\geq\!0,	
	%\\
	\label{eq:thm:SNPL:CI:DF:L}
	&=\int_{-\infty}^{\infty}
	e^{-\tin}\abp{\gcdf{\llr}{\tin}{\m{w}}{\m{q}}-\varepsilon}
	\dif{\tin}
	&
	&\forall\varepsilon\geq\!0.
\end{align}
%The identity \eqref{eq:thm:SNPL:CI:DF:L} will be the starting point
%for all of the approximations we derive in the following.
%We will make additional assumptions on 
%\(\gcdf{\llr}{\cdot}{\m{w}}{\m{q}}\) to bound 
%\(\NPD{\cdot}{\m{q}}{\m{w}}\). 

For example, if \(\m{w}\) is a probability measure 
with a finite \(\EX[\m{w}]{\f{\llr;\m{w},\m{q}}^{2}}\),
then %\subsection{Bounds Under Finite Second Moment Assumption}
the Chebychev-Cantelli inequality and \eqref{eq:thm:SNPL:CI:DF:L} 
lead to the following non-asymptotic Stein's lemma,
which improves similar results reported before in
\cite[Lemma 15]{polyanskiythesis}, 
\cite[Lemma 59]{polyanskiyPV10}, 
\cite[(12)]{burnashev23}.
Proof of Lemma \ref{lem:NPD:ChebyshevCantelli},
which is elementary, is presented in 
Appendix \ref{sec:Proofs:SteinLemmas}
for completeness.

\begin{lemma}\label{lem:NPD:ChebyshevCantelli}
Let \(\m{w}\) be a probability measure and 
\(\m{q}\) be a \(\sigma\)-finite 
measure\begin{subequations}\label{eq:lem:NPD:ChebyshevCantelli}
satisfying \(\EX[\m{w}]{(\f{\llr;\m{w},\m{q}})^{2}}<\infty\), then 
\(\EX[\m{w}]{\f{\llr;\m{w},\m{q}}}\!\in\!\numbers{R}\) and
\begin{align}
	\label{eq:lem:NPD:ChebyshevCantelli:bound}
	\ln\tfrac{\Delta}{1+\Delta}\!-\!\sigma\sqrt{\tfrac{\varepsilon+\Delta}{1-\varepsilon}}	
	\!\leq\! 
	\ln\tfrac{\NPD{\varepsilon}{\m{q}}{\m{w}}}{1-\varepsilon}\!+\!\EX[\m{w}]{\f{\llr;\m{w},\m{q}}}
	&\!\leq\!\sigma\sqrt{\tfrac{1-\varepsilon}{\varepsilon}}
\end{align}		
for all \(\varepsilon\!\in\!(0,1)\), \(\Delta\!>\!0\), 
where
\(\sigma\DEF\sqrt{\EX[\m{w}]{\left(\f{\llr;\m{w},\m{q}}\!-\!\EX[\m{w}]{\f{\llr;\m{w},\m{q}}}\right)^{2}}}\).
%\(\sigma\) is defined as
%\begin{align}
%\label{eq:lem:NPD:ChebyshevCantelli:sigma}		
%\sigma
%&\DEF 
%\sqrt{\int \left(\ln \der{\m{w}}{\m{q}}
	%-\KLD{\m{w}}{\m{q}}\right)^{2}\dif{\m{w}}}.
%&
%&~
%\end{align}			
\end{subequations}	
\end{lemma}	
\begin{remark}
	\(\ln\tfrac{1-\varepsilon}{\NPD{\varepsilon}{\m{q}}{\m{w}}}\) 
	is called \(\varepsilon\)-hypothesis testing divergence and denoted by \(D_{h}^{\varepsilon}(\m{w}\Vert\m{q})\), see 
	\cite[p. 7044]{tomamichelT13}, \cite[(2.6)]{tan14}. 
	Note that for any probability measure \(\m{w}\), 
	monotonicity of divided differences for convex functions 
	and convexity of \(\NPD{\cdot}{\m{q}}{\m{w}}\)
	imply that \(D_{h}^{\varepsilon}(\m{w}\Vert\m{q})\) is 
	decreasing in \(\varepsilon\) and
	\(D_{h}^{\varepsilon}(\m{w}\Vert\m{q})\geq -\ln \norm{\m{q;\sim}}\). 
	Note that inequality holds as an equality 
	for all \(\varepsilon\in(0,1)\) in the 
	trivial case provided that \(\m{w}\AC\m{q}\). 	
\end{remark}

\subsection{Gaussian Mills Ratio}\label{sec:GaussianMillsRatio}
The Gaussian Mills ratio is defined as the ratio of 
Gassian Q-function and Gaussian probability density:
\begin{align}
	\label{eq:def:GMR}
	\GMR{\tau}
	&\DEF\tfrac{\GQF{\tau}}{\GD{\tau}}
	&
	&\forall \tau\!\in\!\numbers{R}.			
\end{align}
Then \(\GMR{\cdot}\!:\!\numbers{R}\!\mapsto\!(0,\infty)\)
is a decreasing strictly convex and invertible function
because 
\(\ln \GMR{\cdot}\!:\!\numbers{R}\!\mapsto\!\numbers{R}\) 
is such a function by \cite{baricz08}.
Strikingly its multiplicative inverse \(\sfrac{1}{\GMR{\cdot}}\) 
is a convex functions as well, see \cite{gasullU14}.

Note that
\(\ln\GMR{\tau\!+\!\delta}\!>\!\ln\GMR{\tau}\!+\!\tfrac{\GMR{\tau}[']}{\GMR{\tau}}\delta\)
for all \(\tau\!\in\!\numbers{R}\) and \(\delta\!\neq\!0\)
where \(\GMR{\tau}[']\!=\!\der{}{\tau}\GMR{\tau}\)
because \(\ln\GMR{\cdot}\) is strictly convex.
Since \(\ln\GMR{\cdot}\) is strictly decreasing 
as well, we have
\begin{align}
	\label{eq:GMRBound:multiplicative:GeneralA}
	\GMR{\tau}\!>\! 
	\GMR{\tau\!+\!\delta}
	&\!>\!\GMR{\tau}e^{\frac{\GMR{\tau}[']}{\GMR{\tau}}\delta}
	&
	&\forall \tau\in\numbers{R},~\delta>0,
\end{align}
On the other hand \(\sfrac{\GMR{\tau}[']}{\GMR{\tau}}\) is
non-decreasing because  \(\ln \GMR{\cdot}\) is convex.
Then using the identity 
\(\GMR{\tau}[']=\tau\GMR{\tau}-1\)
we get
\begin{align}
	\label{eq:GMRBound:multiplicative:GeneralB}
	\GMR{\tau}
	&\!>\!\GMR{\tau\!+\!\delta}
	\!>\!\GMR{\tau}e^{(\dinp-\frac{1}{\GMR{\dinp}})\delta}
	&
	&\forall \tau\!\geq\!\dinp,\delta\!>\!0,\dinp\!\in\!\numbers{R}.
	\intertext{On that other hand 
		\(\sfrac{1}{\GMR{-1.3}}\leq0.19\). Hence}
	\label{eq:GMRBound:multiplicative}
	1.5\delta
	&\!\geq\! 
	\ln\!\tfrac{\GMR{\tau}}{\GMR{\tau+\delta}}
	\!\geq\!0
	&
	&\forall \tau\geq-1.3,~\delta\geq0.	
\end{align}
Furthermore, 
\begin{align}
	\label{eq:GMRBound:positive}
	%\!\!\tfrac{\tau}{1+\tau^{2}}
	%\!\leq\!
	\tfrac{2}{\tau+\sqrt{\tau^2+ 4}}
	%\!\vee\! 
	%\tfrac{\sfrac{\pi}{2}}{\tau+\sqrt{\tau^{2}+ \sfrac{\pi}{2}}}   
	\!\leq\! 
	\GMR{\tau}
	&\!\leq\!
	\tfrac{2}{\tau+\sqrt{\tau^{2}+\sfrac{8}{\pi}}}			
	&
	&\forall\tau\geq 0,
\end{align}	
by \cite[(10), (11), (12)]{borjesson79} and \cite[7.1.13]{AbramowitzStegun}. 
Let the function \(\f{g}[\cdot,\cdot]:\numbers{R}\times(0,1)\mapsto(0,\infty)\) be
\begin{align}
	\label{eq:def:SGMR}
	\f{g}[\dinp,\varepsilon]	
	&\DEF e^{\frac{\dinp^{2}}{2}}
	\GQF{\dinp-\IGQF{\varepsilon}},
	\\
	\label{eq:SGMR:Alternative}
	&=\tfrac{\GQF{\dinp-\IGQF{\varepsilon}}}{\GD{\dinp}\sqrt{2\pi}},
	\\
	\label{eq:SGMR}
	&=\GMR{\dinp\!-\!\IGQF{\varepsilon}}
	\GD{\IGQF{\varepsilon}}e^{\dinp\IGQF{\varepsilon}}
	\intertext{Then}
	\notag
	\pder{}{\varepsilon}\f{g}[\dinp,\varepsilon]	
	&=-e^{\dinp{\IGQF\varepsilon}}
	\\
	\label{eq:SGMR:epsilon:partial}
	&=\tfrac{-\f{g}[\dinp,\varepsilon]}{\GMR{\dinp-\IGQF{\varepsilon}}}\tfrac{1}{\GD{\IGQF{\varepsilon}}}.
	%\\
	%\notag
	%&=\tfrac{-\f{g}[\dinp,\varepsilon]}{\GMR{\dinp-\IGQF{\varepsilon}}}\tfrac{\GMR{\abs{\IGQF{\varepsilon}}}}{\varepsilon\wedge(1-\varepsilon)}.
\end{align}	
%{\color{purple}		
	%%Using \eqref{eq:GMRBound:positive} and
	%%confirming numerically for  \(\tau\in(-1.3,0)\), 
	%%one observe that
	%%\begin{align}
	%%\label{eq:GMRBound:extended}
	%%\tfrac{2}{3+2\tau}
	%%\!\leq\! 
	%%\GMR{\tau}
	%%&\!\leq\!
	%%\tfrac{2}{\tau+\sqrt{\tau^{2}+1.1}}
	%%&
	%%&\forall \tau\geq-1.3.			
	%%\end{align}
	%\begin{align}
	%\label{eq:SGMR:s:partial}
	%\pder{}{\dinp}\f{g}[\dinp,\varepsilon]	
	%&=\dinp\f{g}[\dinp,\varepsilon]	 
	%-e^{\dinp\IGQF{\varepsilon}}\GD{\IGQF{\varepsilon}}
	%\\
	%\notag
	%&=\left(\dinp-\tfrac{1}{\GMR{\dinp\!-\!\IGQF{\varepsilon}}}\right) 
	%\f{g}[\dinp,\varepsilon]	 
	%\end{align}
	%\(\f{g}[\dinp,\varepsilon]\)
	%is decreasing in \(\dinp\) if 
	%\(\varepsilon\geq0.5\).
	%}	
We denote the Gaussian cumulative distribution function 
by \(\GCD{\cdot}\), i.e.,
\begin{align}
	\label{eq:def:GCD}
	\GCD{\dinp}
	&\DEF\int_{-\infty}^{\dinp}\GD{\tau}\dif{\tau}	
	&
	&\forall \dinp\in\numbers{R}.
	\intertext{Note that 
		%definitions of \(\GQF{\cdot}\) in 
		\eqref{eq:def:GQF} and the fact that 
		\(\GD{\tau}\!=\!\GD{-\tau}\) 
		for \(\tau\!\in\!\numbers{R}\) imply}
	\label{eq:GCD}
	\GCD{\dinp}
	&=\GQF{-\dinp}	
	&
	&\forall\dinp\!\in\!\numbers{R}.
\end{align}
\subsection{Kolmogorov Distance to Normality}\label{sec:KolmogorovDistanceToNormality}
\begin{theorem}\label{thm:KDtoN}
If \(\m{w}\) and \(\m{q}\) are \(\sigma\)-finite measures
\begin{subequations}\label{eq:thm:KDtoN}satisfying
\begin{align}
\label{eq:thm:KDtoN:hypothesis}
\sup\nolimits_{\dinp\in\numbers{R}}\abs{\gcdf{\llr}{\xi\!+\!\sigma\dinp}{\m{w}}{\m{q}}-\GCD{\dinp}}	
%\sup\nolimits_{\dinp\in\numbers{R}}\abs{\gcdf{\llr}{\xi\!-\!\sigma\dinp}{\m{w}}{\m{q}}-\GQF{\dinp}}	
&\leq \delta,		
\end{align}
for a \(\xi\!\in\!\numbers{R}\),
a \(\sigma\!>\!0\), and 
a \(\delta\!\in\!(0,1)\), then
\begin{align}\label{eq:thm:KDtoN:upperbound}	
\!\!\!\NPD{\varepsilon}{\m{q}}{\m{w}}		
&\!\leq\! 
e^{-\xi+\frac{\sigma^{2}}{2}}\GQF{\sigma\!-\!\IGQF{\varepsilon\!-\!\delta}}
&
&\forall \varepsilon\!\in\!(\delta,1+\delta),
\\
\label{eq:thm:KDtoN:lowerbound}	
\!\!\!\NPD{\varepsilon}{\m{q}}{\m{w}}		
&\!\geq\!
e^{-\xi+\frac{\sigma^{2}}{2}}\GQF{\sigma\!-\!\IGQF{\varepsilon\!+\!\delta}}
&
&\forall \varepsilon\!\in\!(0,1\!-\!\delta),
\intertext{Furthermore, if \(\varepsilon\!\in\!(\delta,1\!-\!\delta\!-\!\GQF{\sigma})\), then}
\label{eq:thm:KDtoN:Strassen}
\!\!\!\NPD{\varepsilon}{\m{q}}{\m{w}}		
&\!=\!
e^{\!-\!\xi+\sigma\tau}\GMR{\sigma\!-\!\tau}\GD{\tau}\Delta,
%&
%&\forall\varepsilon\!\in\!(\delta,1\!-\!\delta\!-\!\GQF{\!\sigma\!}),\!
\end{align}
where \(\tau\!=\!\IGQF{\varepsilon}\),
and\end{subequations}		
~\(\Delta\)~satisfies
~\(\abs{\ln\Delta}\leq\tfrac{\delta(\sigma+0.5)\sqrt{2\pi}}{(1-\varepsilon-\delta)\wedge(\varepsilon-\delta)}\).			
\end{theorem}
\begin{proof}[Proof of Theorem \ref{thm:KDtoN}]
Note that \(\norm{\m{w;\sim}}\in[1\!-\!\delta,1\!+\!\delta]\) 
by \eqref{eq:thm:KDtoN:hypothesis} because 
\(\norm{\m{w;\sim}}=\lim_{\dinp\uparrow\infty}
\gcdf{\llr}{\dinp}{\m{w}}{\m{q}}\). 
Thus \(\m{w}\) and \(\m{q}\)
satisfy \eqref{eq:assumption:nonvoid}.
\begin{align}
\notag	
\NPD{\varepsilon}{\m{q}}{\m{w}}	
&\mathop{=}^{(a)}
\sigma \int_{-\infty}^{\infty} 
e^{-(\xi+\sigma \dinp)}
\abp{\gcdf{\llr}{\xi\!+\!\sigma\dinp}{\m{w}}{\m{q}}
\!-\!\varepsilon
}\dif{\dinp},
\\
\notag
&\mathop{\leq}^{(b)}
e^{-\xi}\sigma
\int_{-\infty}^{\infty} e^{-\sigma \dinp}
\abp{\GCD{\dinp}\!+\!\delta\!-\!\varepsilon}\dif{\dinp},
\\
\notag
&\mathop{=}^{(c)}
e^{-\xi}\sigma
\int_{\IGCD{\varepsilon-\delta}}^{\infty}
e^{-\sigma \dinp}
(\GCD{\dinp}\!+\!\delta\!-\!\varepsilon)\dif{\dinp},						
\end{align}			
where \((a)\) follows from 	\eqref{eq:thm:SNPL:CI:DF:L}
for \(\tin\!=\!\xi\!+\!\sigma\dinp\), 
\((b)\) follows from the hypothesis in 
\eqref{eq:thm:KDtoN:hypothesis},
\((c)\) holds because 
\(\GCD{\cdot}\!:\!\numbers{R}\!\mapsto\!(0,1)\) is an 
invertible and increasing function.
Then \eqref{eq:thm:KDtoN:upperbound} follows 
from \eqref{eq:GaussianSpectralIntegralIdentityAlternative}
established below for \(\lambda\!=\!\IGCD{\varepsilon\!-\!\delta}\).\vspace{-.1cm}
%\begin{align}
%\notag
%\int_{-\infty}^{\lambda}\!e^{\sigma\dinp}
%\left(\!\GQF{\dinp}\!-\!\GQF{\lambda}\right)\dif{\dinp}
%&\mathop{=}^{}
%\int_{-\infty}^{\lambda}e^{\sigma\dinp}
%\int_{\dinp}^{\lambda}\GD{\tau}
%\dif{\tau}\dif{\dinp},
%\\
%\notag
%&\mathop{=}^{}
%\int_{-\infty}^{\lambda}\GD{\tau}
%\int_{-\infty}^{\tau}e^{\sigma\dinp}
%\dif{\dinp}\dif{\tau},
%\\
%\notag
%&\mathop{=}^{}
%\int_{-\infty}^{\lambda}
%\tfrac{\GD{\tau}e^{\sigma\tau}}{\sigma}
%\dif{\tau},
%\\
%\notag 
%&\!\mathop{=}^{}
%e^{\frac{\sigma^{2}}{2}}
%\int_{-\infty}^{\lambda}
%\tfrac{\GD{\tau-\sigma}}{\sigma}\dif{\tau},
%\\
%%\label{eq:GaussianSpectralIntegralIdentityAlternative}
%&\!\mathop{=}^{}
%\tfrac{\GQF{\sigma-\lambda}}{\sigma\GD{\sigma}\sqrt{2\pi}}.				
%\end{align}	
\begin{align}
\notag
\!\!\!\int_{\lambda}^{\infty}\!e^{-\sigma\dinp}
\left(\!\GCD{\dinp}\!-\!\GCD{\lambda}\right)\dif{\dinp}
&\mathop{=}^{}
\int_{\lambda}^{\infty}e^{-\sigma\dinp}
\int_{\lambda}^{\dinp}\GD{\tau}
\dif{\tau}\dif{\dinp},
\\
\notag
&\mathop{=}^{}
\int_{\lambda}^{\infty}\GD{\tau}
\int_{\tau}^{\infty}e^{-\sigma\dinp}
\dif{\dinp}\dif{\tau},
\\
\notag
&\mathop{=}^{}
\tfrac{1}{\sigma}
\int_{\lambda}^{\infty}
\GD{\tau}e^{-\sigma\tau}
\dif{\tau}
\\
\notag 
&\!\mathop{=}^{}
\tfrac{1}{\sigma}
\int_{\lambda}^{\infty}e^{\frac{\sigma^{2}}{2}}\GD{\tau+\sigma}\dif{\tau},
\\
\label{eq:GaussianSpectralIntegralIdentityAlternative}
&\!\mathop{=}^{}\tfrac{1}{\sigma}e^{\frac{\sigma^{2}}{2}}\GQF{\sigma+\lambda}.		
\end{align}
\eqref{eq:thm:KDtoN:lowerbound} is established using 
the lower bound on \(\gcdf{\llr}{\xi\!-\!\sigma\dinp}{\m{w}}{\m{q}}\)
in \eqref{eq:thm:KDtoN:hypothesis}	by following a similar 
analysis.
It is worth mentioning that that both 
\eqref{eq:thm:KDtoN:upperbound}	and \eqref{eq:thm:KDtoN:lowerbound}	
can be derived using \eqref{eq:thm:SNPL:IOSD:1:L} 
instead of \eqref{eq:thm:SNPL:CI:DF:L}.

To establish \eqref{eq:thm:KDtoN:Strassen}, first note that
\(\phi\!\in\!(\varepsilon\!-\!\delta,\varepsilon\!+\!\delta)\) such 
that \(\NPD{\varepsilon}{\m{q}}{\m{w}}=e^{-\xi}\f{g}[\sigma,\phi]\)
as a result of intermediate value theorem and the definition of 
\(\f{g}[\cdot,\cdot]\) in \eqref{eq:def:SGMR}, 
by \eqref{eq:thm:KDtoN:upperbound} and \eqref{eq:thm:KDtoN:lowerbound}. 
Since \(\ln\f{g}[\cdot,\cdot]\) is continuously differentiable 
in its second argument,  as a result of Taylor's theorem there exists
a \(\phi_{0}\) between \(\phi\) and \(\varepsilon\) such that
\begin{align}
	\label{eq:KolmogorovDistance:3}	
	\ln \f{g}[\sigma,\phi]
	&=\ln \f{g}[\sigma,\varepsilon]
	+(\phi-\varepsilon)\tfrac{\pdimpf[\f{g}]{\varepsilon}[\sigma,\phi_{0}]}{\f{g}[\sigma,\phi_{0}]}.
	\intertext{On the other hand using \eqref{eq:SGMR:epsilon:partial} we get}
	\notag
	\abs{\tfrac{\pdimpf[\f{g}]{\varepsilon}[\sigma,\phi_{0}]}{\f{g}[\sigma,\phi_{0}]}}
	&=\tfrac{1}{\GMR{\sigma-\IGQF{\phi_{0}}}}\tfrac{\GMR{\abs{\IGQF{\phi_{0}}}}}{(1-\phi_{0})\wedge \phi_{0}}
	\\
	\notag
	&\mathop{\leq}^{(a)} \tfrac{1}{\GMR{\sigma-\IGQF{\phi_{0}}}}\tfrac{\sqrt{\sfrac{\pi}{2}}}{(1-\phi_{0})\wedge \phi_{0}},
	\\
	\notag
	&\mathop{\leq}^{(b)} \tfrac{1}{\GMR{\sigma-\IGQF{\varepsilon+\delta}}}\tfrac{\sqrt{\sfrac{\pi}{2}}}{(1-\phi_{0})\wedge \phi_{0}},
	\\
	\notag
	&\mathop{\leq}^{(c)} \tfrac{1}{\GMR{2\sigma}}\tfrac{\sqrt{\sfrac{\pi}{2}}}{(1-\phi_{0})\wedge \phi_{0}},
	\\
	\notag
	&\mathop{\leq}^{(d)} (\sigma+\sqrt{\sigma^{2}+1})\tfrac{\sqrt{\sfrac{\pi}{2}}}{(1-\phi_{0})\wedge \phi_{0}},
	\\
	\label{eq:KolmogorovDistance:4}	
	&\mathop{\leq}^{(e)} (\sigma+0.5)\tfrac{\sqrt{2\pi}}{(1-\phi_{0})\wedge \phi_{0}},
	&
	&~\hspace{1cm}			
\end{align}	
where \((a)\) holds because 
\(\GMR{\tau}\!\leq\!\sqrt{\sfrac{\pi}{2}}\) for all \(\tau\!>\!0\)
by \eqref{eq:GMRBound:positive}%\cite[7.1.13]{AbramowitzStegun}
,
\((b)\) holds  because \(\GMR{\cdot}\) and \(\IGQF{\cdot}\)
are both decreasing functions
and \(\phi_{0}\!\in\!(\varepsilon\!-\!\delta,\varepsilon\!+\!\delta)\),
\((c)\) holds because 
\(\GMR{\cdot}\) and \(\IGQF{\cdot}\) are both decreasing functions,
\(\varepsilon\!+\!\delta\!\leq\! 1\!-\!\GQF{\sigma}\) by hypothesis,
and \(\IGQF{1\!-\!\dinp}\!=\!-\IGQF{\dinp}\)
for all  \(\dinp\!\in\!(0,1)\), 
\((d)\) follows from \eqref{eq:GMRBound:positive}%\cite[7.1.13]{AbramowitzStegun}
,
\((e)\) follows from \(\sqrt{\sigma^{2}\!+\!1}\!\leq\!\sigma\!+\!1\).
Thus 
\eqref{eq:thm:KDtoN:Strassen} follows from
\eqref{eq:KolmogorovDistance:3}, \eqref{eq:KolmogorovDistance:4},
and 
\(\phi_{0}\!\in\!(\varepsilon\!-\!\delta,\varepsilon\!+\!\delta)\).
\end{proof}		

Theorem \ref{thm:KDtoN} assumes the cumulative distribution 
function of the log-likelihood ratio \(\f{\llr;\m{w},\m{q}}\)
in the measure space \((\outS,\outA,\m{w})\) 
is close to that of a normal (Gaussian) distribution 
to bound \(\NPD{\cdot}{\m{q}}{\m{w}}\). 
One can bound \(\NPD{\cdot}{\m{q}}{\m{w}}\) or 
\(\NPD{\cdot}{\m{w}}{\m{q}}\) using Theorem \ref{thm:MeasureChange}
by first bounding \(\NPD{\cdot}{\m{q}}{\m{v}}\) and
\(\NPD{\cdot}{\m{w}}{\m{v}}\) using Theorem \ref{thm:KDtoN}, provided that 
the cumulative distribution function of both
\(\ln\f{f}[e^{\f{\llr;\m{w},\m{q}}}]\) and 
\(\ln\cdual{f}[e^{-\f{\llr;\m{w},\m{q}}}]\) 
in the measure space \((\outS,\outA,\m{v})\)
are close to normal distributions. 
If \(\m{v}=\gamma\m{v;\rno}\) for a positive constant \(\gamma\)
and tilted measure \(\m{v;\rno}\) for an \(\rno\!\in\!\numbers{R}\!\setminus\!\{0,1\}\) 
then these conditions are one and the same; that is the main
idea behind Theorem \ref{thm:TKDtoN} given in the 
following.
\begin{theorem}\label{thm:TKDtoN}
Let \(\m{w}\) and \(\m{q}\) be \(\sigma\)-finite measures  
satisfying \(\ent{1}{\m{w}}{\m{q}}\!\in\!(0,\infty)\)
and assume that an
\(\rno\!\in\!\SGBdomain{\m{w}}{\m{q}}\cap(0,\infty)\)
\begin{subequations}\label{eq:thm:TKDtoN}satisfies
\begin{align}
	\label{eq:thm:TKDtoN:hypothesis}	
	\sup\nolimits_{\dinp\in\numbers{R}}\abs{\gcdf{\llr}{\xi\!+\!\sigma\dinp}{\gamma\m{v;\rno}}{\m{w}}-\GCD{\dinp}}	
	&\leq \delta,		
\end{align}
for a \(\gamma>0\),
a \(\xi\in\numbers{R}\), 
a \(\sigma>0\), and 
a \(\delta\!\in\!(0,\sfrac{1}{8})\).
Let \(\f{\eta}[\cdot]\) be
\begin{align}
	\label{eq:thm:TKDtoN:eta:definition}		
	\f{\eta}[\epsilon]
	&\DEF e^{\frac{\rno\xi-\ln\gamma}{1-\rno}+\frac{\rno^{2} \sigma^{2}}{2(1-\rno)^2}}
	\GQF{\tfrac{\rno \sigma}{\abs{1-\rno}}\!-\!\IGQF{\epsilon}}
\end{align}
for all \(\epsilon\!\in\!(0,1)\) for 
\(\xi\) and \(\sigma\) of \eqref{eq:thm:TKDtoN:hypothesis}.

If \(\rno\!\in\!(0,1)\), then 
for all \(\epsilon\!\in\!(3\delta,1\!-\!3\delta)\) there exists a 
\(\phi\!\in\!(\epsilon\!-\!3\delta,\epsilon\!+\!3\delta)\) such that
\begin{align}
	\label{eq:thm:TKDtoN:zeroone}
	\NPD{\f{\eta}[\epsilon]}{\m{w}}{\m{q}}
	&\!=\!
	e^{-\xi+\frac{\sigma^{2}}{2}}\GQF{\sigma\!-\!\IGQF{1-\phi}}.
	\\
	\intertext{~~If \(\rno\!\in\!(1,\infty)\), then for all 
		\(\epsilon\!\in\!(2\delta,1\!-\!2\delta)\) there exists a 
		\(\phi\!\in\!(\epsilon\!-\!2\delta,\epsilon\!+\!2\delta)\) such that}
	\label{eq:thm:TKDtoN:oneinfinity}
	\NPD{\f{\eta}[\epsilon]}{\m{w}}{\m{q}}
	&\!=\!\norm{\m{w;\sim}}
	\!-\!
	e^{-\xi+\frac{\sigma^{2}}{2}}\GQF{\sigma\!-\!\IGQF{\phi}}.	
\end{align}
Furthermore, if \(\delta\!\leq\!\sfrac{1}{10}\)
and \(\widetilde{\eta}\in[\f{\eta}[\sfrac{3}{5}],\f{\eta}[\sfrac{2}{5}]]\),
then \(\exists\Delta\) satisfying  both
\(\abs{\ln\Delta}\!\leq\!\tfrac{0.43}{1\wedge \rno}\!+\!(\sigma\!+\!2.3)15\sqrt{2\pi}\delta\)
and
\begin{align}
\label{eq:thm:TKDtoN:Strassen}	
\hspace{-.05cm}\NPD{\widetilde{\eta}}{\m{w}}{\m{q}}
&\!=\!
\begin{cases}
\!\f{\theta}[\rno,\sigma]
\widetilde{\eta}^{\frac{\rno-1}{\rno}}
\gamma^{-\frac{1}{\rno}}\Delta\!
&\text{if~}\rno\!\in\!(0,1),\!
\\	
\!\norm{\m{w;\sim}}\!-\!
\f{\theta}[\rno,\sigma]
\widetilde{\eta}^{\frac{\rno-1}{\rno}}
\gamma^{-\frac{1}{\rno}}\Delta\!
&\text{if~}\rno\!>\!1,\!
\end{cases}		
\end{align}
where  
\(\f{\theta}[\rno,\sigma]\!=\!
\left(\sfrac{1}{2\pi}\right)^{\frac{1}{2\rno}}
(\GMR{\sfrac{\rno \sigma}{\abs{\rno-1}}})^{\frac{1-\rno}{\rno}}
\GMR{\sigma}\).		
\end{subequations}						
\end{theorem}
\begin{remark}
	Note that for all \(\rno\!\in\!(0,\infty)\) as \(\sigma\) 
	increases \(\f{\theta}[\rno,\sigma]\) is bounded above
	and below tightly by bounds on the Gaussian Mills ratio %\(\GMR{\cdot}\) 
	given in \eqref{eq:GMRBound:positive}:
	\(\f{\theta}\!\sim\!\tfrac{\rno}{\abs{1-\rno}}\left(\sigma\sqrt{2\pi}\right)^{-\sfrac{1}{\rno}}\)
	as \(\sigma\to\infty\). 
\end{remark}
\begin{proof}[Proof of Theorem \ref{thm:TKDtoN}]
Let us first establish  \eqref{eq:thm:TKDtoN:zeroone}
and \eqref{eq:thm:TKDtoN:oneinfinity}.
To that end we will first invoke Theorem \ref{thm:KDtoN} 
to bound \(\NPD{\cdot}{\m{w}}{\gamma\m{v;\rno}}\)
and \(\NPD{\cdot}{\m{q}}{\gamma\m{v;\rno}}\), and then 
employ Theorem \ref{thm:MeasureChange}.

Note that \eqref{eq:thm:TKDtoN:hypothesis} ensures that 
the hypothesis of Theorem \ref{thm:KDtoN} is satisfied for
\(\NPD{\cdot}{\m{w}}{\gamma\m{v;\rno}}\).
Thus Theorem \ref{thm:KDtoN} and the intermediate value theorem
imply that for all \(\epsilon\!\in\!(\delta,1\!-\!\delta)\) 
there exists a \(\phi\!\in\!(\epsilon\!-\!\delta,\epsilon\!+\!\delta)\) satisfying 
\begin{align}
	\label{eq:TKDtoN:01}
	\NPD{\epsilon}{\m{w}}{\gamma\m{v;\rno}}	
	&\!=\!e^{-\xi+\frac{\sigma^{2}}{2}}\GQF{\sigma\!-\!\IGQF{\phi}}
	&
	&=e^{-\xi}\f{g}[\sigma,\phi].
\end{align}

In order to apply Theorem \ref{thm:KDtoN} to bound
\(\NPD{\cdot}{\m{q}}{\gamma\m{v;\rno}}\), we need to 
first ensure that the hypothesis of Theorem \ref{thm:KDtoN} 
is satisfied for \(\NPD{\cdot}{\m{q}}{\gamma\m{v;\rno}}\). 
To that end first note that \eqref{eq:def:TiltedMeasure} 
and \eqref{eq:TiltedMeasure} imply
\begin{align}
	\notag
	\f{\llr;\gamma\m{v;\rno},\m{w}}	
	&=\ln\gamma+(\rno-1) \f{\llr;\m{w},\m{q}}
	&
	&\m{v;\rno}\text{-a.e.},
	\\
	\notag	
	\f{\llr;\gamma\m{v;\rno},\m{q}}	
	&=\ln\gamma+\rno \f{\llr;\m{w},\m{q}}
	&
	&\m{v;\rno}\text{-a.e.}.	
	\intertext{Thus}
	\notag%\label{eq:TKDtoN:02}
	\f{\llr;\gamma\m{v;\rno},\m{q}}	
	&=\tfrac{\rno}{\rno-1}\f{\llr;\gamma\m{v;\rno},\m{w}}	
	\!+\!\tfrac{1}{1-\rno}\ln\gamma
	&
	&\m{v;\rno}\text{-a.e.}.		
\end{align}
Then \eqref{eq:thm:TKDtoN:hypothesis} 
implies\footnote{Note that Kolmogorov distance between 
a continuous function \(\f{f}\) and a monotonic 
function does not change if the monotonic function 
is replaces by its upper-semicontinuous or
lower-semicontinuous variants. 
Thus for any continuous function \(\f{f}[\cdot]\) 
we have
\begin{align}
	\notag
	\sup\nolimits_{\dinp\in\numbers{R}}
	\abs{\gcdf{\llr}{\dinp}{\m{v}}{\m{q}}-\f{f}[\dinp]}
	&=\sup\nolimits_{\dinp\in\numbers{R}}
	\abs{\lim\nolimits_{\tau\uparrow\dinp}\gcdf{\llr}{\tau}{\m{v}}{\m{q}}-\f{f}[\dinp]}
\end{align}}
\begin{align}
	\notag%\label{eq:TKDtoN:03}
	\sup\nolimits_{\dinp\in\numbers{R}}\abs{\gcdf{\llr}{\tfrac{\ln\gamma-\rno\xi}{1-\rno}
			+\tfrac{\rno\sigma}{\abs{\rno-1}}\dinp}{\gamma\m{v;\rno}}{\m{w}}\!-\!\GCD{\dinp}}
	&\!\leq\!\delta.			
\end{align}	
Thus Theorem \ref{thm:KDtoN} and the intermediate value theorem
imply that for all \(\varkappa\!\in\!(\delta,1\!-\!\delta)\) 
there exists a \(\epsilon\!\in\!(\varkappa\!-\!\delta,\varkappa\!+\!\delta)\) satisfying 
\begin{align}
	\label{eq:TKDtoN:04}
	\NPD{\varkappa}{\m{q}}{\gamma\m{v;\rno}}	
	&\!=\!
	\f{\eta}[\epsilon].
\end{align}
Note that \(\f{\eta}[\cdot]\) defined in 
\eqref{eq:thm:TKDtoN:eta:definition}
is invertible, decreasing, and continuous on \((0,1)\). 
On the other hand 
\(\gamma\norm{\m{v;\rno}}\!\in\!(1\!-\!\delta,1\!+\!\delta)\)
by \eqref{eq:thm:TKDtoN:hypothesis} and 
\(\NPD{\cdot}{\m{q}}{\gamma\m{v;\rno}}\) 
is invertible, decreasing, and continuous on 
\((0,\gamma\norm{\m{v;\rno}})\). 
Thus \eqref{eq:TKDtoN:04} defines an invertible 
increasing and continuous function, i.e., 
\(\f{\zeta}[\cdot]\DEF\f{\eta;;-1}[\NPD{\cdot}{\m{q}}{\gamma\m{v;\rno}}]\), 
on \((\delta,1-\delta)\) satisfying
\(\abs{\f{\zeta}[\varkappa]\!-\!\varkappa}\!\leq\!\delta\)
for all \(\varkappa\) in \((\delta,1-\delta)\).
Then \(\f{\zeta;;-1}[\cdot]\) is increasing on 
its domain and \((2\delta,1-2\delta)\) is a subset of 
the domain of \(\f{\zeta;;-1}[\cdot]\). Hence
\begin{align}
	\label{eq:TKDtoN:05}
	\NPD{\f{\zeta;;-1}[\epsilon]}{\m{q}}{\gamma\m{v;\rno}}
	&\!=\!\f{\eta}[\epsilon]
	&
	&\forall \epsilon\!\in\!(2\delta,1\!-\!2\delta),
	\\
	\label{eq:TKDtoN:06}
	\abs{\epsilon-\f{\zeta;;-1}[\epsilon]}
	&\leq \delta
	&
	&\forall \epsilon\!\in\!(2\delta,1\!-\!2\delta).
\end{align}
Furthermore, \eqref{eq:NDPInversion} and \eqref{eq:TKDtoN:05} imply
\begin{align}
	\label{eq:TKDtoN:07}
	\f{\zeta;;-1}[\epsilon]
	&=\NPD{\f{\eta}[\epsilon]}{\gamma\m{v;\rno}}{\m{q}}
	&
	&\forall \epsilon\!\in\!(2\delta,1\!-\!2\delta).
\end{align}
On the other hand, 
\(\f{\eta}[\epsilon]\in(0,\norm{\m{q;\sim}})\) 
for all \(\epsilon\!\in\!(2\delta,1\!-\!2\delta)\),
thus 
\(\NPD{\f{\eta}[\epsilon]}{\m{w}}{\m{q}}\in(0,\norm{\m{w;\sim}})\)
for all \(\epsilon\) in \((2\delta,1-2\delta)\).
Applying 
\eqref{eq:thm:MeasureChange:Increasing}	 of Theorem \ref{thm:MeasureChange}
for \(\f{f}[\dinp]=\gamma\dinp^{\rno}\) 
for \(\varepsilon=\NPD{\f{\eta}[\epsilon]}{\m{w}}{\m{q}}\)
and invoking \eqref{eq:NDPInversion}, we get
\begin{align}
	\notag	
	\f{\eta}[\epsilon]
	&\!=\!\NPD{\fav{\NPD{\f{\eta}[\epsilon]}{\m{w}}{\m{q}}}{\gamma\m{v;\rno}}}{\m{q}}{\gamma\m{v;\rno}}
	&
	&\forall \epsilon\!\in\!(2\delta,1\!-\!2\delta).
\end{align}
Invoking \eqref{eq:NDPInversion} again, this time for  
\(\NPD{\cdot}{\m{q}}{\gamma\m{v;\rno}}\), we get
\begin{align}
	\notag%\label{eq:TKDtoN:08}
	\!\NPD{\f{\eta}[\epsilon]}{\gamma\m{v;\rno}}{\m{q}}
	&\!=\!\fav{\NPD{\f{\eta}[\epsilon]}{\m{w}}{\m{q}}}{\gamma\m{v;\rno}}
	&
	&\forall \epsilon\!\in\!(2\delta,1\!-\!2\delta).
\end{align}
Thus \eqref{eq:TKDtoN:06} and \eqref{eq:TKDtoN:07} imply
\begin{align}
	\label{eq:TKDtoN:09}	
	\abs{\epsilon-\fav{\NPD{\f{\eta}[\epsilon]}{\m{w}}{\m{q}}}{\gamma\m{v;\rno}}}
	&\leq \delta
	&
	&\forall\epsilon\!\in\!(2\delta,1\!-\!2\delta).	
\end{align}
Since \(\gamma\norm{\m{v;\rno}}\!\in\!(1\!-\!\delta,1\!+\!\delta)\)
by \eqref{eq:thm:TKDtoN:hypothesis}, using triangle inequality 
together with \eqref{eq:MeasureChange}, we get 
\begin{align}
	\label{eq:TKDtoN:10}	
	\abs{1-\epsilon-\mdv{\NPD{\f{\eta}[\epsilon]}{\m{w}}{\m{q}}}{\gamma\m{v;\rno}}}
	&\leq 2\delta
	&
	&\forall\epsilon\!\in\!(2\delta,1\!-\!2\delta).	
\end{align} 
\eqref{eq:NDPInversion}, 
\eqref{eq:thm:MeasureChange:CsiszarDual:Increasing}	
of Theorem \ref{thm:MeasureChange}
for \(\f{f}[\dinp]\!=\!\gamma\dinp^{\rno}\),
\eqref{eq:TKDtoN:01}, and \eqref{eq:TKDtoN:10} 
imply \eqref{eq:thm:TKDtoN:zeroone};
and  \eqref{eq:NDPInversion}, 
\eqref{eq:thm:MeasureChange:CsiszarDual:Decreasing}	
of Theorem \ref{thm:MeasureChange}
for \(\f{f}[\dinp]\!=\!\gamma\dinp^{\rno}\),
\eqref{eq:TKDtoN:01}, and \eqref{eq:TKDtoN:09} 
imply \eqref{eq:thm:TKDtoN:oneinfinity}.

We will establish \eqref{eq:thm:TKDtoN:Strassen}
by invoking \eqref{eq:thm:TKDtoN:zeroone}
and \eqref{eq:thm:TKDtoN:oneinfinity}
together with Taylor's theorem.
Note that \eqref{eq:thm:TKDtoN:eta:definition},
\eqref{eq:thm:TKDtoN:zeroone},
and \eqref{eq:thm:TKDtoN:oneinfinity} can
be\begin{subequations}\label{eq:TKDtoN:11}~expressed 
in terms of \(\f{g}[\cdot,\cdot]\) defined in 
\eqref{eq:def:SGMR} as
\begin{align}
	\f{\eta}[\epsilon]
	&=e^{\frac{\rno\xi-\ln\gamma}{1-\rno}}
	\f{g}[\tfrac{\rno \sigma}{\abs{1-\rno}},\epsilon],
	\\
	\NPD{\f{\eta}[\epsilon]}{\m{w}}{\m{q}}
	&\!=\!
	e^{-\xi}\f{g}[\sigma,1-\phi],
	\\
	\NPD{\f{\eta}[\epsilon]}{\m{w}}{\m{q}}
	&\!=\!\norm{\m{w;\sim}}
	\!-\!
	e^{-\xi}\f{g}[\sigma,\phi].	
\end{align}	
\end{subequations}
Let \(\epsilon\) be the unique solution of 
\(\widetilde{\eta}\!=\!\f{\eta}[\epsilon]\) 
in \([\sfrac{2}{5},\sfrac{3}{5}]\)
and \(\phi\) be the corresponding number 
in \((\epsilon\!-\!3\delta,\epsilon\!+\!3\delta)\)
satisfying \eqref{eq:thm:TKDtoN:zeroone} if \(\rno\!\in\!(0,1)\)
and \eqref{eq:thm:TKDtoN:oneinfinity} if \(\rno\!\in\!(1,\infty)\).
\begin{subequations}\label{eq:TKDtoN:12}
Since \(\ln\f{g}[\cdot,\cdot]\) is continuously differentiable 
in its second argument, there exists \(\epsilon_{0}\) between 
\(\epsilon\) and \(\phi\) satisfying 
\begin{align}
	\label{eq:TKDtoN:12A}	
	\ln\f{g}[\sigma,1\!-\!\phi]
	&\!=\!
	\ln \f{g}[\sigma,1\!-\!\epsilon]
	\!+\!\tfrac{\pdimpf[\f{g}]{\varepsilon}[\sigma,1\!-\!\epsilon_{0}]}{\f{g}[\sigma,1\!-\!\epsilon_{0}]}(\epsilon\!-\!\phi)
	\intertext{if \(\rno\!\in\!(0,1)\) and satisfying}
	\label{eq:TKDtoN:12B}	
	\ln\f{g}[\sigma,\phi]
	&\!=\!\ln\f{g}[\sigma,\epsilon]
	\!+\!\tfrac{\pdimpf[\f{g}]{\varepsilon}[\sigma,\epsilon_{0}]}{\f{g}[\sigma,\epsilon_{0}]}(\phi\!-\!\epsilon),	
\end{align}
for \(\rno\!\in\!(1,\infty)\).	
\end{subequations}
Note that \(\epsilon_{0}\!\in\!(\sfrac{1}{10},\sfrac{9}{10})\) 
in either case. Let us proceed with bounding 
\(\pdimpf[\f{g}]{\varepsilon}[\sigma,\epsilon_{0}]\)
for an \(\epsilon_{0}\!\in\!(\sfrac{1}{10},\sfrac{9}{10})\):
\begin{align}
	\notag	
	\abs{\tfrac{\pdimpf[\f{g}]{\varepsilon}[\sigma,\epsilon_{0}]}{\f{g}[\sigma,\epsilon_{0}]}}
	&=\tfrac{1}{\GMR{\sigma-\IGQF{\epsilon_{0}}}}\tfrac{\GMR{\abs{\IGQF{\epsilon_{0}}}}}{(1-\epsilon_{0})\wedge \epsilon_{0}}
	\\
	\notag
	&\mathop{\leq}^{(a)} 
	\tfrac{1}{\GMR{\sigma+\IGQF{\epsilon_{0}}}}
	\tfrac{\sqrt{\sfrac{\pi}{2}}}{(1-\epsilon_{0})\wedge \epsilon_{0}},
	\\
	\notag
	&\mathop{\leq}^{(b)} 
	\tfrac{5\sqrt{2\pi}}{\GMR{\sigma+\IGQF{\sfrac{1}{10}}}},
	\\
	\notag
	&\mathop{\leq}^{(c)} \tfrac{5\sqrt{2\pi}}{\GMR{\sigma+1.3}},
	\\
	\notag
	&\mathop{\leq}^{(d)}(\sigma\!+\!1.3\!+\!\sqrt{(\sigma\!+\!1.3)^{2}\!+\!4})
	2.5\sqrt{2\pi},
	\\
	\label{eq:TKDtoN:13}	
	&\mathop{\leq}^{(e)} (\sigma+2.3)5\sqrt{2\pi},		
\end{align}
where \((a)\) holds because 
\(\GMR{\tau}\!\leq\!\sqrt{\sfrac{\pi}{2}}\) for all \(\tau\!>\!0\)
by \eqref{eq:GMRBound:positive},%\cite[7.1.13]{AbramowitzStegun}
\((b)\) holds  because \(\GMR{\cdot}\) and \(\IGQF{\cdot}\)
are both decreasing functions
and \(\epsilon_{0}\!\in\!(0.1,0.9)\),
\((c)\) follows from \(\IGQF{\sfrac{1}{10}}\!<\!1.3\) and the monotonicity of 
\(\GMR{\cdot}\),
\((d)\) follows from \eqref{eq:GMRBound:positive},
\((e)\) follows from \(\sqrt{\tau^{2}\!+\!4}\!\leq\!\tau\!+\!2\)
for \(\tau\geq0\).

On the other hand, using
\eqref{eq:GMRBound:multiplicative}, \eqref{eq:SGMR},
and invoking worst case assumptions via
\(\IGQF{\sfrac{2}{5}}\leq 0.26\), 
\begin{subequations}\label{eq:TKDtoN:14}
we get
\begin{align}
	\label{eq:TKDtoN:14A}	
	\!\abs{\tfrac{1-\rno}{\rno}\ln\f{g}[\tfrac{\rno\sigma}{\abs{1-\rno}},\epsilon]
		\!+\!\ln\tfrac{\f{g}[\sigma,1-\epsilon]}{\f{\theta}[\rno,\sigma]}}
	&\!\leq\!\tfrac{0.43}{\rno}
	&
	&\text{if~}\rno\!<\!1,
	\\
	\label{eq:TKDtoN:14B}	
	\!\abs{\tfrac{1-\rno}{\rno}\ln\f{g}[\tfrac{\rno\sigma}{\abs{1-\rno}},\epsilon]
		\!+\!\ln\tfrac{\f{g}[\sigma,\epsilon]}{\f{\theta}[\rno,\sigma]}}
	&\!\leq\!0.43
	&
	&\text{if~}\rno\!>\!1.	
\end{align}		
\end{subequations}
\eqref{eq:thm:TKDtoN:Strassen} follows from 
\eqref{eq:TKDtoN:11},
\eqref{eq:TKDtoN:12},
\eqref{eq:TKDtoN:13},
and
\eqref{eq:TKDtoN:14}.
%{\color{magenta}
%\begin{align}
%\notag
%\varpi
%&\!=\!\tfrac{1-\rno}{\rno}\ln\f{g}[\tfrac{\rno}{1-\rno}\sigma,\varepsilon]
%+\ln\f{g}[\sigma,1\!-\!\varepsilon]-\ln\f{\theta}[\rno,\sigma]
%\\
%\notag
%&\!=\!\tfrac{1-\rno}{\rno}\ln\tfrac{\GMR{\frac{\rno}{1-\rno}\sigma-\IGQF{\varepsilon}}}{\GMR{\frac{\rno}{1-\rno}\sigma}}
%\!+\!\ln\tfrac{\GMR{\sigma+\IGQF{\varepsilon}}}{\GMR{\sigma}}
%\!-\!\tfrac{(\IGQF{\varepsilon})^{2}}{2\rno}
%\intertext{For \(\rno\in(1,\infty)\)}
%\notag
%\varpi
%&\!=\!\tfrac{1-\rno}{\rno}\ln\f{g}[\tfrac{\rno}{\rno-1}\sigma,\varepsilon]
%+\ln\f{g}[\sigma,1\!-\!\varepsilon]
%-\ln\f{\theta}[\rno,\sigma]
%\\
%\notag
%&\!=\!\tfrac{1-\rno}{\rno}\ln\tfrac{\GMR{\frac{\rno}{\rno-1}\sigma-\IGQF{\varepsilon}}}{\GMR{\frac{\rno}{\rno-1}\sigma}}
%\!+\!\ln\tfrac{\GMR{\sigma-\IGQF{\varepsilon}}}{\GMR{\sigma}}
%\!-\!\tfrac{(\IGQF{\varepsilon})^{2}}{2\rno}
%\end{align}
%}
%{\color{blue}
%\begin{align}
%\notag	
%0.2533
%&\leq
%\IGQF{\sfrac{2}{5}}
%&=-\IGQF{\sfrac{3}{5}}
%&%=0.253347
%\leq 0.2534,
%\\	
%\notag	
%0.3186
%&\leq
%\IGQF{\sfrac{3}{8}}
%&=-\IGQF{\sfrac{5}{8}}
%&%=0.31863936
%\leq 0.3187,
%\\	
%\notag	
%0.6744
%&\leq
%\IGQF{\sfrac{1}{4}}
%&=-\IGQF{\sfrac{3}{4}}
%&%=0.67448975
%\leq 
%0.675,
%\\	
%\notag	
%1.150
%&\leq
%\IGQF{\sfrac{1}{8}}
%&=-\IGQF{\sfrac{7}{8}}
%&%=1.15034938
%\leq 
%1.151,
%\\	
%\notag	
%1.281
%&\leq
%\IGQF{\sfrac{1}{10}}
%&=-\IGQF{\sfrac{9}{10}}
%&%=1.281552
%\leq
%1.282.							
%\end{align}
%}	  	 		
\end{proof}	

\section{Applications}\label{sec:Applications}
Under appropriate structural assumptions on \(\m{w}\) and \(\m{q}\),
the hypotheses of Theorems \ref{thm:KDtoN} and \ref{thm:TKDtoN} 
follow from Berry--Essen theorems non-asymptotically 
and from Central Limit Theorem  asymptotically. 
We will explore these implications 
for possibly non-stationary (i.e., time-inhomogeneous) 
memorlyless case in \S\ref{sec:Memoryless}
and for time-homogeneous discrete (i.e., countable state-space) 
Markov case in \S\ref{sec:DiscreteMarkov}.
Then, we will describe the 
asymptotic simple hypothesis testing problem 
and show that the Central Limit Theorem
and Theorem \ref{thm:KDtoN} imply a
second order Stein's lemma in \S\ref{sec:Asymptotic}.

\subsection{Memoryless Case}\label{sec:Memoryless}
In the memoryless case, \(\m{w}\) and \(\m{q}\) 
are assumed to be product measures.
Since both the normalized tilted measure and the likelihood ratio
inherit the product structure, the finite length implications 
of Theorems \ref{thm:KDtoN} and \ref{thm:TKDtoN} 
are immediate even in the non-stationary case,
provided that appropriate moment conditions hold. 
In models with memory, relating tilting operation to the
structural assumptions of the model is the main technical 
challenge, as we will see for the Markov case in 
\S\ref{sec:DiscreteMarkov}.

The product measure is determined by the measure of the product 
events, see \cite[Theorem 3.3.1]{bogachev}.
The product of two \(\sigma\)-finite measures is 
a \(\sigma\)-finite measure because Cartesian product
of two countable sets is also countable. A more detail
discussion products of measures can be found 
in \cite[\S 3.3]{bogachev}.	
\begin{assumption}\label{assumption:Memorylessness}
	\(\m{w}\) and \(\m{q}\) are \(\sigma\)-finite product measures
	of the form \(\m{w}=\otimes_{\tin=1}^{\blx}\m{w;\tin}\)
	and \(\m{q}=\otimes_{\tin=1}^{\blx}\m{q;\tin}\)
	for some positive integer \(\blx\),
	where \(\m{w;\tin}\) and \(\m{q;\tin}\)
	are \(\sigma\)-finite measures
	on the measurable space \((\outS_{\tin},\outA_{\tin})\),
	for each \(\tin\!\in\!\{1,\ldots,\blx\}\).	
\end{assumption}
The memorylessness assumption implies 
the additivity log-likelihood ratios:
if \(\m{w}\) and \(\m{q}\) satisfy
Assumption \ref{assumption:Memorylessness}, then 
\begin{align}
	\label{eq:Memorylessness:LLRAdditivity}
	\f{\llr;\m{w},\m{q}}
	&=\sum\nolimits_{\tin=1}^{\blx}
	\f{\llr;\m{w;\tin},\m{q;\tin}}
	&
	&\m{w;\sim}\text{-a.e.}.		
\end{align}
If \(\m{w;\tin}\)'s are probability measures, 
then \eqref{eq:Memorylessness:LLRAdditivity} and 
Fubini's theorem \cite[Theorem 3.4.4]{bogachev}
imply the additivity of \(\KLD{\m{w}}{\m{q}}\)
and \(\ln\norm{\m{v;\rno;\m{w},\m{q}}}\) 
privided that corresponding summands, i.e.,
\(\KLD{\m{w;\tin}}{\m{q;\tin}}\)'s or 
\(\ln \norm{\m{v;\rno;\m{w;\tin},\m{q;\tin}}}\)'s,
are finite.\footnote{The additivity of \(\ln\norm{\m{v;\rno;\m{w},\m{q}}}\) 
	holds even when \(\m{w;\tin}\)'s are not necessarily probability measures.}

\begin{corollary}\label{cor:Memoryless:KDtoN}
Let \((\m{w},\m{q})\) be a pair of measures satisfying 
Assumption \ref{assumption:Memorylessness},
for which \(\m{w;\tin}\) is a probability measure
\begin{subequations}\label{eq:cor:Memoryless:KDtoN}satisfying 
\(\abs{\KLD{\m{w;\tin}}{\m{q;\tin}}}\!<\!\infty\)
for all \(\tin\), and let \(\xi\), \(\sigma\),  \(\tlx\), 
and \(\delta\) be
\begin{align}
	\label{eq:cor:Memoryless:KDtoN:xi}		
	\xi
	&\!=\!\KLD{\m{w}}{\m{q}}
	\\
	\label{eq:cor:Memoryless:KDtoN:sigma}		
	\sigma^{2}
	&\!=\!\int\!\left(\f{\llr;\m{w},\m{q}}\!-\!\KLD{\m{w}}{\m{q}}\right)^{2}\dif{\m{w}}
	\\
	\label{eq:cor:Memoryless:KDtoN:absolutethirdmoment}		
	\tlx
	&\!=\!\sum\nolimits_{\tin=1}^{\blx}\!
	\int\!\abs{\f{\llr;\m{w;\tin},\m{q;\tin}}\!-\!\KLD{\m{w;\tin}}{\m{q;\tin}}}^{3}
	\dif{\m{w}},	
	\\
	\label{eq:cor:Memoryless:KDtoN:delta}		
	\delta
	&\!=\!0.56\sfrac{\tlx}{\sigma^{3}}.	
\end{align}
If \(\sigma\!>\!0\) and \(\delta\!\in\!(0,\sfrac{1}{2})\),
then \eqref{eq:thm:KDtoN:upperbound},
\eqref{eq:thm:KDtoN:lowerbound},
and \eqref{eq:thm:KDtoN:Strassen}
of Theorem \ref{thm:KDtoN}
hold for \(\xi\), \(\sigma\), and \(\delta\) defined above.	
\end{subequations}		
\end{corollary}
\begin{remark}
	If \((\m{w},\m{q})\) satisfies
	Assumption \ref{assumption:Memorylessness},
	\(\m{w;\tin}\) is a probability measure satisfying 
	\(\abs{\KLD{\m{w;\tin}}{\m{q;\tin}}}\!<\!\infty\)
	for all \(\tin\), and \(\sigma\!=\!0\), then 
	we are in the trivial-non-singular case,
	\(\f{\llr;\m{w},\m{q}}\!=\!\KLD{\m{w}}{\m{q}}\)
	holds \(\m{w}\)-a.s.,
	\(\KLD{\m{w}}{\m{q}}\!=\!-\ln\norm{\m{q;\sim}}\), 
	and \eqref{eq:trivial-non-singular-non-void} imply
	\begin{align}
		\notag
		\NPD{\varepsilon}{\m{q}}{\m{w}}
		&=e^{-\KLD{\m{w}}{\m{q}}}\abp{1-\varepsilon}
		&
		&\forall \varepsilon\geq 0.		
	\end{align}
	Hence \eqref{eq:thm:KDtoN:upperbound},
	\eqref{eq:thm:KDtoN:lowerbound},
	and \eqref{eq:thm:KDtoN:Strassen}
	of Theorem \ref{thm:KDtoN} hold for \(\delta\!=\!0\).
\end{remark}

\begin{proof}[Proof of Corollary \ref{cor:Memoryless:KDtoN}]
	Note that \(\m{w;\tin}\AC\m{q;\tin}\) because 
	\(\abs{\KLD{\m{w;\tin}}{\m{q;\tin}}}\) is finite. 
	Thus \(\f{\llr;\m{w;\tin},\m{q;\tin}}=\ln\der{\m{w;\tin}}{\m{q;\tin}}\)
	holds \(\m{w}\)-a.s. 
	Furhtmore, we can calculate the first moment and the variance of 
	\(\f{\llr;\m{w},\m{q}}\) under \(\m{w}\) using 
	\eqref{eq:Memorylessness:LLRAdditivity} 
	and Fubini's theorem \cite[Theorem 3.4.4]{bogachev}:
	\begin{align}
		\notag
		\KLD{\m{w}}{\m{q}}
		&\!=\!\sum\nolimits_{\tin=1}^{\blx}\!\KLD{\m{w;\tin}}{\m{q;\tin}}
		\\
		\notag
		\sigma^{2}
		&\!=\!\sum\nolimits_{\tin=1}^{\blx}\!
		\int\!\left(\ln \der{\m{w;\tin}}{\m{q;\tin}}\!-\!\KLD{\m{w;\tin}}{\m{q;\tin}}\right)^{2}\!\dif{\m{w;\tin}}.		
	\end{align}
	Furthermore \eqref{eq:thm:KDtoN:hypothesis} holds for \(\xi\), 
	\(\sigma\), and \(\delta\) defined in \eqref{eq:cor:Memoryless:KDtoN} 
	as a result of Berry--Esseen Theorem \cite{berry41,esseen42,shevtsova10}.	
\end{proof}
Implications of \eqref{eq:thm:KDtoN:Strassen} for \(\NPD{\cdot}{\m{q}}{\m{w}}\)
on closed subintervals of \((0,1)\) becomes more apparent
when considered for the stationary memoryless case, i.e., 
when \(\m{w;\tin}\!=\!\m{w;1}\) and \(\m{q;\tin}\!=\!\m{q;1}\)
for all \(\tin\) for some \(\m{w;1}\) and \(\m{q;1}\).
Note that \(\KLD{\m{w}}{\m{q}}=\blx\KLD{\m{w;1}}{\m{q;1}}\),
\(\sigma=\sigma_{1}\sqrt{\blx}\), and
\(\delta=\sfrac{\delta_{1}}{\sqrt{\blx}}\) for
some \(\sigma_{1}\) and \(\delta_{1}\) in the stationary case. 
Thus\footnote{To convert the hypothesis of \eqref{eq:thm:KDtoN:Strassen}
	to a lower bound on \(\sqrt{\blx}\), we employ the fact that
	\(\GQF{\dinp}\leq\tfrac{1}{\dinp\sqrt{2\pi}}\)
	for all \(\dinp\geq0\) .} 
for all \(\blx\geq (\tfrac{\sigma_{1}\delta_{1}+\sfrac{1}{\sqrt{2\pi}}}{\epsilon_{1} \sigma_{1}})^{2}\)
the following equality holds for all  \(\varepsilon\in[\epsilon_{1},1\!-\!\epsilon_{1}]\)
\begin{align}
	\NPD{\varepsilon}{\m{q}}{\m{w}}		
	&\!=\!\tfrac{\f{\Delta}[\varepsilon]}{\sqrt{2\pi}}
	e^{\!-\!\blx\KLD{\m{w;1}}{\m{q;1}}+\sqrt{\blx}\sigma_{1}\tau+\frac{\tau^{2}}{2}}
	\GMR{\sqrt{\blx}\sigma_{1}\!-\!\tau},
\end{align}
where \(\tau\!=\!\IGQF{\varepsilon}\), 
\(\abs{\ln\f{\Delta}[\varepsilon]}\!\leq\!\tfrac{(\sigma_{1}\sqrt{\blx}+0.5)\delta_{1}\sqrt{2\pi}}{\epsilon_{1}\sqrt{\blx}-\delta_{1}}\)
and the Gaussian Mills ratio  \(\GMR{\sqrt{\blx}\sigma_{1}\!-\!\tau}\) 
is bounded tightly by \eqref{eq:GMRBound:positive}
because \(\blx\geq (\tfrac{\sigma_{1}\delta_{1}+\sfrac{1}{\sqrt{2\pi}}}{\epsilon_{1} \sigma_{1}})^{2}\)
ensures \(\sqrt{\blx}\sigma_{1}\!\geq\!\IGQF{\epsilon_{1}}\!\geq\!\tau\). 

\begin{corollary}\label{cor:Memoryless:TKDtoN}
Let \((\m{w},\m{q})\) be a pair of measures satisfying 
Assumption \ref{assumption:Memorylessness}
for \((\m{w;\tin},\m{q;\tin})\) 
pairs\begin{subequations}\label{eq:cor:Memoryless:TKDtoN}~
satisfying 
\(\norm{\m{v;\rno;\m{w;\tin},\m{q;\tin}}}\!<\!(0,\infty)\) 
for all \(\tin\) 
for some  \(\rno\!\in\!(0,\infty)\!\setminus\!\{1\}\),
and let \(\xi\), \(\sigma\),  \(\tlx\), 
and \(\delta\) be
\begin{align}
	\label{eq:cor:Memoryless:TKDtoN:gamma}		
	\gamma
	&\!=\!\sfrac{1}{\norm{\m{v;\rno;\m{w},\m{q}}}}
	\\
	\label{eq:cor:Memoryless:TKDtoN:xi}		
	\xi
	&\!=\!\KLD{\m{v}}{\m{w}}
	\\
	\label{eq:cor:Memoryless:TKDtoN:sigma}		
	\sigma^{2}
	&\!=\! (\rno\!-\!1)^{2}\int\!\left(\f{\llr;\m{w},\m{q}}
	\!-\!\EX[\m{v}]{\f{\llr;\m{w},\m{q}}}\right)^{2}\dif{\m{v}},
	\\
	\label{eq:cor:Memoryless:TKDtoN:absolutethirdmoment}		
	\tlx
	&\!=\! \abs{\rno\!-\!1}^{3}
	\sum\nolimits_{\tin=1}^{\blx}
	\int\abs{\f{\llr;\m{w;\tin},\m{q;\tin}}-\EX[\m{v}]{\f{\llr;\m{w;\tin},\m{q;\tin}}}}^{3}\dif{\m{v}},	
	\\
	\label{eq:cor:Memoryless:TKDtoN:delta}		
	\delta
	&\!=\!0.56\sfrac{\tlx}{\sigma^{3}},	
\end{align}		
where \(\m{v}\!=\!\gamma\m{v;\rno;\m{w},\m{q}}\)\end{subequations}.
If \(\abs{\xi}\!<\!\infty\), \(\sigma\!>\!0\),
and \(\delta\!\in\!(0,\sfrac{1}{8})\)
then \eqref{eq:thm:TKDtoN:zeroone},
\eqref{eq:thm:TKDtoN:oneinfinity},
and \eqref{eq:thm:TKDtoN:Strassen}
of Theorem \ref{thm:TKDtoN}
hold for \(\gamma\), \(\xi\), \(\sigma\), \(\delta\) defined above.	
\end{corollary}
If the hypotheses\begin{subequations}\label{eq:SGB:identity} of 
Corollary \ref{cor:Memoryless:TKDtoN} hold, then
\begin{align}
	\label{eq:SGB:identity:W}
	\KLD{\m{v}}{\m{w}}
	&\!=\!(\rno\!-\!1)
	\EX[\m{v}]{\f{\llr;\m{w},\m{q}}}\!+\!\ln\gamma, 
	\\
	\label{eq:SGB:identity:Q1}
	\KLD{\m{v}}{\m{q}}
	&\!=\!\rno\EX[\m{v}]{\f{\llr;\m{w},\m{q}}}\!+\!\ln\gamma,
	\\
	\label{eq:SGB:identity:Q2}
	&\!=\!\tfrac{\rno\xi-\ln\gamma}{\rno-1}
\end{align}
for \(\m{v}\!=\!\gamma\m{v;\rno;\m{w},\m{q}}\).
The identities in \eqref{eq:SGB:identity} are present 
in \cite[Theorem 5]{shannonGB67A} 
for\end{subequations}~the case when \(\outS\) is a
finite set and \(\m{w}\) and \(\m{q}\) are 
probability measures.				
Note that the expression on the right hand side 
of \eqref{eq:SGB:identity:Q2} shows up in 
\eqref{eq:thm:TKDtoN:eta:definition} defining 
\(\f{\eta}[\cdot]\).

\begin{remark}
	If \((\m{w},\m{q})\) satisfies
	Assumption \ref{assumption:Memorylessness}
	for pairs \((\m{w;\tin},\m{q;\tin})\) satisfying
	\(\norm{\m{v;\rno;\m{w;\tin},\m{q;\tin}}}\!\in\!(0,\infty)\) 
	for all \(\tin\) for some  \(\rno\!\in\!(0,\infty)\),
	and \(\abs{\xi}\!<\!\infty\), \(\sigma\!=\!0\),
	then we are in the trivial-non-singular case,
	\(\f{\llr;\m{w},\m{q}}
	\!=\!\ln\norm{\m{w;\sim}}\!-\!\ln\norm{\m{q;\sim}}\)
	holds \(\m{w}\)-a.s.,
	\(\KLD{\m{v}}{\m{w}}\!=\!-\ln\norm{\m{w;\sim}}\),
	\(\KLD{\m{v}}{\m{q}}\!=\!-\ln\norm{\m{q;\sim}}\),
	and \eqref{eq:trivial-non-singular-non-void} imply
	\begin{align}
		\NPD{\tau}{\m{w}}{\m{q}}
		&=e^{-\KLD{\m{v}}{\m{w}}}\abp{1-\tau e^{\KLD{\m{v}}{\m{q}}}}
		&
		&\forall \tau\geq0.	
	\end{align}
	Hence, \eqref{eq:thm:TKDtoN:zeroone}
	\eqref{eq:thm:TKDtoN:oneinfinity},
	and \eqref{eq:thm:TKDtoN:Strassen}
	of Theorem \ref{thm:TKDtoN} holds for \(\delta\!=\!0\).
\end{remark}
\begin{proof}[Proof of Corollary \ref{cor:Memoryless:TKDtoN}]
	Since \(\norm{\m{v;\rno;\m{w;\tin},\m{q;\tin}}}\!<\!(0,\infty)\) 
	for all \(\tin\), \eqref{eq:Memorylessness:LLRAdditivity}  
	and Fubini's theorem \cite[Theorem 3.4.4]{bogachev} imply
	\begin{align}
		\notag	
		\norm{\m{v;\rno;\m{w},\m{q}}}
		&\!=\!\prod\nolimits_{\tin=1}^{\blx}
		\norm{\m{v;\rno;\m{w;\tin},\m{q;\tin}}},
		\\
		\notag
		\f{\llr;\m{v},\m{w}}
		&\!=\!\sum\nolimits_{\tin=1}^{\blx}
		\left[(\rno-1)\f{\llr;\m{w;\tin},\m{q;\tin}}-\ln\norm{\m{v;\rno;\m{w;\tin},\m{q;\tin}}}\right]
		&
		&\m{v}\text{-a.s.}.
		\intertext{Consequently, \(\m{v}\) inherits the product 
			structure of \(\m{w}\) and \(\m{q}\), i.e.,}
		\notag	
		\m{v}	
		&\!=\!\bigotimes\nolimits_{\tin=1}^{\blx}\tfrac{\m{v;\rno;\m{w;\tin},\m{q;\tin}}}{\norm{\m{v;\rno;\m{w;\tin},\m{q;\tin}}}}.
	\end{align}
	Thus \(\f{\llr;\m{v},\m{w}}\) is a sum of
	random variables that are independent under \(\m{v}\).
	Hence  \eqref{eq:thm:TKDtoN:hypothesis} holds for \(\gamma\), 
	\(\xi\), \(\sigma\), \(\delta\) defined in \eqref{eq:cor:Memoryless:TKDtoN} 
	as a result of Berry--Esseen Theorem \cite{berry41,esseen42,shevtsova10}.	
\end{proof}
Implications of \eqref{eq:thm:TKDtoN:Strassen} for \(\NPD{\cdot}{\m{q}}{\m{w}}\) 
in the memoryless case becomes more apparent when considered for the stationary
case:
If \(\m{w;\tin}\!=\!\m{w;1}\) and \(\m{q;\tin}\!=\!\m{q;1}\) for all \(\tin\) 
for some \(\m{w;1}\) and \(\m{q;1}\) satisfying 
\(\norm{\m{v;\rno;\m{w;1},\m{q;1}}}\!<\!(0,\infty)\) for some 
\(\rno\!\in\!(0,\infty)\!\setminus\!\{1\}\), then 
\(\ln\norm{\m{v;\rno;\m{w},\m{q}}}\!=\!\blx\ln\norm{\m{v;\rno;\m{w;1},\m{q;1}}}\),
\(\KLD{\m{v}}{\m{w}}\!=\!\blx\KLD{\m{v;1}}{\m{w;1}}\)
%\(\KLD{\m{v}}{\m{q}}\!=\!\blx\KLD{\m{v;1}}{\m{q;1}}\) 
for \(\m{v;1}\!=\!\sfrac{\m{v;\rno;\m{w;1},\m{q;1}}}{\norm{\m{v;\rno;\m{w;1},\m{q;1}}}}\), 
\(\sigma\!=\!\sigma_{1}\sqrt{\blx}\) and \(\delta\!=\!\sfrac{\delta_{1}}{\sqrt{\blx}}\). 
Let \(\f{\tau}[\cdot]\) be
\begin{align}
	\notag	
	%\label{eq:thm:TKDtoN:eta:definition}		
	\f{\tau}[\epsilon]
	&\DEF \tfrac{\rno^{2} \sigma^{2}}{2(1-\rno)^2}+\ln \GQF{\tfrac{\rno \sigma}{\abs{1-\rno}}\!-\!\IGQF{\epsilon}}.
\end{align}
For all \(\blx\geq (10 \delta_{1})^{2}\) and 
\(\tau_{1}\!\in\![\f{\tau}[\sfrac{3}{5}],\f{\tau}[\sfrac{2}{5}]]\), then \(\exists\Delta\) satisfying  both
\(\abs{\ln\Delta}\!\leq\!\tfrac{0.43}{1\wedge \rno}\!+\!(\sigma_{1}\!+\!(\sfrac{2.3}{\blx}))15\sqrt{2\pi}\delta_{1}\)
\begin{align}
	\notag	
	%\label{eq:thm:TKDtoN:Strassen}	
	&\!\NPD{e^{\tau_{1}-\blx\KLD{\m{v;1}}{\m{q;1}}}}{\m{w}}{\m{q}}
	\\
	\notag	
	&\hspace{.5cm}\!=\!
	\begin{cases}
		\Delta\f{\theta}_{\blx}
		e^{\frac{\rno-1}{\rno}\tau_{1}-\blx\KLD{\m{v;1}}{\m{w;1}}}\!,
		&\text{if~}\rno\!\in\!(0,1),\!
		\\	
		\norm{\m{w;1,\sim}}^{\blx}\!-\!
		\Delta\f{\theta}_{\blx}
		e^{\frac{\rno-1}{\rno}\tau_{1}-\blx\KLD{\m{v;1}}{\m{w;1}}}\!
		&\text{if~}\rno\!>\!1,\!
	\end{cases}	
\end{align}
where  
\(\f{\theta}_{\blx}\!=\!
\left(\sfrac{1}{2\pi}\right)^{\frac{1}{2\rno}}
(\GMR{\sfrac{\rno \sigma_{1}\sqrt{\blx}}{\abs{\rno-1}}})^{\frac{1-\rno}{\rno}}
\GMR{\sigma_{1}\sqrt{\blx}}\). As a result of \eqref{eq:GMRBound:positive}:
\(\f{\theta}\!\sim\!\tfrac{\rno}{\abs{1-\rno}}\left(\sigma_{1}\sqrt{2\pi \blx}\right)^{-\sfrac{1}{\rno}}\)
as \(\blx\to\infty\). 	
\subsection{Discrete Markov Case}\label{sec:DiscreteMarkov}
Tests between time-homogeneous \(\sigma\)-finite Markov measures 
provide a relatively simple model with memory 
for the simple hypothesis testing problem. 
To investigate the problem for Markov measures at a level of generality 
commensurate with our framework thus far, we must ---at the very least--- 
consider \(\sigma\)-bounded transition measures.
We defer this generalization to \S\ref{sec:Asymptotic}
and consider only the discrete models,
i.e., models with a countable state-space,
which are rich enough to illustrate some
of the most important subtleties of the problem and of ensuring 
the hypotheses of Theorems \ref{thm:KDtoN} and \ref{thm:TKDtoN}.

Let us recall standard notational conventions about
discrete Markov measures, first. 
For countable sets \(\inpS\) and \(\outS\), a discrete 
transition measure \(\tm{W}\) from \(\inpS\) into \(\outS\), 
is a function of the from
\(\tm{W}\!:\!\inpS\!\times\!\sss{\outS}\!\mapsto\![0,\infty]\),
satisfying 
\(\tm{W}[\dinp;\oev]\!=\!\sum_{\dout\in\oev}\tm{W}[\dinp;\{\dout\}]\)
for all \(\dinp\!\in\!\inpS\) and \(\oev\!\subset\!\outS\). 
With customary abuse of notation, we use \(\tm{W}[\dinp;\dout]\), 
rather than \(\tm{W}[\dinp;\{\dout\}]\) for singletons and 
employ \(\tm{W}\) to denote 
both the transition measure described above
and 
the corresponding countably infinite dimensional matrix with 
non-negative entries whose rows and columns correspond to 
elements of \(\inpS\) and \(\outS\), respectively.
With this interpretation, we define the support \(\supp{\tm{W}}\) 
of the discrete transition measure \(\tm{W}\) as the set of 
\((\dinp,\dout)\) pairs for which \(\tm{W}[\dinp;\dout]\) is positive:
\begin{align}
	\label{eq:def:Support}	
	\supp{\tm{W}}
	&\DEF\{(\dinp,\dout)\!\in\!\inpS\!\times\!\outS:\tm{W}[\dinp;\dout]\!>\!0\}.
\end{align} 
A discrete transition measure \(\tm{W}\) is called
\begin{itemize}
	\item \(\sigma\)-finite iff  
	\(\tm{W}[\dinp;\dout]\!<\!\infty~\forall\dinp\!\in\!\inpS,\dout\!\in\!\outS\),
	\item \(\sigma\)-bounded iff
	\(\sup_{\dinp\in\inpS,\dout\in\outS}\tm{W}[\dinp;\dout]\!<\!\infty\),
	\item bounded iff  
	\(\sup_{\dinp\in\inpS}\sum_{\dout}\tm{W}[\dinp;\dout]\!<\!\infty\),
	\item a transition probability  iff  
	\(\sum_{\dout}\!\tm{W}[\dinp;\dout]\!=\!1~\forall\dinp\!\in\!\inpS\).
\end{itemize}  
%\begin{definition}\label{def:DiscreteMarkovKernel}
\begin{subequations}\label{eq:def:DiscreteMarkovKernel}
	For any discrete \(\sigma\)-finite transition measure
	\(\tm{W}\) from \(\outS\) into \(\outS\)
	and \(\blx\!\in\!\numbers{Z}[+]\),
	discrete \(\sigma\)-finite transition measure
	\(\tm{W;;\otimes\blx\!}\) 
	from \(\outS\) into \(\outS_{1}^{\blx}\)
	and discrete transition measure 
	\(\tm{W;;\blx\!}\) from \(\outS\) into \(\outS\)
	are defined via their values for singletons as
	\begin{align}
		\label{eq:def:DiscreteMarkovKernel:Path}	
		\tm{W;;\otimes\blx\!}[\dout_{0};\dout_{1}^{\blx}]	
		&\DEF \prod\nolimits_{\jnd=1}^{\blx}
		\tm{W}[\dout_{\jnd-1};\dout_{\jnd}]
		&
		&\forall \dout_{0}^{\blx}\!\in\!\outS_{0}^{\blx},
		\\
		\label{eq:def:DiscreteMarkovKernel:KolmogorovChapman}
		\tm{W;;\blx\!}[\dout_{0};\dout_{\blx}]	
		&\DEF \sum\limits_{\dout_{1}^{\blx-1}\!\in\outS_{1}^{\blx-1}}
		\!\!\tm{W;;\otimes\blx\!}[\dout_{0};\dout_{1}^{\blx}]	
		&
		&\forall \dout_{0},\dout_{\blx}\!\in\!\outS.\!		
	\end{align}	
	For \(\sigma\)-bounded \(\tm{W}\)'s, 
	\(\tm{W;;\blx\!}\)'s are not necessarily, even, \(\sigma\)-finite.
	If \(\tm{W}\) is bounded, then \(\tm{W;;\blx\!}\) is
	also bounded and it is the \(n^{\text{th}}\) power of 
	\(\tm{W}\) as a countably infinite dimensional matrix. 
\end{subequations}	
%\end{definition} 
If \(\tm{W}\) is a transition probability, 
then \(\tm{W;;\otimes\blx\!}[\dout_{0};\cdot]\) 
and \(\tm{W;;\blx\!}[\dout_{0};\cdot]\) 
specify the probabilities of \(\blx\)-step trajectories (i.e., paths)  
and \(\blx\)-step transitions, respectively, for the corresponding
discrete Markov chain for the initial state \(\dout_{0}\).

The equivalence classes and related concepts about discrete Markov chains, 
e.g., \cite[\S 4.2]{gallagerSP}, \cite[\S 8.4]{shiryaev2019probability2}, 
extend naturally  to \(\sigma\)-finite discrete transition measures, 
\cite{vere-Jones67,seneta,Nummelin}, because they are in essence 
determined by the support of the transition probabilities as 
defined \eqref{eq:def:Support}.
Let \(\tm{W}\) be a \(\sigma\)-finite discrete transition measure 
from \(\outS\) to \(\outS\) and \(\supp{\tm{W}}\) be its support. 
Then a \(\dout\!\in\!\outS\) is accessible from an 
\(\dinp\!\in\!\outS\), i.e., \(\dinp\!\rightarrow\!\dout\), 
iff \(\exists\blx\!\in\!\numbers{Z}[+]\) such that 
\((\dinp,\dout)\!\in\!\supp{\tm{W;;\blx}}\).
States \(\dinp\) and \(\dout\) are equivalent, 
i.e. \(\dinp\!\leftrightarrow\!\dout\), iff both 
\(\dinp\!\rightarrow\!\dout\), 
and \(\dout\!\rightarrow\!\dinp\) hold. 
\(\supp{\tm{W}}\) partitions the set of states \(\outS\) into 
disjoint equivalence classes.
Each equivalence class \(\set{S}\) is either a singleton of 
the form \(\{\dinp\}\) satisfying \(\nexists\dout\in\outS\) 
such that \(\dinp\leftrightarrow\dout\), or it is a set 
satisfying for each of its elements \(\dinp\) 
the following condition: \(\dout\in\set{S}\) iff 
\(\dinp\leftrightarrow\dout\).
An equivalence class \(\set{S}\) is closed 
(i.e., essential according to \cite{seneta}) iff 
\((\dinp,\dout)\notin\supp{\tm{W}}\)
for all \(\dinp\!\in\!\set{S}\) and 
\(\dout\!\notin\!\set{S}\).
An equivalence class \(\set{S}\) is periodic iff 
\(\exists\blx_{0}\!\geq\!2\) such that 
\((\dinp,\dinp)\notin \supp{\tm{W;;\blx}}\) for all 
\(\dinp\!\in\!\set{S}\) unless \(\blx\) is an intiger 
multiple of \(\blx_{0}\). 
An equivalence class \(\set{S}\) is  aperiodic iff it is 
not periodic.
A discrete transition measure \(\tm{W}\) is said to be 
irreducible if it has a unique equivalence class
and 
indecomposable if it has a unique closed equivalence class. 	
%Generalization of these concept to transition measures 
%without discreteness is possible; but that discussion 
%is more nuanced \cite{Nummelin}

Both for the simple hypothesis testing problem and 
lossless source coding problem Markov transition measures of interest 
have typically been the ones associated with a time-homogeneous 
irreducible aperiodic Markov chains on a finite state-space. 
Such Markov chains have a unique stationary distributions 
and the convergence of \(\blx\)-step transition probabilities to the
stationary distribution is geometric \cite[Theorem 4.3.5]{gallagerSP}.
For Markov chains with countably infinite state-spaces 
the convergence of \(\blx\)-step transition probabilities to 
the stationary distribution is ensured for all initial states for 
positive-recurrent irreducible aperiodic Markov chains, 
see \cite[6.26]{gallagerSP}, 
\cite[Theorem 3, \S 8.6]{shiryaev2019probability2}. 
Even then, the convergence to the stationary distribution
is neither necessarily uniform over the states 
\cite[\S 6.3.3]{gallagerSP}, nor geometric.
For transition probabilities both 
Dobrushin--Doeblin condition 
and Doeblin's minorization condition
are necessary-and-sufficient conditions for 
the uniform geometric convergence of 
\(\blx\)-step transition probabilities to the stationary distribution 
of the Markov chain in total variation distance over the initial states, 
see \cite{Nummelin,nagaev57,nagaev61};
we state these conditions for the discrete case. 
Dobrushin--Doeblin condition immediately 
leads to specific convergence parameters.
Doeblin's minorization condition, on the other hand, 
can be extended without modification to transition measures
that are not transition probabilities.

A discrete transition probability \(\tm{W}\) 
from \(\outS\) to \(\outS\) satisfies
Dobrushin--Doeblin condition for \((k_{0},\delta_{0})\) 
iff constants \(k_{0}\!\in\!\numbers{Z}[+]\)
and \(\delta_{0}\!\in\!(0,1)\) satisfy,
see \cite[(1.1)--(1.4)]{nagaev57}, 
\begin{align}
	\label{eq:DobrushinDoeblin}	
	\norm{\tm{W;;k_{0}\!}[\dinp;\cdot]\!-\!\tm{W;;k_{0}\!}[\dout;\cdot]}
	&\!\leq\!2\delta_{0}
	&
	&\forall\dinp,\dout\!\in\!\outS.
	\intertext{If Dobrushin--Doeblin condition is satisfied, then
		there exist  constants 
		\(\gamma\!\in\!(0,\infty)\) and \(\rho\!\in\!(0,1)\) satisfying}
	\label{eq:DobrushinDoeblin:Alternative}	
	\norm{\m{p;\!\tm{W}}[\cdot]\!-\!\tm{W;;\blx\!}[\dout;\cdot]}	
	&%\!\leq\!\delta_{0}^{\lfloor\frac{\blx}{k_{0}}\rfloor}
	\!<\!2\gamma \rho^{\blx}
	&
	&\forall\dout\!\in\!\outS,
	\blx\!\in\!\numbers{Z}[+],
\end{align}		
where \(\m{p;\!\tm{W}}[\cdot]\) is the stationary distribution 
of \(\tm{W}\), e.g., \eqref{eq:DobrushinDoeblin:Alternative} holds 
for \(\gamma=\sfrac{1}{\delta_{0}}\) and
\(\rho=\delta_{0}^{\sfrac{1}{k_{0}}}\),
see \cite[(0.2)]{nagaev61}.

A discrete transition measure \(\tm{W}\) 
from \(\outS\) to \(\outS\) satisfies 
Doeblin's minorization condition for 
\((\mlx,\gamma,\m{\rfm})\) iff \(\mlx\!\in\!\numbers{Z}[+]\), 
\(\gamma\!\in\!(0,\infty)\),
and \(\m{\rfm}\) is a probability measure on \(\outS\) satisfying,
see \cite[Theorem 6.15, Remarks 6.1]{Nummelin},
\begin{align}
	\label{eq:DoeblinMinorization}	
	\tm{W;;\mlx}[\dinp;\dout]
	&\geq \gamma\m{\rfm}[\dout]
	&
	&\forall \dinp,\dout\in\outS.	
	\intertext{
		Evidently, \eqref{eq:DoeblinMinorization} holds iff the following condition
		holds
	}
	\label{eq:DoeblinMinorization:Alternative}	
	\ent{\gamma}{\tm{W;;\mlx}[\dinp;\cdot]}{\m{\rfm}}	
	&=\gamma
	&
	&\forall \dinp\in\outS.	
\end{align}

For any discrete transition measure \(\tm{W}\) from \(\inpS\) to \(\outS\)
and \(\sigma\)-finite measure \(\m{p}\) on \(\inpS\),
the  measure  \(\m{p}\!\otimes\!\tm{W}\) on \(\inpS\!\times\!\outS\)
is defined via its value on singletons as
\begin{align}
	\m{p\!\otimes\!W}[\{(\dinp,\dout)\}]
	&\!=\!\m{p}[\dinp]\tm{W}[\dinp;\dout]	
	&
	&\forall\dinp\!\in\!\inpS,\dout\!\in\!\outS.		
\end{align}
Evidently, if  \(\m{p}\) is \(\sigma\)-finite 
and \(\tm{W}\) is \(\sigma\)-bounded,
then \(\m{p}\!\otimes\!\tm{W}\) is \(\sigma\)-finite.
If \(\m{p}\) is a probability measure
and \(\tm{W}\) is a transition probability,
then \(\m{p}\!\otimes\!\tm{W}\) is a probability measure. 

The functional Berry--Esseen Theorem of Nagaev
in \cite{nagaev61} implies the following lemma 
through a standard state-space 
enlargement argument and Jensen's inequality.
\begin{lemma}\label{lem:CSSMFBE}
	Let \(\tm{W}\) be a discrete transition probability 
	satisfying \eqref{eq:DobrushinDoeblin}	 
	with the stationary distribution \(\m{p;\!\tm{W}}\) 
	and \(\tm{Q}\) be a \(\sigma\)-finite discrete transition measure
	\begin{subequations}
		satisfying \(\f{m;4}<\infty\), where 
		\begin{align}
			\label{eq:lem:CSSMFBE:MomentSupremum}	
			\f{m;k}
			&\!=\!\sup\nolimits_{\dinp}
			\sum\nolimits_{\dout}
			\abs{\ln\tfrac{\tm{W}[\dinp;\dout]}{\tm{Q}[\dinp;\dout]}}^{k}
			\tm{W}[\dinp;\dout]
			&
			&\forall k\!\in\!\numbers{Z}[+].
		\end{align}
		For all positive integers \(\blx\), 
		let the random vector \(\out_{0}^{\blx+1}\) be distributed as 
		\(\m{p;\tm{W}}\!\otimes\!\tm{W;;\otimes\blx+1}\).
		Then the following expectations and limit, 
		defining \(\overline{\xi}\) and \(\overline{\sigma}\), 
		exist and are finite and \(\rv{S;\blx}\) is defined as
		\begin{align}
			\label{eq:lem:CSSMFBE:xibar}		
			\!\overline{\xi}
			&\!=\!
			\EX{\ln\tfrac{\tm{W}[\out_{0};\out_{1}]}{\tm{Q}[\out_{0};\out_{1}]}},
			\\
			\label{eq:lem:CSSMFBE:sigmabar}		
			\!\overline{\sigma}^{2}
			&\!=\!\lim\limits_{\blx\to\infty}\EX{\tfrac{1}{\blx}\left(\! 
				\ln\tfrac{\tm{W;;\otimes\blx\!}[\out_{0};\out_{1}^{\blx}]}{\tm{Q;;\otimes\blx}[\out_{0};\out_{1}^{\blx}]}
				-\blx\overline{\xi}\right)^{\!2}},
			\\
			\label{eq:lem:CSSMFBE:sum}
			\!\rv{S;\blx}
			&\!=\!\tfrac{1}{\overline{\sigma}\sqrt{\blx}}
			\left[\!\ln\!\tfrac{\tm{W;;\otimes\blx+1}[\out_{0};\out_{1}^{\blx+1}]}{\tm{Q;;\otimes\blx+1}[\out_{0};\out_{1}^{\blx+1}]}
			-(\blx\!+\!1)\overline{\xi}\right].
		\end{align}
		If \(\overline{\sigma}>0\),
		then \(\rv{S;\blx}\) satisfies
		\begin{align}
			\abs{\PX{\rv{S;\blx}
					\!\leq\! \tau }[\out_{0}\!=\!\dout]
				\!-\!\GCD{\tau}}
			&\!\leq\! \overline{\delta}
			&
			&\forall \tau\!\in\!\numbers{R}, \dout\in\outS,
		\end{align}
		where \(\overline{\delta}\) is defined in terms of 
		\(\rho\!=\!(\delta_{0})^{\frac{1}{1+k_{0}}}\), \(\f{M}[\rho]\), \(\f{c}[\rho]\), as
		\begin{align}
			\label{eq:lem:CSSMFBE:Mrho}
			\f{M}[\rho]
			&\!=\!3\tfrac{3^{k_{0}}(1+k_{0})(2+4\rho)+1}{1-\rho},
			\\
			\label{eq:lem:CSSMFBE:crho}
			\f{c}[\rho]
			&\!=\!136(\f{M}[\rho])^{3}+\tfrac{32}{3}(\f{M}[\rho])^{2}+1,
			\\
			\label{eq:lem:CSSMFBE:deltabar}
			\overline{\delta}
			&\!=\!c_{1} \tfrac{\f{m;3} \f{c}[\rho]}{\overline{\sigma}^{3}\sqrt{\blx}}
			\!+\!c_{2}\tfrac{\f{m;1}}{\overline{\sigma}\sqrt{\blx}}
			\!+\!c_{3}(\tfrac{1\!+\!2\rho}{3})^{\blx}(1\!+\!\tfrac{\f{m;1}}{\overline{\sigma}}),
		\end{align}
		for universal constants \(c_{1}\), \(c_{2}\), \(c_{3}\) 
		that do not depend on \(\tm{W}\) or \(\tm{Q}\).
		Furthermore, \(\exists(\f{\blx;0},\f{A})\)
		determined by \((k_{0},\delta_{0})\) such that
		\begin{align}
			\label{eq:lem:CSSMFBE:deltabar-bound}
			\overline{\delta}
			&\leq \tfrac{\f{A}}{\sqrt{\blx}}\left(1+\tfrac{\f{m;1}}{\overline{\sigma}}+\tfrac{\f{m;3}}{\overline{\sigma}^{3}}\right)
			&
			&\forall \blx\geq\blx_{0}.
		\end{align}
	\end{subequations}
\end{lemma}
Constants \(c_{1}\), \(c_{2}\), \(c_{3}\) are universal, 
but Lemma \ref{lem:CSSMFBE} is not universal 
because the coefficients of the ratios of the moments depend 
on \(\tm{W}\) via \(\delta_{0}\) and \(k_{0}\). 
\begin{proof}[Proof of Lemma \ref{lem:CSSMFBE}]
	Note that \(\f{m;1}\), \(\f{m;2}\), and \(\f{m;3}\) are finite 
	as a result of Jensen's inequality and 
	the convexity of \(\abs{\cdot}^{\rno}\) 
	for \(\rno\geq1\) because \(\f{m;4}\) is finite by assumption.
	To establish the lemma, we cannot invoke \cite[Theorem 1]{nagaev61}
	to the Markov chain of \(\out_{\tin}\)'s 
	because summands of \(\rv{S;\blx}\) are 
	\(\ln\tfrac{\tm{W}[\out_{\tin-1};\out_{\tin}]}{\tm{Q}[\out_{\tin-1};\out_{\tin}]}\)
	hence not of the the form \(\f{f}[\out_{\tin}]\). 
	We will invoke \cite[Theorem 1]{nagaev61} to the Markov chain 
	of \((\out_{\tin-1},\out_{\tin})\) pairs.
	
	Let \(\tm{U}\) be a discrete transition measure 
	from \(\outS\times\outS\) into \(\outS\times\outS\),
	defined as \(\tm{U}[(\dinp_{1},\dinp_{2});(\dout_{1},\dout_{2})]	
	\!=\!\IND{\dinp_{2}=\dout_{1}}
	\tm{W}[\dinp_{2};\dout_{2}]\). Then
	\begin{align}
		\notag	
		\tm{U;;\otimes\blx}[\dsta_{0};\dsta_{1}^{\blx}]		
		&=\IND{\dsta_{0,2}^{\blx-1}=\dsta_{1,1}^{\blx}}		
		\tm{W;;\otimes\blx}[\dsta_{0,2};\dsta_{1,2}^{\blx}],
		\\
		\notag	
		\tm{U;;\blx}[\dsta_{0};\dsta_{\blx}]		
		&=\tm{W;;\blx-1}[\dsta_{0,2};\dsta_{\blx,1}]		
		\tm{W}[\dsta_{\blx,1};\dsta_{\blx,2}].
	\end{align}
	Thus \(\m{p;\tm{U}}\!=\!\m{p;\!\tm{W}}\!\otimes\!\tm{W}\). 
	Furthermore, if \(\tm{W}\) satisfies Dobrushin--Doeblin condition 
	for \((k_{0},\delta_{0})\),
	then \(\tm{U}\) satisfies Dobrushin--Doeblin condition 
	for \((k_{0}\!+\!1,\delta_{0})\).
	If \(\sta_{1}^{\blx}\!\sim\!\m{p}\!\otimes\!\tm{U;;\blx-1}\)
	for \(\m{p}\!=\!\IND{\dout=\cdot}\!\otimes\!\tm{W}\),
	and \(\out_{\tin}\!=\!\sta_{\tin,2}\), then 
	\(\out_{1}^{\blx}\!\sim\!\tm{W;;\otimes\blx\!}[\dout;\cdot]\).
	Then for \(\f{m;11}\) defined below, rather than \(\f{m;1}\),
	Lemma \ref{lem:CSSMFBE} follows from 
	\cite[(0.4) and Theorem 1]{nagaev61}
	for 
	initial distribution \(\m{p}[\dsta]\!=\!\IND{\dout=\dout_{1}}\!\tm{W}[\dout;\dout_{2}]\), 
	transition probability \(\tm{U}\), 
	and
	function \(\f{f}[\dsta]\!=\!\ln\!\tfrac{\tm{W}[\dout_{1};\dout_{2}]}{\tm{Q}[\dout_{1};\dout_{2}]}\), 
	where \(\dsta=(\dout_{1},\dout_{2})\).
	\begin{align}
		\notag	
		\f{m;11}
		&\!=\!\EX{\abs{\ln\tfrac{\tm{W}[\out_{0};\out_{1}]}{\tm{Q}[\out_{0};\out_{1}]}}}[\out_{0}\!=\!\dout].
	\end{align}
	Lemma \ref{lem:CSSMFBE} holds for  \(\f{m;1}\)
	because \(\f{m;11}\leq \f{m;1}\) by definition.
\end{proof}

The conditional Kullaback-Leibler divergence is defined as 
\begin{align}
	\label{eq:def:CKLD}
	\CKLD{\tm{W}}{\tm{Q}}{\m{p}}
	&\DEF 
	\KLD{\m{p}\otimes\tm{W}}{\m{p}\otimes\tm{Q}}.	
\end{align}
Thus \(\overline{\xi}\) defined in \eqref{eq:lem:CSSMFBE:xibar}
is equal to \(\CKLD{\tm{W}}{\tm{Q}}{\m{p;\tm{W}}}\).
\begin{assumption}\label{assumption:DiscreteMarkov}
\(\m{w}\) and \(\m{q}\) are \(\sigma\)-finite measures
on \(\outS_{0}^{\blx+1}\) satisfying
\(\m{w}=\m{w;0}\otimes\tm{W;;\otimes\blx+1}\)
and
\(\m{q}=\m{q;0}\otimes\tm{Q;;\otimes\blx+1}\)
for some countable set \(\outS\),
\(\sigma\)-finite measures \(\m{w;0}\) and \(\m{q;0}\)
on \(\outS\),
and \(\sigma\)-finite transition measures \(\tm{W}\) and \(\tm{Q}\) 
from \(\outS\) to \(\outS\).
\end{assumption}
\begin{remark}
Note that any \(k\)-step discrete transition measure 
\(\tm{W}\) from \(\outS_{1}^{k}\) to \(\outS\) can be described via 
\(1\)-step discrete transition measure 
\(\tm{U}\) from \(\outS_{1}^{k}\) to \(\outS_{1}^{k}\) of form
\begin{align}
	\notag	
	\tm{U}[\ama{1}{k};\bma{1}{k}]	
	&=\IND{\ama{2}{k}=\bma{1}{k-1}}	\tm{W}[\amn{k};\bmn{k}]
\end{align}
Thus, we only consider one-step discrete transition 
measures in Corollaries \ref{cor:DiscreteMarkov:KDtoN}
and \ref{cor:DiscreteMarkov:TKDtoN} without loss of generality. 
\end{remark}	

\begin{corollary}\label{cor:DiscreteMarkov:KDtoN}
Let \((\m{w},\m{q})\) be a pair of measures satisfying 
Assumption \ref{assumption:DiscreteMarkov},
for a probability measure \(\m{w;0}\) and
transition probability \(\tm{W}\)
satisfying \eqref{eq:DobrushinDoeblin},
for which \(\f{m;4}\), defined in 
\eqref{eq:lem:CSSMFBE:MomentSupremum},
and \(\EX[\m{w;0}]{\left(\ln\der{\m{w;0}}{\m{q;0}}\right)^{2}}\)
are\begin{subequations}\label{eq:cor:DiscreteMarkov:KDtoN}~finite.
Let \(\xi\), \(\sigma\),  and \(\delta\) be
\begin{align}
	\label{eq:cor:DiscreteMarkov:KDtoN:xi}			
	\xi
	&\!=\! 
	(\blx+1) \CKLD{\tm{W}}{\tm{Q}}{\m{p;\tm{W}}}+\KLD{\m{w;0}}{\m{q;0}},
	\\
	\label{eq:cor:DiscreteMarkov:KDtoN:sigma}		
	\sigma
	&\!=\!
	\overline{\sigma}\sqrt{\blx},
	\\
	\label{eq:cor:DiscreteMarkov:KDtoN:delta}		
	\delta
	&\!=\!
	\overline{\delta}+\tfrac{1}{ \overline{\sigma}^{2}\blx \sqrt{8\pi e}}
	\EX[\m{w;0}]{\ln^{2}\der{\m{w;0}}{\m{q;0}}},	
\end{align}
for \(\overline{\sigma}\) and \(\overline{\delta}\) 
are defined \eqref{eq:lem:CSSMFBE:sigmabar} and
\eqref{eq:lem:CSSMFBE:deltabar}.	
\end{subequations}Then \eqref{eq:thm:KDtoN:upperbound},
\eqref{eq:thm:KDtoN:lowerbound},
and \eqref{eq:thm:KDtoN:Strassen}
of Theorem \ref{thm:KDtoN}
hold for \(\xi\), \(\sigma\), and \(\delta\) defined above
whenever \(\sigma\!>\!0\) 
and \(\delta\!\in\!(0,\sfrac{1}{2})\).
\end{corollary}			
If \(\m{w;0}\!=\!\m{q;0}\), then 
Corollary \ref{cor:DiscreteMarkov:KDtoN} follows from 
Lemma \ref{lem:CSSMFBE} and Theorem \ref{thm:KDtoN}. 
To prove Corollary \ref{cor:Memoryless:KDtoN} for 
\(\m{w;0}\!\neq\!\m{q;0}\) case 
without assuming \(\abs{\ln\m{w;0}[\dout]-\ln\m{q;0}[\dout]}\)
to be \(\m{w;0}\)-a.s. bounded,
we invoke the following lemma.
\begin{lemma}\label{lem:KolmogorovDistanceBound}
Let \(\inp\) and \(\sta\) be a random variable satisfying
\begin{align}
	\notag	
	\sup\nolimits_{\tau}
	\abs{\PX{\sta\leq \tau}[\inp]-\GCD{\sfrac{\tau}{\sigma}}}
	&\leq \overline{\delta }
	&
	&\text{a.s.}.
\end{align}	
Let \(\f{f}\) be a function such that
\(\EX{\f{f}[\inp]}\!=\!0\). Then
\begin{align}
	\notag	
	\sup\nolimits_{\tau}
	\abs{\PX{\sta\!+\!\f{f}[\inp]\!\leq\!\tau}-\GCD{\sfrac{\tau}{\sigma}}}
	&\leq \overline{\delta}+\tfrac{1}{ \sigma^{2} \sqrt{8\pi e } }\EX{(\f{f}[\inp])^{2}}.
\end{align}	
\end{lemma}
Proof of Lemma \ref{lem:KolmogorovDistanceBound} is presented 
in Appendix \ref{sec:Proof:lem:KolmogorovDistanceBound}.
\begin{proof}[Proof of Corollary \ref{cor:DiscreteMarkov:KDtoN}]
If \(\out_{0}^{\blx+1}\!\sim\!\m{w}\), then by
Lemma \ref{lem:CSSMFBE}, the hypothesis of 
Lemma \ref{lem:KolmogorovDistanceBound} holds for  
\(\inp=\out_{0}\), 
\(\overline{\delta}\) defined in \eqref{eq:lem:CSSMFBE:deltabar},
\(\sigma=\overline{\sigma}\sqrt{\blx}\), 
\(\f{f}[\dinp]=\ln\tfrac{\m{w;0}[\dinp]}{\m{q;0}[\dinp]}
-\KLD{\m{w;0}}{\m{q;0}}\), 
and
\begin{align}
	\notag
	\sta
	&=\ln\!\tfrac{\tm{W;;\otimes\blx+1}[\out_{0};\out_{1}^{\blx+1}]}{\tm{Q;;\otimes\blx+1}[\out_{0};\out_{1}^{\blx+1}]}
	\!-\!(\blx\!+\!1)\CKLD{\tm{W}}{\tm{Q}}{\m{p;\tm{W}}}.
\end{align}	
Then the hypothesis of Theorem \ref{thm:KDtoN}
follows from Lemma \ref{lem:KolmogorovDistanceBound} 
for \(\xi\), \(\sigma\), and \(\delta\) defined
in \eqref{eq:cor:DiscreteMarkov:KDtoN:xi},
\eqref{eq:cor:DiscreteMarkov:KDtoN:sigma},
and
\eqref{eq:cor:DiscreteMarkov:KDtoN:delta},
respectively, because
\begin{align}
	\notag
	\f{\llr;\m{w},\m{q}}
	&=\sta+\f{f}[\out_{0}]+\KLD{\m{w;0}}{\m{q;0}}.	
\end{align}
Then Corollary \ref{cor:DiscreteMarkov:KDtoN}
follows from  Theorem \ref{thm:KDtoN}.		
\end{proof}
	
For the simple hypothesis testing problem,
Markov structure ---like memorylessness--- is hereditary 
under the tilting operation in the sense that
if \((\m{w},\m{q})\) satisfies Assumption \ref{assumption:DiscreteMarkov},
then both  \((\m{v;\rno},\m{q})\) and \((\m{v;\rno},\m{w})\) satisfy 
Assumption \ref{assumption:DiscreteMarkov} for the same integer \(\blx\). 
To demonstrate this property more explicitly, 
let us introduce the concept of tilted transition measure. 
Let \(\tm{W}\) and \(\tm{Q}\) be \(\sigma\)-finite discrete transition 
measures from \(\inpS\) to \(\outS\). 
Then for any \(\rno\!\in\!\numbers{R}\) 
the order \(\rno\) tilted transition measure \(\tm{V;\rno;\!\tm{W}\!,\!\tm{Q}\!}\) 
from \(\inpS\) to \(\outS\) is defined as
\begin{subequations}
	\label{eq:def:TiltedTransitionMeasure}
	\begin{align}
		\hspace{-.1cm}\supp{\tm{V;\rno;\!\tm{W}\!,\!\tm{Q}}}
		&\DEF \supp{\tm{W}}\cap\supp{\tm{Q}},
		\\
		\hspace{-.1cm}\tm{V;\rno;\!\tm{W}\!,\!\tm{Q}\!}[\dinp;\dout]
		&\DEF\tm{W}[\dinp;\dout]^{\rno}\!\tm{Q}[\dinp;\dout]^{1-\rno}\!
		&
		&\forall(\dinp,\dout)\!\in\!\supp{\tm{V;\rno;\!\tm{W}\!,\!\tm{Q}}}.
	\end{align}
\end{subequations}
Then for any \((\m{w},\m{q})\) pair satisfying Assumption \ref{assumption:DiscreteMarkov}, 
the tilted measure \(\m{v;\rno;\m{w},\m{q}}\) defined in \eqref{eq:def:TiltedMeasure}
satisfies
\begin{align}
	\label{eq:DiscreteMarkov:TiltedTransitionMeasure}	
	\m{v;\rno;\m{w},\m{q}}
	&\!=\!\m{v;\rno;\m{w;0},\m{q;0}}\otimes\tm{V;\rno;\!\tm{W}\!,\!\tm{Q},\otimes(\blx+1)}.
	&
	~
\end{align}
Unlike memorylessness, however, the Markov structure is not 
preserved with the scaling of \(\m{v;\rno;\m{w},\m{q}}\) with 
\(\norm{\m{v;\rno;\m{w},\m{q}}}\). 
Nevertheless, one can construct a probability measure with 
the Markov structure using Doob's \(\f{h}\)-transform. 

Let \(\tm{V}\) be a \(\sigma\)-finite discrete transition 
measure from \(\outS\) to \(\outS\) and 
\(\f{h}\!:\!\outS\!\mapsto\!(0,\infty)\) be an eigenfunction of 
\(\tm{V}\) for a positive eigenvalue \(\lambda\) satisfying 
\(\tm{V}\f{h}=\lambda\f{h}\). Then Doob's \(\f{h}\)-transform of 
\(\tm{V}\), denoted by \(\tm{V;\!\f{h}}\), is defined pointwise as 
\begin{align}
	\label{eq:def:DoobsTransform}	
	\tm{V;\!\f{h}}[\dinp;\dout]
	&\DEF\tfrac{\tm{V}[\dinp;\dout] \f{h}[\dout]}{\lambda \f{h}[\dinp]}
	&
	&\forall \dinp,\dout\!\in\!\outS.
\end{align}
Note that \(\tm{V;\!\f{h}}\) is a discrete transition probability 
from \(\outS\) to \(\outS\) by construction. 
Furthermore,\begin{subequations}\label{eq:DoobsTransform}	
corresponding probabilities of \(\blx\)-step trajectories 
(i.e., paths)  and \(\blx\)-step transitions satisfy
\begin{align}
	\label{eq:DoobsTransform:npath}	
	\!\tm{V;\!\f{h};\otimes\blx}[\dout_{0};\dout_{1}^{\blx}]
	&\DEF\tfrac{\tm{V;;\otimes\blx}[\dout_{0};\dout_{1}^{\blx}] \f{h}[\dout_{\blx}]}{\lambda^{\blx} \f{h}[\dout_{0}]}
	&
	&\forall \dout_{0}^{\blx}\!\in\!\outS_{0}^{\blx},
	\blx\!\in\!\numbers{Z}[+\!],
	\\
	\label{eq:DoobsTransform:nstep}	
	\!\tm{V;\!\f{h};\blx}[\dout_{0};\dout_{\blx}]
	&\DEF\tfrac{\tm{V;;\blx}[\dout_{0};\dout_{\blx}] \f{h}[\dout_{\blx}]}{\lambda^{\blx} \f{h}[\dout_{0}]}
	&
	&\forall \dout_{0},\dout_{\blx}\!\in\!\outS,\blx\!\in\!\numbers{Z}[+\!].
\end{align}
In\end{subequations}~the following, we will invoke 
\eqref{eq:def:DiscreteMarkovKernel:Path},
\eqref{eq:DiscreteMarkov:TiltedTransitionMeasure},
and \eqref{eq:DoobsTransform} 
to express \(\m{v;\rno;\m{w},\m{q}}\) for a \((\m{w},\m{q})\) 
pair satisfying Assumption \ref{assumption:DiscreteMarkov} in terms of 
\(\tm{V;\!\f{h};\otimes\blx+1}\) for the case when 
\(\tm{V}\!=\!\tm{V;\rno;\!\tm{W}\!,\!\tm{Q}\!}\). 
Thus the existence of Doob's \(\f{h}\)-transform
---at least for the approriate closed equivalence class 
of \(\tm{V;\rno;\!\tm{W}\!,\!\tm{Q}\!}\)--- 
is necessary for our analysis to go through. 

If \(\tm{V}\) is irreducible, aperiodic, and finite
and \(\outS\) is a finite set, 
then there exists a unique \(\f{h}\), of the form 
\(\f{h}\!:\!\outS\!\mapsto\!(0,\infty)\), satisfying 
\(\tm{V}\f{h}\!=\!\lambda \f{h}\) for the maximum 
modulus eigen value \(\lambda\)  by Perron-Frobenius theorem, 
\cite[Theorem 8.4.4]{hornJohnson},
as previously noted in 
\cite{watanabeH17,NakagawaK93B,Natarajan85,Vasek80,WolferW23}. 
If \(\tm{V}\) is irreducible, aperiodic, and \(\sigma\)-finite
and \(\outS\) is a countably infinite set, 
then a more nuanced picture emerges. 
For the case when \(\tm{V;;\blx}\) is \(\sigma\)-finite
for all positive integers \(\blx\), 
e.g., when \(\tm{V}\) is bounded, 
\(R\)-recurrence of \(\tm{V}\), as defined in \cite{vere-Jones67},
implies the existence of a unique \(\f{h}\), of the form 
\(\f{h}\!:\!\outS\!\mapsto\!(0,\infty)\),  satisfying 
\(\tm{V}\f{h}=\lambda \f{h}\) by \cite[Theorem 4.1]{vere-Jones67}.
Let us summarize some of the fundamental findings of 
the spectral theory of infinite non-negative matrices 
from \cite{vere-Jones67,seneta}.
\begin{lemma}\label{lem:CPclassification}
For any irreducible aperiodic bounded discrete 
transition measure \(\tm{V}\) from \(\outS\) to \(\outS\)
there exists a \(\lambda\!\in\!(0,\infty)\) satisfying 
\begin{align}
	\label{eq:lem:CPclassification}	
	\lim\nolimits_{\blx\to\infty}	
	\sqrt[\blx]{\tm{V;;\blx}[\dinp;\dout]}
	&=\lambda 
	&
	&\forall \dinp,\dout\in\outS.
\end{align}
Either \(\sum_{\blx=1}^{\infty}(\sfrac{1}{\lambda})^{\blx}\tm{V;;\blx}[\dinp;\dout]\!<\!\infty\)
for all \( \dinp\) and \(\dout\) in \(\outS\)
and \(\tm{V}\) is said to be \(R\)-transient, 
or 
\(\sum_{\blx=1}^{\infty}(\sfrac{1}{\lambda})^{\blx}\tm{V;;\blx}[\dinp;\dout]\!=\!\infty\)
for all \( \dinp\) and \(\dout\) in \(\outS\)
and \(\tm{V}\) is said to be \(R\)-recurrent.
Similarly, for an \(R\)-recurrent \(\tm{V}\), either
\(\lim\nolimits_{\blx\to\infty}(\sfrac{1}{\lambda})^{\blx}\tm{V;;\blx}[\dinp;\dout]\!=\!0\)
for all \( \dinp\) and \(\dout\) in \(\outS\)
and \(\tm{V}\) is said to be \(R\)-null, 
or 
\(\lim\nolimits_{\blx\to\infty}(\sfrac{1}{\lambda})^{\blx}\tm{V;;\blx}[\dinp;\dout]\!>\!0\)
for all \( \dinp\) and \(\dout\) in \(\outS\)
and \(\tm{V}\) is said to be \(R\)-positive. 
Furthermore, if \(\tm{V}\) is \(R\)-recurrent, 
then there exists 
a unique (upto a scaling) \(R\)-invariant function 
\(\f{h}\!:\!\outS\!\mapsto\!(0,\infty)\) satisfying 
\(\tm{V}\f{h}\!=\!\lambda\f{h}\) 
and a unique (upto a  scaling) \(R\)-invariant measure 
\(\m{\pi}\!:\!\outS\!\mapsto\!(0,\infty)\) satisfying 
\(\m{\pi}\tm{V}\!=\!\lambda\m{\pi}\). 
\end{lemma}	
\begin{proof}[Proof of Lemma \ref{lem:CPclassification}]
There exists a non-negative \(\lambda\)  
satisfying \eqref{eq:lem:CPclassification}	
by\footnote{The existence of positive, possibly infinite, 
	\(\lambda\) satisfying \eqref{eq:lem:CPclassification} 
	also follows from \cite[Theorem 6.1]{seneta} 
	and Hadamard formula, \cite[Lemma 1.1.8]{krantzparks}.} 
\cite[Theorem A]{vere-Jones67}.  
\(\lambda\) is positive because 
\(\supp{\tm{V;;\blx}}\!=\!\outS\!\times\!\outS\) 
for all \(\blx\) large enough and 
\(\tm{V;;\mlx+\blx}[\dinp;\dinp] 
\!\geq\!\tm{V;;\mlx}[\dinp;\dinp] 
\tm{V;;\blx}[\dinp;\dinp]\)
for all \(\mlx,\blx\!\in\!\numbers{Z}[+]\) 
and \(\dinp\!\in\!\outS\).
Furthermore, \(\lambda\) is finite
because boundedness of \(\tm{V}\) implies the
existence of a real number \(M\) 
satisfying \(\tm{V}\f{\ones}\!\leq\!\f{\ones} M\); 
hence satisfying
\(\tm{V;;\blx}\f{\ones}\!\leq\!\f{\ones} M^{\blx}\) 
for all positive integers \(\blx\).
Dichotomies about
\(\sum_{\blx=1}^{\infty}(\sfrac{1}{\lambda})^{\blx}\tm{V;;\blx}[\dinp;\dout]\)
and 
\(\lim\nolimits_{\blx\to\infty}(\sfrac{1}{\lambda})^{\blx}\tm{V;;\blx}[\dinp;\dout]\)
are established in \cite[Theorem B]{vere-Jones67}. 
The existence of a unique  invariant function \(\f{h}\) 
and measure \(\m{\pi}\) for \(R\)-recurrent \(\tm{V}\)
are established in \cite[Theorem 4.1]{vere-Jones67}.
\end{proof}

Evidently, if \(\tm{V;\!\f{h}}\) is the Doob's \(\f{h}\)-transform of 
\(\tm{V}\) for the \(\lambda\) in \eqref{eq:lem:CPclassification},
then \(\tm{V}\) is  \(R\)-transient/\(R\)-null/\(R\)-positive
iff \(\tm{V;\!\f{h}}\) is transient/null-recurrent/positive-recurrent as characterized
in \cite[Ch. 6]{gallagerSP}, \cite[\S8.5 Theorem 5]{shiryaev2019probability2}.
%\cite[Theorem 6.2.6, 6.3.5]{gallagerSP} 
Since we are building on \cite{nagaev61}, via Lemma \ref{lem:CSSMFBE}, 
we will only discuss the case when \(\tm{V;\!\f{h}}\) satisfies 
the Doeblin's minorization condition. 
Lemma \ref{lem:CPpositivity} in the following, 
presents the characterization of \(R\)-positivity
in terms of \(R\)-invariant measure and \(R\)-invariant 
function of \cite{vere-Jones67,seneta,Nummelin}, recalled in
Lemma \ref{lem:CPclassification}, and 
investigates implications of this characterization 
for Doeblin's minorization condition.
\begin{lemma}\label{lem:CPpositivity}
Let \(\tm{V}\) be an irreducible aperiodic bounded
discrete transition measure from \(\outS\) to \(\outS\),
then \(\tm{V}\) is \(R\)-positive and 
\eqref{eq:lem:CPclassification} holds
iff there exist 
\(\f{h}\!:\!\outS\!\mapsto\!(0,\infty)\) 
and
\(\m{\pi}\!:\!\outS\!\mapsto\!(0,\infty)\)
satisfying 
\(\tm{V}\f{h}\!=\!\lambda\f{h}\),
\(\m{\pi}\tm{V}\!=\!\lambda\m{\pi}\),
\(\m{\pi}\f{h}\!=\!\sum_{\dinp}\m{\pi}[\dinp]\f{h}[\dinp]\!<\!\infty\)
for some \(\lambda\!\in\!(0,\infty)\).
Furthermore, let \(\tm{V}\) be \(R\)-positive,
then 
\(\tm{V}\) satisfies Doeblin's minorization condition 
and \(\sup_{\dout}\f{h}[\dout]\!<\!\infty\)
iff
\(\tm{V;\!\f{h}}\) satisfies Doeblin's minorization condition 
and \(\inf_{\dout}\f{h}[\dout]\!>\!0\).	
\end{lemma}	
\begin{proof}[Proof of Lemma \ref{lem:CPpositivity}]
The characterization of \(R\)-positivity is established in
\cite[Criterion III]{vere-Jones67}, %\cite[Theorem D]{vere-Jones67}
see\footnote{The same characterization extends to the
	general transition measures for 
	\(\m{\pi}\f{h}\!=\!\int\!\f{h}\dif{\m{\pi}}\),	
	as well, see \cite[Definition 5.3]{Nummelin}
	and \cite[Proposition 5.8]{Nummelin}.}
also \cite[Theorem 6.4]{seneta}.

For any \(R\)-positive \(\tm{V}\), the \(R\)-invariant function
\(\f{h}\) satisfies \(\f{h}\!=\!\lambda^{-\mlx}\tm{V;;\mlx}\f{h}\)
for any \(\mlx\!\in\!\numbers{Z}[+]\). 
Then Doeblin's minorization condition for \(\tm{V}\)
implies \(\f{h}\!\geq\!(\m{\rfm}\f{h})\lambda^{-\mlx}\gamma \f{\ones}\).
Thus \(\inf_{\dout}\f{h}[\dout]\!>\!0\). If 
\(\sup_{\dout}\f{h}[\dout]\!<\!\infty\), then 
\(\tm{V;\!\f{h}}\) satisfies Doeblin's minorization condition 
as a result of 
\eqref{eq:def:DiscreteMarkovKernel:KolmogorovChapman} and
\eqref{eq:DoobsTransform}.

For any \(R\)-positive \(\tm{V}\), let us define 
\(\f{g}\!:\!\outS\!\mapsto\!(0,\infty)\) to be the pointwise
multiplicative inverse of \(\f{h}\), i.e.,  
\(\f{g}[\dout]\!=\!\sfrac{1}{\f{h}[\dout]}\) for all 
\(\dout\!\in\!\outS\). 
Then the Doeblin's minorization condition
for \(\tm{V;\!\f{h}}\) and \eqref{eq:DoobsTransform} implies
\(\f{g}[\dout_{0}]\tm{V;;\mlx}[\dout_{0};\dout_{\mlx}]
\!\geq\!\lambda^{\mlx}\gamma \m{\rfm}[\dout_{\mlx}]\f{g}[\dout_{\mlx}]\)
for all \(\dout_{0}\), \(\dout_{\mlx}\) in \(\outS\).  
On the other hand, 
\(\tm{V}\f{\ones}\!<\!M\f{\ones}\)
and hence \(\tm{V;;\mlx}\f{\ones}\!<\!M^{\mlx}\f{\ones}\)
hold for a finite \(M\)
because \(\tm{V}\) is bounded. Thus
\(\inf_{\dout}\f{g}[\dout]\!
\geq\!(\m{\rfm}\f{g})(\sfrac{\lambda}{M})^{\mlx}
\gamma\!>\!0\). 
Hence \(\sup_{\dout}\f{h}[\dout]\!<\!\infty\).
If  we also have \(\inf_{\dout}\f{h}[\dout]\!>\!0\), then 
\(\tm{V}\) satisfies Doeblin's minorization condition 
as a result of 
\eqref{eq:def:DiscreteMarkovKernel:KolmogorovChapman} 
and \eqref{eq:DoobsTransform}.	
\end{proof}	
\begin{remark}\label{rem:IndependenceOfRPositivtyAndDoeblinHypothesis}
Common limit value \(\lambda\) in \eqref{eq:lem:CPclassification} 
is \(1\) for any recurrent transition probability \(\tm{V}\). 
Thus \(R\)-positivity of \(\tm{V}\) does 
not imply Doeblin's minorization condition for \(\tm{V}\), 
because there are positive-recurrent transition
probabilities that do not satisfy Doeblin's minorization,
e.g., birth--death Markov chains \cite[\S6.4]{gallagerSP}. 
One might be tempted to think that Doeblin's minorization 
condition for \(\tm{V}\) implies  the \(R\)-positivity of 
\(\tm{V}\), because Doeblin's minorization condition 
for transition probabilities implies positive-recurrence.
This, however, is not true there are 
\(R\)-transient\footnote{The transition measure
	\(\tm{V}[\dinp;\dout]\!=\!\tfrac{1}{8}\IND{\!\dout=1\!}
	\!+\!\tfrac{\dinp^{2}+2\dinp}{2(\dinp+2)^{2}}\IND{\!\dout=\dinp+1\!}\)
	from \(\numbers{Z}[+]\) to \(\numbers{Z}[+]\) 
	satisfies \(\tm{V}\f{h}\!=\!(\sfrac{1}{2})\f{h}\) for
	\(\f{h}[\dinp]\!=\!(\dinp\!+\!1)^{2}\) and  \(\tm{V}\) is \(R\)-transient because 
	\(\tm{V;\!\f{h}\!}[\dinp;\dout]\!=\!\tfrac{1}{(\dinp+1)^{2}}\IND{\!\dout=1\!}
	\!+\!\tfrac{\dinp^{2}+2\dinp}{(\dinp+1)^{2}}\IND{\!\dout=\dinp+1\!}\) 
	is transient.}
and \(R\)-null\footnote{The transition measure
	\(\tm{V}[\dinp;\dout]\!=\!\tfrac{1}{4}\IND{\!\dout=1\!}
	\!+\!\tfrac{\dinp}{2(\dinp+2)}\IND{\!\dout=\dinp+1\!}\)
	from \(\numbers{Z}[+]\) to \(\numbers{Z}[+]\) 
	satisfies 
%	\(\m{\pi}\tm{V}\!=\!(\sfrac{1}{2})\m{\pi}\) 
%	for \(\m{\pi}[\dinp]\!=\!\tfrac{1}{(\dinp+1)\dinp}\) and 
	\(\tm{V}\f{h}\!=\!(\sfrac{1}{2})\f{h}\) for \(\f{h}[\dinp]\!=\!\dinp\!+\!1\) 
 	and \(\tm{V}\) is \(R\)-null because
  	\(\tm{V;\!\f{h}}[\dinp;\dout]\!=\!\tfrac{1}{\dinp+1}\IND{\!\dout=1\!}
	\!+\!\tfrac{\dinp}{\dinp+1}\IND{\!\dout=\dinp+1\!}\) is null-recurrent.} 
discrete transition measures satisfying Doeblin's minorization condition.
A set of sufficient conditions for \(R\)-positivity 
of discrete substochatic transition measures is 
presented in \cite[Theorem 1]{FerrariKM96}; 
corresponding conditions for bounded discrete transition 
measures can be obtained using that fact that any bounded
transition measure \(\tm{V}\) satisfies
\(\tm{V}\!=\!\gamma\tm{U}\) for a substochastic transition measure \(\tm{U}\)
and \(\gamma\!\in\![1,\infty)\).
\end{remark}

Doeblin--Dobbrushin condition assumed for 
Corollary \ref{cor:DiscreteMarkov:KDtoN}
implies the uniqueness of the closed equivalence class,
i.e., indecomposablity, for the Markov chain. 	
The tilting operation described in 
\eqref{eq:def:TiltedTransitionMeasure}
can, in principle, change the equivalence classes,
i.e., the equivalence classes of \(\tm{V;\rno;\tm{W},\tm{Q}}\) 
and \(\tm{W}\) may be different. 
One can refrain from coping with the subtleties arising 
from this issue explicitly and ensure that
supports of \(\tm{V;\rno;\tm{W},\tm{Q}}\) and \(\tm{W}\) 
are the same, i.e., \(\supp{\tm{V;\rno;\tm{W},\tm{Q}}}\!=\!\supp{\tm{W}}\),
by assuming either \( \supp{\tm{W}}\!=\!\supp{\tm{Q}}\),
see \cite{nagaevA02,nagaevA03,watanabeH17},
or \(\supp{\tm{W}}\!=\!\supp{\tm{Q}}\!=\!\outS\!\times\!\outS\),
see \cite{Natarajan85,NakagawaK93B}.
Evidently, this issue does not show up in 
the lossless source coding problem 
considered in \cite{Vasek80,DavissonLS81}  
because \(\supp{\tm{Q}}\!=\!\outS\!\times\!\outS\).
One can, however, avoid making such assumptions
in the simple hypothesis testing problem as well
by considering the restrictions of 
the transition measures.

For any discrete transition measure \(\tm{V}\) 
from \(\outS\) to \(\outS\)  and \(\set{S}\subset\outS\),
the restriction \(\tm{V;\!|\set{S}}\) of \(\tm{V}\) to \(\set{S}\) 
is a discrete transition measure from \(\set{S}\) to \(\set{S}\) 
defined pointwise  as
\begin{align}
\label{eq:def:Restriction}	
\tm{V;\!|\set{S}}[\dinp;\dout]
&=\tm{V}[\dinp;\dout]
&
&\forall \dinp,\dout\in\set{S}.
\end{align} 
If either \(\set{S}\) or its complement 
\(\outS\setminus\set{S}\) is closed, i.e., 
\((\dinp,\dout)\notin\supp{\tm{V}}\) either 
for all \(\dinp\!\in\!\set{S}\) and \(\dout\!\notin\!\set{S}\)
or 
for all \(\dinp\!\notin\!\set{S}\) and \(\dout\!\in\!\set{S}\),
then
\begin{align}
\label{eq:Restriction}		
\tm{V;;\blx}[\dinp;\dout]
&=\tm{V;\!|\set{S};\blx}[\dinp;\dout]
&
&\forall \dinp,\dout\in\set{S}.
\end{align} 
Thus if either \(\set{S}\) or its complement is closed then 
the restriction of \(\tm{V;;\blx}\) to \(\set{S}\) 
is equal to the \(\blx^{\text{th}}\) power of \(\tm{V;\!|\set{S}}\),
i.e., \(\tm{V^{\blx};\!|\set{S}}=\tm{V;\!|\set{S};\blx}\).

\begin{corollary}\label{cor:DiscreteMarkov:TKDtoN}
Let \((\m{w},\m{q})\) be a pair of measures satisfying 
Assumption \ref{assumption:DiscreteMarkov},
\(\set{S}\) be a closed equivalence class of 
\(\supp{\tm{W}}\cap\supp{\tm{Q}}\)
satisfying \(\m{w;0}[\dout]\m{q;0}[\dout]\!=\!0\)
for all \(\dout\!\notin\!\set{S}\)
and \(\norm{\m{v;\rno;\m{w;0},\m{q;0}}}\!\in\!(0,\infty)\)
for some \(\rno\!\in\!(0,\infty)\!\setminus\!\{1\}\). 
If the restriction of \(\tm{V;\rno;\tm{W},\tm{Q}}\) to \(\set{S}\)
is a bounded \(R\)-positive transition measure satisfying 
Doeblin's minorization condition 
with a bounded \(R\)-invariant function 
\(\f{h}\!:\!\set{S}\!\mapsto\!(0,\infty)\),
i.e., \(\sup_{\dout\in\set{S}}\f{h}[\dout]\!<\!\infty\),
then \(\inf_{\dout\in\set{S}}\f{h}[\dout]\!>\!0\),
and\begin{subequations}\label{eq:cor:DiscreteMarkov:TKDtoN}
there exist
\(\lambda\!\in\!(0,\infty)\),
\(k_{0}\!\in\!\numbers{Z}[+]\),
\(\delta_{0}\!\in\!(0,1)\) satisfying
\begin{align}
	\label{eq:cor:DiscreteMarkov:TKDtoN:CPclassification}	
	\lim\nolimits_{\blx\to\infty}	
	\sqrt[\blx]{\tm{V;\rno;\tm{W},\tm{Q},\blx}[\dinp;\dout]}
	&=\lambda 
	&
	&\forall\dinp,\dout\!\in\!\set{S},
	\\
	\label{eq:cor:DiscreteMarkov:TKDtoN:DobrushinDoeblin}	
	\norm{\tm{V;\!\f{h};k_{0}\!}[\dinp;\cdot]\!-\!\tm{V;\!\f{h};k_{0}\!}[\dout;\cdot]}
	&\!\leq\!2\delta_{0}
	&
	&\forall\dinp,\dout\!\in\!\set{S},
\end{align}
where \(\tm{V;\!\f{h}}\) is a transition probability 
from \(\set{S}\) to \(\set{S}\) defined as
\begin{align}
	\label{eq:cor:DiscreteMarkov:TKDtoN:DoobsTransform}		
	\tm{V;\!\f{h}}[\dinp;\dout]
	&=\tfrac{\f{h}[\dout]}{\lambda \f{h}[\dinp]}\tm{V;\rno;\tm{W},\tm{Q}}[\dinp;\dout]
	&
	&\forall\dinp,\dout\!\in\!\set{S},
	\\
	\label{eq:cor:DiscreteMarkov:TKDtoN:RinvariantRatio}
	\varsigma
	&=\left(\sup\nolimits_{\dinp,\dout\in\set{S}}\tfrac{\f{h}[\dout]}{\f{h}[\dinp]}\right)^{\frac{1}{\abs{1-\rno}}}.
\end{align}
If \(\f{m;4}\), defined below in 
\eqref{eq:cor:DiscreteMarkov:TKDtoN:MomentSupremum}, 
is finite, then the expectations and limit, 
defining \(\overline{\xi}\) and \(\overline{\sigma}\) 
exist and are finite
\begin{align}
	\label{eq:cor:DiscreteMarkov:TKDtoN:MomentSupremum}	
	\f{m;k}
	&\!=\!\sup\limits_{\dinp\in\set{S}}
	\sum\limits_{\dout\in\set{S}}
	\abs{\ln\tfrac{\tm{V;\!\f{h}}[\dinp;\dout]}{\tm{W}[\dinp;\dout]}}^{k}
	\tm{V;\!\f{h}}[\dinp;\dout]
	\qquad~\forall k\!\in\!\numbers{Z}[+],
	\\
	\label{eq:cor:DiscreteMarkov:TKDtoN:xibar}		
	\!\overline{\xi}
	&\!=\!
	\EX{\ln\tfrac{\tm{V;\!\f{h}}[\out_{0};\out_{1}]}{\tm{W}[\out_{0};\out_{1}]}},
	\\		
	\label{eq:cor:DiscreteMarkov:TKDtoN:sigmabar}		
	\!\overline{\sigma}^{2}
	&\!=\!\lim\limits_{\blx\to\infty}\EX{\tfrac{1}{\blx}\left(\!
		\!\ln\!\tfrac{\tm{V;\!\f{h};\otimes\blx\!}[\out_{0};\out_{1}^{\blx}]}{\tm{W;;\otimes\blx}[\out_{0};\out_{1}^{\blx}]}
		\!-\!\blx\overline{\xi}\!\right)^{\!2}},	
\end{align}
where \(\out_{0}^{\blx+1}\) is distributed as 
\(\m{p;\!\tm{V;\!\f{h}}}\!\otimes\!\tm{V;\!\f{h};\otimes\blx+1}\)
for all \(\blx\). If \(\overline{\sigma}\!>\!0\)
\(\EX[\m{v;0}]{\!\ln^{2}\der{\m{v;0}}{\m{w;0}}\!}\!<\!\infty\) 
for \(\m{v;0}\!=\!\tfrac{\m{v;\rno;\m{w;0},\m{q;0}}}{\norm{\m{v;\rno;\m{w;0},\m{q;0}}}}\),
and \(\sqrt{\blx}\!\geq\!\tfrac{\abs{1-\rno}\varsigma}{\rno\IGQF{0.4}\overline{\sigma}}\)
then \eqref{eq:thm:TKDtoN:Strassen}
of Theorem \ref{thm:TKDtoN} hold for all 
\(\tilde{\eta}\in[\varsigma\f{\eta}[\sfrac{3}{5}],\sfrac{\f{\eta}[\sfrac{2}{5}]}{\varsigma}]\)
for a \(\Delta\) satisfying
\(\abs{\ln\Delta}\!\leq\!\tfrac{0.43}{1\wedge \rno}\!+\!(\sigma\!+\!2.3)15\sqrt{2\pi}\delta\!+\!\varsigma\)
for \(\gamma\), \(\xi\), \(\sigma\), \(\delta\) defined in the following
whenever \(\delta\!\in\!(0,\sfrac{1}{10})\):
\begin{align}
	\label{eq:cor:DiscreteMarkov:TKDtoN:gamma}
	\gamma
	&\!=\!\sfrac{1}{(\lambda^{\blx+1}\norm{\m{v;\rno;\m{w;0},\m{q;0}}})}
	\\
	\label{eq:cor:DiscreteMarkov:TKDtoN:xi}
	\xi
	&\!=\! 
	(\blx+1)\CKLD{\tm{V;\!\f{h}}}{\tm{W}}{\m{p;\!\tm{V;\!\f{h}}}}+\KLD{\m{v;0}}{\m{w;0}},
	\\
	\label{eq:cor:DiscreteMarkov:TKDtoN:sigma}
	\sigma
	&\!=\!
	\overline{\sigma}\sqrt{\blx},
	\\
	\label{eq:cor:DiscreteMarkov:TKDtoN:delta}
	\delta
	&\!=\!\overline{\delta}
	\!+\!\tfrac{1}{ \overline{\sigma}^{2}\blx \sqrt{8\pi e}}
	\EX[\m{v;0}]{\ln^{2}\der{\m{v;0}}{\m{w;0}}},
\end{align}
where \(\overline{\delta}\) is given in \eqref{eq:lem:CSSMFBE:deltabar} 
for \(\f{m;k}\) in \eqref{eq:cor:DiscreteMarkov:TKDtoN:MomentSupremum},	
\(\f{M}[\rho]\) in \eqref{eq:lem:CSSMFBE:Mrho},
\(\f{c}[\rho]\) in \eqref{eq:lem:CSSMFBE:crho},
and \(\rho\!=\!(\delta_{0})^{\frac{1}{1+k_{0}}}\)
for \(\delta_{0}\) and \(k_{0}\) 
in \eqref{eq:cor:DiscreteMarkov:TKDtoN:DobrushinDoeblin}.	
\end{subequations}	
\end{corollary}	
It is worth mentioning that if closed equivalence class
\(\set{S}\) of Corollary  \ref{cor:DiscreteMarkov:TKDtoN} 
is a finite subset of \(\outS\), then the restriction of 
\(\tm{V;\rno;\tm{W},\tm{Q}}\) to \(\set{S}\)
is always an  \(R\)-positive, bounded transition measure 
satisfying Doeblin's minorization condition and 
\(\EX[\m{v;0}]{\left(\ln\m{v;0}\!-\!\ln\m{w;0}\right)^{2}}\)
is always finite. 	
\begin{proof}[Proof of Corollary \ref{cor:DiscreteMarkov:TKDtoN}]
Note that \(\set{S}\) is a closed equivalence class of 
the order \(\rno\) tilted transition measure 
\(\tm{V;\rno;\tm{W},\tm{Q}}\), as well. As a result 
\(\tm{V;\rno|\set{S};\tm{W},\tm{Q}}\), i.e.,
the restriction of \(\tm{V;\rno;\tm{W},\tm{Q}}\) to \(\set{S}\), 
is irreducible. The restriction is also aperiodic  
because it satisfies Doeblin's minorization condition.
Since the equivalence class  \(\set{S}\) is closed,
\begin{align}
	\notag
	\tm{V;\rno|\set{S};\tm{W},\tm{Q},\blx}[\dinp;\dout]
	&=\tm{V;\rno;\tm{W},\tm{Q},\blx}[\dinp;\dout]
	&
	&\forall \dinp,\dout\in\set{S},\blx\!\in\!\numbers{Z}[+].
\end{align}
Then \eqref{eq:cor:DiscreteMarkov:TKDtoN:CPclassification} holds for a 
\(\lambda\!\in\!(0,\infty)\) by Lemma \ref{lem:CPclassification},
and \(\tm{V;\!\f{h}}\) defined in \eqref{eq:cor:DiscreteMarkov:TKDtoN:DoobsTransform} 
is transition measure satisfying 
Doeblin's minorization condition and \(R\)-invariant function \(\f{h}\) 
satisfies \(\inf_{\dout\in\set{S}}\f{h}[\dout]\!>\!0\) 
by Lemma \ref{lem:CPpositivity}. 
Then \(\tm{V;\!\f{h}}\) satisfies Dobrushin--Doeblin condition 
and there exist \(k_{0}\!\in\!\numbers{Z}[+]\)
and \(\delta_{0}\!\in\!(0,1)\) satisfying
\eqref{eq:cor:DiscreteMarkov:TKDtoN:DobrushinDoeblin}
because these two conditions are equivalent for transition 
probabilities.	 

Note that \(\m{w;\sim}[\outS_{0}^{\blx+1}\!\setminus\!\set{S}_{0}^{\blx+1}]\!=\!0\). 
Let the measure \(\m{s}\AC\m{w;\sim}\) be  
\begin{align}
	%\label{eq:cor:DiscreteMarkov:1}
	\notag
	\m{s}[\dout_{0}^{\blx+1}]
	&\!=\!\m{q}[\dout_{0}^{\blx+1}]
	\left(\tfrac{\f{h}[\dout_{\blx+1}]}{\f{h}[\dout_{0}]}\right)^{\frac{1}{1-\rno}}
	&
	&\forall \dout_{0}^{\blx+1}\in\set{S}_{0}^{\blx+1}.
\end{align}
Then 
\begin{align}
	\label{eq:cor:DiscreteMarkov:2}	
	\sfrac{\m{s}}{\varsigma}
	\leq\m{q}
	&\leq \varsigma \m{s}
	&
	&\m{w;\sim}\text{-a.e..}
\end{align}
Note that \(\gamma\m{v;\rno;\m{w},\m{s}}=\m{v;0}\m{V;\!\f{h};\otimes(\blx+1)}\).
If \(\out_{0}^{\blx+1}\!\sim\!\gamma\m{v;\rno;\m{w},\m{s}}\), then 
by Lemma \ref{lem:CSSMFBE}, the hypothesis of 
Lemma \ref{lem:KolmogorovDistanceBound} holds for  
\(\inp=\out_{0}\), 
\(\overline{\delta}\) defined in \eqref{eq:lem:CSSMFBE:deltabar},
\(\sigma=\overline{\sigma}\sqrt{\blx}\), 
\(\f{f}[\dinp]=\ln\tfrac{\m{v;0}[\dinp]}{\m{w;0}[\dinp]}
-\KLD{\m{v;0}}{\m{w;0}}\), 
and
\begin{align}
	\notag
	\sta
	&=\ln\!\tfrac{\tm{V;\!\f{h};\otimes\blx+1}[\out_{0};\out_{1}^{\blx+1}]}{\tm{W;;\otimes\blx+1}[\out_{0};\out_{1}^{\blx+1}]}
	\!-\!(\blx\!+\!1)\CKLD{\tm{V;\!\f{h}}}{\tm{W}}{\m{p;\tm{V;\!\f{h}}}}.
\end{align}	
Then for \((\m{w},\m{s})\) the hypothesis of Theorem \ref{thm:TKDtoN}
follows from Lemma \ref{lem:KolmogorovDistanceBound} 
for \(\gamma\), \(\xi\), \(\sigma\), and \(\delta\) 
defined in \eqref{eq:cor:DiscreteMarkov:TKDtoN:gamma},
\eqref{eq:cor:DiscreteMarkov:TKDtoN:xi},
\eqref{eq:cor:DiscreteMarkov:TKDtoN:sigma},
and
\eqref{eq:cor:DiscreteMarkov:TKDtoN:delta},
respectively, because
\begin{align}
	\notag
	\f{\llr;\gamma\m{v;\rno;\m{w},\m{s}},\m{w}}
	&=\sta+\f{f}[\out_{0}]+\KLD{\m{v;0}}{\m{w;0}}.	
\end{align}
Thus \eqref{eq:thm:TKDtoN:zeroone},
\eqref{eq:thm:TKDtoN:oneinfinity},
and \eqref{eq:thm:TKDtoN:Strassen}
of Theorem \ref{thm:TKDtoN}
hold for \(\NPD{\cdot}{\m{w}}{\m{s}}\), rather than 
\(\NPD{\cdot}{\m{w}}{\m{q}}\).
On the other hand as result of 
\eqref{eq:lem:NPD:Elementary:ScalarMultiplication:mW},
\eqref{eq:lem:NDP:Elementary:Monotonicity:mW},
of Lemma \ref{lem:NPD:Elementary}
and \eqref{eq:cor:DiscreteMarkov:2},
we have
\begin{align}
	%\label{eq:cor:DiscreteMarkov:3}
	\notag
	\NPD{\tfrac{\cdot}{\sfrac{1}{\varsigma}}}{\m{w}}{\m{s}}
	=\NPD{\cdot}{\m{w}}{\sfrac{\m{s}}{\varsigma}}
	&\!\leq\!\NPD{\cdot}{\m{w}}{\m{q}}\!\leq\!
	\NPD{\cdot}{\m{w}}{\varsigma\m{s}}=
	\NPD{\tfrac{\cdot}{\varsigma}}{\m{w}}{\m{s}}.
\end{align}
Then \eqref{eq:thm:TKDtoN:Strassen}
of Theorem \ref{thm:TKDtoN} hold for all 
\(\tilde{\eta}\in[\varsigma\f{\eta}[\sfrac{3}{5}],\sfrac{\f{\eta}[\sfrac{2}{5}]}{\varsigma}]\)
for a \(\Delta\) satisfying
\(\abs{\ln\Delta}\!\leq\!\tfrac{0.43}{1\wedge \rno}\!+\!(\sigma\!+\!2.3)15\sqrt{2\pi}\delta\!+\!\varsigma\)
as a result of \eqref{eq:thm:TKDtoN:Strassen}
of Theorem \ref{thm:TKDtoN} for \(\NPD{\cdot}{\m{w}}{\m{s}}\).
Note that 
\(\sqrt{\blx}\!\geq\!\tfrac{\abs{1-\rno}\ln\varsigma}{\rno\IGQF{0.4}\overline{\sigma}}\)
ensures
\(\sfrac{\f{\eta}[\sfrac{2}{5}]}{\varsigma}\geq \varsigma\f{\eta}[\sfrac{3}{5}]\)
because
\begin{align}
	\notag	
	\ln \tfrac{\f{\eta}[\sfrac{2}{5}]}{\varsigma^{2}\f{\eta}[\sfrac{3}{5}]}
	&=\tfrac{2\rno\sigma \IGQF{0.4}}{\abs{1-\rno}}
	-2\ln\varsigma+\ln\tfrac{\GMR{\frac{\rno \sigma}{\abs{1-\rno}}-\IGQF{\sfrac{2}{5}}}}{\GMR{\frac{\rno \sigma}{\abs{1-\rno}}+\IGQF{\sfrac{2}{5}}}}
	\\
	\notag
	&\geq \tfrac{2\rno\IGQF{0.4}}{\abs{1-\rno}}
	\left[\overline{\sigma}\sqrt{\blx}
	-\tfrac{\abs{1-\rno}\ln\varsigma}{\rno\IGQF{0.4}}
	\right].
\end{align}	
\end{proof}

\subsection{Asymptotic Case}\label{sec:Asymptotic}
To pose the asymptotic simple hypothesis testing problem
and describe the asymptotic bounds in a generality 
that is commensurate with the framework we have employed thus 
far, let us first define the transition measures 
(i.e., the measure kernels), 
\cite[\S 10.7]{bogachev}, \cite[\S I.6]{cinlar}.
\begin{definition}\label{def:TransitionMeasure}
	Let \((\inpS,\inpA)\) and \((\outS,\outA)\) be measurable spaces
	and \(\tm{W}\) be a mapping from \(\inpS\times\outA\) to \([0,\infty]\).
	Then \(\tm{W}\) is called a transition measure
	from \((\inpS,\inpA)\) into \((\outS,\outA)\) iff 
	the following two conditions hold:
	\begin{enumerate}[(i)]
		\item For every \(\dinp\!\in\!\inpS\), 
		the function \(\tm{W}[\dinp;\cdot]\) is
		a measure on \(\outA\)\!.
		\item  For every \(\oev\!\in\!\outA\), 
		the function \(\tm{W}[\cdot;\oev]\) is
		\(\inpA\)-measurable.	
	\end{enumerate}
	If \(\tm{W}[\dinp;\cdot]\) is a probability measure for all 
	\(\dinp\!\in\!\inpS\), then 
	\(\tm{W}\) is called a 
	transition probability (probability kernel).
\end{definition}
A transition measure \(\tm{W}\) from \((\inpS,\inpA)\) into \((\outS,\outA)\)
is said to be finite if 
\(\tm{W}[\dinp;\outS]\!<\!\infty\) for all \(\dinp\!\in\!\inpS\) 
and \(\sigma\)-finite if 
\(\tm{W}[\dinp;\cdot]\) is \(\sigma\)-finite for all \(\dinp\!\in\!\inpS\).
Similarly, \(\tm{W}\) is said to be bounded if
\(\tm{W}[\cdot;\outS]\) is a bounded function
and \(\sigma\)-bounded if there exists a (countable)
\(\outA\)-measurable partition \(\{\oev_{\blx}\}_{\blx\in\numbers{Z}}\) 
of \(\outS\) such that \(\tm{W}[\cdot;\oev_{\blx}]\) is a 
bounded\footnote{We refrain from introducing concepts
	of \(\Sigma\)-boundedness and \(\Sigma\)-finiteness
	introduced in \cite[\S I.6]{cinlar} for conceptual parsimony.
	Note that every \(\sigma\)-finite transition measure 
	is also a \(\Sigma\)-finite transition measure 
	and every \(\sigma\)-bounded transition measure 
	is also a \(\Sigma\)-bounded transition measure
	for the definitions employed in  in \cite[\S I.6]{cinlar}.} 
function for each \(\blx\).

\begin{lemma}\label{lem:TransitionMeasure}
	Let \(\tm{W}\) be a \(\sigma\)-bounded 
	transition measure from \((\inpS,\inpA)\) into \((\outS,\outA)\) 
	and \(\rfm\) be a \(\sigma\)-finite measure on \(\inpA\). Then 
	there exists a unique \(\sigma\)-finite measure \(\mu\) on 
	\(\inpA\otimes\outA\) with 
	\begin{align}
		\label{eq:lem:TransitionMeasure:Joint}
		\m{\mu}[\oev_{\dinp}\!\times\!\oev_{\dout}]	
		&\!=\!
		\int_{\oev_{\dinp}}\!\tm{W}[\dinp;\oev_{\dout}]\m{\rfm}[\dif{\dinp}]
		&
		&\forall\oev_{\dinp}\!\in\!\inpA,\oev_{\dout}\!\in\!\outA.	
	\end{align}
	In addition for any non-negative 
	\(\inpA\otimes\outA\)-measurable function 
	\(\f{f}\) the mapping 
	\(\dinp\mapsto\int_{\outS} 
	\f{f}[\dinp,\dout]
	\tm{W}[\dinp;\dif{\dout}]\) defines
	a \(\inpA\)-measurable non-negative function
	satisfying
	\begin{align}
		\label{eq:lem:TransitionMeasure:Fubini:nonnegative}
		\!\!\int_{\inpS\times\outS}\!\f{f}[\dinp,\dout] \m{\mu}[\dif{(\dinp,\dout)}]	
		&\!=\!\int_{\inpS}\!\int_{\outS}\! 
		\f{f}[\dinp,\dout]
		\tm{W}[\dinp;\dif{\dout}]
		\m{\rfm}[\dif{\dinp}].\!	
	\end{align}	
\end{lemma}
Lemma \ref{lem:TransitionMeasure} is a generalization 
of \cite[Theorem 10.7.2]{bogachev} 
for probability measures to \(\sigma\)-finite measures,
which follows from 
\cite[Proposition 6.9 and Theorem 6.11]{cinlar}.

\begin{remark}
	With the construction  given in Assumption \ref{assumption:DiscreteMarkov}, 	
	\(\m{w}\) and \(\m{q}\) are guaranteed to be \(\sigma\)-finite 
	for \(\sigma\)-finite discrete transition measures 
	\(\tm{W}\) and \(\tm{Q}\) because \(\outS\) is countable. 
	When we apply the same construction to transitions measures 
	of the form described in Definition \ref{def:TransitionMeasure},
	we rely on \cite[Theorem 6.11]{cinlar} to 
	ensure that \(\m{w}\) and \(\m{q}\) are \(\sigma\)-finite;
	hence we will assume \(\tm{W}\) and \(\tm{Q}\)
	to be \(\sigma\)-bounded rather merely \(\sigma\)-finite. 	
\end{remark}

Invoking the sign based decomposition
\(\f{f}\!=\!\f{f;;+}\!-\!\f{f;;-}\), we see that
\eqref{eq:lem:TransitionMeasure:Fubini:nonnegative}
holds for any \(\mu\)-integrable function \(\f{f}\).
Furthermore, the mapping 
\(\dinp\!\mapsto\!\int_{\outS} 
\f{f}[\dinp,\dout]\tm{W}[\dinp;\dif{\dout}]\) 
defines a \(\inpA\)-measurable function that is finite 
\(\m{\rfm}\)-a.e. for any \(\m{\mu}\)-integrable function \(\f{f}\). 
We invoke these assertions about \(\m{\mu}\)-integrable functions 
only for the case of when \(\tm{W}\) is a 
transition probability and \(\m{\rfm}\) is probability measure, 
which is covered by \cite[Theorem 10.7.2]{bogachev}.

We adopt the notation similar to 
\cite[(6.23)]{cinlar} and denote \(\m{\mu}[\cdot]\)
defined in \eqref{eq:lem:TransitionMeasure:Joint} 
by \(\m{\rfm}\otimes\tm{W}\).
Now we are ready to pose the asymptotic simple 
hypothesis testing problem. 

For all positive integers \(\blx\),
let \((\outS_{\blx},\outA_{\blx})\) be a measurable space,
and let \(\outS_{1}^{\blx}\) and \(\outA_{1}^{\blx}\) be corresponding
(ordered)\begin{subequations}\label{eq:def:ASHTP}
	Cartesian product set and (ordered) product 
	\(\sigma\)-algebra, 
	\begin{align}
		\label{eq:def:ASHTP:SampleSpace}
		\outS_{1}^{\blx}
		&=\outS_{1}^{\blx-1}\times\outS_{\blx}
		&
		&\forall \blx\!\in\!\{2,3,\ldots\},
		\\ 
		\label{eq:def:ASHTP:SigmaAlgebra}
		\outA_{1}^{\blx}
		&=\outA_{1}^{\blx-1}\otimes\outA_{\blx}
		&
		&\forall \blx\!\in\!\{2,3,\ldots\}.		
	\end{align}
	Let \(\m{w;1}\) and \(\m{q;1}\) be \(\sigma\)-finite
	measures on \(\outA_{1}\) and for each \(\blx\) in 
	\(\{2,3,\ldots\}\), let \(\tm{W;\blx}\) 
	and \(\tm{Q;\blx}\) be \(\sigma\)-bounded 
	transition measures from \((\outS_{1}^{\blx-1},\outA_{1}^{\blx-1})\) 
	into \((\outS_{\blx},\outA_{\blx})\), then \(\m{w;1;\blx}\) and
	\(\m{q;1;\blx}\) are \(\sigma\)-finite
	measures on \(\outA_{1}^{\blx}\) defined as
	\begin{align}
		\label{eq:def:ASHTP:NullHypothesis}
		\m{w;1;\blx}
		&=\m{w;1;\blx-1}\otimes\tm{W;\blx}
		&
		&\forall \blx\!\in\!\{2,3,\ldots\},
		\\ 
		\label{eq:def:ASHTP:AlternativeHypothesis}
		\m{q;1;\blx}
		&=\m{q;1;\blx-1}\otimes\tm{Q;\blx}
		&
		&\forall \blx\!\in\!\{2,3,\ldots\}.	
	\end{align}
	The asymptotic simple hypothesis testing problem is the study 
	of the asymptotic behavior of 
	\(\NPD{\cdot}{\m{q;1;\blx}}{\m{w;1;\blx}}\) 
	as \(\blx\) tends to infinity, 
	under various structural assumptions on 
	\(\{(\m{w;1;\blx},\m{q;1;\blx})\}_{\blx\in\numbers{Z}[+]}\). 
\end{subequations} 
\begin{definition}\label{def:SASHTP}
	Asymptotic simple hypothesis testing problem for
	\(\{(\m{w;1;\blx},\m{q;1;\blx})\}_{\blx\in\numbers{Z}[+]}\)
	described in \eqref{eq:def:ASHTP} is called standard 
	iff \(\m{w;1}\) is a probability measure
	and \(\tm{W;\blx}\) is a transition probability
	for all \(\blx\!\in\!\{2,3,\ldots\}\). 
\end{definition}
In the following, we will only be interested 
in the standard case, for which probability measures 
\(\{\m{w;1;\blx}\}_{\blx\in\numbers{Z}[+]}\)
uniquely extend to a probability measure on 
\(\outA_{1}^{\infty}\), as demonstrated by
Ionescu Tulcea, see \cite[Theorem 10.7.4]{bogachev}. 
We denote the resulting probability measure
and corresponding expectation by \(\PX[\m{w}]{\cdot}\) and 
\(\EX[\m{w}]{\cdot}\), 
respectively.\footnote{We do not need the existence of the unique extension 
	\(\PX[\m{w}]{\cdot}\), either to define \(\overline{D}\) and \(\overline{\sigma}\), 
	or to obtain the results discussed in the following. 
	However, \(\PX[\m{w}]{\cdot}\) makes the connection to classical limit theorems 
	for sums of random variables more apparent than it would be otherwise.} 
Furthermore, \(\{\outA_{1}^{\blx}\}_{\blx\in\numbers{Z}[+]}\) form a 
filtration in \(\outA_{1}^{\infty}\) and 
\(\{(\f{\llr;\m{w;1;\blx},\m{q;1;\blx}},\outA_{1}^{\blx})\}_{\blx\in\numbers{Z}[+]}\) 
form stochastic sequence in  \(\outA_{1}^{\infty}\)
by construction, see \cite[\S 7.1]{shiryaev2019probability2}. %\cite[\S VII.1]{shiryaev} 
Thus
\begin{align}
	\notag	
	\EX[\m{w}]{\f{\llr;\m{w;1;\blx},\m{q;1;\blx}}}
	&=\KLD{\m{w;1;\blx}}{\m{q;1;\blx}}
	&
	&\forall \blx\in\numbers{Z}[+],
\end{align}
whenever the expectation exists.
For a standard asymptotic simple hypothesis testing problem,
when %\begin{subequations}\label{eq:def:SASHTP}
the defining limit exists,
we define \(\overline{D}\) as
\begin{align}
	\label{eq:def:SASHTP:mean}		
	\overline{D}
	&\DEF\lim\nolimits_{\blx\to\infty}\tfrac{1}{\blx}
	\EX[\m{w}]{\f{\llr;\m{w;1;\blx},\m{q;1;\blx}}}.
	%\\
	%\label{eq:def:SASHTP:sigma}		
	%\overline{\sigma}
	%&\DEF\lim\nolimits_{\blx\to\infty}
	%\sqrt{\tfrac{1}{\blx}
		%	\EX[\m{w}]{\left(\f{\llr;\m{w;1;\blx},\m{q;1;\blx}}\!-\!
			%		\KLD{\m{w;1;\blx}}{\m{q;1;\blx}}\right)^{2}}}.
\end{align}
%\end{subequations}	
The log-likelihood-ratio of a standard asymptotic simple 
hypothesis testing problem is said to satisfy the weak 
law of large numbers iff\footnote{Note that
	\(\overline{D}\!\in\!\numbers{R}\) excludes both
	\(\overline{D}\!=\!\infty\) and
	\(\overline{D}\!=\!-\infty\). Since 
	\(\norm{\m{q;1;\blx}}\) can be infinite,
	not only \(\overline{D}\), but also 
	\(\KLD{\m{w;1;\blx}}{\m{q;1;\blx}}\)'s 
	can be \(-\infty\), in principle.} 
\(\overline{D}\in\numbers{R}\), and
\begin{align}\label{eq:assumption:WeakLawOfLargeNumbers}
	\lim\limits_{\blx\to\infty}
	\PX[\m{w}]{\abs{\tfrac{1}{\blx}\f{\llr;\m{w;1;\blx},\m{q;1;\blx}}
			\!-\!\overline{D}}>\varsigma}
	&\!=\!0
	&
	&\forall\varsigma\!>\!0.
\end{align}
\begin{lemma}[The First Order Asymptotic Stein's Lemma]\label{lem:NPD:WeakLawOfLargeNumbers}
	If the log-likelihood-ratio of a standard asymptotic simple 
	hypothesis testing problem satisfies weak law of large 
	numbers, i.e. \eqref{eq:assumption:WeakLawOfLargeNumbers}, 
	then 
	\begin{align}
		\label{eq:lem:NPD:WeakLawOfLargeNumbers}
		\lim\nolimits_{\blx\to\infty}\tfrac{1}{\blx}\ln 
		\NPD{\varepsilon}{\m{q;1;\blx}}{\m{w;1;\blx}}
		&\!=\!-\overline{D}
		&
		&\forall \varepsilon\!\in\!(0,1).		
	\end{align}		
\end{lemma}
Lemma \ref{lem:NPD:WeakLawOfLargeNumbers} 
follows from \eqref{eq:thm:SNPL:CI:DF:L}, i.e., \eqref{eq:Beta}, 
see Appendix \ref{sec:Proofs:SteinLemmas} for a proof.

The log-likelihood-ratio of a standard asymptotic simple 
hypothesis testing problem is said to satisfy the central 
limit theorem iff \(\overline{D}\in\numbers{R}\),
and there exist a \(\overline{\sigma}>0\) satisfying
\begin{align}\label{eq:assumption:CentralLimitTheorem}
\lim\limits_{\blx\to\infty}
\PX[\m{w}]{\tfrac{\f{\llr;\m{w;1;\blx},\m{q;1;\blx}}-\blx\overline{D}}{\sqrt{\blx}}\!\leq\!
	\overline{\sigma}\dinp}
&\!=\!1\!-\!\GQF{\dinp}
&
&\forall\dinp\!\in\!\numbers{R}.
\end{align}
\begin{theorem}[The Second Order Asymptotic Stein's Lemma]\label{thm:NPD:CentralLimitTheorem}
If the log-likelihood-ratio of a standard asymptotic simple 
hypothesis testing problem satisfies Central Limit Theorem, 
i.e., \eqref{eq:assumption:CentralLimitTheorem}, then 
\begin{align}
\label{eq:thm:NPD:CentralLimitTheorem}
\lim\limits_{\blx\to\infty}\tfrac{1}{\overline{\sigma}\sqrt{\blx}}
\left(\ln \NPD{\varepsilon}{\m{q;1;\blx}}{\m{w;1;\blx}}
\!+\!\blx \overline{D}\right)
&\!=\!\IGQF{\varepsilon}
&
&\forall \varepsilon\!\in\!(0,1).		
\end{align}	
\end{theorem}
\begin{proof}[Proof of Theorem \ref{thm:NPD:CentralLimitTheorem}]
First note that \eqref{eq:assumption:CentralLimitTheorem} is nothing 
but the convergence of distribution functions of the sequence random variables
\(\left\{\tfrac{1}{\overline{\sigma}\sqrt{\blx}}(\f{\llr;\m{w;1;\blx},\m{q;1;\blx}}\!-\!\blx\overline{D})\right\}_{\blx\in\numbers{Z}[+]}\) 
to standard Gaussian cumulative distribution function,
see \cite[Definition 1, \S3.1]{shiryaev2016probability1}. 
This is equivalent to the weak convergence of corresponding 
sequence of probability measures to the standard Gaussian measure 
on \(\numbers{R}\) by \cite[Theorem 2]{shiryaev2016probability1}.
On the other hand the weak convergence convergence to 
the Gaussian measure is metrized by the Kolmogorov distance
by \cite[Theorem 6]{gibbsS02} because Gaussian measure is 
absolutely continuous in the Lebesgue measure. 
Thus \eqref{eq:assumption:CentralLimitTheorem} implies
that for all \(\delta\!>\!0\) there exists an \(\blx_{0}\) 
such that
\begin{align}
\notag
\sup\nolimits_{\dinp\in\numbers{R}}
\abs{\PX[\m{w}]{\tfrac{\f{\llr;\m{w;1;\blx},\m{q;1;\blx}}-\blx\overline{D}}{ \overline{\sigma} \sqrt{\blx}}\!>\!\dinp}
-\GQF{\dinp}}
&\!\leq\!\delta
&
&\forall \blx\!\geq\!\blx_{0}.
\end{align}	
Consequently, for all \(\delta\!>\!0\) there exists 
an \(\blx_{0}\) such that for all \(\blx\!\geq\!\blx_{0}\),
probability measure \(\m{w;1;\blx}\) and
\(\sigma\)-finite measure \(\m{q;1;\blx}\) 
satisfies \eqref{eq:thm:KDtoN:hypothesis}
for \(\xi=\blx\overline{D}\),
\(\sigma=\overline{\sigma} \sqrt{\blx}\),
and \(\delta\). Then 
for all \(\varepsilon\!\in\!(0,1)\), 
\(\delta\in(0,\varepsilon\wedge(1\!-\!\varepsilon))\),
Theorem \ref{thm:KDtoN} and 
\eqref{eq:GMRBound:positive} imply
\begin{align}
	\notag
	\limsup\limits_{\blx\to\infty}
	\tfrac{1}{\overline{\sigma}\sqrt{\blx}}
	\left(\ln \NPD{\varepsilon}{\m{q;1;\blx}}{\m{w;1;\blx}}
	\!+\!\blx \overline{D}\right)
	&\!\leq\!\IGQF{\varepsilon}\!+\!\tfrac{\delta\sqrt{2\pi}}{(1-\delta-\varepsilon)\wedge(\varepsilon-\delta)},
	\\
	\notag
	\liminf\limits_{\blx\to\infty}
	\tfrac{1}{\overline{\sigma}\sqrt{\blx}}
	\left(\ln \NPD{\varepsilon}{\m{q;1;\blx}}{\m{w;1;\blx}}
	\!+\!\blx \overline{D}\right)
	&\!\geq\!\IGQF{\varepsilon}\!-\!
	\tfrac{\delta\sqrt{2\pi}}{(1-\delta-\varepsilon)\wedge(\varepsilon-\delta)}.	
\end{align}	
Thus \(\liminf\) expression is greater than or equal 
to \(\IGQF{\varepsilon}\),
and \(\limsup\) expression less than or equal to 
\(\IGQF{\varepsilon}\).
Hence \eqref{eq:thm:NPD:CentralLimitTheorem} holds 
for all \(\varepsilon\!\in\!(0,1)\).	
\end{proof}				

It is worth mentioning that Theorem \ref{thm:NPD:CentralLimitTheorem} 
is instructive even in the stationary and memoryless
case because the absolute third moment of a random variable 
can be divergent even when Lindenberg condition's are satisfied 
and Central Limit Theorem holds. 
Likewise in the time-homogeneous Markovian case 
and non-stationary memoryless case
there are incidences in which 
hypotheses of  Corollaries \ref{cor:Memoryless:KDtoN}
and \ref{cor:DiscreteMarkov:KDtoN} fail but
the hypothesis of Theorem \ref{thm:NPD:CentralLimitTheorem}
holds. 
Theorem \ref{thm:NPD:CentralLimitTheorem} demonstrates that 
Central Limit Theorem implies a second order asymptotic Stein's lemma.
Using Theorem \ref{thm:KDtoN} one can also establish third order 
order asymptotic Stein's lemma tying the rate of convergence to
normal distribution in the Central Limit Theorem to the third order
term in asymptotic Stein's lemma.

\section{Discussion}\label{sec:Discussion}
Non-asymptotic formulations of information transmission problems 
enjoyed a considerable attention in the recent past 
\cite{hayashi09B,polyanskiythesis,polyanskiyPV10,polyanskiy13,
	moulin17,yavasKE24,tan14,wangCR09,wangR12,tomamichelH13,tomamichelT13}. 
One-shot bounds, such as those in 
\eqref{eq:MetaConverse}, \eqref{eq:MaximalCode}, 
\eqref{eq:DependenceTestingBound}, and \eqref{eq:DependenceTestingBound2},
were previously interpreted merely as preliminary results for
the ultimate asymptotic results.
With the surge of interest in non-asymptotic formulations, 
one-shot bounds have become ends in themselves, 
heavily studied and cataloged as part of recent efforts 
to obtain non-asymptotic bounds. 
However, the one-shot quantities appearing in these bounds
---such as \(\NPD{\cdot}{\m{q}}{\m{w}}\),
\(\ent{\cdot}{\m{q}}{\m{w}}\),
\(\scq{\cdot}{\m{w}}{\m{q}}\),
\(\lscq{\cdot}{\m{w}}{\m{q}}\),
and
\(\uscq{\cdot}{\m{w}}{\m{q}}\)---
are rarely interpreted explicitly as functions of their 
real parameters; instead, they are customarily studied 
as functions of the measures \(\m{w}\) and \(\m{q}\) 
for a fixed value of the real parameter.
Analyzing these quantities as functions of their real 
parameters is the paradigm shift that led to 
Theorems \ref{thm:SNPL} and \ref{thm:MeasureChange}. 
Studying these one-shot quantities as functions of 
their real parameters is significantly more tractable 
than one might initially assume. 
The convexity of \(\NPD{\cdot}{\m{q}}{\m{w}}\),
which is well-known, proves to be the key tool in 
relating 
\(\NPD{\cdot}{\m{q}}{\m{w}}\),
\(\ent{\cdot}{\m{q}}{\m{w}}\),
\(\scq{\cdot}{\m{w}}{\m{q}}\),
\(\lscq{\cdot}{\m{w}}{\m{q}}\),
and
\(\uscq{\cdot}{\m{w}}{\m{q}}\)
to one another as functions.

Theorems \ref{thm:SNPL} and \ref{thm:MeasureChange} can be 
interpreted as non-asymptotic information spectrum results
that establish relationships between one-shot information spectrum 
quantities for the simple hypothesis testing problem. 
These relationships have immediate consequences for known one-shot bounds, 
such as the meta-converse in \eqref{eq:MetaConverse},
the maximum-code-theorem in \eqref{eq:MaximalCode},
and dependence testing bounds in \eqref{eq:DependenceTestingBound}
and \eqref{eq:DependenceTestingBound2}.  
More importantly, we believe similar characterizations 
can be established for other information transmission problems using 
the non-asymptotic information spectrum method.
Following the initial arXiv submission of this work \cite{nakiboglu26A} 
and  its subsequent conference appearances \cite{nakiboglu26B,nakiboglu26C},
the authors of \cite{elkayamF16} proposed 
a pairwise error probability framework for one-shot 
information theory in three arXiv submissions 
\cite{elkayamF26A,elkayamF26B,elkayamF26C}.

Exact characterizations of optimal performance for 
an information transmission problem are of limited use 
if they are difficult to compute. 
Theorems \ref{thm:KDtoN}, \ref{thm:TKDtoN}, and 
\ref{thm:NPD:CentralLimitTheorem} establish 
tractable and tight approximations by assuming 
the Kolmogorov distance to normality for the 
log-likelihood ratio distribution is bounded
by \(\delta\).  
This approach allows us not only to derive 
the non-asymptotic approximations presented in 
the corollaries for the non-stationary memoryless  
and the time-homogeneous Markov cases, 
but also to demonstrate the distinction between 
the implications of the Central Limit Theorem and 
the Berry--Esseen Theorem for the simple hypothesis testing
problem.

Bounded Kolmogorov distance hypothesis for the log-likelihood 
ratio distributions naturally complements the characterization 
given in \eqref{eq:thm:SNPL:CI:DF:L}. Furthermore, the 
approximating distribution need not be Gaussian, as pointed out 
in \cite[\S III A]{elkayamF16}. 
One can obtain \(\smallo{1}\) approximations for  
\(\ln\NPD{\cdot}{\mQ}{\mW}\), such as those in
\cite[Theorem 1.1]{strassen62},
\cite[Theorem 18]{moulin17},
\cite{nagaevA02}, 
by using more accurate approximations for 
\(\gcdf{\llr}{\cdot}{\m{w}}{\m{q}}\) in \eqref{eq:thm:SNPL:CI:DF:L}.  
Applying these more accurate approximation with tilting, 
the corresponding improvements for the large deviations regime
in \cite{csiszarL71,vazquezFKL18} can also be recovered.

Han's Information Spectrum Method \cite{han} is a powerful, robust, 
and unifying approach for studying information transmission problems. 
Its primary drawback is that it is an asymptotic method 
by design (e.g., compare the present manuscript with \cite{han00}). 
Non-asymptotic applications of information spectrum ideas have 
proliferated following the surge of non-asymptotic formulations 
sparked by \cite{hayashi09B,polyanskiythesis,polyanskiyPV10}. 
By interpreting one-shot quantities as functions of 
their real parameters, we are able to demonstrate previously 
unknown relations and significantly streamline known ones. 
We believe these results are not isolated. 
A non-asymptotic variant of the Information Spectrum Method 
may prove more analytically tractable than previously assumed.

The broader implications of one-shot spectral representations 
remain largely unexplored. We conclude by outlining
several specific research problems raised by this work.
Although we confined our analysis to discrete Markov chains 
in \S\ref{sec:DiscreteMarkov}, our results can be readily extended 
to Markov chains with abstract state-spaces. 
The results from \cite{vere-Jones67,seneta} that we employed 
have already been established for general transition measures 
in \cite{Nummelin}, and the Dobrushin--Doeblin condition was 
originally proposed for Markov chains on measurable spaces 
without the discreteness assumption 
\cite[(1.1)]{nagaev57}, \cite[(0.1)]{nagaev61}. 
Such a generalization would remove some of the superficial
assumptions employed in \cite{Koopmans60}.

An arguably more interesting generalization of the results 
in \S\ref{sec:DiscreteMarkov} is the removal of 
Doeblin's minorization condition. 
While a straightforward proof relying on 
Lemma \ref{lem:NPD:Elementary}
and \eqref{eq:cor:DiscreteMarkov:2}
would not be possible in that case, 
similar results could likely be obtained using 
Bolthausen's functional Berry--Esseen Theorem 
\cite{Bolthausen80}. 
Extending this to a time-inhomogeneous Markov framework 
is also of interest, though likely more challenging.

Our primary focus in this work was \(\NPD{\cdot}{\m{q}}{\m{w}}\).
When \(\NPDd{\cdot}{\m{q}}{\m{w}}\) and \(\NPD{\cdot}{\m{q}}{\m{w}}\)
differ due to a probability mass in \(\m{w;\sim}\) (and hence in \(\m{q;\sim}\)),
the analysis of \(\NPDd{\cdot}{\m{q}}{\m{w}}\) becomes far more nuanced
and likelihood-ratio tests are not necessarily optimal for all \(\varepsilon\) 
values for \(\NPDd{\cdot}{\m{q}}{\m{w}}\). 
Unlike, likelihood-ratio tests, general detectors 
lack an inherent total order and are only partially ordered, 
see \cite[Appendix B]{polyanskiythesis}.
Semicontinuous inverses of monotonic functions appear to be 
the appropriate mathematical tool to analyze \(\NPDd{\cdot}{\m{q}}{\m{w}}\)
in this setting. However, it remains unclear to what extent the results in 
\cite[Appendix B]{polyanskiythesis} can be improved.

\section*{Acknowledgment}
Author would like to thank Hao-Chung Cheng for 
informing him about \cite[Proposition 3.2]{AudenaertMF12} 
and \cite[Proposition 6.1]{chengL26}.\\
{\it LLM use disclosure:}
The author primarily used Google's Gemini 3.1-Pro 
(and, to a lesser extent, free versions of DeepSeek, 
OpenAI's ChatGPT, and Anthropic's Claude) 
to explore new concepts, paraphrase text, 
conduct literature searches, construct counterexamples, 
and test proof ideas regarding Markov chains. 
The only direct use of an LLM for a proof was 
Lemma \ref{lem:KolmogorovDistanceBound}. 
Gemini generated this proof from a single prompt
containing the statement of the lemma, 
which the author then verified and translated 
into the paper's notation. 
The author independently verified every 
mathematical statement, proof, and reference 
in the final manuscript and assumes full 
responsibility for its contents.

\appendix%%\appendices
\subsection{Proofs Of The Lemmas About LRQs}\label{sec:Proofs:LikelihoodRatioQuantiles}
\begin{proof}[Proof of Lemma \ref{lem:LRQ}]
First, note that 
\(\lscqs{\cdot}{\m{w}}{\m{q}}\) and \(\uscqs{\cdot}{\m{w}}{\m{q}}\),
defined in \eqref{eq:def:LRQ:halfline},
are both non-increasing in \(\varepsilon\) 
by \eqref{eq:LRQ:halfline:Monotonicity},
i.e.,
\begin{align}
	\notag
	\scqs[-1]{\varepsilon}{\m{w}}{\m{q}}
	&\subset\scqs[-1]{\tau}{\m{w}}{\m{q}}	
	&
	&\forall\varepsilon,\tau\in\numbers{R}:\varepsilon>\tau.	
\end{align}
In \eqref{eq:def:LRQ}, as \(\varepsilon\) increases 
the set that we are taking the infimum over, 
i.e., \(\scqs[-1]{\varepsilon}{\m{w}}{\m{q}}\), 
can only become smaller or stay the same. 
Thus both \(\lscq{\cdot}{\m{w}}{\m{q}}\)
and \(\uscq{\cdot}{\m{w}}{\m{q}}\) are 
non-decreasing on \(\numbers{R}\).

Furthermore, as an immediate consequence of \eqref{eq:def:LRQ:halfline}
\begin{subequations}\label{eq:LRQ:halfline}and
\eqref{eq:LRQ:halfline:Monotonicity},
\begin{align}
	\label{eq:LRQM:halfline}	
	\lscqs{\varepsilon}{\m{w}}{\m{q}}
	&=\bigcap\nolimits_{\tau:\tau<\varepsilon}\scqs[-1]{\tau}{\m{w}}{\m{q}},
	\\	
	%\notag	
	%\lscqs{\varepsilon}{\m{w}}{\m{q}}
	%&=\bigcap\nolimits_{\blx=1}^{\infty}\scqs[-1]{\varepsilon_{\blx}}{\m{w}}{\m{q}}
	%&
	%&\text{if~}\varepsilon_{\blx}\!\uparrow\!\varepsilon,
	%\\	
	%\notag	
	%\lscqs{\varepsilon}{\m{w}}{\m{q}}
	%&=\bigcap\nolimits_{\blx\in\numbers{Z}[+]}\scqs[-1]{\varepsilon_{\blx}}{\m{w}}{\m{q}}
	%&
	%&\forall
	%\{\varepsilon_{\blx}\}_{\blx\in\numbers{Z}[+]}\!:\!
	%\varepsilon_{\blx}\!\uparrow\!\varepsilon,
	%\\
	\label{eq:LRQP:halfline}	
	\uscqs{\varepsilon}{\m{w}}{\m{q}}
	&=\bigcup\nolimits_{\tau:\tau>\varepsilon}\scqs[-1]{\tau}{\m{w}}{\m{q}},
	%\\
	%\notag	
	%\uscqs{\varepsilon}{\m{w}}{\m{q}}
	%&=\bigcup\nolimits_{\blx=1}^{\infty}\scqs[-1]{\varepsilon_{\blx}}{\m{w}}{\m{q}}
	%&
	%&\text{if~}\varepsilon_{\blx}\!\downarrow\!\varepsilon,
	%\\	
	%\notag	
	%\uscqs{\varepsilon}{\m{w}}{\m{q}}
	%&=\bigcup\nolimits_{\blx\in\numbers{Z}[+]}\scqs[-1]{\varepsilon_{\blx}}{\m{w}}{\m{q}}
	%&
	%&\forall
	%\{\varepsilon_{\blx}\}_{\blx\in\numbers{Z}[+]}\!:\!
	%\varepsilon_{\blx}\!\downarrow\!\varepsilon,
\end{align}	
where \end{subequations}	
\(\scqs[-1]{\cdot}{\m{w}}{\m{q}}\) in \eqref{eq:LRQ:halfline} can either 
be\footnote{Note that \(\{\scqs[-1]{\varepsilon_{\blx}}{\m{w}}{\m{q}}\}_{\blx\in\numbers{Z}[+]}\)
	is a non-decreasing sequence of set whenever  \(\varepsilon_{\blx}\!\downarrow\!\varepsilon\)
	and \(\{\scqs[-1]{\varepsilon_{\blx}}{\m{w}}{\m{q}}\}_{\blx\in\numbers{Z}[+]}\)
	is a non-increasing sequence of set whenever  \(\varepsilon_{\blx}\!\uparrow\!\varepsilon\),
	by \eqref{eq:LRQ:halfline:Monotonicity}. Thus
	\(\scqs[-1]{\varepsilon_{\blx}}{\m{w}}{\m{q}}\uparrow\uscqs{\varepsilon}{\m{w}}{\m{q}}\) 
	whenever
	\(\varepsilon_{\blx}\!\downarrow\!\varepsilon\)
	and 
	\(\scqs[-1]{\varepsilon_{\blx}}{\m{w}}{\m{q}}\downarrow\lscqs{\varepsilon}{\m{w}}{\m{q}}\)
	whenever  
	\(\varepsilon_{\blx}\!\uparrow\!\varepsilon\).} 	
\(\uscqs{\cdot}{\m{w}}{\m{q}}\) or \(\lscqs{\cdot}{\m{w}}{\m{q}}\).
\begin{align}
	\notag	
	\lscq{\varepsilon}{\m{w}}{\m{q}}
	&\mathop{=}^{(a)} 
	\sup [0,\infty)
	\setminus\left(\bigcap\nolimits_{\tau:\tau<\varepsilon}\scqs[-1]{\tau}{\m{w}}{\m{q}}\right),
	\\
	\notag	
	&\mathop{=}^{(b)} 
	\sup\bigcup\nolimits_{\tau:\tau<\varepsilon}\left([0,\infty)\setminus\scqs[-1]{\tau}{\m{w}}{\m{q}}\right),
	\\
	\notag	
	&\mathop{=}^{(c)} 
	\sup\nolimits_{\tau:\tau<\varepsilon}
	\sup \left([0,\infty)\setminus\scqs[-1]{\tau}{\m{w}}{\m{q}}\right),
	\\
	\notag	
	&\mathop{=}^{(d)} 
	\sup\nolimits_{\tau:\tau<\varepsilon}
	\scq[-1]{\tau}{\m{w}}{\m{q}}			
\end{align}		
where 
\((a)\) follows from
\eqref{eq:def:lr},	
\eqref{eq:def:LRQ:LSC},	
\eqref{eq:def:LRQM:halfline},	
\eqref{eq:LRQM:halfline:CB},	
and \eqref{eq:LRQM:halfline};	
\((b)\) follows from De Morgan's law;
\((c)\) holds because
\(\sup\left(\bigcup\nolimits_{\tau\in\set{T}}\set{B}_{\tau}\right)
\!=\!\sup\nolimits_{\tau\in\set{T}}\left(\sup\set{B}_{\tau}\right)\)
for any collection \(\{\set{B}_{\tau}\}_{\tau\in\set{T}}\) of sets 
of the real numbers;
\((d)\) follows from
\eqref{eq:def:LRQ},
\eqref{eq:def:LRQ:halfline},  and	
\eqref{eq:LRQ:halfline:CB}.	
Then \eqref{eq:lem:LRQ:leftlimit} 
holds because both \(\uscq{\tau}{\m{w}}{\m{q}}\) and
\(\lscq{\tau}{\m{w}}{\m{q}}\) non-decreasing in \(\tau\).
\begin{align}
	\notag	
	\uscq{\varepsilon}{\m{w}}{\m{q}}
	&\mathop{=}^{(a)} 
	\inf \bigcup\nolimits_{\tau:\tau>\varepsilon}\scqs[-1]{\tau}{\m{w}}{\m{q}},
	\\
	\notag	
	&\mathop{=}^{(b)} 
	\inf\nolimits_{\tau:\tau>\varepsilon} \inf \scqs[-1]{\tau}{\m{w}}{\m{q}},
	\\
	\notag	
	&\mathop{=}^{(c)} 
	\inf\nolimits_{\tau:\tau>\varepsilon}\scq[-1]{\tau}{\m{w}}{\m{q}},		
\end{align}
where 
\((a)\) follows from
\eqref{eq:def:lr},	
\eqref{eq:def:LRQ:USC},	
\eqref{eq:def:LRQP:halfline},	
and \eqref{eq:LRQP:halfline};	
\((b)\) holds because 
\(\inf\left(\bigcup\nolimits_{\tau\in\set{T}}\set{B}_{\tau}\right)
\!=\!\inf\nolimits_{\tau\in\set{T}}\left(\inf\set{B}_{\tau}\right)\)
for any collection \(\{\set{B}_{\tau}\}_{\tau\in\set{T}}\) of sets 
of the real numbers;
\((c)\) follows from
\eqref{eq:def:LRQ} and	
\eqref{eq:def:LRQ:halfline}.
Then \eqref{eq:lem:LRQ:rightlimit}
holds because both \(\uscq{\cdot}{\m{w}}{\m{q}}\) and
\(\lscq{\cdot}{\m{w}}{\m{q}}\) are non-decreasing.

We are left with confirming 
\eqref{eq:lem:LRQUpperBound}, \eqref{eq:lem:LRQLowerBound}, 
\eqref{eq:lem:LRQ:AEequality}, and
\eqref{eq:lem:LRQ:StrictlyMonone}. 
Note that \eqref{eq:lem:LRQLowerBound} for \(\varepsilon\!>\!\norm[1]{\m{w;\sim}}\)
and \eqref{eq:lem:LRQUpperBound} for 
\(\varepsilon\!<\!0\) follow from the definition 
of \(\f{\ylr;\m{w},\m{q}}\) in \eqref{eq:def:lr}. 
Thus we need to establish 
\eqref{eq:lem:LRQUpperBound} and
\eqref{eq:lem:LRQLowerBound} only  
for  \(\varepsilon\in[0,\norm[1]{\m{w;\sim}}]\).
Let us consider \eqref{eq:lem:LRQUpperBound}, first.
If \(\lscq{\varepsilon}{\m{w}}{\m{q}}\!=\!\infty\), 
then \eqref{eq:lem:LRQUpperBound} 
follows from  \(\m{w;\sim}[\f{\ylr;\m{w},\m{q}}\!\leq\!\infty]
\!=\!\norm[1]{\m{w;\sim}}\).
If \(\lscq{\varepsilon}{\m{w}}{\m{q}}\!<\!\infty\) and
\(\m{w;\sim}[\f{\ylr;\m{w},\m{q}}\!\leq\!\lscq{\varepsilon}{\m{w}}{\m{q}}]\!=\!\infty\), 
then \eqref{eq:lem:LRQUpperBound} holds trivially.
If both \(\lscq{\varepsilon}{\m{w}}{\m{q}}\!<\!\infty\) and
\(\m{w;\sim}[\f{\ylr;\m{w},\m{q}}\!\leq\!\lscq{\varepsilon}{\m{w}}{\m{q}}]\!<\!\infty\) hold, then 
\(\exists\{\gamma_{\blx}\}_{\blx\in\numbers{Z}[+]}\!\subset\!(0,\infty)\)
satisfying \(\gamma_{\blx}\!\downarrow\!\lscq{\varepsilon}{\m{w}}{\m{q}}\)
and the continuity of  \(\m{w;\sim}\) from above (i.e.,  at \(\emptyset\)), 
implies
\begin{align}
	\notag	
	\m{w;\sim}[\f{\ylr;\m{w},\m{q}}\!
	\leq\!\lscq{\varepsilon}{\m{w}}{\m{q}}]
	&\!=\!\lim\nolimits_{\blx\to\infty}
	\m{w;\sim}[\f{\ylr;\m{w},\m{q}}\!
	\leq\!\gamma_{\blx}],
	\\
	\notag
	&\!\geq\!\varepsilon, 				
\end{align}
where the inequality holds because 
\(\m{w;\sim}[\f{\ylr;\m{w},\m{q}}\!\leq\!\gamma]\!\geq\!\varepsilon\)
for all \(\gamma\) in \((\lscq{\varepsilon}{\m{w}}{\m{q}},\infty)\)
by \eqref{eq:def:LRQ:LSC}. 
Thus \eqref{eq:lem:LRQUpperBound} holds
for the case when both
\(\lscq{\varepsilon}{\m{w}}{\m{q}}\) and 
\(\m{w;\sim}[\f{\ylr;\m{w},\m{q}}\!\leq\!\lscq{\varepsilon}{\m{w}}{\m{q}}]\)
are finite, as well.

Let us proceed with establishing \eqref{eq:lem:LRQLowerBound}.
If \(\uscq{\varepsilon}{\m{w}}{\m{q}}\!=\!-\infty\), then 
\(\m{w}[\f{\ylr;\m{w},\m{q}}\!<\!\uscq{\varepsilon}{\m{w}}{\m{q}}]\!=\!0\) 
and \eqref{eq:lem:LRQLowerBound} holds trivially.
If \(\uscq{\varepsilon}{\m{w}}{\m{q}}\!>\!-\infty\), then 
\(\exists\{\gamma_{\blx}\}_{\blx\in\numbers{Z}[+]}\!\subset\!\numbers{R}\)
satisfying \(\gamma_{\blx}\!\uparrow\!\uscq{\varepsilon}{\m{w}}{\m{q}}\)
and the continuity of  \(\m{w}\) from below 
(i.e., \(\sigma\)-additivity) implies
\begin{align}
	\notag	
	\m{w}[\f{\ylr;\m{w},\m{q}}\!<\!\lscq{\varepsilon}{\m{w}}{\m{q}}]
	&\!=\!\lim\nolimits_{\blx\to\infty}
	\m{w}[\f{\ylr;\m{w},\m{q}}\!\leq\!\gamma_{\blx}],
	\\
	\notag
	&\!\leq\!\varepsilon,	
\end{align}
where the inequality holds because 
\(\m{w}[\f{\ylr;\m{w},\m{q}}\!\leq\!\gamma]\!\leq\!\varepsilon\)
for all \(\gamma\) in \((-\infty,\uscq{\varepsilon}{\m{w}}{\m{q}})\)
by \eqref{eq:LRQ:USC}.
Thus \eqref{eq:lem:LRQLowerBound} holds
for \(\uscq{\varepsilon}{\m{w}}{\m{q}}\!>\!-\infty\) case,
as well.

\(\lscq{\cdot}{\m{w}}{\m{q}}\) and \(\uscq{\cdot}{\m{w}}{\m{q}}\) 
are the right and left limits of \(\lscq{\cdot}{\m{w}}{\m{q}}\) 
by \eqref{eq:lem:LRQ:leftlimit} and \eqref{eq:lem:LRQ:rightlimit}.
Those left and right limits are equal and \(\lscq{\cdot}{\m{w}}{\m{q}}\) 
is continuous everywhere except for a countable set 
because monotonic functions have 
at most countably many discontinuities by 
\cite[Theorem 4.30]{rudin} and
\(\lscq{\cdot}{\m{w}}{\m{q}}\) 
is a non-decreasing function.
The points of discontinuity should 
lie between zero and 
\(\norm[1]{\m{w;\sim}}\) because 
\(\lscq{\varepsilon}{\m{w}}{\m{q}}\!=\!\uscq{\varepsilon}{\m{w}}{\m{q}}\!=\!-\infty\) 
for all \(\varepsilon\!<\!0\)
and
\(\lscq{\varepsilon}{\m{w}}{\m{q}}\!=\!\uscq{\varepsilon}{\m{w}}{\m{q}}\!=\!\infty\) 
for all \(\varepsilon\!>\!\norm[1]{\m{w;\sim}}\).
Thus \eqref{eq:lem:LRQ:AEequality} holds 
for some countable subset \(\set{D}\) of the closed
interval \([0,\norm[1]{\m{w;\sim}}]\).

\eqref{eq:lem:LRQ:StrictlyMonone} is void for \(\m{w}\!\perp\!\m{q}\) 
case because \(\m{w}\!\perp\!\m{q}\) iff \(\norm[1]{\m{w;\sim}}\!=\!0\).
If \(\m{w}\!\not\perp\!\m{q}\), i.e., \(\norm[1]{\m{w;\sim}}\!>\!0\),
then:
\begin{itemize}
	\item \(\lscq{\varepsilon}{\m{w}}{\m{q}}\) ---and hence \(\uscq{\varepsilon}{\m{w}}{\m{q}}\)---
	is positive for any \(\varepsilon\) in \((0,\norm[1]{\m{w;\sim}})\)
	by \eqref{eq:lem:LRQUpperBound}, because 
	\(\m{w;\sim}[\f{\ylr;\m{w},\m{q}}\!\leq\!0]=0\).
	\item  \(\uscq{\varepsilon}{\m{w}}{\m{q}}\) ---and hence \(\lscq{\varepsilon}{\m{w}}{\m{q}}\)---
	is finite for any \(\varepsilon\) in \((0,\norm[1]{\m{w;\sim}})\)
	by \eqref{eq:lem:LRQLowerBound}, because 
	\(\m{w}[\f{\ylr;\m{w},\m{q}}\!<\!\infty]
	=\norm[1]{\m{w;\sim}}\).	
\end{itemize}	
\end{proof}		

\subsection{Proofs Of The Lemmas About PES}\label{sec:Proofs:PrimitiveEntropySpectrum}
\begin{proof}[Proof of Lemma \ref{lem:PES}]

	Both \(\f{\ylr;\m{w},\m{q}}\!\wedge\!\gamma\!\to\!0\) 
	as \(\gamma\downarrow0\) and
	\(\f{\ylr;\m{w},\m{q}}\!\wedge\!\gamma\!\leq\!\f{\ylr;\m{w},\m{q}}\!\wedge\!1\)
	hold \(\m{q}\)-a.e. for \(\gamma\in(0,1)\) and 
	\(\int \left(\f{\ylr;\m{w},\m{q}}\!\wedge\!1\right)\dif{\m{q}}<\infty\),
	i.e., \(\ent{1}{\m{w}}{\m{q}}\) is finite, by hypothesis.
	Then \eqref{eq:lem:PES:zero} 
	follows from the dominated convergence theorem
	\cite[Corollary 2.8.6]{bogachev}.
	On the other hand 
	\(\m{w}[\f{\ylr;\m{w},\m{q}}\!\leq\!\gamma]
	\!\leq\!\ent{\gamma}{\m{w}}{\m{q}}\!\leq\!\norm[1]{\m{w;\sim}}\!\)
	for all \(\gamma\!\in\!(0,\infty)\)
	by \eqref{eq:def:EntropySpectrum}
	and
	\(\lim\nolimits_{\gamma\uparrow\infty}
	\m{w}[\f{\ylr;\m{w},\m{q}}\!\leq\!\gamma]
	\!=\!\norm[1]{\m{w;\sim}}\)
	by the countable additivity, i.e., the continuity from below, 
	of \(\m{w}\).
	
	Let us proceed with establishing \eqref{eq:lem:PES:leftderivative}.
	First using \eqref{eq:def:lr}, \eqref{eq:lem:PES:integral:r}, 
	and the definition of left derivative; then invoking 
	the additivity of the measure \(\m{q}\) for disjoint events
	\(\{\dout:\f{\ylr;\m{w},\m{q}}\!\geq\!\gamma\}\) and 
	\(\{\dout:\gamma\!>\!\f{\ylr;\m{w},\m{q}}\!>\!\tau\}\)
	for \(\tau\!\in\!(0,\gamma)\), we get
	\begin{align}
		\notag
		\lder{\ent{\gamma}{\m{w}}{\m{q}}}
		&\mathop{=}^{}\lim\nolimits_{\delta\downarrow0}\tfrac{1}{\delta}
		\int_{\gamma-\delta}^{\gamma}\!\m{q}[\f{\ylr;\m{w},\m{q}}>\tau]\dif{\tau},
		\\
		\label{eq:lem:PES:leftderivative1}
		&\mathop{=}^{}\m{q}[\f{\ylr;\m{w},\m{q}}\!\geq\!\gamma]
		\!+\!\lim\nolimits_{\delta\downarrow0}
		\int_{\gamma-\delta}^{\gamma}\!
		\tfrac{\m{q}[\gamma>\f{\ylr;\m{w},\m{q}}>\tau]}{\delta}\dif{\tau}.
	\end{align}
	On the other hand for all \(\delta\in[0,\gamma)\) and \(\tau\in[\gamma\!-\!\delta,\gamma]\) 
	we have
	\begin{align}
		\label{eq:lem:PES:leftderivative2}	
		0
		\!\leq\!\m{q}[\gamma\!>\!\f{\ylr;\m{w},\m{q}}\!>\!\tau]
		&\!\leq\!\m{q}[\gamma\!>\!\f{\ylr;\m{w},\m{q}}\!>\!\gamma-\delta].
	\end{align}
	The continuity of \(\m{w}\) from above (i.e., at \(\emptyset\)) imply
	\begin{align}
		\label{eq:lem:PES:leftderivative3}	
		\lim\nolimits_{\delta\downarrow0}
		\m{q}[\gamma\!>\!\f{\ylr;\m{w},\m{q}}\!>\!\gamma-\delta]
		&=0.	
	\end{align}
	\eqref{eq:lem:PES:leftderivative} follows from 
	\eqref{eq:def:lr},
	\eqref{eq:lem:PES:leftderivative1},
	\eqref{eq:lem:PES:leftderivative2}, and
	\eqref{eq:lem:PES:leftderivative3}.

	Let us proceed with establishing \eqref{eq:lem:PES:rightderivative}.
	First using \eqref{eq:def:lr}, \eqref{eq:lem:PES:integral:r}, 
	and the definition of right derivative; then invoking 
	the additivity of the measure \(\m{q}\) for disjoint events
	\(\{\dout:\f{\ylr;\m{w},\m{q}}\!>\!\tau\}\) and 
	\(\{\dout:\tau\!\geq\!\f{\ylr;\m{w},\m{q}}\!>\!\gamma\}\)
	for \(\tau\!\in\!(\gamma,\infty)\), we get
	\begin{align}
		\notag
		\rder{\ent{\gamma}{\m{w}}{\m{q}}}
		&\mathop{=}^{}\lim\nolimits_{\delta\downarrow0}\tfrac{1}{\delta}
		\int_{\gamma}^{\gamma+\delta}\!\m{q}[\f{\ylr;\m{w},\m{q}}>\tau]\dif{\tau},
		\\
		\label{eq:lem:PES:rightderivative1}
		&\mathop{=}^{}\m{q}[\f{\ylr;\m{w},\m{q}}\!>\!\gamma]
		\!-\!\lim\limits_{\delta\downarrow0}
		\int_{\gamma}^{\gamma+\delta}\!\!
		\tfrac{\m{q}[\tau\geq\f{\ylr;\m{w},\m{q}}>\gamma]}{\delta}\dif{\tau}.
	\end{align}
	On the other hand for all \(\delta\!\in\![0,\infty)\) and \(\tau\!\in\![\gamma,\gamma\!+\!\delta]\) 
	we have
	\begin{align}
		\label{eq:lem:PES:rightderivative2}	
		0
		\!\leq\!\m{q}[\tau\!\geq\!\f{\ylr;\m{w},\m{q}}\!>\!\gamma]
		&\!\leq\!\m{q}[\gamma+\delta\!\geq\!\f{\ylr;\m{w},\m{q}}\!>\!\gamma].
	\end{align}
	The continuity of \(\m{w}\) from above (i.e., at \(\emptyset\)) imply
	\begin{align}
		\label{eq:lem:PES:rightderivative3}	
		\lim\nolimits_{\delta\downarrow0}
		\m{q}[\gamma+\delta\!\geq\!\f{\ylr;\m{w},\m{q}}\!>\!\gamma]
		&=0.	
	\end{align}
	\eqref{eq:lem:PES:leftderivative} follows from 
	\eqref{eq:def:lr},
	\eqref{eq:lem:PES:rightderivative1},
	\eqref{eq:lem:PES:rightderivative2}, and
	\eqref{eq:lem:PES:rightderivative3}.
\end{proof}		
\begin{proof}[Proof of Lemma \ref{lem:PES:Elementary}]
	\(\ent{\gamma}{\m{w}}{\m{q}}\) is  non-decreasing and concave 
	in the pair \((\m{w},\gamma\m{q})\) 
	because the expression in side the infimum in 
	\eqref{eq:EntropySpectrum:Variational} is
	a non-decreasing and linear (hence concave) 
	function in the pair \((\m{w},\gamma\m{q})\)
	for all \(\oev\!\in\!\outA\).
	Then both \eqref{eq:lem:PES:Elementary:Monotonicty} 
	and \eqref{eq:lem:PES:Elementary:Concavity} hold 
	because pointwise infimum of non-decreasing functions is nondecreasing
	and pointwise infimum of concave functions is concave.
	\eqref{eq:lem:PES:Elementary:ScalarMultiplication:mW} and
	\eqref{eq:lem:PES:Elementary:ScalarMultiplication:mQ} follow from 
	the definition of primitive entropy spectrum in \eqref{eq:def:EntropySpectrum}.
	%as well as \eqref{eq:lrf:Elementary:ScalarMultiplication} and \eqref{eq:lem:PES:integral:p:DF}.
\end{proof}		

\subsection[Proofs Of \(1^{\text{st}}\!\)-Order Stein Lemmas]{Proofs of First-Order Stein Lemmas}\label{sec:Proofs:SteinLemmas}
\begin{proof}[Proof of Lemma \ref{lem:NPD:ChebyshevCantelli}] 
First note that \(\EX[\m{w}]{\abs{\llr_{\m{w},\m{q}}}}\) is finite 
because \(\EX[\m{w}]{(\llr_{\m{w},\m{q}})^{2}}\) is finite 
and \(\EX{\abs{\inp}}^{2}\leq\EX{\inp^{2}}\)
for any extended real valued random variable \(\inp\).
Then \(\EX[\m{w}]{\llr_{\m{w},\m{q}}}\), and hence \(\KLD{\m{w}}{\m{q}}\) 
exists and it is a real number.

Chebyshev-Cantelli inequality implies
\begin{align}
\notag	
\gcdf{\llr}{\tau}{\m{w}}{\m{q}}
&\!\leq\!\tfrac{\sigma^{2}}{\sigma^{2}+(\KLD{\m{w}}{\m{q}}-\tau)^{2}}
&
&\forall\tau\!\leq\!\KLD{\m{w}}{\m{q}}.		
\end{align}
Then \(\gcdf{\llr}{\tau}{\m{w}}{\m{q}}\leq \varepsilon\)
for all \(\tau\leq \KLD{\m{w}}{\m{q}}-\sigma\sqrt{\tfrac{1-\varepsilon}{\varepsilon}}\).
Thus as a result of \eqref{eq:thm:SNPL:CI:DF:L}
\begin{align}	
%\notag
%\gcdf{\llr}[\wedge]{\varepsilon}{\m{w}}{\m{q}}
%&\geq \KLD{\m{w}}{\m{q}}-\sigma\sqrt{\tfrac{1-\varepsilon}{\varepsilon}},
%\\
\notag
\NPD{\varepsilon}{\m{q}}{\m{w}}
&\!=\!\int_{\KLD{\m{w}}{\m{q}}-\sigma\sqrt{\frac{1-\varepsilon}{\varepsilon}}}^{\infty}
e^{-\tau}\abp{\gcdf{\llr}{\tau}{\m{w}}{\m{q}}\!-\!\varepsilon}\dif{\tau}.
\end{align}
Then the upper bound in \eqref{eq:lem:NPD:ChebyshevCantelli:bound} 
follows from \(\gcdf{\llr}{\tau}{\m{w}}{\m{q}}\!\leq\!1\).
Note that Chebyshev-Cantelli inequality also implies
\begin{align}
\notag	
1-\gcdf{\llr}{\tau}{\m{w}}{\m{q}}
&\!\leq\!\tfrac{\sigma^{2}}{\sigma^{2}+(\tau-\KLD{\m{w}}{\m{q}})^{2}}
&
&\forall\tau\!\geq\!\KLD{\m{w}}{\m{q}}.	
\end{align}
Thus 
\(1\!-\!\gcdf{\llr}{\tau}{\m{w}}{\m{q}}\!\leq\!\tfrac{1-\varepsilon}{1+\Delta}\)
for all \(\tau\geq \KLD{\m{w}}{\m{q}}+\sigma\sqrt{\tfrac{\varepsilon+\Delta}{1-\varepsilon}}\).
Thus as a result of \eqref{eq:thm:SNPL:CI:DF:L}
\begin{align}
%\notag
%\gcdf{\llr}[\vee]{\varepsilon}{\m{w}}{\m{q}}
%&\leq \KLD{\m{w}}{\m{q}}+\sigma\sqrt{\tfrac{\varepsilon}{1-\varepsilon}},
%\\
%\notag
%\gcdf{\llr}[\vee]{\tfrac{\varepsilon+\Delta}{1+\Delta}}{\m{w}}{\m{q}}
%&\leq \KLD{\m{w}}{\m{q}}+\sigma\sqrt{\tfrac{\varepsilon+\Delta}{1-\varepsilon}},		
%\\
\notag
\NPD{\varepsilon}{\m{q}}{\m{w}}
&\!\geq \!
\int_{\KLD{\m{w}}{\m{q}}+\sigma\sqrt{\frac{\varepsilon+\Delta}{1-\varepsilon}}}^{\infty}
e^{-\tau}\abp{\gcdf{\llr}{\tau}{\m{w}}{\m{q}}\!-\!\varepsilon}
\!\dif{\tau}
\\
\notag
&\!\geq\!
\int_{\KLD{\m{w}}{\m{q}}+\sigma\sqrt{\frac{\varepsilon+\Delta}{1-\varepsilon}}}^{\infty}
e^{-\tau}\abp{1\!-\!\tfrac{1-\varepsilon}{1+\Delta}\!-\!\varepsilon}
\!\dif{\tau}
\\
\notag
&\!=\!e^{-\KLD{\m{w}}{\m{q}}-\sigma\sqrt{\frac{\varepsilon+\Delta}{1-\varepsilon}}}\tfrac{\Delta(1-\varepsilon)}{1+\Delta}.
\end{align}
\end{proof}					

\begin{proof}[Proof of Lemma \ref{lem:NPD:WeakLawOfLargeNumbers}] 
As a result of \eqref{eq:assumption:WeakLawOfLargeNumbers} 
for any \(\varsigma>0\) and \(\delta>0\) there exists a 
\(\blx_{0}(\varsigma,\delta)\)  such that
\begin{align}
\notag
%\label{eq:WeakLawOfLargeNumbers-1}
\PX[\m{w}]{\llr_{\m{w;1;\blx},\m{q;1;\blx}}\!\leq\! (\overline{D}\!-\!\varsigma)\blx}
&\!\leq\!\delta
&
&\forall\blx\!\geq\! \blx_{0}(\varsigma,\delta).	
\end{align}
Thus as a result of \eqref{eq:thm:SNPL:CI:DF:L}	 for any 
\(\blx\!\geq\! \blx_{0}(\varsigma,\varepsilon)\) have
\begin{align}
\notag
\NPD{\varepsilon}{\m{q;1;\blx}}{\m{w;1;\blx}}
%&\!=\!\int_{-\infty}^{\infty} e^{-\tin}
%\abp{\PX[\m{w}]{\llr_{\m{w;1;\blx},\m{q;1;\blx}}\leq\tin}
%	\!-\!\varepsilon}\!\dif{\tin},
%\\
%\notag
&\!=\!\int_{(\overline{D}-\varsigma)\blx}^{\infty} e^{-\tin}
\abp{\PX[\m{w}]{\llr_{\m{w;1;\blx},\m{q;1;\blx}}\leq\tin}
	\!-\!\varepsilon}\!\dif{\tin},
\\
\notag
&\!\leq\!(1\!-\!\varepsilon)
e^{-(\overline{D}-\varsigma)\blx}.
\end{align}
Thus 
\begin{align}
\label{eq:WeakLawOfLargeNumbers-2}
\limsup\nolimits_{\blx\to\infty}\tfrac{1}{\blx}\ln 
\NPD{\varepsilon}{\m{q;1;\blx}}{\m{w;1;\blx}}
&\!\leq\!-\!\overline{D}\!+\!\varsigma
&
&\forall\varsigma\!>\!0.
\end{align}
As a result of \eqref{eq:assumption:WeakLawOfLargeNumbers} 
for any \(\varsigma>0\) and \(\delta>0\) there exists a 
\(\blx_{1}(\varsigma,\delta)\)  such that
\begin{align}
%\label{eq:WeakLawOfLargeNumbers-3}
\notag
\PX[\m{w}]{\llr_{\m{w;1;\blx},\m{q;1;\blx}}\!>	\! (\overline{D}\!+\!\varsigma)\blx}
&\!\leq\!\delta
&
&\forall\blx\!\geq\!\blx_{1}(\varsigma,\delta).	
\end{align}
Thus as a result of \eqref{eq:thm:SNPL:CI:DF:L} for any 
\(\blx\!\geq\! \blx_{1}(\varsigma,\tfrac{1-\varepsilon}{2})\) have
\begin{align}
\notag
\NPD{\varepsilon}{\m{q;1;\blx}}{\m{w;1;\blx}}
&\!=\!\int_{-\infty}^{\infty} e^{-\tin}
\abp{1\!-\!\varepsilon\!-\!
	\PX[\m{w}]{\llr_{\m{w;1;\blx},\m{q;1;\blx}}>\tin}}\!\dif{\tin},
\\
\notag
&\!\geq\!\int_{-(\overline{D}+\varsigma)\blx}^{\infty} e^{-\tin}
\abp{1\!-\!\varepsilon
	\!-\!\PX[\m{w}]{\llr_{\m{w;1;\blx},\m{q;1;\blx}}>\tin}}\!\dif{\tin},
\\
\notag
&\!\geq\!\tfrac{1-\varepsilon}{2}e^{-(\overline{D}+\varsigma)\blx}.
\end{align}	
Thus 
\begin{align}
\label{eq:WeakLawOfLargeNumbers-4}
\liminf\nolimits_{\blx\to\infty}\tfrac{1}{\blx}\ln 
\NPD{\varepsilon}{\m{q;1;\blx}}{\m{w;1;\blx}}
&\!\geq\!-\!\overline{D}\!-\!\varsigma
&
&\forall\varsigma\!>\!0.
\end{align}	
\eqref{eq:lem:NPD:WeakLawOfLargeNumbers} follows from 
\eqref{eq:WeakLawOfLargeNumbers-2} and
\eqref{eq:WeakLawOfLargeNumbers-4}. 
\end{proof}

\subsection[Proof of Lemma \ref{lem:KolmogorovDistanceBound}]{Proof of Lemma \ref{lem:KolmogorovDistanceBound}}\label{sec:Proof:lem:KolmogorovDistanceBound}
\begin{proof}[Proof of Lemma \ref{lem:KolmogorovDistanceBound}]
Let us denote the random variable \(\f{f}[\inp]\) by \(\rv{V}\).
For any random variable \(\out\), the triangle inequality for Kolmogorov 
distance imply 
\begin{align}
	\notag
	\metric{d}[\cdf{\rv{V}+\sta}[\cdot];\!\GCD{\tfrac{\cdot}{\sigma}}]	
	&\!\leq\! 
	\metric{d}[\cdf{\rv{V}+\sta}[\cdot];\!\cdf{\rv{V}+\out}[\cdot]]
	\!+\!
	\metric{d}[\cdf{\rv{V}+\out}[\cdot];\!\GCD{\tfrac{\cdot}{\sigma}}].	
\end{align}	
Let \(\out\) be a zero-mean Gaussian random variable 
with variance \(\sigma^{2}\) independent of \(\inp\), 
and hence \(\rv{V}\).
Then
\begin{align}
	\notag
	\cdf{\rv{V}+\out}[\tau]
	&\!=\!\EX{\GCD{\tfrac{\tau-\rv{V}}{\sigma}}}.
\end{align}		
On the other hand, the law of iterated expectations asserts
\begin{align}
	\notag
	\cdf{\rv{V}+\sta}[\tau]
	&=\EX{\EX{\IND{\sta+\rv{V}\leq \tau}}[\inp]}.
\end{align}	
Thus using first the triangle inequality for expectations,
and then the hypothesis of the lemma we get
\begin{align}
	\notag
	\abs{\cdf{\rv{V}+\out}[\tau]\!-\!\cdf{\rv{V}+\sta}[\tau]}
	&\leq\EX{\abs{\GCD{\tfrac{\tau-\rv{V}}{\sigma}}
			\!-\!\EX{\IND{\sta+\rv{V}\leq \tau}}[\inp]
	}}
	\\
	\notag
	&\leq \overline{\delta}
	&
	&\forall \tau.
\end{align}
Thus \(\metric{d}[\cdf{\rv{V}+\sta}[\cdot];\cdf{\rv{V}+\out}[\cdot]]
\leq \overline{\delta}\).

On the other hand a result of Taylor's theorem, 
for all \(\dinp\in\numbers{R}\), there exists a \(\xi_{\dinp}\) between \(0\)
and \(\dinp\) satisfying
\begin{align}
	\notag	
	\GCD{\tfrac{\tau-\dinp}{\sigma}}
	&=\GCD{\tfrac{\tau}{\sigma}}-\tfrac{\dinp}{\sigma}\GD{\tfrac{\tau}{\sigma}}
	-\tfrac{\dinp^{2}}{2\sigma^{2}}
	\left.\left(\tin\GD{\tin}\right)\right\vert_{\tin=\tfrac{\tau-\xi_{\dinp}}{\sigma}}.
\end{align}
On the other hand \(\sup_{\tin}\abs{\tin\GD{\tin}}=\tfrac{1}{\sqrt{2\pi e}}\). Thus
using \(\EX{\rv{V}}=0\), we get
\begin{align}
	\notag	
	\EX{\GCD{\tfrac{\tau-\rv{V}}{\sigma}}}
	&=\GCD{\tfrac{\tau}{\sigma}}
	\pm\tfrac{1}{2\sigma^{2}\sqrt{2\pi e}}\EX{\rv{V}^{2}}
\end{align}
Thus \(\metric{d}[\cdf{\rv{V}+\out\!~}[\cdot];\GCD{\tfrac{\cdot}{\sigma}}]
\!\leq\!\tfrac{\EX{\rv{V}^{2}}}{2\sigma^{2}\sqrt{2\pi e}}\) and lemma follows.	
\end{proof}	
\iffreshbib
%\IEEEtriggeratref{97}
\bibliographystyle{IEEEtran}
\bibliography{main.bib}
\else  
%\IEEEtriggeratref{44}
\newcommand{\noopsort}[1]{} \newcommand{\printfirst}[2]{#1}
  \newcommand{\singleletter}[1]{#1} \newcommand{\switchargs}[2]{#2#1}

\fi
\ifnullhyperlink\end{NoHyper}\fi
\end{document}